\documentclass[3pd]{elsarticle}
\usepackage{hyperref}
\usepackage{xcolor}

\usepackage{graphicx,lscape,rotating}

\usepackage{amssymb}
\usepackage{amsmath}
\usepackage{multirow}
\newcommand\EatDot[1]{}

\usepackage{lineno}

\begin{document}  

\def \cebaf {{\textsc{cebaf}}}
\def \GX    {{\textsc{GlueX}}}
\def \gx    {{\textsc{GlueX}}}
\def \jlab  {{JLab}}

 
\begin{frontmatter} 


\title{The \gx~ Beamline and Detector}

\author[odu]{S.~Adhikari}
\author[fsu]{C.S.~Akondi}
\author[fsu]{H.~Al~Ghoul}
\author[gsi]{A.~Ali\fnref{f0}}
\fntext[f0]{Also at Goethe University Frankfurt, 60323 Frankfurt am Main, Germany.}
\author[odu]{M.~Amaryan}
\author[uofa]{E.G.~Anassontzis}
\author[cmu]{A.~Austregesilo}
\author[jlab]{F.~Barbosa}
\author[fsu]{J.~Barlow}
\author[cmu]{A.~Barnes}
\author[fsu]{E.~Barriga}
\author[iu]{R.~Barsotti}
\author[uofr]{T.D.~Beattie}
\author[jlab]{J.~Benesch}
\author[cua,mephi]{V.V.~Berdnikov}
\author[jlab]{G.~Biallas}
\author[uncw]{T.~Black}
\author[fiu]{W.~Boeglin}
\author[jlab,odu]{P.~Brindza}
\author[gwu]{W.J.~Briscoe}
\author[jlab]{T.~Britton}
\author[jlab]{J.~Brock}
\author[usm]{W.K.~Brooks}
\author[fsu]{B.E.~Cannon}
\author[jlab]{C.~Carlin}
\author[jlab]{D.S.~Carman}
\author[jlab]{T.~Carstens\fnref{f1}}
\fntext[f1]{Current address: 103 Riverside Dr, Yorktown, VA 23692.}
\author[ihep]{N.~Cao}
\author[itep]{O.~Chernyshov}
\author[jlab]{E.~Chudakov}
\author[asu]{S.~Cole}
\author[gwu]{O.~Cortes}
\author[jlab]{W.D.~Crahen}
\author[fsu]{V.~Crede}
\author[jlab]{M.M.~Dalton}
\author[uncw]{T.~Daniels}
\author[jlab]{A.~Deur}
\author[jlab]{C.~Dickover}
\author[fsu]{S.~Dobbs}
\author[itep]{A.~Dolgolenko}
\author[fiu]{R.~Dotel}
\author[asu]{M.~Dugger}
\author[gsi]{R.~Dzhygadlo}
\author[iu]{A.~Dzierba}
\author[jlab]{H.~Egiyan}
\author[fiu]{T.~Erbora}
\author[fsu]{A.~Ernst}
\author[fsu]{P.~Eugenio}
\author[mit]{C.~Fanelli}
\author[gwu]{S.~Fegan\fnref{f2}}
\fntext[f2]{Current address: University of York, York YO10 5DD, United Kingdom.}
\author[uofr]{A.M.~Foda}
\author[iu]{J.~Foote}
\author[iu]{J.~Frye}
\author[jlab]{S.~Furletov}
\author[uncw]{L.~Gan}
\author[ncatsu]{A.~Gasparian}
\author[itep]{A.~Gerasimov}
\author[yerevan]{N.~Gevorgyan}
\author[iu]{C.~Gleason}
\author[gsi]{K.~Goetzen}
\author[fsu]{A.~Goncalves}
\author[itep]{V.S.~Goryachev}
\author[fiu]{L.~Guo}
\author[usm]{H.~Hakobyan}
\author[gsi]{A.~Hamdi\fnref{f0}}
\author[mit]{J.~Hardin}
\author[uofr]{C.L.~Henschel\fnref{f4}}
\fntext[f4]{Current address:Department of Physics and Astronomy, University of Calgary, Calgary, AB, T2N 1N4, Canada.}
\author[uofr]{G.M.~Huber}
\author[jlab]{C.~Hutton}
\author[wm]{A.~Hurley}
\author[uofa]{P.~Ioannou\fnref{f5}}
\fntext[f5]{Deceased.}
\author[glasgow]{D.G.~Ireland}
\author[jlab]{M.M.~Ito}
\author[cmu]{N.S.~Jarvis}
\author[uconn]{R.T.~Jones}
\author[yerevan]{V.~Kakoyan}
\author[uofr]{S. ~Katsaganis}
\author[cua]{G.~Kalicy}
\author[fiu]{M.~Kamel}
\author[jlab]{C.D.~Keith}
\author[cua]{F.J.~Klein\fnref{f6}}
\fntext[f6]{Current address: Office of Academic Computing Services, University of Maryland, College Park, MD 20742.}
\author[gsi]{R.~Kliemt}
\author[uofr]{D.~Kolybaba}
\author[uofa]{C.~Kourkoumelis}
\author[uofr]{S.T.~Krueger\fnref{f7}}
\fntext[f7]{Current address: iQMetrix, 311 Portage Avenue, Winnipeg, MB, R3B 2B9, Canada.}
\author[usm]{S.~Kuleshov}
\author[umass,itep]{I.~Larin}
\author[jlab]{D.~Lawrence}
\author[iu]{J.P.~Leckey}
\author[fsu]{D.I.~Lersch}
\author[uofr]{B.D.~Leverington\fnref{f8}}
\fntext[f8]{Current address: Heidelberg Universitaet, Physikalisches Institut 3.406, 69120 Heidelberg, Germany.}
\author[cmu]{W.I.~Levine}
\author[wm]{W.~Li}
\author[ihep]{B.~Liu}
\author[glasgow]{K.~Livingston}
\author[uofr]{G.J.~Lolos}
\author[tomsk,tomskb]{V.~Lyubovitskij}
\author[jlab]{D.~Mack}
\author[yerevan]{H.~Marukyan}
\author[jlab]{P.T.~Mattione\fnref{f9}}
\fntext[f9]{Current address: Deep Silver Volition, 1 E Main St., Champaign, IL 61820.}
\author[itep]{V.~Matveev}
\author[jlab]{M.~McCaughan}
\author[cmu]{M.~McCracken}
\author[cmu]{W.~McGinley}
\author[uconn]{J.~McIntyre}
\author[jlab]{D.~Meekins}
\author[usm]{R.~Mendez}
\author[cmu]{C.A.~Meyer}
\author[umass]{R.~Miskimen}
\author[iu]{R.E.~Mitchell}
\author[uconn]{F.~Mokaya}
\author[asu]{K.~Moriya}
\author[gsi]{F.~Nerling\fnref{f0}}
\author[fsu]{L.~Ng}
\author[gwu]{H.~Ni}
\author[fsu]{A.I.~Ostrovidov}
\author[uofr]{Z.~Papandreou}
\author[mit]{M.~Patsyuk}
\author[fiu]{C.~Paudel}
\author[glasgow]{P.~Pauli}
\author[ncatsu]{R.~Pedroni}
\author[jlab]{L.~Pentchev}
\author[gsi]{K.J.~Peters\fnref{f0}}
\author[gwu]{W.~Phelps}
\author[jlab]{J.~Pierce\fnref{f12}}
\fntext[f12]{Current address: Oak Ridge National Laboratory, Oak Ridge, TN 37831.}
\author[jlab]{E.~Pooser}
\author[jlab]{V.~Popov}
\author[uconn]{B.~Pratt}
\author[jlab]{Y.~Qiang\fnref{f13}}
\fntext[f13]{Current address: Toshiba Medical Research Institute USA, Inc., 706 N Deerpath Dr, Vernon Hills, IL 60061.}
\author[nw]{N.~Qin}
\author[jlab]{V.~Razmyslovich\fnref{f14}}
\fntext[f14]{Current address:  660 E Raven way, Gilbert, AZ 85297.}
\author[fiu]{J.~Reinhold}
\author[asu]{B.G.~Ritchie}
\author[juelich]{J.~Ritman}
\author[nw]{L.~Robison}
\author[mephi]{D.~Romanov}
\author[usm]{C.~Romero}
\author[nsu]{C.~Salgado}
\author[jlab]{N.~Sandoval}
\author[jlab]{T.~Satogata}
\author[wm]{A.M.~Schertz}
\author[juelich]{S.~Schadmand}
\author[umass]{A.~Schick}
\author[cmu]{R.A.~Schumacher}
\author[gsi]{C.~Schwarz}
\author[gsi]{J.~Schwiening}
\author[uofr]{A.Yu.~Semenov}
\author[uofr]{I.A.~Semenova}
\author[nw]{K.K.~Seth}
\author[ihep]{X.~Shen}
\author[iu]{M.R.~Shepherd}
\author[jlab]{E.S.~Smith\corref{cor}}
\cortext[cor]{Corresponding author: Tel.: +1 757 269 7625.}\ead{elton@jlab.org}
\author[cua]{D.I.~Sober}
\author[jlab]{A.~Somov}
\author[mephi]{S.~Somov}
\author[usm]{O.~Soto}
\author[asu]{N.~Sparks}
\author[cmu]{M.J.~Staib}
\author[jlab]{C.~Stanislav}
\author[wm]{J.R.~Stevens}
\author[jlab]{J.~Stewart\fnref{f16}}
\fntext[f16]{Current address: Brookhaven National Laboratory, Upton, New York 11973.}
\author[gwu]{I.I.~Strakovsky}
\author[asu]{B.C.L.~Sumner}
\author[uofr]{K.~Suresh}
\author[itep]{V.V.~Tarasov}
\author[jlab]{S.~Taylor}
\author[uofr]{L.A.~Teigrob\fnref{f17}}
\fntext[f17]{Current address: Tykans Group Inc., 3412 25 St. NE, Calgary, AB, T1Y 6C1.}
\author[uofr]{A.~Teymurazyan}
\author[glasgow]{A.~Thiel}
\author[mephi]{I.~Tolstukhin\fnref{f18}}
\fntext[f18]{Current address: Argonne National Laboratory, Argonne, Illinois 60439.}
\author[nw]{A.~Tomaradze}
\author[usm]{A.~Toro}
\author[fsu]{A.~Tsaris}
\author[cmu]{Y.~Van~Haarlem\fnref{f19}}
\fntext[f19]{Current address: Commonwealth Scientific and Industrial Research Organisation, Lucas Heights, NSW~2234, Australia.}
\author[uofa]{G.~Vasileiadis}
\author[usm]{I.~Vega}
\author[iu]{G.~Visser}
\author[uofa]{G.~Voulgaris}
\author[cua]{N.K.~Walford\fnref{f20}}
\fntext[f20]{Current address: AbbVie Deutschland GmbH, Knollstrasse 67061, Ludwigshafen, Germany.}
\author[glasgow]{D.~Werthm\"uller\fnref{f21}}
\fntext[f21]{Current address: University of York, York YO10 5DD, United Kingdom.}
\author[jlab]{T.~Whitlatch}
\author[odu]{N.~Wickramaarachchi}
\author[mit]{M.~Williams}
\author[jlab]{E.~Wolin\fnref{f22}}
\fntext[f22]{Current address: 2808 Linden Ln, Williamsburg, VA 23185.}
\author[nw]{T.~Xiao}
\author[mit]{Y.~Yang}
\author[iu,uofr]{J.~Zarling}
\author[wuhan]{Z.~Zhang}
\author[ihep]{Q.~Zhou}
\author[wuhan]{X.~Zhou}
\author[jlab]{B.~Zihlmann}
%
%
\address[asu]{Arizona State University, Tempe, Arizona 85287, USA}
\address[uofa]{National and Kapodistrian University of Athens, 15771 Athens, Greece}
\address[cmu]{Carnegie Mellon University, Pittsburgh, Pennsylvania 15213, USA}
\address[cua]{Catholic University of America, Washington, D.C. 20064, USA}
\address[uconn]{University of Connecticut, Storrs, Connecticut 06269, USA}
\address[fiu]{Florida International University, Miami, Florida 33199, USA}
\address[fsu]{Florida State University, Tallahassee, Florida 32306, USA}
\address[gwu]{The George Washington University, Washington, D.C. 20052, USA}
\address[glasgow]{University of Glasgow, Glasgow G12 8QQ, United Kingdom}
\address[gsi]{GSI Helmholtzzentrum f\"ur Schwerionenforschung GmbH, D-64291 Darmstadt, Germany}
\address[iu]{Indiana University, Bloomington, Indiana 47405, USA}
\address[ihep]{Institute of High Energy Physics, Beijing 100049, People's Republic of China}
\address[itep]{Alikhanov Institute for Theoretical and Experimental Physics NRC (Kurchatov Institute), Moscow, 117218, Russia}
\address[jlab]{Thomas Jefferson National Accelerator Facility, Newport News, Virginia 23606, USA}
\address[juelich]{Nuclear Physics Institute, Forschungszentrum Juelich, 52428 Juelich, Germany}
\address[umass]{University of Massachusetts, Amherst, Massachusetts 01003, USA}
\address[mit]{Massachusetts Institute of Technology, Cambridge, Massachusetts 02139, USA}
\address[mephi]{National Research Nuclear University Moscow Engineering Physics Institute, Moscow 115409, Russia}
\address[nsu]{Norfolk State University, Norfolk, Virginia 23504, USA}
\address[ncatsu]{North Carolina A\&T State University, Greensboro, North Carolina 27411, USA}
\address[uncw]{University of North Carolina at Wilmington, Wilmington, North Carolina 28403, USA}
\address[nw]{Northwestern University, Evanston, Illinois 60208, USA}
\address[odu]{Old Dominion University, Norfolk, Virginia 23529, USA}
\address[uofr]{University of Regina, Regina, Saskatchewan, Canada S4S 0A2}
\address[usm]{Universidad T\'ecnica Federico Santa Mar\'ia, Casilla 110-V Valpara\'iso, Chile}
\address[tomsk]{Tomsk State University, 634050 Tomsk, Russia}
\address[tomskb]{Tomsk Polytechnic University, 634050 Tomsk, Russia}
\address[yerevan]{A. I. Alikhanian National Science Laboratory (Yerevan Physics Institute), 0036 Yerevan, Armenia}
\address[wm]{College of William and Mary, Williamsburg, Virginia 23185, USA}
\address[wuhan]{Wuhan University, Wuhan, Hubei 430072, People's Republic of China}

\begin{abstract}
The \gx~ experiment at Jefferson Lab has been designed to study photoproduction reactions with a 9-GeV linearly polarized photon beam. The energy and arrival time of beam photons are tagged using a scintillator
hodoscope and a scintillating fiber array. The photon flux is determined using a pair spectrometer, while the linear polarization of the photon beam is determined using a polarimeter based on triplet photoproduction.
Charged-particle tracks from interactions in the central target are analyzed in a solenoidal field using a central straw-tube
drift chamber and six packages of planar chambers with cathode strips and drift wires. Electromagnetic showers are reconstructed in a cylindrical scintillating fiber calorimeter inside the magnet and
a lead-glass array downstream. Charged particle identification is achieved by measuring energy loss in the wire chambers and using the flight time of particles between the target and detectors outside the magnet. The signals from all detectors are recorded with flash ADCs and/or pipeline TDCs into memories allowing trigger decisions with a latency of 3.3\,$\mu$s. The detector operates routinely at
trigger rates of 40~kHz and data rates of 600 megabytes per second. We describe the photon beam, the \gx~ detector components, electronics, data-acquisition and monitoring systems, and the performance of the experiment during the first three 
years of operation.
\end{abstract}   


\end{frontmatter}



\tableofcontents

\section[The \gx{} experiment]{\label{sec:gluexexperiment} The \gx{} experiment}
The search for Quantum ChromoDynamics (QCD) exotics uses data from a wide range of experiments and production mechanisms. Historically, the searches have looked for the gluonic excitations of mesons, searching for states of pure glue, glueballs, and hybrid mesons where the gluonic field binding the quark-anti-quark pair has been excited. Most experiments searching for glueballs looked for scalar mesons~\cite{Crede:2008vw}, where the searches relied on over-population of nonets, as well as unusual meson decay patterns. In the search for hybrid mesons~\cite{Meyer:2010ku,Meyer:2015eta}, efforts have focused on particles with exotic quantum numbers, i.e. systems beyond simple quark-anti-quark configurations. Good evidence exists for an isospin $1$ state, the $\pi_{1}(1600)$. Looking collectively at past studies, data from high-statistics photoproduction experiments in the energy range above $6$~GeV are lacking. 

\begin{figure}[hbt]\centering
\includegraphics[width=0.75\textwidth]{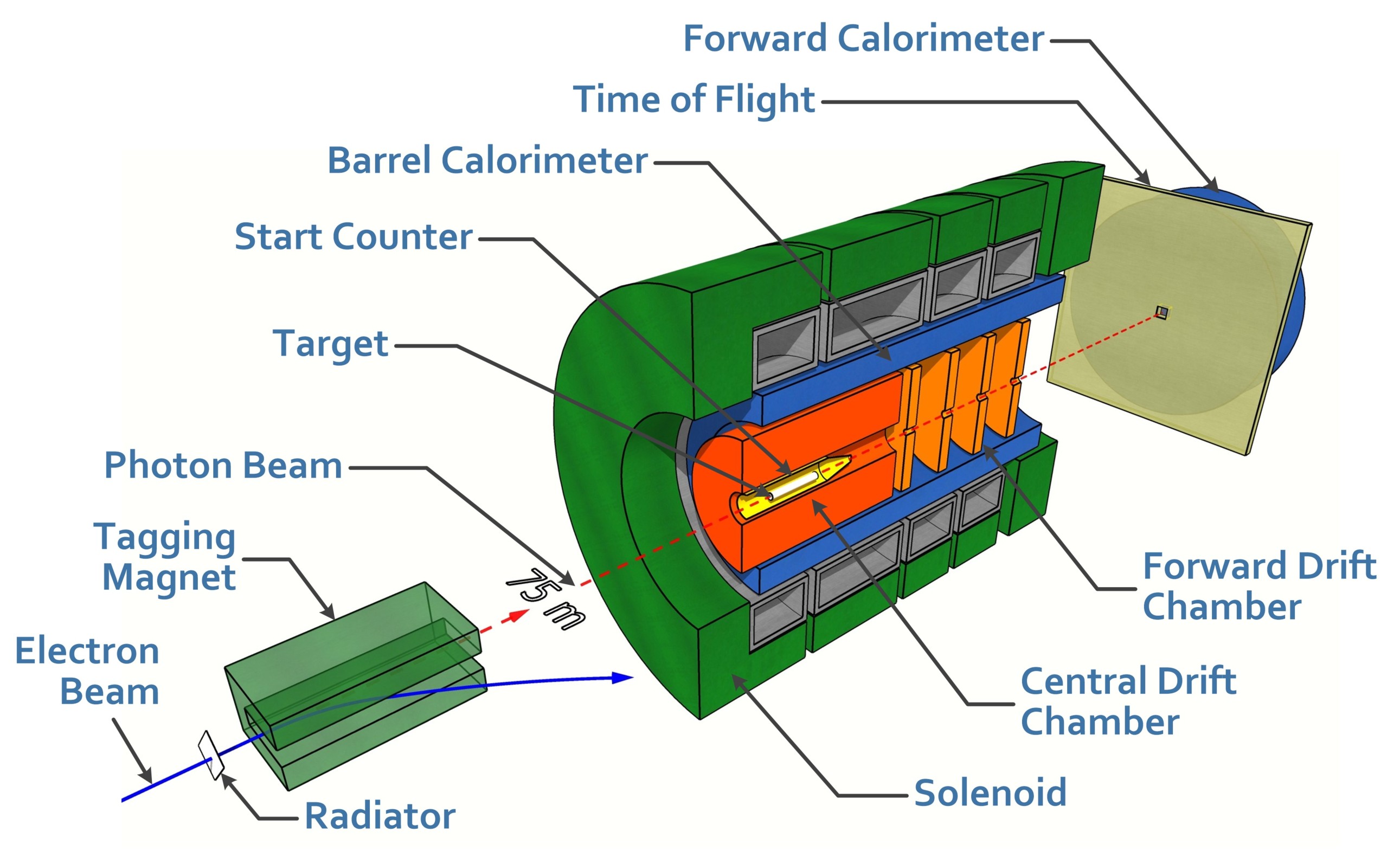}
\caption[]{\label{fig:gluex_cut-away}(Color online)A cut-away drawing of the \GX{} detector in Hall D, not to scale.}
\end{figure}
The \emph{Glu}onic \emph{Ex}citation (\gx{}) experiment at the 
US Department of Energy's Thomas Jefferson National Accelerator Facility (JLab)\footnote{Thomas Jefferson National Accelerator Facility, 12000 Jefferson Ave., Newport News, VA 23606, https://www.jlab.org.} has been built to search for and map out the spectrum of exotic hybrid mesons using a 9-GeV linearly-polarized photon beam incident on a proton target\cite{gluex-ref}. The \gx{} detector and beamline are shown schematically in Figure~\ref{fig:gluex_cut-away}. The detector is nearly hermetic for both charged particles and photons arising from reactions in the cryogenic target at the center of the detector, allowing for reconstruction of exclusive final states. A 2-T solenoidal magnet surrounds the drift chambers used for charged-particle tracking. Two electromagnetic calorimeters cover the central and forward regions, and a scintillation detector downstream provides particle-identification capability through time-of-flight measurements.

\subsection[The Hall-D complex]{The Hall-D complex \label{sec:gluexexperiment:complex}}
\begin{figure}[tbp]\centering
\includegraphics[width=0.75\textwidth]{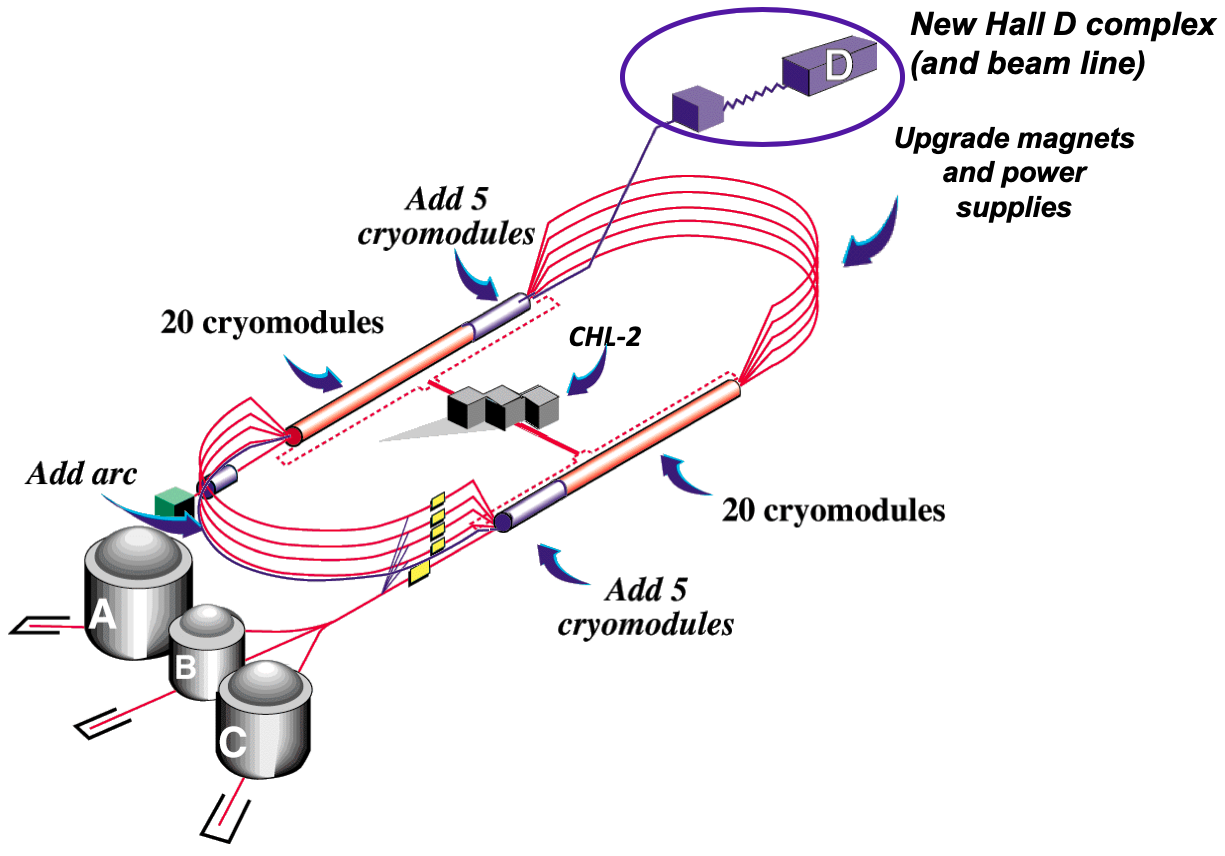}
\caption[]{\label{fig:CEBAF-graphic}(Color online) Schematic of the CEBAF accelerator showing the additions made during the 12-GeV project. The Hall-D complex is located at the north-east end.}
\end{figure}

The \gx{} experiment is housed in the Hall-D complex at JLab (see Fig.\ref{fig:CEBAF-graphic}). This new facility starts with an extracted electron beam at the north end of the Continuous Electron Beam Accelerator Facility (CEBAF) \cite{Leemann:2001dg,CEBAF12GeV}. The electron beam is delivered to the Tagger Hall, where the maximum energy is 12~GeV, due to one more pass through the north linac than the other experimental halls (A, B and C).  Here, linearly-polarized photons are produced through coherent bremsstrahlung off a 50~$\mu$m thick diamond crystal radiator.
The scattered electrons pass through a tagger magnet and are bent into tagging detectors. A high-resolution scintillating-fiber tagging array covers the 8 to 9~GeV energy range, and a tagger hodoscope covers photon energies both from 9~GeV to the endpoint, and from 8~GeV to 3~GeV. Electrons not interacting in the diamond are directed into a 60 kW electron beam dump. The tagged photons travel to the Hall-D experimental hall. The distance from the radiator to the primary collimator is 75~m. The collimator of 5~mm diameter removes off-axis incoherent photons. The front face of the collimator is instrumented with an active collimator to aid in beam tuning.  The beamline and tagging system are described below in Section\,\ref{sec:beamline}.

Downstream of the primary collimator is a thin beryllium radiator used by both the Triplet Polarimeter, which measures the linear polarization of the photons, and a Pair Spectrometer, which is used to measure the flux of the photons. More information on the production, tagging and monitoring of the photon beam can be found in Section~\ref{sec:beamline}. 
The photon beam continues through to a liquid hydrogen target at the heart of the \gx{} detector, and then to the end of the experimental hall where it enters the photon beam dump.

\begin{figure*}[tbp]
\centering
  \includegraphics[angle=0,viewport=95 115 628 500,clip,width=1.0\linewidth]{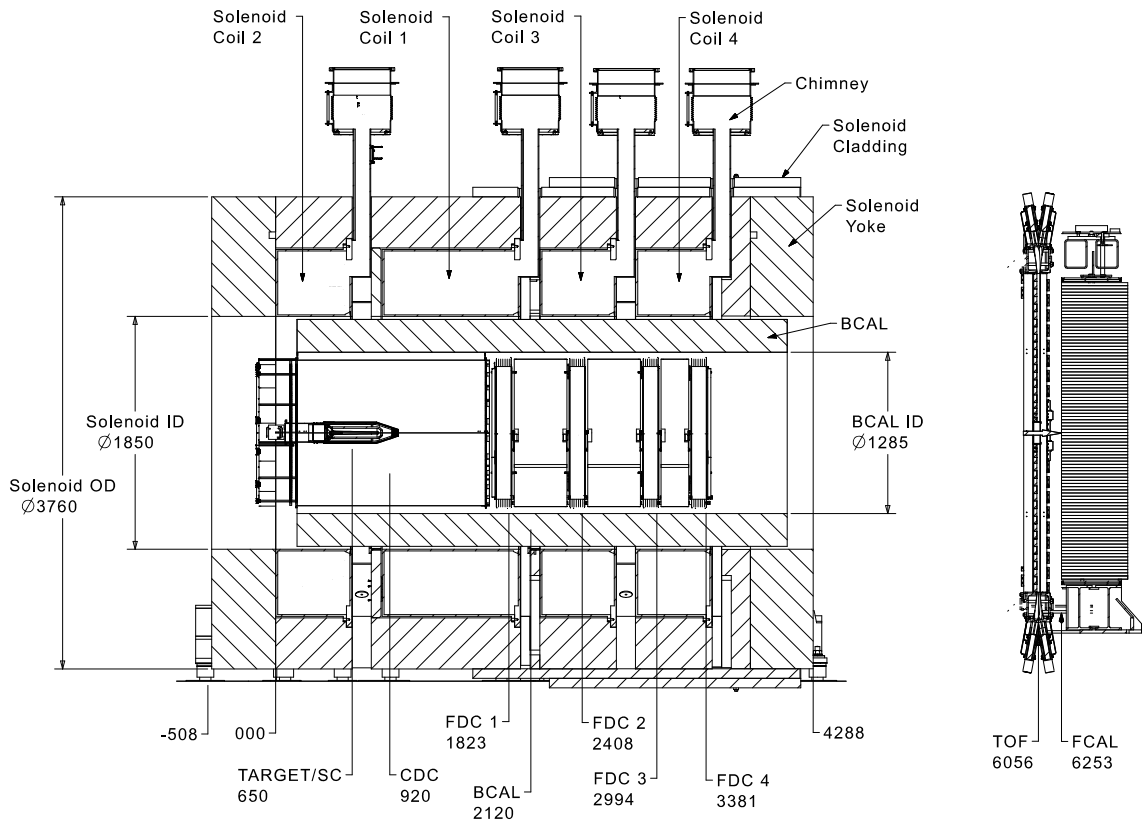}%
  \caption[layout]{GlueX spectrometer layout. Dimensions are given in mm. The
    numbers show the Z-coordinates of the detectors' centers, or of
    the front face of the FCAL modules.
    Glossary: 
              SC  - Start Counter (Section \ref{sec:st}), 
              CDC - Central Drift Chamber (Section \ref{sec:cdc}), 
              FDC - Forward Drift Chamber (Section \ref{sec:fdc}),
              BCAL - Barrel Calorimeter (Section \ref{sec:bcal}), 
              TOF -  Time-of-Flight hodoscope (Section \ref{sec:tof}), 
              FCAL - Forward Calorimeter (Section \ref{sec:fcal}).
%
    \label{fig:layout_spectrometer}
  }
\end{figure*}

The layout of the \gx{} detector is shown in Fig.~\ref{fig:layout_spectrometer}. The spectrometer is based on a 4-m-long solenoidal magnet that is operated at a maximum field of 2~T, see Section~\ref{sec:solenoid}. The liquid-hydrogen target is located inside the upstream bore of the magnet. The target consists of a 2-cm-diameter, 30-cm-long volume of hydrogen, as described in Section~\ref{sec:target}. Surrounding the target is the Start Counter, which consists of 30 thin scintillator paddles that bend to a nose on the down-stream end of the hydrogen target. The Start Counter is the primary detector that registers the time coincidence of the radio-frequency (RF) bunch containing the incident electron and the tagged photon producing the interaction. More information on this scintillator detector can be found in Section~\ref{sec:scintillators}. 


The Central Drift Chamber, a cylindrical straw-tube detector, starts at a radius of 10~cm from the beam line. The active volume of the chamber extends from 48~cm upstream to 102~cm downstream of the target center, and from 10~cm to 56~cm in radius. The Central Drift Chamber consists of 28 layers of straw tubes in axial and two stereo orientations. Downstream of the central tracker is the Forward Drift Chamber, which consists of four packages, each containing 6 planar layers in alternating $u$-$y$-$v$ orientations. Both cathodes and anodes in the Forward Drift Chamber are read out, providing three-dimensional space point measurements. More details on the tracking system are provided in Sections~\ref{sec:tracking} and \ref{sec:trackingperformance}. 

Downstream of the magnet is the Time-of-Flight wall. This system consists of two layers of scintillator paddles in a crossed pattern, and, in conjunction with the Start Counter, is used to measure the flight time of charged particles. More information on the time-of-flight system is provided in Section~\ref{sec:scintillators}. 
Photons arising from interactions within the \gx{} target are detected by two calorimeter systems. The Barrel Calorimeter, located inside the solenoid, consists of layers of scintillating fibers alternating with lead sheets. The Forward Calorimeter is downstream of the Time-of-Flight wall, and consists of $2800$ lead-glass blocks. More information on the the calorimeters can be found in Section~\ref{sec:calorimeters}.

\subsection[Experimental requirements]{Experimental requirements \label{sec:intro:requirements}}
The physics goals of the GlueX experiment require the reconstruction of exclusive final states. Thus, the \gx{} detector must be able to reconstruct both charged particles ($\pi^{\pm}$, $K^{\pm}$ and $p/\bar{p}$) and particles decaying into photons ($\pi^{\circ}$, $\eta$, $\omega$ and $\eta^{\prime}$). For this capability, the charged particles and photons must be reconstructed with good momentum and energy resolution. The experiment must also be able to reconstruct the energy of the incident photon (8 to 9~GeV) with high accuracy ($0.1$\%) and have knowledge of the linear polarization (maximum $\sim$40\%) of the photon beam to an absolute precision of 1\%. Finally, many interesting final states involve more than five particles. Thus, the \gx{} detector must also be nearly hermetic for both charged particles and photons, with an acceptance that is reasonably uniform, well understood, and accurately modeled in simulation.

In practice, the typical momentum resolution for charged particles is $1$--$3\%$, while the resolution is 8-9\% for very-forward high-momentum particles.  For most charged particles, the tracking system has nearly hermetic acceptance for polar angles from $1^\circ-2^{\circ}$ to $150^{\circ}$. However, protons with momenta below about 250~MeV/c are absorbed in the hydrogen target and not detected. A further challenge is the reconstruction of tracks from charged pions with momenta under 200~MeV/c due to spiraling trajectories in the magnetic field.
The measurement of energy loss ($dE/dx$) in the Central Drift Chamber enables the separation of pions and protons up to about 800~MeV/c, while time-of-flight determination allows separation of forward-going pions and kaons up to about 2~GeV/c.

For photons produced from the decays of reaction products, the typical energy resolution is 5 to 6\%$/\sqrt{E_{\gamma}}$. Photons above 60 MeV can be detected in the Barrel Calorimeter, with some variation depending on the incident angle.
The interaction point along the beam direction is determined by comparing the information from the readouts on the upstream and downstream ends of the detector. In the Forward Calorimeter, photons with energies larger than 100~MeV can be detected with uniform resolution across the face of the detector. There is a gap between the calorimeters at around $11^{\circ}$, where energy can be lost due to shower leakage. Both photon detection efficiency and energy resolution are degraded in this region. 
 
\subsection{Data requirements \label{sec:intro:data_requirements}}
The physics analyses need to be carried out in small bins of energy and momentum transfer, necessitating not only the ability to reconstruct exclusive final states but also to collect sufficient statistics. 
While exact cross sections are not known, the cross sections of interest will be in the 10~nb to 1~$\mu$b range. 

This paper describes the operation of \gx{} Phase I.
During this initial phase, the \GX{} experiment has run with a data acquisition system capable of collecting data using photon beams of a few $10^{7}~\gamma/$s in the coherent peak (8.4-9 GeV), with an expectation to run with 2.5 times higher rates in the future. 
The data acquisition system ran routinely at 40\,kHz with raw event sizes of 15-20 kilobytes, collecting about 600~megabytes of data per second. With firmware improvements, future running is expected at 90 kHz and 1~gigabyte per second. Details of the trigger and data acquisition are presented in Sections~\ref{sec:trig} and \ref{sec:daq}.

\subsection{Coordinate system \label{sec:intro:coordinates}}
For reference, we introduce here the overall experiment coordinate system, which is used in this document and throughout the analysis. The z-axis is defined along the nominal beamline increasing downstream. The coordinate system 
is right-handed with the y-axis pointing vertically up and the x-axis pointing approximately north. 
The origin is located 50.8\,cm (20 inches) downstream of the upstream side of the upstream endplate of the solenoid, placing the nominal center of the target at (0,0,65\,cm).

\section[The coherent photon source and beamline]{The coherent photon source and beamline \label{sec:beamline}}

\begin{table}[tbp]
\begin{center}
\caption
[Design properties of the electron beam]
{Electron beam parameters. 
The emittance, energy spread and       
related parameters are estimates
based on a model of the transport line from
the accelerator to the Hall D radiator.
The dimensions of the beam spot at the position of 
the radiator are directly measured, and vary around the
stated values by $\pm 30\%$
depending on beam conditions. 
Values for image size at collimator,
obtained by projection of the electron beam
spot convergence forward to the position of
the primary photon collimator, have relative
uncertainties of 50\%.
\label{tab:elecprop}}  
\begin{tabular}{|l|c|}
\hline\hline
parameter & design results \\
\hline
energy & 12~GeV \\ 
energy spread, RMS & $2.2$~MeV \\
transverse $x$ emittance & 2.7~mm$\cdot\mu$rad \\
transverse $y$ emittance & 1.0~mm$\cdot\mu$rad \\
$x$ spot size at radiator, RMS & 1.1~mm \ \\
$y$ spot size at radiator, RMS & 0.7~mm \ \\
$x$ image size at collimator, RMS & 0.5~mm \\
$y$ image size at collimator, RMS & 0.5~mm \\
image offset from collimator axis, RMS & 0.2~mm \\
distance radiator to collimator & 75.3~m \\
\hline\hline
\end{tabular}
\end{center}
\end{table}

\subsection{CEBAF electron beam \label{sec:ebeam}}
CEBAF has a race track configuration with two parallel linear accelerators based on superconducting radio frequency (RF) technology \cite{Leemann:2001dg}. The machine operates at 1.497 GHz and delivers beam to Hall D at 249.5~MHz.\footnote{Hall D beam at 499 MHz is possible, but not the norm.} Precise timing signals for the accelerator beam bunches are available to the experiment and are used to determine the time that individual photon bunches pass through the target. The nominal properties for the CEBAF electron beam to the Tagger Hall are listed in Table\,\ref{tab:elecprop}.
\begin{figure}[t]
\begin{center}
 \includegraphics[clip=true,width=0.98\linewidth]{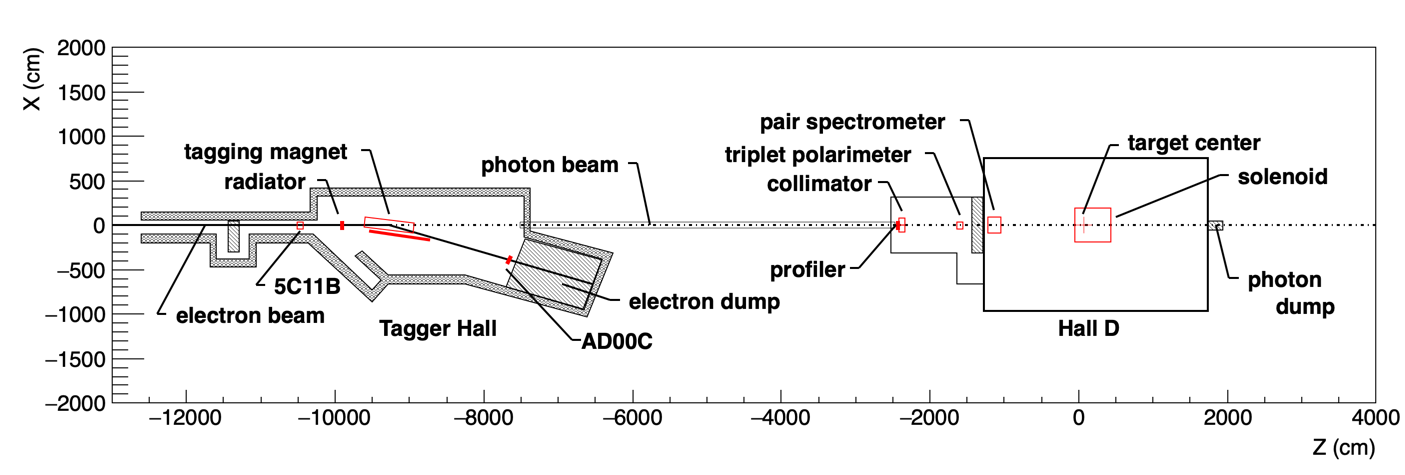}
\end{center}
\caption{Schematic layout of the Hall-D complex, showing the Tagger Hall, Hall D, and
several of the key beamline devices.
Also indicated are the locations of the 5C11B and AD00C beam monitors.
        }
\label{fig:beam:Draw_beamline} 
\end{figure}

\subsection{Hall-D photon beam \label{sec:gbeam}}
\begin{figure}[tbp]
\begin{center}
 \includegraphics[clip=true,width=0.49\linewidth]{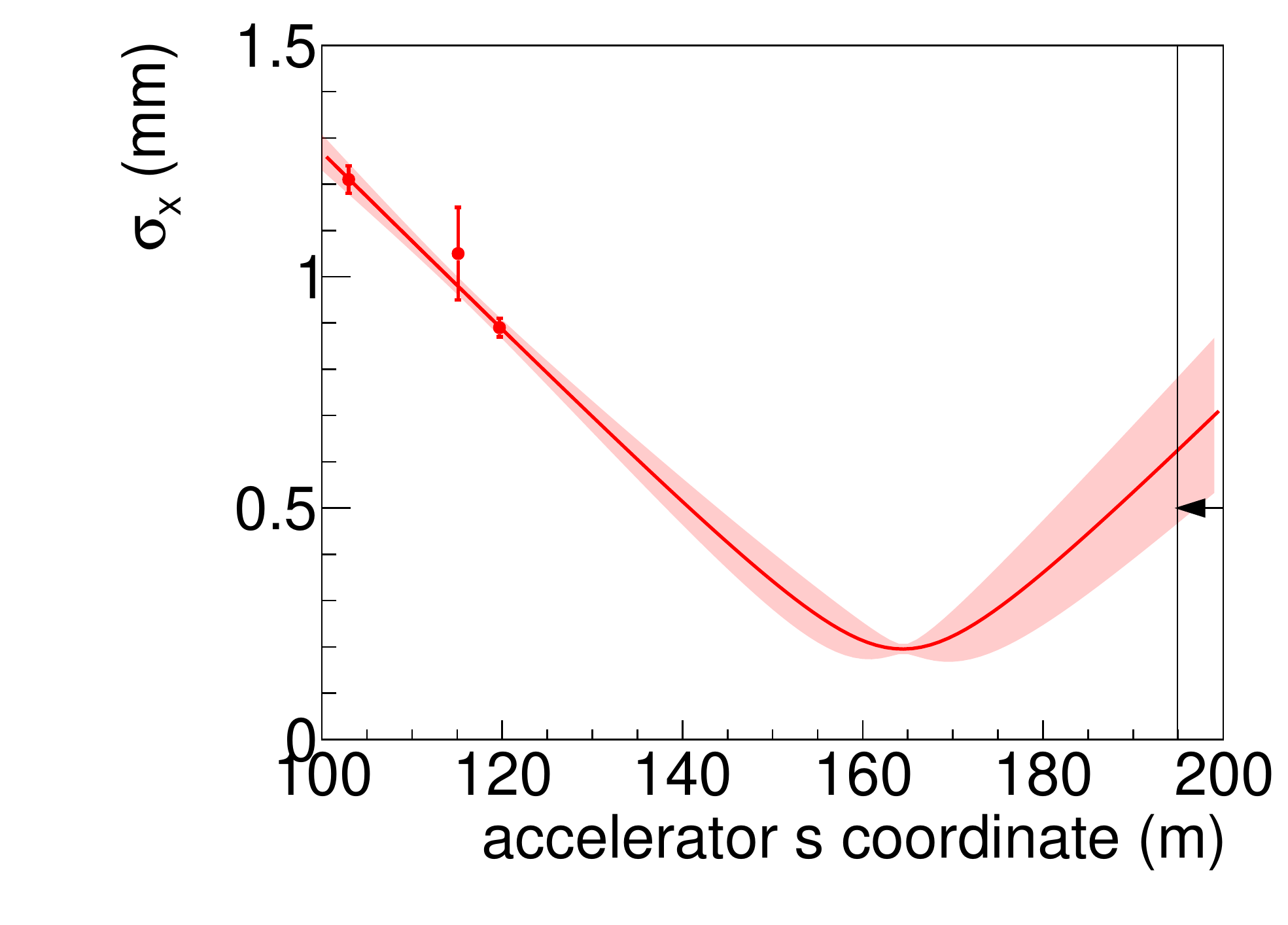}
 \includegraphics[clip=true,width=0.49\linewidth]{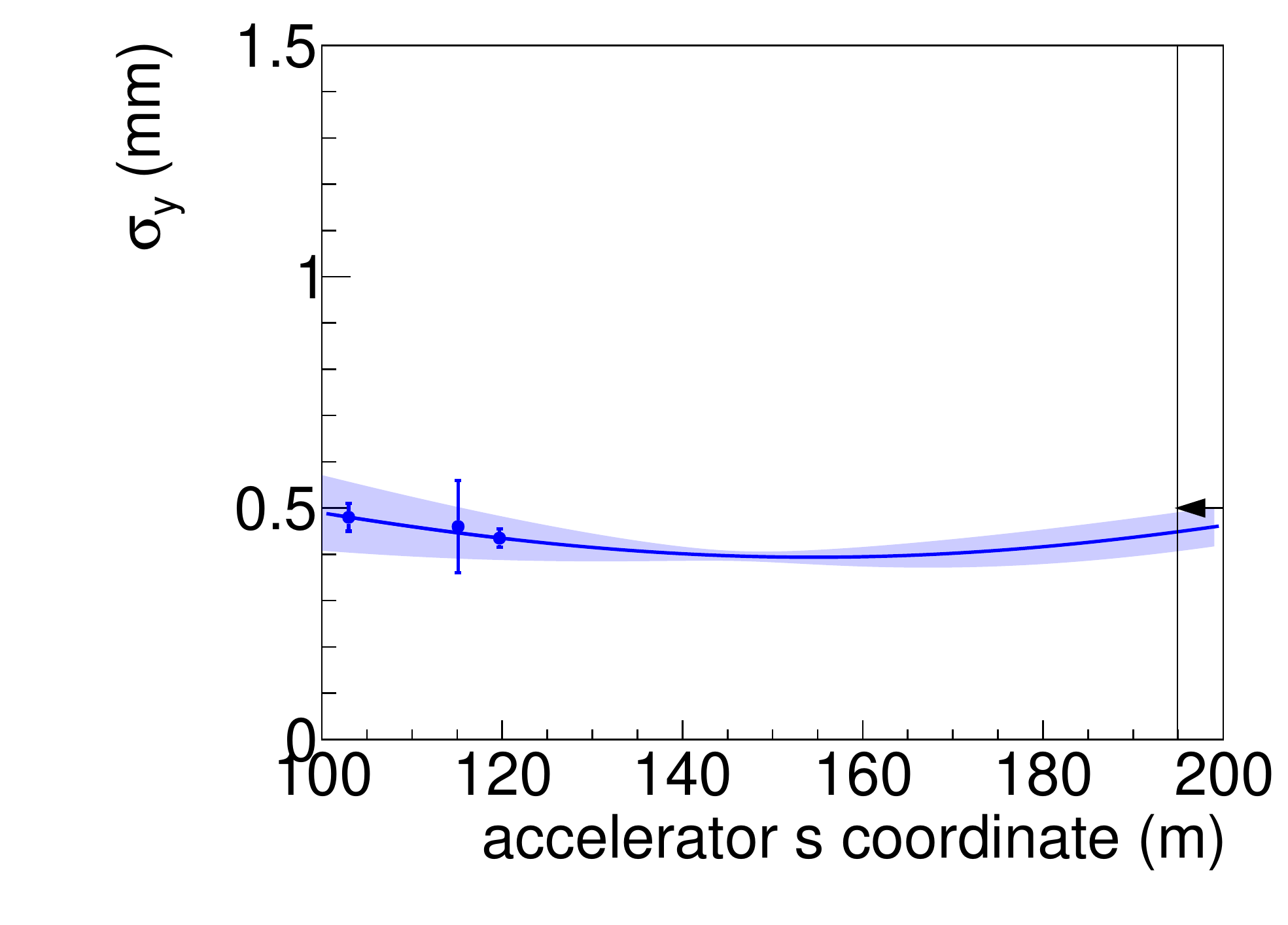}
\end{center}
\caption{(Color online) Measurements of the root-mean-square width of the electron beam 
in horizontal (left)
and vertical (right) projections as a function of position along the beamline, based on
harp scans (data points) of the electron beam. The radiator position is just upstream
of the third data point. The primary collimator position is marked by the vertical line
indicated by the arrow. The curve downstream of the radiator is an extrapolation from
the measured data points, with extrapolation uncertainty indicated by the shaded regions.
        }
\label{fig:beam:convergence} 
\end{figure}

The Hall-D complex, described in Section \ref{sec:gluexexperiment:complex} and shown schematically in Fig.~\ref{fig:beam:Draw_beamline}, includes a dedicated Tagger Hall, an associated collimator cave, and Experimental Hall D itself. A linearly-polarized photon beam is created using the process of coherent bremsstrahlung \cite{timm1969,LIVINGSTON2009205} when the electron beam passes through an oriented diamond radiator at the upstream end of the Tagger Hall.
The electron beam position at the radiator is monitored and controlled using beam position monitors (5C11 and 5C11B) which are located at the end of the accelerator tunnel just upstream of the Tagger Hall (see Fig.\,\ref{fig:beam:Draw_beamline}.)
The CEBAF electron beam is tuned to converge as it passes through the radiator, ideally
so that the electron beam forms a virtual focus at the collimator 
located 75\,m downstream of the radiator.
At the collimator, the virtual spot size of 0.5\,mm is small compared to the cm-scale size of the photon beam on the front face of the collimator,
such that a cut on photon position at the collimator is effectively a cut on photon emission angle at the radiator. 
The convergence properties of the electron beam are measured by scanning the beam profile with vertical and horizontal wires. The wire scanners are referred to as "harps."
Examples of the horizontal and vertical convergence of the electron beam envelope (undeflected by the tagger magnet)
measured using harp scans and projected downstream along the beamline are shown in Fig.\,\ref{fig:beam:convergence}.

The photon beam position on the collimator is monitored using an active collimator positioned just upstream of the primary photon beam collimator
(described below in section \ref{sec:coll}). 
The position stability of the photon beam is maintained during normal
operation by a feedback system that locks the position of the electron
beam at the 5C11B beam position monitor and, consequently, the photon beam at the active collimator. The stability of the electron
beam current and position is monitored using an independent beam monitor
(AD00C in Fig.\,\ref{fig:beam:Draw_beamline}) 
located immediately upstream of the electron dump.

\begin{figure}[tbp]
\begin{center}
 \includegraphics[clip=true,width=0.5\linewidth]{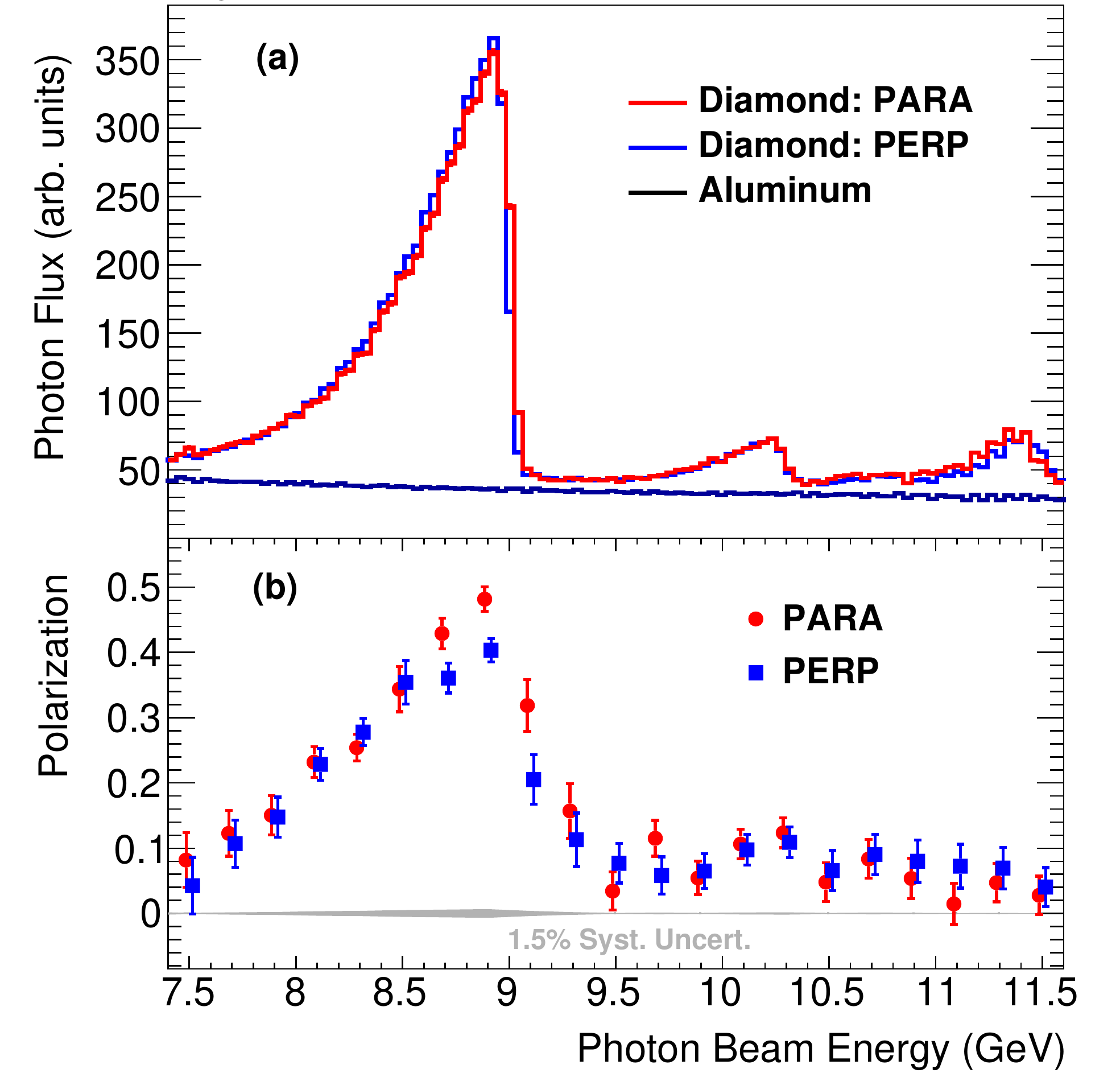}
\end{center}
\caption{(color online) (a) Collimated photon beam intensity versus energy as measured by the Pair Spectrometer.
(b) Collimated photon beam polarization as a function of beam energy,
as measured by the Triplet Polarimeter, with data points offset horizontally by $\pm0.015$~GeV for clarity.
The labels PARA and PERP refer to orientations of the diamond radiator that result in polarization
planes that are parallel and perpendicular to the horizontal, respectively.
        }
\label{fig:beam:fig0_beam} 
\end{figure}

The linearly-polarized photon beam is produced via a radiator placed in the electron beam just upstream of the
Tagger (section \ref{sec:tag}). A properly aligned 20--60\,$\mu$m thick diamond crystal
radiator produces
linearly polarized photons via coherent brems\-strah\-lung in enhancements \cite{timm1969,LIVINGSTON2009205},
that appear as peaks at certain energies in the collimated brems\-strah\-lung intensity spectrum (Fig.\,\ref{fig:beam:fig0_beam}), superimposed upon the ordinary  continuum brems\-strah\-lung spectrum from an aluminum radiator.
The energies of the coherent photon peaks and the degree of polarization in each of those peaks depend on the crystal orientation with respect to the incident electron beam.
Adjustment of the orientation of the diamond crystal with respect to the incoming
electron beam permits production of essentially any coherent photon peak energy up to that of the energy of the incident electron beam, as well as the
degree or direction of linear polarization.
A choice of 9 GeV for the primary peak energy, corresponding to 40\% peak linear polarization,
was found to be optimum for the \GX{} experiment with a 12-GeV incident electron beam.

The degree of polarization for a coherent bremsstrahlung beam is greatest for photons emitted at small
angles with respect to the incident electron direction. Collimation of the photon beam to a fraction
of the characteristic brems\-strah\-lung angle exploits this correlation to significantly enhance
the average polarization of the beam. 
In the nominal \GX{} beamline configuration, a 5.0-mm-diameter collimator
\footnote{A 3.4\,mm collimator is also available, and has been used for some physics production runs
with the thinnest (20 $\mu m$) diamond.}
positioned 75\,m downstream of the radiator is used, corresponding to a cut at approximately
1/2 $m/E$ in characteristic angle, where $m$ is the electron rest mass and
$E$ is the energy of the incident electron. 
The photon beam energy spectrum and photon flux after collimation are measured
by the Pair Spectrometer (section \ref{sec:ps}), located downstream of the collimator in Hall D. 

An example of the measured photon spectrum and degree of polarization with a 12-GeV electron beam is
shown in Fig.\,\ref{fig:beam:fig0_beam}. The spectrum labeled ``Aluminum" in 
Fig.\,\ref{fig:beam:fig0_beam}(a) shows the spectrum of ordinary (incoherent) brems\-strah\-lung,
normalized to the approximate thickness of the diamond radiator in terms of radiation lengths.
The expected degree of linear polarization in the energy range of 8.4--9.0~GeV is $\sim$40\% after
collimation.  The photon beam polarization is directly measured by the triplet polarimeter (section \ref{sec:tpol})
located just upstream of the pair spectrometer. The stability of the beam polarization is independently
monitored via the observed azimuthal asymmetry in various photoproduction reactions, particularly that for $\rho$ photoproduction \cite{gx3076}.

Typical values for parameters and properties of the photon beam are given in Table\,\ref{tab:operates}.
In the sections that follow, we describe in more detail how the linearly-polarized photon beam
is produced, how the photon energy is determined using the tagging spectrometer, how the photon beam polarization
spectrum and flux are measured with the Pair Spectrometer and Triplet Polarimeter, and how the photon
flux is calibrated using the Total Absorption Counter.

\begin{table}[btp]
\begin{center}
\caption[Typical parameters for the \GX{} photon beam]{\label{tab:operates}
Typical parameters for the \GX{} photon beam,
consistent with the electron beam properties listed in Table \ref{tab:elecprop},
a diamond radiator of thickness 50~$\mu$m, and the standard primary collimator of diameter 5.0~mm
located at the nominal position. The electron beam current incident on the radiator is taken to  be $150$~nA.
The hadronic rates are calculated for the \GX{} 30~cm liquid hydrogen target.}

\begin{tabular}{|l|r|}
\hline\hline
$E$ upper edge of the coherent peak & 9~GeV \\
Coherent peak effective range       & 8.4 - 9.0~GeV\\
Net tagger rate in the coherent peak range & 45~MHz  \\
$N_{\gamma}$ in the peak range after collimator & 24~MHz  \\
Maximum polarization in the peak, after collimator & 40\% \\
Mean polarization in the peak range, after collimator & 35\% \\
Power absorbed on collimator & 0.60~W \\
Power incident on target & 0.23~W \\
Total hadronic rate & 70 kHz \\
Hadronic rate in the peak range & 3.7 kHz \\
\hline\hline
\end{tabular}
\end{center}
\end{table}

\subsection{Goniometer and radiators \label{sec:radiators}}
For the linearly-polarized photon beam normally used in \GX{} production running, diamond radiators 
are used to produce a coherent bremsstrahlung beam. This requires precise alignment of the diamond
radiator, in order to produce a single dominant coherent peak\footnote{Defined as 0.6 GeV below the coherent edge (nominally 9 GeV). The position of the edge scales approximately with the primary incident electron beam energy.}  with the desired energy and polarization
by scattering the beam electrons from the crystal planes associated with a particular reciprocal lattice
vector.
A multi-axis goniometer, manufactured by Newport Corporation, precisely
adjusts the relative orientation of the
diamond radiator with respect to the incident electron beam horizontally, vertically and rotationally about the $X$, $Y$ and $Z$ axes, respectively.
The Hall-D goniometer holds several radiators, any of which may be moved into the beam for use at any time
according to the requirements of the experiment.

In addition to the diamond radiators, several aluminum radiators of thicknesses ranging from 1.5 to 40~$\mu$m are used to normalize the rate spectra measured in the Pair Spectrometer, correcting for its acceptance.
A separate rail for these amorphous radiators is 
positioned 615~mm downstream of the goniometer.

\subsubsection{Diamond selection and quality control \label{sec:diamonds}}
The properties of diamond are uniquely suited for coherent brems\-strah\-lung radiators.
The small lattice constant and high Debye temperature of diamond result in an exceptionally high probability
for coherent scattering in the brems\-strah\-lung process \cite{Bilokon:1983}.
Also, the high coherent scattering probability is a consequence of the small atomic number of carbon (Z = 6). At the dominant crystal momentum (9.8 keV) corresponding to the leading (2,2,0) reciprocal lattice vector, the small atomic number results in minimal screening of the nuclear charge by inner shell electrons.
Diamond is the best known material in terms of its coherent radiation
fraction, and its unparalleled thermal conductivity and radiation hardness make it
well-suited for use in a high-intensity electron beam environment.

The position of the coherent edge in the photon beam intensity spectrum is a simple monotonic
function of the angle between the incident electron beam direction and the normal to the (2,2,0)
crystal plane. The 12-GeV-electron beam entering the radiator has a divergence less than
10 $\mu$rad, corresponding to a broadening of the coherent edge in
Fig.\,\ref{fig:beam:fig0_beam} by just 7~MeV. However, if the 
incident electron beam had 
to travel through 100~$\mu$m of diamond material prior to radiating, the
resulting electron beam emittance would
increase by a factor of 10 due to multiple Coulomb scattering, resulting in a proportional increase
in the width of the coherent edge. Such broadening of the coherent peak diminishes both the degree of polarization in the coherent peak as well as the collimation efficiency in the forward direction.
Hence, diamond radiators for \GX{} must be significantly
thinner than 100 microns. 

The cross-sectional area of a diamond target must also be large enough to completely contain the electron beam so that the beam does not overlap with the material of the target holder. Translated to the beam spot dimensions from Table\,\ref{tab:elecprop}, \gx{}
requires a target with transverse size 5~mm or greater. Uniform single-crystal diamonds of
this size are now available as slices cut from natural gems, HPHT (high-pressure, 
high-temperature) synthetics, and CVD (chemical vapor deposition) single crystals. Natural gems are ruled out due to cost. HPHT crystals had been thought
to be far superior to CVD single crystals in terms of their diffraction widths, but our
experience did not bear this out. \GX{} measurements of the
x-ray rocking curves of CVD crystals obtained from the commercial vendor Element Six\footnote{Element Six, https://www.e6.com/en.} routinely
showed widths that were within a factor 2 of the theoretical Darwin width,
similar to the results we found for the best HPHT diamonds that were available to us
\cite{YANG2010719,YANG2012}.

\begin{figure}[tbp]
\begin{center}
 \includegraphics[clip=true,width=0.7\linewidth]{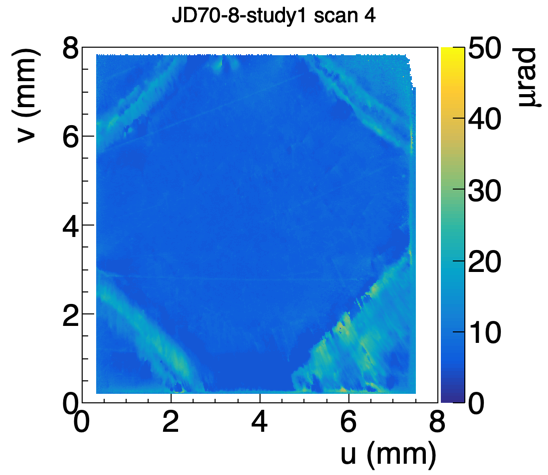}
\end{center}
\caption{(color online) Rocking curve RMS width topograph taken of the (2,2,0) reflection
from a CVD diamond crystal using 15~keV X-rays at the C-line at CHESS.
The bright diagonal lines in the corners
indicate regions of increased local strain, coinciding with growth boundaries radiating
outward from the seed crystal used in the CVD growth process. 
        }
\label{fig:diamond_rocking_curve_rms} 
\end{figure}

Fig.~\ref{fig:diamond_rocking_curve_rms} shows a rocking curve topograph of a diamond
radiator taken with 15~keV x-rays at the 
Cornell High Energy Synchrotron Source (CHESS). The instrumental
resolution of this measurement is of the same order as the Darwin width for this
diffraction peak, approximately 5 $\mu$rad. During operation, the electron beam spot would
be confined to the relatively uniform central region. Any region in
this figure with a rocking curve root-mean-square width of 20~$\mu$rad or less is indistinguishable
from a perfect crystal for the purposes of \GX{}.
Regardless of whether or not better HPHT diamonds exist, these Element Six CVD diamonds have sufficiently narrow  diffraction widths for our application.  This, coupled with their lower cost relative to HPHT material, made
them the obvious choice for the Hall-D photon source.

The diamond radiator fabrication procedure began with procurement of the raw
material in the form of $7\times 7\times 1.2$~mm$^3$ CVD single-crystal plates from the
vendor. After x-ray rocking curve scans of the raw material were taken to verify crystal
quality, the acceptable diamonds were shipped to a second vendor, Delaware Diamond Knives (DDK). At DDK, the
1.2-mm-thick samples were sliced into three samples of 250 $\mu$m thickness each, then
each one was polished on both sides down to a final thickness close to 50 $\mu$m. The
samples, now of dimensions $7\times 7\times 0.05$~mm$^3$ were fixed to a small aluminum
mounting tab using a tiny dot of conductive epoxy placed in one corner.
These crystals were then returned to the synchrotron light source
for final x-ray rocking curve measurements prior to final
approval for use in the \GX{} photon source.

The useful lifetime of a diamond radiator in the \GX{} beamline is limited by the 
degradation in the sharpness of the coherent edge due to accumulation of radiation damage.
Experience during the early phase of \GX{} running showed that after exposure to
about 0.5 C of integrated electron beam charge, the width of the coherent edge 
increased enough that the entire coherent peak was no longer contained within the energy
window of the tagger microscope. When a crystal reached this degree of degradation, the
radiator was regarded as no longer usable, and a new crystal was installed.

During Phase 1 of \GX{}, radiator crystals were replaced three times due
to degradation, twice with 50~$\mu$m radiators and once with a 20~$\mu$m radiator. The 20-$\mu$m
diamond was introduced to test whether the reduced multiple Coulomb scattering 
might result in an
observable increase in peak polarization. This turned out not to be the case, for
two reasons. The first is that to take full advantage of the reduced multiple
scattering in the radiator for increased peak polarization, the collimator size 
must be reduced proportionally. A 3.4-mm-diameter collimator was available for
this purpose, but variability observed in the convergence properties of the electron
beam at the radiator overruled running with any collimator smaller than 5~mm,
even when a thinner radiator was in use.

The second reason is that any improvements
from reduced multiple scattering that came with the smaller radiator thickness
were more than offset by strong indications of radiation
damage that appeared not long after the 20~$\mu$m crystal was put into production.
The rapid appearance of radiation damage 
was partly due to the larger beam current (factor 2.5) that was needed to
produce the same photon flux as with a 50~$\mu$m crystal, but that factor alone
did not fully
explain what was seen. Subsequent x-ray measurements showed that a large buckling of
the 20~$\mu$m crystal had occurred in the region of the incident electron beam spot, 
evidently due to  local differential expansion of the diamond lattice arising from
radiation damage. Once the crystal buckled, the energy of the coherent
peak varied significantly across the electron beam spot, effectively broadening
the peak. Fortunately, the greater stiffness of a 50~$\mu$m crystal
appears to suppress this local buckling under similar conditions of radiation damage.

Based on these observations, 50~$\mu$m was selected as the
optimum thickness for \GX{} diamond radiators: thin enough to limit the effects
of multiple scattering and thick enough to suppress buckling from internal stress
induced by radiation damage. The effective useful lifetime of a 50~$\mu$m radiator
in the photon source is about 0.5 C integrated incident electron charge. 
This lifetime might be extended somewhat by the use of thermal annealing to partially remove the effects of radiation damage. 
This possibility will be explored when the pace of diamond replacement increases with the start of full-intensity running (\gx{} Phase 2) and the number of spent radiators starts to accumulate.

\begin{figure}[tbp]
\begin{center}
   \includegraphics[width=0.95\linewidth,viewport=80 200 750 400]{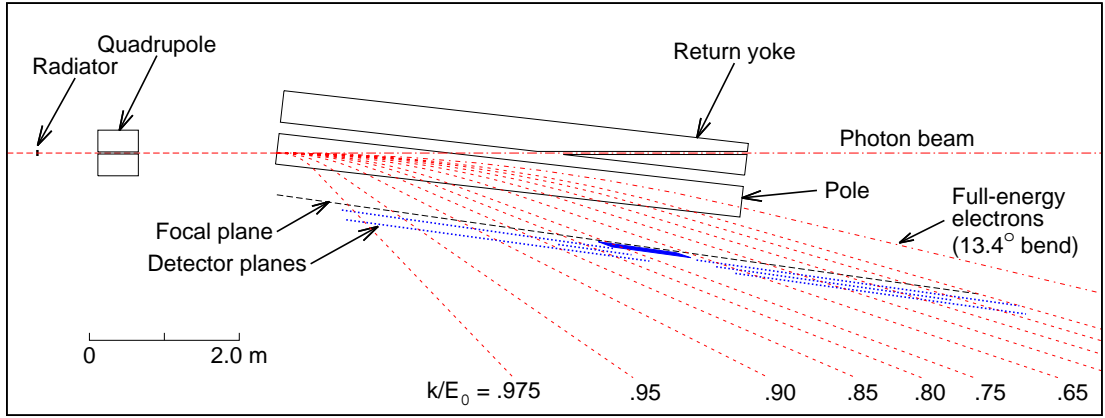}
\caption{Schematic diagram of the tagging spectrometer, showing the paths of the electron
and photon beams. Dotted lines indicate post-radiation electron trajectories identified by
the energy the electron gave up to an associated radiated photon, as a fraction of the beam energy E$_0$.
The Tagger focal plane detector arrays TAGH and TAGM are described in the text.
       \label{fig:beam:BEAM_taggerplot}  }
\end{center}
\end{figure}

\subsection{Photon tagging system \label{sec:tag}}
After passing through the radiator, the combined photon and electron beams enter
the photon tagging spectrometer (Tagger). The full-energy electrons are swept out of
the beamline by a dipole magnet and redirected into a shielded beam dump. The
subset of beam electrons that radiated a significant fraction of their energy in
the radiator are deflected to larger angles by the dipole field. 
These post-brems\-strah\-lung electrons exit through a thin window
along the side of the magnet, and are detected in a highly segmented
array of scintillators called the Tagger Hodoscope, as shown in
Fig.\,\ref{fig:beam:BEAM_taggerplot}. The TAGH counters span
the full range in energy from 25\% to 97\% of the full electron beam energy. A high-energy-resolution device known as the Tagger Microscope (TAGM) covers the
energy range corresponding to the primary coherent peak, indicated by the denser
portion of the focal plane in Fig.\,\ref{fig:beam:BEAM_taggerplot}. 
The quadrupole magnet upstream of the Tagger dipole provides a weak vertical focus, optimizing the efficiency of the Tagger Microscope for tagging collimated photons.
A 0.8~Tm permanent dipole magnet is installed downstream of the Tagger magnet on the photon beam line, in order to prevent the electron beam from reaching Hall D should the Tagger magnet trip.

Both the TAGM and TAGH devices are used to determine the energy of individual
photons in the photon beam via coincidence, using
the relation $E_{\gamma} = E_{0} - E_{e}$, where $E_{0}$ is the primary electron
beam energy before interaction with the radiator, and $E_{e}$ is the
energy of the post-brems\-strah\-lung electron determined by its detected position at the
focal plane. Multiple radiative interactions in a 50 $\mu$m diamond radiator
($3\times 10^{-4}$ radiation lengths) produce uncertainties in
$E_{\gamma}$ of the same order as the intrinsic energy spread of the incident
electron beam.

\subsubsection{Tagger magnet \label{sec:tagMagnet}}
The Hall-D Tagger magnet deflects electrons in the horizontal plane, allowing the
brems\-strah\-lung-produced photons to continue to the experimental hall while
bending the electrons that produced them into the focal plane detectors.
Electrons that lose little or no energy in the
radiator are deflected by 13.4$^\circ$ into the electron beam dump.

The Hall-D Tagger magnet is an Elbek-type room temperature dipole magnet, similar
to the JLab Hall-B tagger magnet \cite{BORGGREEN19631, Sober2000263}. 
The magnet is 1.13~m wide, 1.41~m high and 6.3~m long, weighing 80~metric tons,
with a normal operating field of 1.5~T for a  12-GeV incident electron beam, 
a maximum field of 1.75 T, and a pole gap of 30 mm. 
The magnet
design was optimized using the detailed magnetic field calculation  provided by the
TOSCA simulation package and ray tracing of electron beam trajectories~\cite{DIPOLE_YANG,DIPOLE_SOMOV}.

The \gx{} experiment requirements mandate that the scattered
electron beam be measured with an accuracy of 12~MeV (0.1\% of the incident electron
energy). This requires that the magnetic field integrals along all useful electron
trajectories be known to 0.1\%. The magnetic field was mapped at Jefferson Lab and
the detailed field maps were augmented by detailed TOSCA calculations, which have
allowed us to meet these goals. Details of the magnet mapping and uniformity are
found in Ref.\,\cite{gx4271}.

\subsubsection{Tagger Microscope \label{sec:TAGM}}
The Tagger Microscope (TAGM) is a high-resolution hodoscope that counts post-~\!\!brems\-strah\-lung electrons corresponding to the primary coherent peak.
Normally the TAGM is positioned to cover between 8.2 and 9.2~GeV in photon energy, but the TAGM is designed to be movable should a different peak energy be desired.
The microscope is segmented along the horizontal axis into 102 energy bins (columns) of approximately equal width. Each column is segmented 
in five sections (rows) along the vertical axis. The vertical segmentation allows the possibility of scattered electron
collimation, which gives a significant increase in photon polarization when used in
combination with photon collimation. The purpose of the quadrupole magnet upstream
of the dipole is to provide the vertical focus needed to make the
double-collimation scheme work efficiently. Summed signals are also available for
each column for use in normal operation when electron collimation is not desired.

The Tagger Microscope consists of a two-dimensional array of square scintillating
fibers packed in a dense array of dimensions $102\times 5$. The fibers are multi-clad
BCF-20 with a $2\times 2$ mm$^2$ square transverse profile, manufactured by Saint-Gobain\footnote{Saint-Gobain, https://www.saint-gobain.com/en}.
The cladding varies in thickness from 100 microns near the corners to 70 microns in
the middle of the sides, with an active area of $1.8\times 1.8$~mm$^2$ per fiber.
Variations at the level of 5\% in the transverse size of the fibers impose a practical
lower bound of 2.05~mm on the pitch of the array. The detection efficiency of the TAGM
averages 75\% across its full energy range, in good agreement with the geometric
factor of 77\%.

Each scintillating fiber is 10~mm long, fused at its downstream end to
a clear light guide of matching dimensions (Saint-Gobain BCF-98) that
transmits the scintillation light from the focal plane to a shielded box where
a silicon photomultiplier (SiPM) converts light pulses into electronic signals. The
scintillators are oriented so that the electron trajectories are parallel to the
fiber axis, providing large signals for electrons from the radiator, in contrast
to the omni-directional electromagnetic background in the tagger hall.

\begin{figure}[tbh]
\begin{center}
 \includegraphics[clip=true,width=0.95\linewidth]{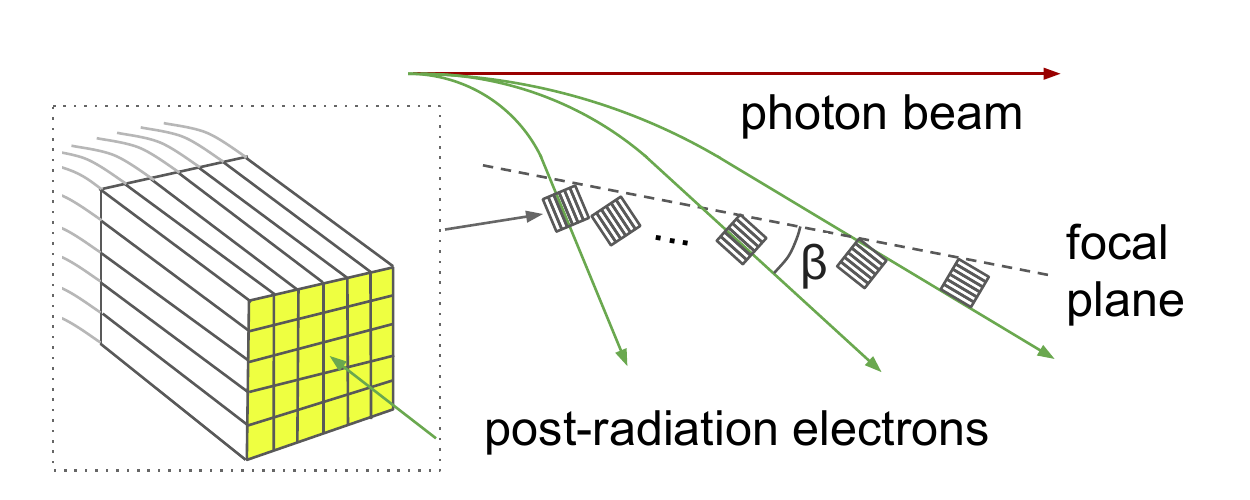}
\end{center}
\caption{
Conceptual overview of the tagger microscope design, showing the fiber bundles and
light guides (left), and the orientation of these bundles aligned with the incoming
electron beam direction in the tagger focal plane (right). The variation of the
crossing angle $\beta$ is exaggerated for the sake of illustration.
        }
\label{fig:TAGM_conceptual}
\end{figure}

Because the electron trajectories do not cross the focal plane at right angles, the
fiber array must be staggered along the dispersion direction. A staggering step
occcurs every 6 columns, as illustrated in Fig.~\ref{fig:TAGM_conceptual}. The slight
variation of the crossing angle $\beta$ is taken into account by a carefully adjusted
fan-out that is implemented by small evenly-distributed gaps at the rear ends of 
adjacent 6-column groups (bundles). A total of 17 such bundles comprise the full
Tagger Microscope.

The far ends of the scintillation light guides are coupled to Hamamatsu S10931-050P SiPMs. The SiPMs are mounted on a custom-built two-stage preamplifier board, with 15 SiPMs per board. In addition to the 15 individual signals generated
by each preamplifier, the boards also produce three analog sum outputs, each the sum
of five adjacent SiPMs corresponding to the five fibers in a single column. All 510
SiPMs are individually biased by custom bias control boards, one for every two
preamplifier boards. The control boards connect to the
preamplifiers over a custom backplane, and communicate with the
experimental slow controls system over ethernet. Each control board has the
capability to electronically select between two gain modes for the preamplifiers
on that board:
a low gain mode used during regular tagging operation, and a high gain
mode used for triggering on single-pixel pulses during bias calibration.
Each bias control board manages the control and biasing for two preamplifiers.
The control board also measures live values for environmental parameters
(voltage levels and temperatures) in the TAGM electronics, so that alarms can
be generated by the experimental control system whenever any of these parameters
stray outside predefined limits.

Pulse height and timing information for 122 channels from the TAGM is provided by analog-to-digital converters (ADCs) and time-to-digital converters (TDCs). These 122
signals include the 102 column sums plus the individual fiber signals from
columns 7, 27, 81, and 97. Here, each channel goes through a 1:1
passive splitter, with one output going to an ADC and the other through
discriminators to a TDC. The ADCs are 250-MHz flash ADCs with 12-bit
resolution and a full-scale pulse amplitude of 1 V. The TDCs are based
on the F1 TDC chip \cite{Fischer:2000zu}, with a least-count of 62~ps. Pulse thresholds in both
the ADC and discriminator modules are programmable over the range 1-1000 mV
on an individual channel basis, covering the full dynamic range of the TAGM
front end. The TAGM preamplifier outputs (before splitting) saturate at around
2~V pulse amplitude.

The mean pulse charge in units of SiPM pixels corresponding to a
single high-energy electron varies from 150 to 300 pC, depending on the fiber,
with an average of 220 pC and standard deviation of 25 pC. During calibration,
this yield is measured individually for each fiber by selectively biasing
the SiPMs on each row of fibers, one row at a time, and reading out the column
sums. Once all 510 individual fiber yields have been measured, the bias voltages
within each column are adjusted to compensate for yield variations, so
that the mean pulse height in a given column is the same regardless of which
fiber in the column detected the electron. The ADC readout and discriminator
thresholds are set individually for each column, for optimum efficiency and
noise rejection.

The ADC firmware provides an approximate time for each pulse, in addition to the
pulse amplitude. During offline reconstruction, this time information is used to
associate ADC and TDC pulse information from the same channel, so that a
time-walk correction can be applied to the TDC time. 
Once this correction
has been applied, a time resolution of 230~ps is achieved for the TAGM.
This resolution is based on data collected at rates on the order of 1~MHz
per column, while the typical rate in the tagger microscope is about 0.5~MHz.
The readout was designed to operate at rates up to 4~MHz per column.
A brief test above 2~MHz per column allowed visual
inspection of the pulse waveforms from the TAGM, without change in the
pulse shape or amplitude. 

\subsubsection{Broadband tagging hodoscope}\label{sec:TAGHIntro}
\begin{figure}[tbp]
\begin{center}
      \includegraphics[width=0.95\linewidth,viewport=80 200 750 400]{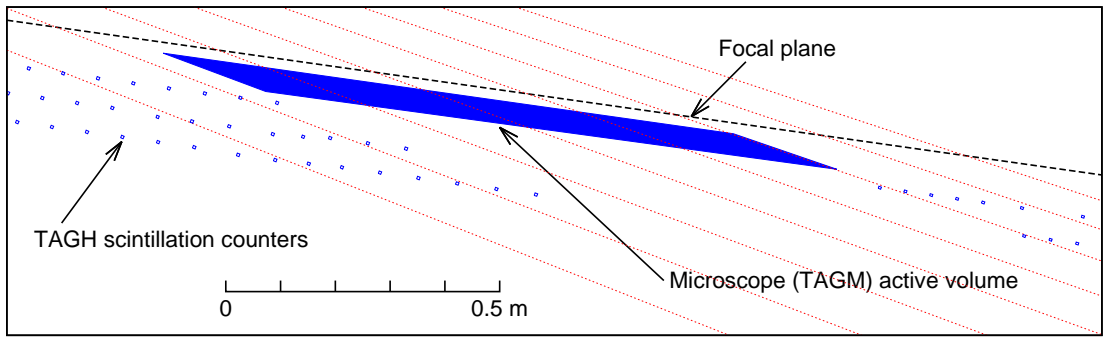}
\caption{Schematic of electron trajectories in the region of the microscope. Shown are the three layers of hodoscope counters on either side of the microscope and the 
               region covered by the microscope.
       \label{fig:beam:BEAM_taggerdetectors}  }

\end{center}
\end{figure}

The Tagger Hodoscope (TAGH) consists of 222 scintillator counters distributed over a length of 9.25~m and mounted just behind the focal plane of the tagger magnet.
The function of this hodoscope is to tag the full range of photon energy from 25\% to 97\%
of the incident electron energy. A gap in the middle of that range is left open for the registration of the primary coherent peak by the Tagger Microscope. The geometry of the counters in the
vicinity of the microscope is shown in Fig.\,\ref{fig:beam:BEAM_taggerdetectors}. 
This broad coverage aids in alignment of the diamond radiator and expands the \GX{} physics program reach to photon energies outside the range of the coherent peak.
The coverage of the hodoscope counters in the region below 60\% drops to half,
with substantial gaps in energy between the counters. This was done because
the events of primary interest to \GX{} come from interactions of photons
within and above the coherent peak; 
within and above the coherent peak the coverage is 100\% up to the 97\% $E_0$ cutoff.

Each counter in the hodoscope is a sheet of EJ-228 scintillator, 6 mm thick and
40 mm high. The counter widths vary along the focal plane, from 21~mm near the
end-point region down to 3~mm at the downstream end. The scintillators are
coupled to a Hamamatsu R9800 photomultiplier tube (PMT) via a cylindrical acrylic (UVT-PMMA) light
guide 22.2 mm in diameter and 120 mm long. Each PMT is wrapped in $\mu$-metal
to shield the tube from the fringe field of the tagger magnet.

Each PMT is instrumented with a custom designed active base~\cite{tagh:base},
consisting of a high-voltage divider and an amplifier powered by current
flowing through the divider. The base provides two signal outputs, one going
to a flash ADC and the other through a discriminator to a TDC.
Operating the amplifier with a gain factor of 8.5 allows the PMT to operate at a
lower voltage of 900~V and reduce the PMT anode current, therefore improving
the rate capability. The energy bite of each counter ranges between 8.5 and
30 MeV for a 12 GeV incident electron beam.  Typical rates during production
running are 1 MHz above the coherent peak and 2 MHz per counter below the
coherent peak. The maximum sustainable rate per counter is about 4~MHz.

The counters are mounted with their faces normal to the path of the
scattered electrons in two or three rows slightly downstream of the focal
plane, as shown in Fig.\,\ref{fig:beam:BEAM_taggerdetectors}.
This allows the counters to be positioned without horizontal gaps in
the dispersion direction, enabling complete coverage of the entire
tagged photon energy range.

The mounting frame of the hodoscope is suspended from the ceiling of the Tagger Hall
to provide full flexibility for positioning TAGH. The frame is constructed
to also support the addition of counters to fill in the energy range currently
occupied by the microscope when the TAGM location is changed.

A similar procedure to that described in Section\,\ref{sec:TAGM} for the TAGM is used to apply
a time-walk correction to the TDC times from the TAGH counters. Once this
time-walk correction
is applied, the time resolution of the TAGH is 200~ps. No significant
degradation of this resolution is expected at the operating rates planned
for Phase 2 running, which are on the order of 2 MHz per counter above
the coherent peak. Under these conditions, the rates in the TAGH counters
below the coherent peak would average around 4~Mhz, which is at the top
of their allowed range. These counters will be turned off when
running at full intensity.  

\subsection{Tungsten keV filter}

\begin{figure}[tbp]
\begin{center}
\includegraphics[width=0.5\linewidth]{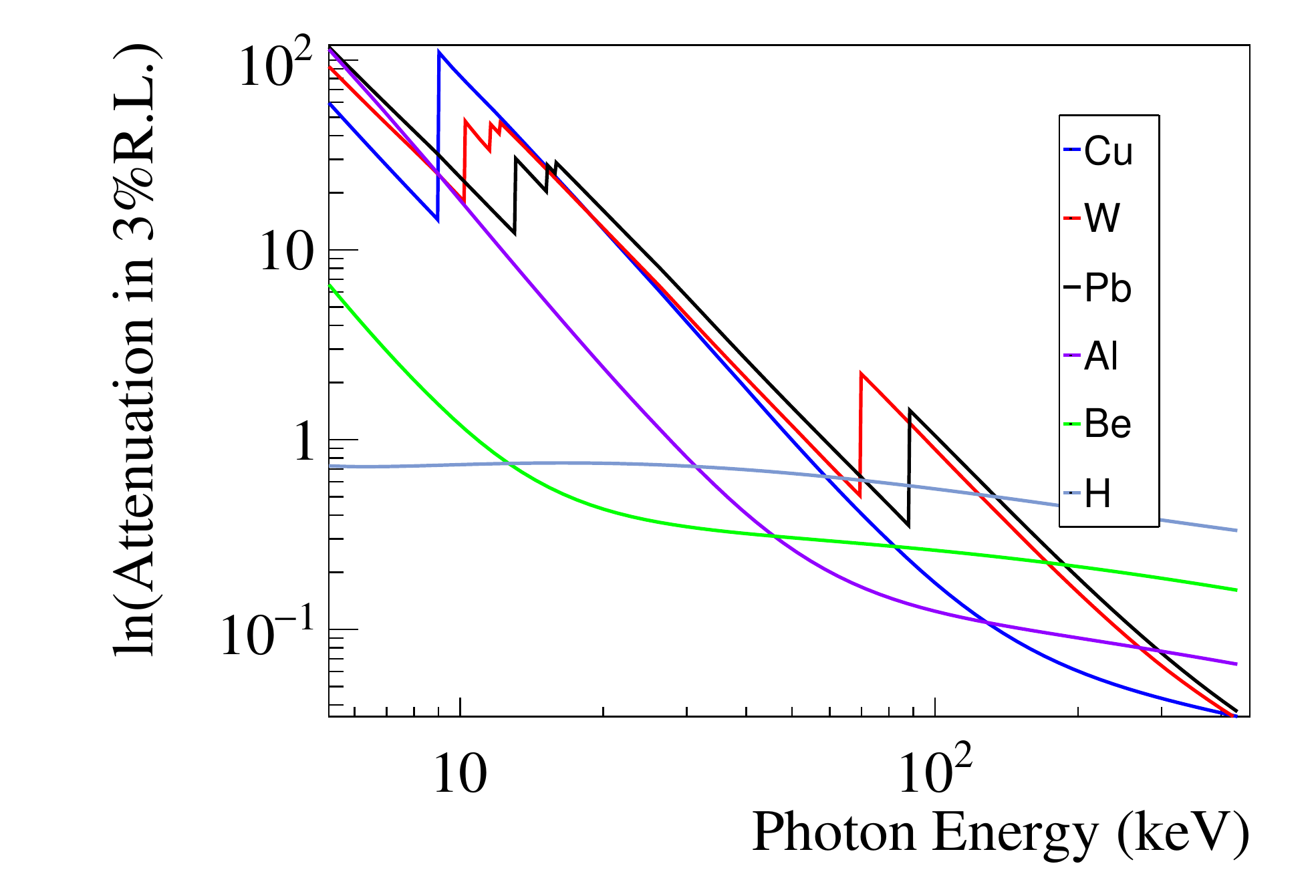}
\caption{Attenuation of low-energy photons in foils with a thickness of  $3\%$ of a radiation length for different materials as a function of photon energy. The W foil was selected to reduce the random background hits in the detector drift chambers. The attenuation coefficients  are taken from Ref.\,\cite{nist_xrays}.
\label{fig:attenuation}  }
\end{center}
\end{figure}

To reduce the photon flux in the $10-100$~keV range, a $100$~$\mu$m tungsten foil ($3\%$ of a radiation length) was installed in the beam line at the entrance of the collimator cave.  We have studied the effect of different foil materials on the anode currents and  random hits in the drift chambers (see Section\,\ref{sec:tracking}), as these factors limit the high-intensity operation of the experiment. By comparing the effect of different materials (Al, Cu, W) with fixed radiation lengths (see Fig.\ref{fig:attenuation})  we learned that the drift chambers are mostly affected by photons in the 70-90 keV range. The analysis of the pulse shape  of the random hits in the CDC confirmed that these photons directly produce hits in the inner layers of the chamber. The insertion of the tungsten foil reduced the number of random hits in the inner CDC layers by a factor of up to 8 and the anode current by $55\%$. The reduction of the current in the FDC was more moderate, about $25\%$. Note that the FDC sense wires are as close as $3$~cm to the beam, while in the CDC the closest wires are at $10$~cm.

\subsection{Beam profiler}
The beam profiler is located immediately upstream of the collimator (see Fig.\,\ref{fig:beam:Draw_beamline}) and is
used to measure the photon beam intensity in a plane normal to the incident
photon beam. The profiler consists of two planes
of scintillating fibers, giving information on the photon beam profile
in the X and Y projections. Each plane consists of 64 square fibers,
2 mm in width, read out by four 16-channel multi-anode PMTs. The beam profiler
is only used during beam setup until the photon beam is centered on the active collimator.

\subsection{Active collimator \label{sec:coll}}
The active collimator monitors the photon beam position and provides
feedback to micro-steering magnets in the electron beamline, for the
purpose of suppressing drifts in photon beam position. 
The
design of the active collimator for \GX{} is based on a device 
developed at SLAC for monitoring
the coherent bremsstrahlung beam there \cite{Miller:1973yi}.
The \GX{} active collimator is located on
the upstream face of the primary collimator, and consists of a dense
array of tungsten pins attached to tungsten base plates. The tungsten
plate intercepts off-axis beam photons before they enter the collimator,
creating an electromagnetic shower that cascades through the array
of pins. High-energy delta rays created by the
shower in the pins (known as ``knock-ons") 
are emitted forward into the primary collimator. The resulting net current between the tungsten plates and the collimator is proportional to the intensity of the photon beam on the plate.
The tungsten plates are mounted on an insulating support, and the plate
currents are monitored by a preamplifier with pA sensitivity. 

The tungsten plate is segmented radially into two rings, and each ring is
segmented azimuthally into four quadrants. The asymmetry of the induced 
currents on the plates in opposite quadrants indicates the degree of
displacement of the photon beam from the intended center position. Typical
currents on the tungsten sectors are at the level of 1.4~nA (inner ring)
and 0.85~nA (outer ring) when running with a 50~$\mu$m diamond crystal
and a 200-nA incident electron beam current. The current-sensitive preamplifiers
used with the active collimator are PMT-5R devices manufactured by
ARI Corporation\footnote{Advanced Research Instruments Corporation, http://aricorp.com.}. The PMT-5R has six remotely selectable gain settings
ranging from $10^{12}$~V/A to $10^6$~V/A, selectable by powers of 10.
This provides an excellent dynamic
range for operation of the beam over a wide range of intensities, from
1~nA up to several $\mu$A. The preamplifier input stage exhibits a fixed
gain-bandwidth product of about 2~Hz-V/pA which limits its bandwidth at
the higher gain settings, for example 2~Hz at $10^{12}$~V/A, 20~Hz at
$10^{11}$~V/A.

In-situ electronic noise on the individual wedge currents is measured to
be 1.5~pA/$\sqrt{\mbox{Hz}}$ on the inner ring, and 15~pA/$\sqrt{\mbox{Hz}}$
on the outer ring. The sensitivity of the current asymmetry to position is
0.160/mm for the inner ring and 0.089/mm for the outer. 
With a 50~micron diamond and 200~nA beam current, operating the active
collimator at a bandwidth of 1~kHz yields a measurement error in the
position of the beam centroid of 150~$\mu$m for the inner ring and
450~$\mu$m for the outer ring.
The purpose of the outer ring is to help locate the beam when the beam location
has shifted more than 2~mm from the collimator axis, where the response
of the inner ring sectors becomes nonlinear.

The maximum deviation allowed for the Hall D photon beam position 
relative to the collimator axis is 200~$\mu$m. The active collimator
readout was designed with kHz bandwidth so that use in a
fast feedback loop would suppress motion of the beam at 60~Hz and harmonics
that might exceed this limit. Experience with the Hall-D beam has shown
that the electron beam feedback system already suppresses this motion
to less than 100~$\mu$m amplitude, so that fast feedback using the active
collimator is not required during normal operation. Instead, the active collimator is
used in a slow feedback loop which locks the photon beam position at
the collimator with a correction time constant of a few seconds. This
slow feedback system
is essential for preventing long-term drifts in the photon beam
position that would otherwise occur on the time scale of hours or
days. The active collimator can achieve 200~$\mu$m position resolution down to beam currents as low as 2~nA when operated in this mode with noise averaging over a 5~s interval.

\subsection{Collimator}
The photon beam produced at the diamond radiator contains both incoherent
and coherent bremsstrahlung components. In the region of the coherent peak,
where photon polarization is at its maximum, the angular spread of coherent
bremsstrahlung photons is less than that of incoherent bremsstrahlung.
The characteristic emission angle for incoherent bremsstrahlung is
$m/E = 43$~$\mu$rad at 12~GeV, whereas the coherent flux within the
primary peak is concentrated below 15~$\mu$rad with respect to the beam
direction. Collimation increases the degree of linear polarization in
the photon beam by suppressing the incoherent component relative to the
coherent part.

The Hall-D primary collimator provides apertures of 3.4 mm and 5.0 mm in a
tungsten block mounted on an X-Y table. The 5.0~mm collimator is used
under normal \GX{} running conditions.
The tungsten collimator is surrounded by lead shielding.
The collimator may also be positioned to block the beam to prevent
high-intensity beam from entering the experimental hall during tuning
of the electron beam. Downstream of the primary collimator, a
sweeping magnet and shield wall, followed by a secondary collimator
with its sweeping magnet and shield wall, suppress charged
particles and photon background around the photon beam that are
generated in the primary collimator. The photon beam exiting the
collimation system then passes through a thin pair conversion target. The resulting $e^+e^-$ pairs are used to continuously monitor the photon beam flux and polarization.

\subsection{Triplet Polarimeter \label{sec:tpol}}
The Triplet Polarimeter (TPOL) is used to measure the degree of polarization
of the linearly-polarized photon beam \cite{DUGGER2017115}.
The polarimeter uses the process of $e^+e^-$ pair production on atomic electrons 
in a beryllium target foil, with the scattered atomic electrons
measured using a silicon strip detector.
Information on the degree of polarization of the photon beam is
obtained by analyzing the azimuthal distribution of the scattered
atomic electrons.

\subsubsection{Determination of photon polarization \label{sec:polarization}}
Triplet photoproduction occurs when the polarized photon beam interacts
with the electric field of an atomic electron within a target material
and produces a high energy $e^+e^-$ pair. When coupled with
trajectory and energy information of the $e^+e^-$ pair, the azimuthal
angular distribution of the recoil electron provides a measure of
the photon beam polarization. The cross section for triplet photoproduction
can be written as $\sigma_t = \sigma_0 [ 1 - P \Sigma \cos(2\varphi)]$
for a polarized photon beam, where $\sigma_0$ is the unpolarized triplet
cross section, $P$ the photon beam polarization, $\Sigma$ the beam
asymmetry for the process, and $\varphi$ the azimuthal angle of the
recoil electron trajectory with respect to the plane of polarization
for the incident photon beam. To determine the photon beam polarization,
the azimuthal distribution of the recoil electrons is recorded and fit
to the function $A [ 1- B \cos(2\varphi)]$  where the variables $A$ and
$B$ are parameters of the fit, with $B = P \Sigma$. The value of
$\Sigma$ depends on the beam photon energy, the
thickness of the converter target, and the geometry of the setup.
The value of $\Sigma$ was determined to be $0.1990 \pm 0.0008$ at 9~GeV for the \GX{} beamline
and a 75~micron Be converter~\cite{DUGGER2017115}.

The TPOL detects the recoil electron arising
from triplet photoproduction. 
This system consists of a converter tray and positioning assembly, which holds
and positions a beryllium foil converter where the triplet photoproduction
takes place.  A silicon strip detector (SSD) detects the recoil electron
from triplet photoproduction, providing energy and azimuthal angle information
for that particle. A vacuum housing, containing the pair production target and
SSD, supplies a vacuum environment minimizing multiple Coulomb scattering
between target and SSD. Preamplier and signal filtering electronics are placed
within a Faraday-cage housing.

The preamplifier enclosure is lined with a layer of copper foil to reduce exterior electromagnetic signal interference.
Signals from the downstream (azimuthal sector) side of the SSD are fed to a charge-sensitive preamplifier located outside the vacuum.
In operation, the TPOL vacuum box is coupled directly to the evacuated
beamline through which the polarized photon beam passes. 

Upon entering TPOL, the photon beam passes into the
beryllium converter, triplet photoproduction takes place, an
$e^+e^-$ pair is emitted from the target in the forward direction,
and a recoil electron ejected from the target at large angles with
respect to the beam is detected by the SSD within the TPOL vacuum chamber.
The recoil electron is ejected at large angles and detected by the SSD.
The $e^+e^-$ pair, together with any beam photons that did not
interact with the converter material, pass through the downstream port of
the TPOL vacuum box into the evacuated beamline, which in turn passes
through a shielding wall into the Hall-D experimental area. 
The $e^+e^-$ pair then enters the vacuum box and magnetic field of the \GX{}
Pair Spectrometer, while photons continue through an evacuated beamline to
the target region of the \gx{} detector. Accounting for all sources of
uncertainty from this setup, the total estimated systematic error in
the TPOL asymmetry $\Sigma$ is 1.5\% \cite{DUGGER2017115}.

\subsection{Pair Spectrometer  \label{sec:ps}}
The main purpose of the Pair Spectrometer (PS) \cite{BARBOSA2015376} 
is to measure the spectrum of the
collimated photon beam and determine the fraction of linearly polarized
photons in the coherent peak energy region. The TPOL relies on the PS
to trigger on pairs in coincidence with hits in the recoil detector.
The PS is also used to monitor the photon beam flux, and for 
energy calibration of the tagging hodoscope and microscope detectors.

The PS, located at the entrance to Hall D, 
reconstructs the energy of a beam photon by detecting
the $e^+e^-$ pair produced by the photon in a thin converter.
The converter used is typically the beryllium target housed within TPOL; 
otherwise the PS has additional converters that may
be inserted into the beam with thicknesses ranging between 0.03\%
and 0.5\% of a radiation length.
The produced $e^+e^-$ leptons are deflected in a modified 18D36 dipole magnet
with an effective field length of about 0.94~m and detected in two
layers of scintillator detectors: a high-granularity hodoscope and
a set of coarse counters, referred to as PS and PSC counters, respectively.
The detectors are partitioned into two identical arms positioned symmetrically on
opposite sides of the photon beam line. The PSC consists of sixteen
scintillator counters, eight in each detector arm. Each PSC counter is
4.4~cm wide and 2~cm thick in the direction along the lepton trajectory
and 6~cm high. Light from the PSC counters is detected using Hamamatsu
R6427-01 PMTs. The PS hodoscope consists of 145 rectangular tiles
(1 mm and 2 mm wide) stacked together. Hamamatsu SiPMs were chosen for
readout of the PS counters
~\cite{Barbosa:2017zzw,Somov:2017kif,Tolstukhin:2014zsa}.

Each detector arm covers an $e^\pm$ momentum range between 3.0 GeV/c
and 6.2 GeV/c,  corresponding to reconstructed photon energies between
6~GeV and 12.4~GeV. The relatively large acceptance of the hodoscope
enables energy determination for photons with energies from below the coherent peak
to the beam endpoint energy near 12~GeV.

The pair energy resolution of the PS hodoscope is about 25 MeV.
The time resolution of the PSC counters is 120~ps, which allows coincidence measurements between the tagging detectors and the PS within an electron beam bunch. Signals from the PS detector are delivered to the trigger system,
as described in Section~\ref{sec:trig}. The typical rate of PS double-arm coincidences
is a few kHz. Details about the performance of the spectrometer are given in~\cite{Somov:2017vhp,Somov:2016bgb}.

\subsubsection{Determination of photon flux         \label{sec:ps_flux}}
The intensity of beam photons incident on the \GX{} target is
important for the extraction of cross sections. The photon flux
is determined by converting a known fraction of the photon beam to
$e^\pm$ pairs and counting them in the PS as     a function
of energy. Data from the PS are  collected using a PS trigger, which
runs in parallel to the  main \GX{} physics trigger, as described in
Section~\ref{sec:trig}. The number of beam photons integrated over
the run period is obtained individually for each tagger counter (TAGH
and TAGM), i.e., for each photon beam energy bin. 

The PS calibration parameter used in the flux determination, a product
of the  converter thickness, acceptance, and the detection efficiency
for leptons, is determined using calibration runs with the Total 
Absorption Counter (TAC)~\cite{somov_flux}. The TAC is a small calorimeter
(see Section~\ref{sec:tac}) inserted directly into the photon
beam immediately upstream of the photon beam dump to count the number of beam photons as a function of energy.
These absolute-flux calibration runs are performed at reduced beam intensities in
order to limit the rate of accidental tagging coincidences.
Data are acquired simultaneously from the PS and TAC.
These data enable an absolute flux calibration for the PS
by measuring the number of reconstructed $e^+e^-$ pairs for a given
number of photons of the same energy seen by the TAC. 
Uncertainties on the photon flux determinations are currently being
investigated. The expected precision of the flux determination is on
the level of $1\%$.

\subsection{Total Absorption Counter \label{sec:tac}}
The TAC is a high-efficiency lead-glass calorimeter, used at low beam currents ($<$ 5nA) to determine the overall normalization of the flux from the \gx{} coherent bremsstrahlung facility.  
This device is intended to count all beam photons above a certain energy threshold, which
have a matching hit in the tagger system.
There would be a very large number of overlapping pulses in the TAC if it is used
with the production photon flux, resulting in low detection efficiency and therefore large
systematic uncertainties. Therefore, the TAC is only inserted into the beam during dedicated runs at very low intensities when the detector can run with near 100\% efficiency.
The TAC was originally developed for and deployed in Hall B, for photon beam operations with CLAS \cite{clasnote1992014, clasnote1993011,clasnote1999002}.

Only a certain fraction of the photons produced at the radiator reach the
target and causes an interaction that is seen in the \gx{} detector.
The count of tagged photons reaching the \gx{} target is
determined as a function of energy from individual TAC coincidence
measurements with each tagging counter. Simultaneous with these
counts, the coincidences between each of the tagging counters and
converted pairs detected in the pair spectrometer are also
recorded. The ratio between the count of tagged pairs and tagged
TAC events thus determined for each tagging counter are used to
convert the tagged rate in the pair spectrometer that is observed
during normal operation into a total count of
tagged photons for each tagging counter that were incident on the
\gx{} target.




\section[Solenoid Magnet]{Solenoid magnet 
  \label{sec:solenoid}
}

\subsection[Overview]{Overview \label{sec:sol:overview}
}

The core of the \gx{} spectrometer is a superconducting
solenoid with a bore diameter and overall yoke length of approximately 2~m and 4.8~m, respectively. The photon beam passes along the axis of
the solenoid.  At the nominal current of 1350~A, the magnet provides a magnetic field along the axis of about 2~T.

The magnet was designed and built at SLAC in the early
1970's~\cite{Alcorn-confer-1972} for the LASS
spectrometer~\cite{Aston:1987uc}. The solenoid employs a cryostatically
stable design with cryostats designed to be opened and
serviced with hand tools. The magnet was refurbished and modified%
\footnote{
  The front plate of the flux return yoke was modified, leading to a
  swap of the two front coils and modifications of the return flux
  yoke in order to keep the magnetic forces on the front coil under
  the design limit.  The original gaps between the yoke's rings were
  filled with iron. The Cryogenic Distribution Box was designed and
  built for \gx{}.
} 
for the \gx{} experiment~\cite{Ballard:2011tm, Ballard:2015wma}. 

The magnet is constructed of four separate superconducting coils and
cryo\-stats. The flux return yoke is made of several iron rings.  The
coils are connected in series. A common liquid helium tank is located
on top of the magnet, providing a gravity feed of the liquid to the
coils. The layout of the coil cryostats and the flux return iron
yoke is shown in Fig.~\ref{fig:layout_spectrometer}.
Table~\ref{tab:sol:summary} summarizes the salient parameters of the  magnet.


\begin{table}[thp]
 \begin{center}
   \small
   \begin{tabular}{lr}
     \hline
     \hline
       Inside diameter of coils        & 2032 mm \\
       Clear bore diameter             & 1854 mm \\
       Overall length along iron       & 4795~mm \\
       Inside iron diameter            & 2946~mm \\
       Outside iron diameter           & 3759~mm \\
       Original yoke, cast and annealed - steel & AISI 1010 \\
       Added filler plates - steel     & ASTM A36 \\
       Full weight                     & 284~t      \\
       Full number of turns            & 4608 \\
       Number of separate coils        & 4 \\
       Turns per coil 2                &  928 \\   
       Turns per coil 1                & 1428 \\   
       Turns per coil 3                &  776 \\   
       Turns per coil 4                & 1476 \\   
       Total conductor weight          & 13.15 t \\
       Coil resistance at $\sim$300~K & 15.3~$\Omega$ \\   
       Coil resistance at  $\sim$10~K & $\sim$0.15~$\Omega$ \\   
       Design operational current      & 1500 A \\
       Nominal current (actual)        & 1350 A \\
       Maximal central field at 1350~A & 2.08 T \\   
       Inductance at 1350~A            & 26.4 H \\
       Stored energy at 1350~A         & 24.1 MJ  \\
       Protection circuit resistor     & 0.061 $\Omega$  \\
       Coil cooling scheme             & helium bath \\   
       Total liquid helium volume      & 3200 $\ell$ \\
       Operating temperature (actual)  &  4.5~K \\   
       Refrigerator liquefaction rate at 0~A        & 1.7~g/s    \\
       Refrigerator liquefaction rate at 1350~A     & 2.7~g/s    \\
     \hline
   \end{tabular}
   \normalsize
 \end{center}
  \caption{
    Key parameters of the \gx{} solenoid. The
    coils are listed in order along the beam direction.
    \label{tab:sol:summary}
  }
\end{table}

\subsection[Conductor and Coils]{Conductor and Coils
 \label{sec:sol:coils}
}

The superconductor composite is made of niobium--titanium filaments
in a copper substrate, twisted and shaped into a
$\sim$7.62$\times$1~mm$^2$ rectangular band. The laminated conductor
is made by soldering the superconductor composite band between two
copper strips
to form a rectangular cross section of 7.62$\times$5.33~mm$^2$.
The measured residual resistivity ratio of the conductor at $\sim{}300$K and
$\sim{}15$K is $\approx{}$100.  

As the coil was wound, a 0.64~mm-thick stainless steel support band
and two 0.2~mm-thick Mylar insulating strips were wound together with it
for pre-tensioning and insulation%
. 
The liquid helium is in contact with the shorter (5.33~mm) sides of
the cable.

Each of the coils consists of a number of subcoils. Each subcoil
contains a number of ``double pancakes'' with the same number of
turns.
Each double pancake is made from a single piece of conductor. The
voltage across the subcoils is monitored using special wires. These pass
through vertical cryostats, called chimneys, along with the helium supply pipes and the
main conductor.

The cold helium vessel containing the coil is supported within the
warm cryostat vacuum vessel by a set of columns designed to provide
sufficient thermal insulation. The columns are equipped with strain
gauges for monitoring the stresses on the columns. The helium vessel
is surrounded by a nitrogen-cooled thermal shield made of copper and stainless-steel panels.
Super-insulation is placed between the vacuum vessel and the nitrogen
shield.  The vacuum vessels are attached to the matching iron rings of
the yoke.

The power supply%
\footnote{Danfysik System 8000 Type 854.}
provides up to 10~V DC for establishing the operating current while ramping. The supply also
includes a protection circuit, which can be engaged by a quench
detector as well as by other signals. During trips, a small dump resistor
of 0.061~$\Omega$ limits the maximum voltage on the magnet to 100~V. The dumping time constant of $L/R \approx 7$~min is
relatively long, but safe according to the original design of the
magnet. A large copper mass and the helium bath are able to absorb a
large amount of energy during a quench without overheating the solder
joints. This permits the use of an ``intelligent''  quench detector with low noise sensitivity and a relatively slow decision time of 0.5~s. The quench detector compares the measured voltages on different subcoils
in order to detect a resistive component.
While ramping the current, such a voltage is proportional
to the subcoil inductance.  Relative values of inductance of various
subcoils depend on the value of the current because of saturation
effects in the iron yoke. Transient effects are also present at changes
of the slew rate caused by Foucault currents in the yoke.
The system includes two redundant detectors: one uses analog signals
and a simplified logic, another is part of the PLC control system (see
Section~\ref{sec:sol:controls}) which uses digitized signals. The PLC
digital programmable device is more sensitive since this monitoring system takes into
account the dependence of the coils' inductance on the current and
provides better noise filtering.  The ramping slew rate is
limited by the transient imbalance of the voltages on subcoils that
may trigger the quench detector. Additionally,
unexplained voltage spikes of 1~ms duration
have been observed in coil 2 at high
slew rates, which can  trigger the quench detector. Powering up the
magnet to 1350~A takes about 8~h.

For diagnostic purposes two 40-turn pickup coils are installed on the bore
surface of the vacuum vessel of each of the coils. 

\subsection[Cooling System]{
   Cooling System
   \label{sec:sol:cryo}
}

The cooling system is described in detail in Ref.~\cite{Lavendure:2014:refrig}.
A stand-alone helium refrigerator located
in a building adjacent to Hall D provides liquid helium and nitrogen
via a transfer line to the Cryogenic Distribution Box above the
magnet. The transfer line delivers helium at 2.6~atm, and 6~K to a Joule-Thomson
(JT) valve providing liquid to a cylindrical common helium tank in the
Distribution Box. The level of liquid helium in the tank is measured
with a superconducting wire probe;%
\footnote{
  American Magnetics Model 1700 with HS-1/4-RGD-19"/46"-4LDCP-LL6-S sensor
}
 the liquid level is kept at about half of the tank diameter. The cold helium gas
 from the tank is returned to the refrigerator, which keeps the
 pressure at the top of the tank at 1.2~atm corresponding to about
 4.35~K at the surface of the liquid.%
\footnote{
  The original implementation at SLAC did not recycle the helium and
  operated at atmospheric pressure.
}
Each coil is connected to the common helium tank by two vertical 2-inch
pipes.  One pipe is open at the bottom of the tank while the other one
is taller than the typical level of helium inside the tank. The main
conductor and the wires for voltage monitoring pass through the former
pipe. Additionally, two $\sim$6~m long, 3/8~inch ID pipes go outside
the coil's helium vessel, from the Distribution Box to the bottom of
the coil. One of those pipes, connected to a JT valve in the box, is used to fill the coil initially, but is not used during operation.
The other pipe reaches the bottom of
the common helium tank in order to provide 
a thermo-syphon effect essential
for the proper circulation of helium in the coil. The main current is
delivered into the helium tank via vapor-cooled leads, and is
distributed to the coils by a superconducting cable. After cooling the
leads, the helium gas is warmed and returned to the refrigeration
system. The gas flow through the leads is regulated based on the
current in the magnet; at 1350~A, the flow is about 0.25~g/s. The
coils and the Distribution Box are equipped with various sensors for
temperature, pressure, voltage, and flow rates.

\subsection[Measurements and Controls]{
         Measurements and Controls
        \label{sec:sol:controls}
}

The control system for the superconducting solenoid, power supply, 
and cryogenic system, is based on Programmable Logic
Controllers (PLC)%
\footnote{
  Allen-Bradley Programmable Logic Controllers
  \url{http://ab.rockwellautomation.com/Programmable-Controllers}.
}%
.
The PLC system digitizes the signals from various sensors, communicates with
other devices, reads out the data into a programmable unit for
analysis, and sends commands to various devices. Additionally, the PLC is
connected to EPICS\footnote{Experimental Physics and 
Industrial Control System, https://epics.anl.gov.} in order to display and archive the data (see
Section \ref{sec:controls}).  The practical sampling limit for the readout of the sensor
is a few Hz, which is too low for
detection of fast voltage spikes on the coils due to motion, shorts,
or other effects. Therefore, the voltage taps from the coils and the pickup coils are read
out by a PXI system\footnote{
  National Instruments, PXI Platform, \url{http://www.ni.com/pxi/}.
}, which provides a sampling rate of about 100~kHz. The
PXI system also reads out several accelerometers attached to the
coils' chimneys, which can detect motion inside the coils. The PXI
CPU performs initial integration and arranges the data in time-wise
rows with a sampling rate of 10~kHz.  The PLC system reads out the
data from the PXI system. Additionally, the PXI data are read out by
an EPICS server at the full 10~kHz sampling rate and are recorded for further analysis.

\subsection[Field calculation and measurement]{
         Field calculation and measurement
        \label{sec:sol:field}
}

The momentum resolution of the \gx{} spectrometer is larger than
1\% and is dominated by multiple scattering and the spatial
resolution of the coordinate detectors.  Thus, a fraction of a percent is
sufficient accuracy for the field determination.  The coils are
axially symmetric, while the flux return yoke is nearly axially
symmetric, apart from the holes for the chimneys. The field was
calculated using a 2-dimensional field calculator {\it
  Poisson/Superfish}%
\footnote{
   Poisson/Superfish developed at LANL,
   \url{https://laacg.lanl.gov/laacg/services/serv_codes.phtml\#ps}.
} 
, assuming axial symmetry.  The model of the magnet included the
fine structure of the subcoils and the geometry of the yoke
iron. Different assumptions about the magnetic properties of the yoke
iron have been used: the {\it Poisson} default AISI 1010 steel, the
measurements of the original yoke iron made at SLAC, and the 1018
steel used for the filler plates. Since the results of the field
calculations differ by less than 0.1\%, the default {\it Poisson} AISI
1010 steel properties were used for the whole yoke iron in the final
field map calculations.

\begin{figure}[!htb]
  \begin{center}
     \includegraphics[angle=0,width=1.0\linewidth]{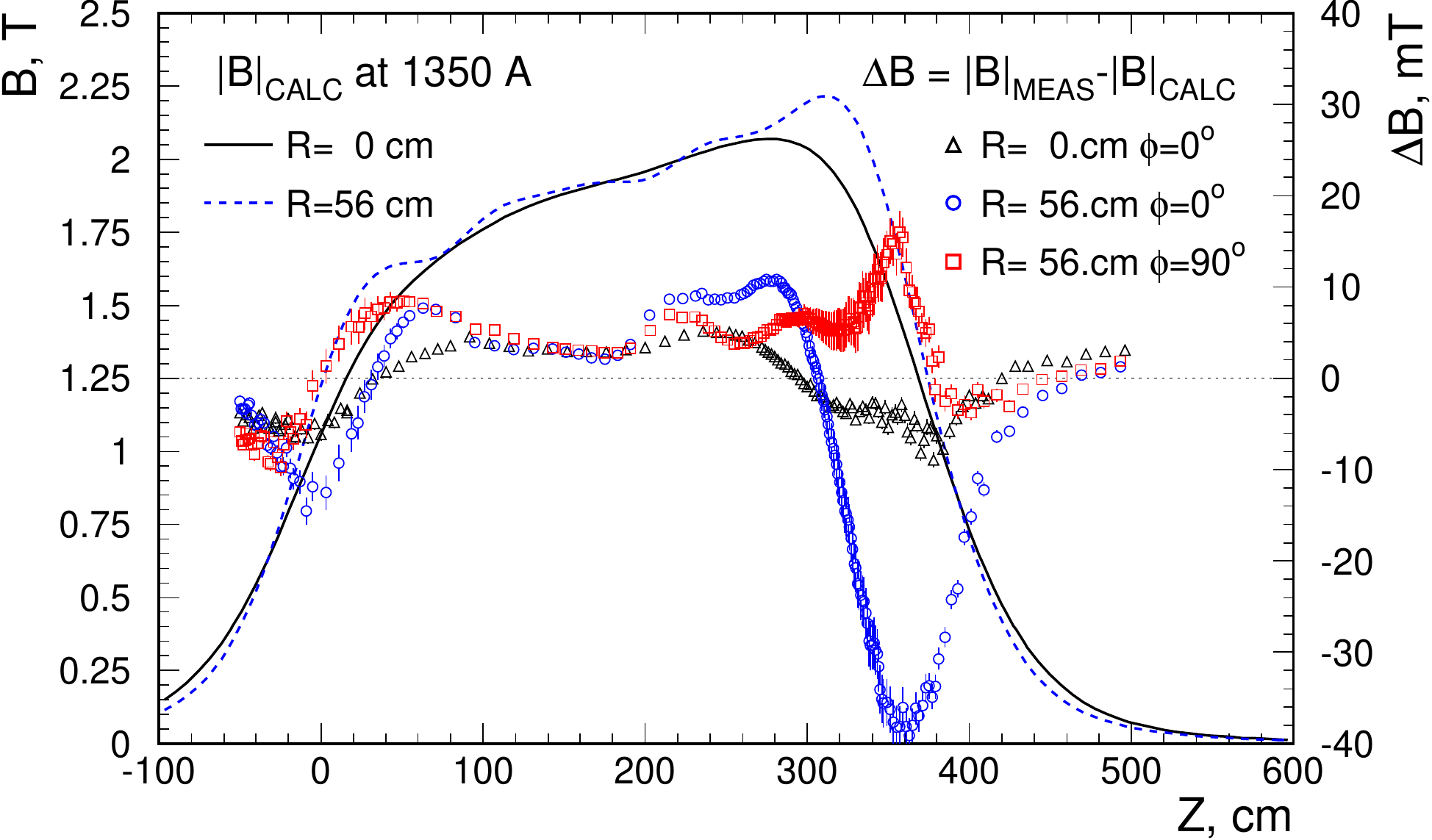}%
  \end{center}
  \caption{
    The full field at 1350~A calculated with {\it Poisson} (left
    scale) on the axis and at the edge of the tracking fiducial volume (R=56~cm). The deviations of the measurements from the
    calculations are shown (right scale) on the axis, and at R=56~cm. The measurements were made at 6 azimuthal
    angles. We show the angles (0$^\circ$ and 90$^\circ$) with the largest deviations from the
    calculations.
    \label{fig:sol:field_comparison}
  }
\end{figure}

The three projections of the magnetic field have been measured along lines
parallel to the axis, at four values of the radius and at up to six values
of the azimuthal angle. The calculated field and the measured deviations are shown in Fig.~\ref{fig:sol:field_comparison}. The
tracking detectors occupy the volume of $R<56$~cm and $45<Z<340$~cm. In
this volume the field deviation at $R=0$ does not exceed 0.2\%. The
largest deviation of 1.5\% is observed at the downstream edge of the
fiducial volume and at the largest radius. Such a field uncertainty in
that region does not noticeably affect the momentum resolution. In
most of the fiducial volume the measured field is axially symmetric to
$\approx$0.1\% and deviates from this symmetry by $\approx$2\% at the
downstream edge and the largest radius.

The calculated field map is used for track reconstruction and
physics analyses.
   

\section[Target]{Target \label{sec:target} }
\begin{figure*}
\begin{center}
\includegraphics[width=4.5in]{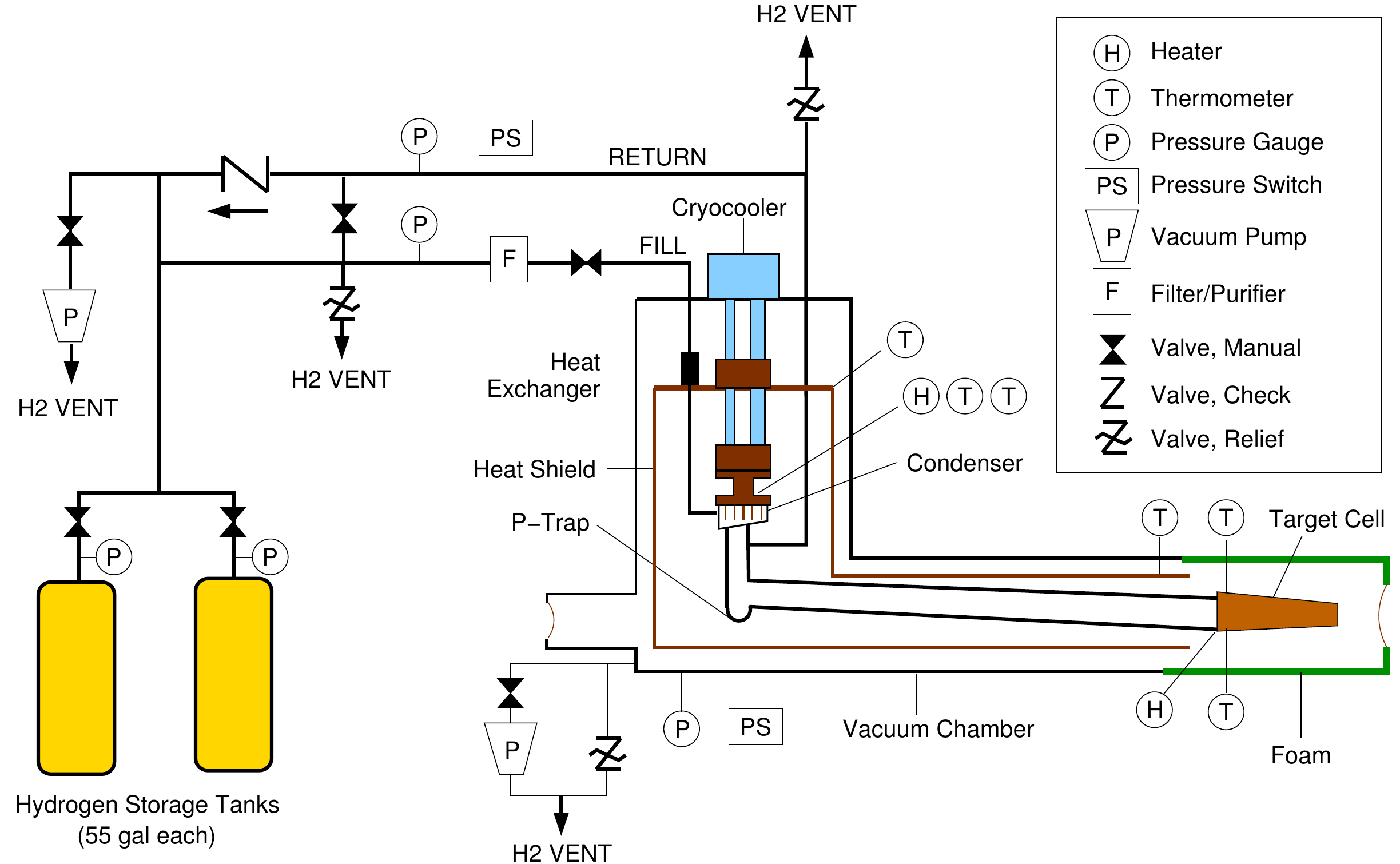}
\end{center}
\caption{Simplified process and instrumentation diagram for the GlueX liquid hydrogen target (not to scale).
In the real system, the P-trap is above the level of the target cell and is used to
promote convective cooling of the target cell from room temperature.}
\label{fig:Target}
\end{figure*}

A schematic diagram of the \gx{} liquid hydrogen cryotarget is shown in Fig.~\ref{fig:Target}. The major components of the system are a pulse tube cryocooler,\footnote{Cryomech model PT415.} a condenser, and a target cell.  These items are contained within an aluminum and stainless steel `L'-shaped vacuum chamber with an extension of closed-cell foam\footnote{Rohacell 110XT, Evonik Industries AG.} surrounding the target cell. In turn, the \gx{} Start Counter (Sec.~\ref{sec:st}) surrounds the foam chamber and is supported by the horizontal portion of the vacuum chamber. Polyimide foils, 100~$\mu$m thick, are used at the upstream and downstream ends of the chamber as beam entrance and exit windows. The entire system, including the control electronics, vacuum pumps, gas-handling system, and tanks for hydrogen storage, is mounted on a small cart that is attached to a set of rails for insertion into the \gx{} solenoid.  To satisfy flammable gas safety requirements, the system is connected at multiple points to a nitrogen-purged ventilation pipe that extends outside Hall D.

Hydrogen gas is stored inside two 200~l tanks and
is cooled and condensed into a small copper and stainless steel container,
the condenser, that is thermally anchored to the second cooling stage of the cryocooler. 
The first stage of the cryocooler is used to
cool the H$_2$ gas to about 50~K before it enters the condenser.
The first stage also cools a copper thermal shield that surrounds all
lower-temperature components of the system except for the
target cell itself, which is wrapped in a few layers of aluminized-mylar/cerex insulation.

The condenser is comprised of a copper C101 base
sealed to a stainless steel can with an indium O-ring.  Numerous vertical 
fins are cut into the copper base, giving a large surface area for condensing hydrogen gas.
A heater and a pair of calibrated Cernox thermometers\footnote{Cernox, Lake Shore Cryotronics.}
are attached outside the condenser, and are used to regulate the heater temperature when the
system is filled with liquid hydrogen.

\begin{figure}
\includegraphics[width=3.5in]{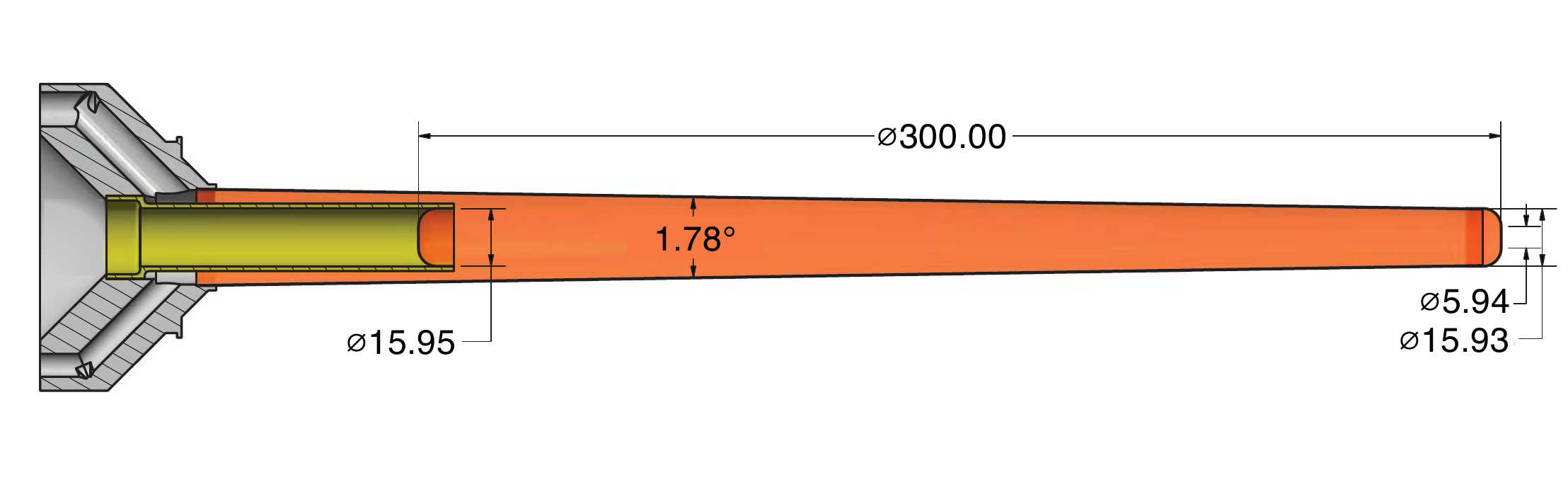}
\caption{Target cell for the liquid hydrogen target.  Dimensions are in mm.  }
\label{fig:TargetCell}
\end{figure}

The target cell, shown in Fig.~\ref{fig:TargetCell}, is similar to
designs used in Hall B at JLab~\cite{HAKOBYAN2008218}.  
The cell walls are made from 100-$\mu$m-thick aluminized
polyimide sheet wrapped in a conical shape and glued along the edge,
overlapping into a 2~mm wide scarf joint.  
The conical shape prevents bubbles from collecting inside the cell, while the
scarf joint reduces the stress riser at the glue joint.  This conical
tube is glued to an aluminum base, 
along with stainless steel fill and return tubes leading to the condenser, a feed-through for two calibrated Cernox thermometers inside the cell, and a
polyamide-imide support for the reentrant upstream beam window.  
Both the upstream and downstream beam
windows are made of non-aluminized,
100~$\mu$m thick polyimide films that have been extruded into the
shapes indicated in Fig.~\ref{fig:TargetCell}. These windows are clearly
visible in Fig.~\ref{fig:z-vertex} where reconstructed vertex positions are shown. All items are glued together using
a two-part epoxy\footnote{3M Scotch-Weld epoxy adhesive DP190 Gray.}
that has been in reliable use at cryogenic temperatures for long periods. 
A second  heater, attached to the aluminum base,
is used to empty the cell for background measurements.
The base is attached to a kinematic mount, which is in turn
supported inside the vacuum chamber using a system of carbon fiber rods.    
The mount is used to correct the pitch and yaw
of the cell, while $X$, $Y$, and $Z$ adjustments 
are accomplished using positioning screws on the target cart.

During normal operation, a sufficient amount of hydrogen gas is condensed from the storage tanks
until the target cell, condenser, and interconnecting piping are filled with liquid hydrogen
and an equilibrium pressure of about 19~psia is achieved.  
The condenser temperature is regulated at 18~K, while the
liquid in the cell cools to about 20.1~K. The latter temperature is 1~K below the saturation
temperature of H$_2$, which eliminates boiling within the cell and permits a more
accurate determination of the fluid density, 
$71.2 \pm 0.3$~mg/cm$^3$.  
The system can be cooled from room temperature and filled with liquid hydrogen in
approximately six hours.  Prior to measurements using an empty target cell, the liquid hydrogen is boiled back into the storage tanks in about five minutes.  H$_2$ gas continues to condense and drain towards the target cell, but the condensed hydrogen is immediately 
evaporated by the cell heater.  In this way, the cell does not warm above 40~K and
can be re-filled with liquid hydrogen in about twenty minutes.

Operation of the cryotarget is highly automated, requires minimal user intervention,
and has operated in a very reliable and predictable manner throughout the
experiment. 
The target controls\footnote{The control logic uses National Instruments CompactRIO 9030.} are handled by a LabVIEW program, 
while a standard EPICS softIOC running in Linux provides a
bridge between the controller and JLab's EPICS enviroment (see Section\,\ref{sec:controls}).     
Temperature readback and control of the condenser and target cell thermometers
are managed by a four-input temperature
controller\footnote{Lake Shore Model 336.} with PID control loops of 50 and 100~W.
Strain gauge pressure sensors measure the fill and return pressures with 0.25\% 
accuracy.  When filled with subcooled liquid, 
the long-term temperature ($\pm 0.2$~K) and pressure ($\pm 0.1$~psi)
stability of the liquid hydrogen enable a determination of the density to better than 0.5\%.

\section{Tracking detectors \label{sec:tracking}}
\subsection[Central drift chamber]{Central drift chamber \label{sec:cdc}}

\begin{figure}[tbp]
\begin{center}
\includegraphics[width=0.7\textwidth]{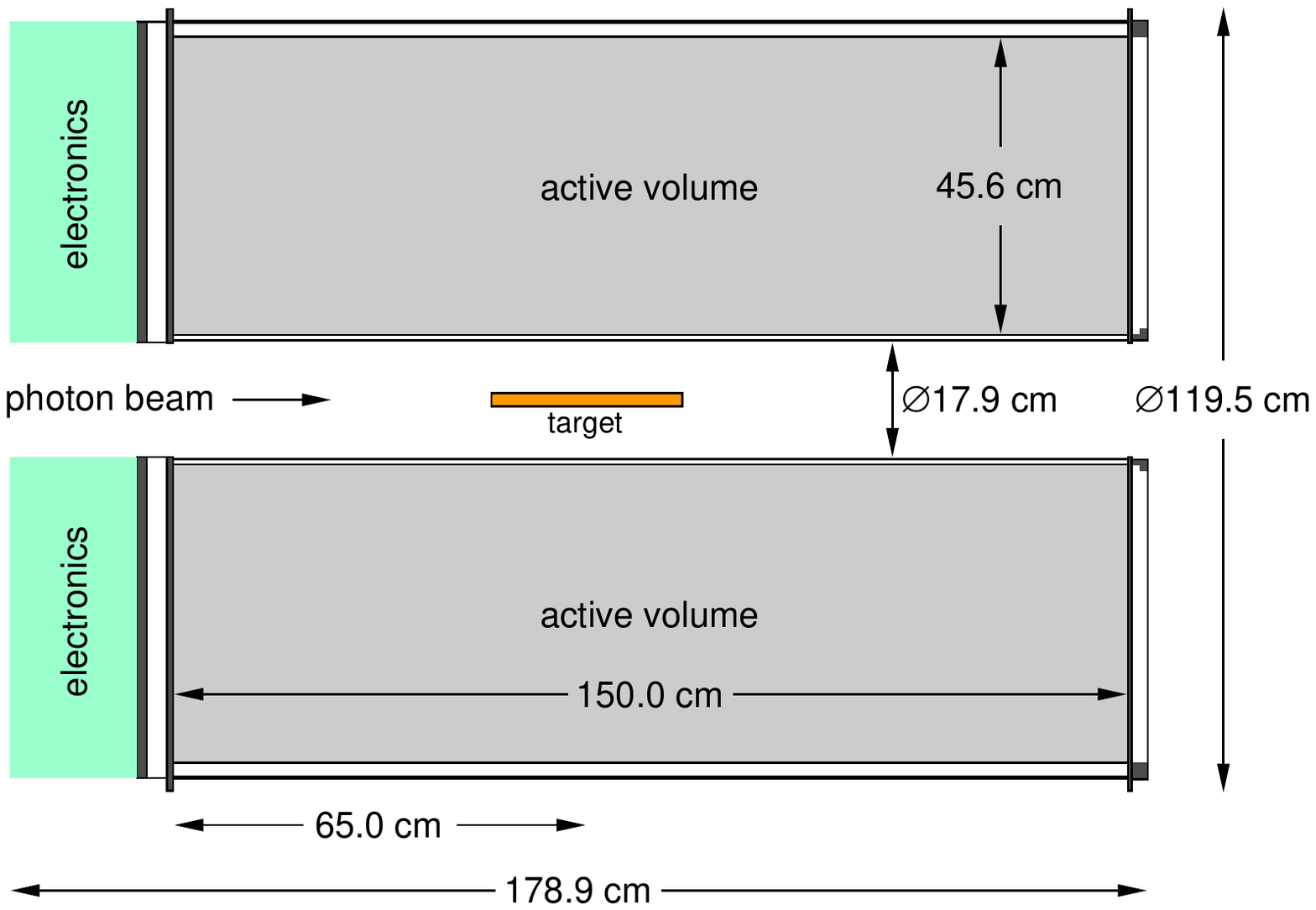}  
\caption{\label{fig:CDC_schematic}          
  Cross-section through the cylindrically symmetric Central Drift Chamber, along the beamline.}  
\end{center}
\end{figure}

The Central Drift Chamber (CDC) is a cylindrical straw-tube drift chamber which is used to track charged particles by providing position, timing and energy loss measurements~\cite{VanHaarlem:2010yq,GlueXCDCNIM}.
The CDC is situated inside the Barrel Calorimeter, surrounding the target and Start Counter. 
The active volume of the CDC is traversed
by particles coming from the hydrogen target with polar angles between $6^{\circ}$ and $168^{\circ}$, with optimum 
coverage for polar angles between $29^{\circ}$ and $132^{\circ}$.  
The CDC contains 3522 anode wires of 20~$\mu$m diameter gold-plated tungsten inside Mylar\footnote{www.mylar.com} straw tubes of diameter 1.6~cm in $28$ layers,
located in a cylindrical volume which is 1.5~m long, with an inner radius of 10~cm and outer radius of 56~cm, as measured from the beamline.  
Readout is from the upstream end. 
Fig.\,\ref{fig:CDC_schematic} shows a schematic diagram of the detector.

\begin{figure}[tbp]
\begin{center}
\includegraphics[width=0.7\textwidth]{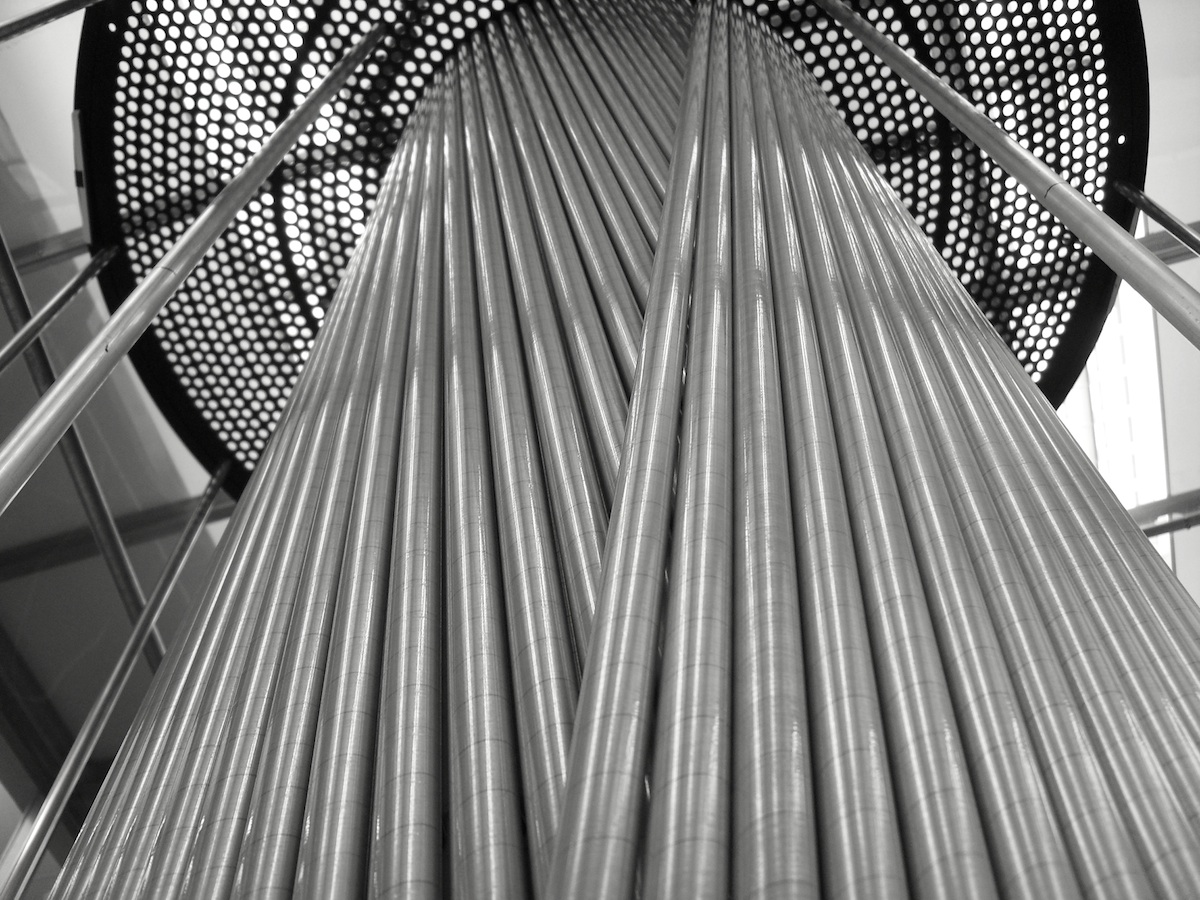}  
\caption{\label{fig:CDC_stereotubes}          
  The Central Drift Chamber during construction. A partially completed layer of stereo straw tubes is shown, surrounding a layer of straw tubes at the opposite stereo angle. Part of the carbon fiber endplate, two temporary tension rods and some of the 12 permanent support rods linking the two endplates can also be seen.}  
\end{center}
\end{figure}

The straw tubes are arranged in 28 layers; 12 layers are axial, and 16 layers are at stereo angles of $\pm 6^{\circ}$ to provide position information along the beam direction.
The stereo angle was chosen to balance the extra tracking information provided by the unique combination of stereo and axial straws along a trajectory against the size of the unused volume inside the chamber at each transition between stereo and axial layers. 
Fig.\,\ref{fig:CDC_stereotubes} shows the CDC during construction. 

The volume surrounding the straws is enclosed by an inner cylindrical wall of 0.5\,mm G10 fiberglass, an outer cylindrical wall of 1.6\,mm aluminum, and two circular endplates. 
The upstream endplate is made of aluminum, while the downstream endplate is made of carbon fiber. The endplates are connected by 12 aluminum support rods. 
Holes milled through the endplates support the ends of the straw tubes, which were glued into place using several small components per tube, described more fully in~\cite{GlueXCDCNIM}.  
These components also support the anode wires, which were installed with 30~g tension.
At the upstream end, these components are made of aluminum and were glued in place using conductive epoxy\footnote{TIGA 920-H, www.loctite.com}. 
This attachment method provides a good electrical connection to the inside walls of the straw tubes, which are coated in aluminum.
The components at the downstream end are made of Noryl plastic\footnote{www.sabic.com} and were glued in place using conventional non-conductive epoxy\footnote{3M Scotch-Weld DP460NS, www.3m.com}.
The materials used for the downstream end were chosen to be as lightweight as feasible so as to minimize the energy loss of charged particles passing through them. 

At each end of the chamber, a cylindrical gas plenum is located outside the endplate.  
The gas supply runs in 12 tubes through the volume surrounding the straws into the downstream plenum. 
There the gas enters the straws and flows through them into the upstream plenum. From the upstream plenum the gas flows into the volume surrounding the straws, and from there the gas exhausts to the outside, bubbling through small jars of mineral oil.
The gas mixture used is 50$\%$ argon and 50$\%$ carbon dioxide at atmospheric pressure. 
This gas mixture was chosen since its drift time characteristics provide good position resolution~\cite{VanHaarlem:2010yq}.
A small admixture (approximately 1$\%$) of isopropanol is used to prevent loss of performance due to aging\cite{KADYK1991436,VAVRA20031}. 
Five thermocouples are located in each plenum and used to monitor the temperature of the gas.
The downstream plenum is 2.54~cm deep, with a sidewall of ROHACELL\footnote{www.rohacell.com} and a final outer wall of aluminized Mylar film, and the upstream plenum is 3.18~cm deep, with a polycarbonate sidewall and a polycarbonate disc outer wall. 

The readout cables pass through the polycarbonate disc and the upstream plenum to reach the anode wires. 
The cables are connected in groups of 20 to 24 to transition boards mounted onto the polycarbonate disc; the disc also supports the connectors for the high-voltage boards. 
Preamplifiers~\cite{hdnote2515} are mounted on the high-voltage boards. The aluminum endplate, outer cylindrical wall of the chamber, aluminum components connecting the straws to the aluminum endplate and the inside walls of the straws are all connected to a common electrical ground. 
The anode wires are held at +2.1~kV during normal operation.

\subsection[Forward Drift Chamber]{Forward Drift Chamber
\label{sec:fdc} }

The Forward Drift Chamber (FDC) consists of 24 disc-shaped planar drift chambers of 1~m diameter \cite{FDC_NIM}.
They are grouped into four packages inside the bore of the spectrometer magnet.
Forward tracking requires good multi-track separation due to the
high particle density in the forward region.
This is achieved via additional cathode strips on both sides of the wire plane allowing for a  reconstruction of a space point on the track from each chamber. 
The FDC registers particles emitted into polar angles as low as $1^\circ$ and up to $10^\circ $
with all the chambers, while having partial coverage up to $20^\circ$.

\begin{figure}[tbp]
\begin{center}
\includegraphics[width=0.95\textwidth]{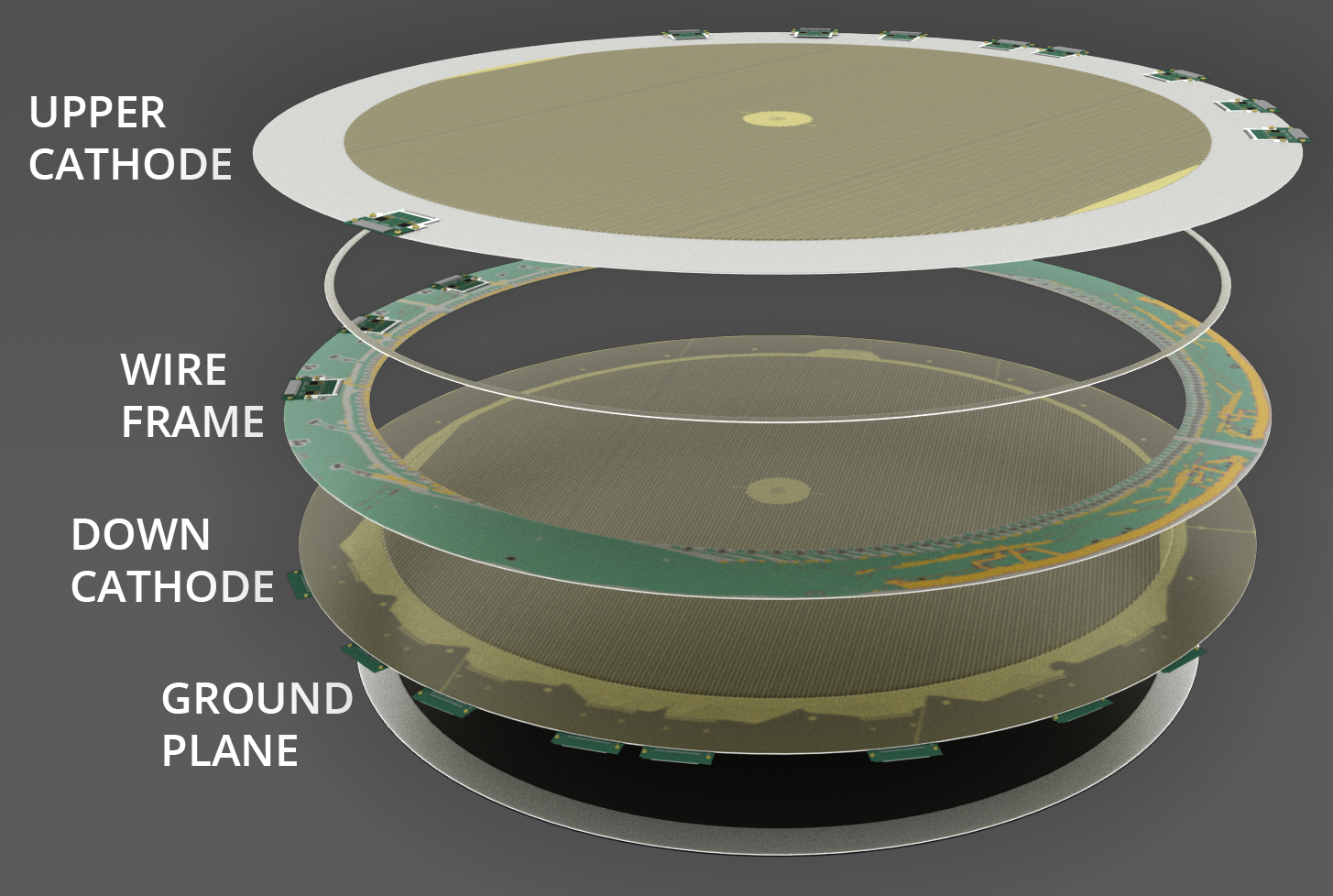} 
\caption{\label{FDC_OneCell}
Artist rendering of one FDC chamber showing components. From top to bottom: upstream cathode, wire frame, downstream cathode, ground plane that separates the chambers. The diameter of the active area is $1$~m.
}
\end{center}
\end{figure}

One FDC chamber consists of a wire plane with cathode planes on either sides at a distance of $5$~mm from the wires (Fig.~\ref{FDC_OneCell}).
The frame that holds the wires is made out of ROHACELL with a thin
G10 fiberglass skin in order to minimize the material and
allow low energy photons to be detected in the outer electromagnetic calorimeters.

The wire plane has sense ($20~\mu$m diameter) and field ($80$~$\mu$m) wires $5$~mm apart, forming a field cell of $10\times 10$~mm$^2$. 
To reduce the effects of the magnetic field, 
a ``slow" gas mixture of $40\%$~Ar and $60\%$~CO$_2$ is used.
A positive high voltage of about $2.2$~kV is applied to the sense wires and a negative high voltage of $0.5$~kV to the field wires. 
The cathodes are made out of $2$-$\mu$m-thin copper strips on Kapton foil with a pitch of $5$~mm, and are held at ground potential. The strips on the two cathodes are arranged at $30^\circ $ relative to each other and at angles of $75^\circ $ and $105^\circ $ angle with respect to the wires.

The six chambers of a package are separated by thin aluminized Mylar.
Each chamber is rotated relative to the previous one by $60^\circ $.
The total material of a package in the sensitive area corresponds to $0.43\%$ radiation lengths, with about half of that in the area along the beam line that has no copper on the cathodes.
The sense wires in the inner area of $6-7.8$~cm diameter (depending on the distance of the package to the target) are increased in thickness from $20$~$\mu$m to $\sim 80$~$\mu$m, which makes them insensitive to the high rates along the beam.
The distance between the first and last package is $1.69$~m. 
All chambers are supplied with gas in parallel. 
In total, $2,304$ wires and $10,368$ strips are read using charge preamplifiers with $10$~ns peaking time, with a gain of $0.77$~mV/fC for the wires and $2.6$~mV/fC for the strips.

\subsection{Electronics \label{sec:dcelectronics}}
The high voltage (HV) supply units used are CAEN A1550P\footnote{www.caen.it}, with noise-reducing filter modules added to each crate chassis. 
The low voltage (LV) supplies are Wiener MPOD MPV8008\footnote{www.wiener-d.com}. 
The preamplifiers are a custom JLab design based on an ASIC~\cite{hdnote2515}
with 24 channels per board; the preamplifiers are charge-sensitive, capacitively coupled to the wires in the CDC and FDC, and directly coupled to strips in the FDC. 

Pulse information from the CDC anode wires and FDC cathode strips are obtained and read out using 72-channel 125 MHz flash ADCs (FADCs) \cite{Visser2008,5873864}. These use Xilinx\footnote{www.xilinx.com} Spartan-6 FPGAs (XC6SLX25) for signal digitization and data processing with 12 bit resolution.
Each FADC receives signals from three preamplifiers. 
The signal cables from different regions of the drift chambers are distributed between the FADCs in order to share out the processing load as evenly as possible.  

The FADC firmware is activated by a signal from the \gx{} trigger. The firmware then computes the following quantities for pulses observed above a given threshold within a given time window: pulse number, arrival time, pulse height, pulse integral, pedestal level preceding the pulse, and a quality factor indicating the accuracy of the computed arrival time. 
Signal filtering and interpolation are used to obtain the arrival time to the nearest 0.8~ns. 
The firmware performs these calculations both for the CDC and FDC alike, and uses different readout modes to provide the data with the precision required by the separate detectors. 
For example, the CDC electronics read out only one pulse but require both pulse height and integral, while the FDC electronics read out up to four pulses and do not require a pulse integral.  

The FDC anode wires are read out using the JLab pipeline F1 TDC\cite{hdnote1021} with a nominal least count of 120~ps. 

\subsection[Gas system]{Gas system \label{sec:gas}}
Both the CDC and FDC operate with the same gases, argon and CO$_{2}$. Since the relative mixture of
the two gases is slightly different for the two tracking chambers, the gas system has two separate but identical mixing stations. There is one gas supply of argon and CO$_{2}$ for both mixing stations. A limiting opening in the supply
lines provides over-pressure protection to the gas system, and filters in the gas lines provide protection against potential
pollution of the gas from the supply. Both gases are mixed using mass flow controllers (MFCs) that can be 
configured
to provide the desired mixing ratio of argon and CO$_{2}$.  MFCs and control electronics from
BROOKS Instruments\footnote{BROOKS Instruments, https://www.brooksinstrument.com/en/products/mass-flow-controllers.} are used throughout.

The mixed gas is filled into storage tanks, with one tank for the CDC and another for the FDC. The pressures are
regulated by controlling the operation of the MFCs with a logic circuit based on an Allen-Bradley ControlLogix system\footnote{Allen-Bradley, https://ab.rockwellautomation.com/}
 that keeps
the pressure in the tank between 10 and 12~psi. The tank serves both as a reservoir and a buffer.
A safety relief valve on each tank
provides additional protection against over-pressure. While the input pressure to the MFC is at 40~psi, the pressure after
the MFC is designed to always be less than 14~psi above atmospheric pressure. After the mixing tank, a provision is
built into the system to allow the gas to pass through an alcohol bath to add a small amount of alcohol gas to the gas mixture.
This small admixture of alcohol protects the wire chambers from aging effects caused by radiation exposure from the beam.
This part of the gas system is located above ground in a separate gas shed, before the gas mixture is transported
to the experimental hall via polyethylene pipes.

Additional MFCs in the hall allow the exact amount of gas provided to the chambers to be specified: one MFC for the CDC and another 
four MFCs for the individual FDC packages. The CDC is operated with a flow of 1.0~l/m, while each FDC package is operated with
a flow of 0.1~l/m. To protect the chambers from over-pressure, there is a bypass line at the input to the detectors that
is open to the atmosphere following a bubbler containing mineral oil. The height of the oil level determines the maximum possible gas pressure at
the input to the chambers. There is a second bubbler at the output to protect against possible air back-flow into
the chamber. The height of the oil above the exhaust line determines the operating pressure inside the chambers.

Valves are mounted at many locations in the gas system to monitor various pressures with a single pressure sensor. The pressures of all six FDC chambers are monitored, as well as the CDC gas at the input, downstream gas plenum and the exhaust. 
A valve in the exhaust line can be used to divert some gas from the chamber to an oxygen sensor. Trace quantities of oxygen will reduce the gas gain and reduce tracking efficiency. The oxygen levels in the chamber are below 100~ppm.

\subsection{Calibration, performance and monitoring \label{sec:dccalib}}
Time calibrations for the drift chambers are used to remove the time offset due to the electronics, so that after calibration the earliest possible arrival time of the pulse signals is at 0~ns. These offsets and the function parameters used to describe the relationship between the pulse arrival time and the closest distance between the track and the anode wire are obtained for each session of data taking. 

The CDC measures the energy loss, $dE/dx$, of tracks over a wide range of polar angles, including recoiling target protons as well as more forward-going tracks. Gain calibrations are made to ensure that $dE/dx$ is consistent between tracking paths through different straws and stable over time. 
The procedure entails matching the position of the minimum ionizing peak for each of the 3522 straws, and then matching the $dE/dx$ at 1.5~GeV/c to the calculated value of 2.0~ keV/cm. This takes place during the early stages of data analysis. Gain calibration for the individual wires is performed each time the HV is switched on and whenever any electronics modules are replaced. Gain calibration for the chamber as a whole is performed for each session of data taking; these sessions are limited to two hours as the gain is very sensitive to the atmospheric pressure. Position calibrations were necessary to describe the small deflection of the straw tubes midway along their length; these were performed in 2016 and repeated in 2017, with no significant difference found between the two sets of results.  Position resolution from the CDC is of the order of 130~$\mu$m and its detection efficiency per straw is over 98\% for tracks up to 4~mm from the CDC wire. The efficiency decreases as the distance between the track and the wire increases, but the close-packing arrangement of the straw tubes and the large number of straws traversed by each track compensate for this. 

\begin{figure}[tbp]
\begin{center}
\includegraphics[width=0.95\textwidth]{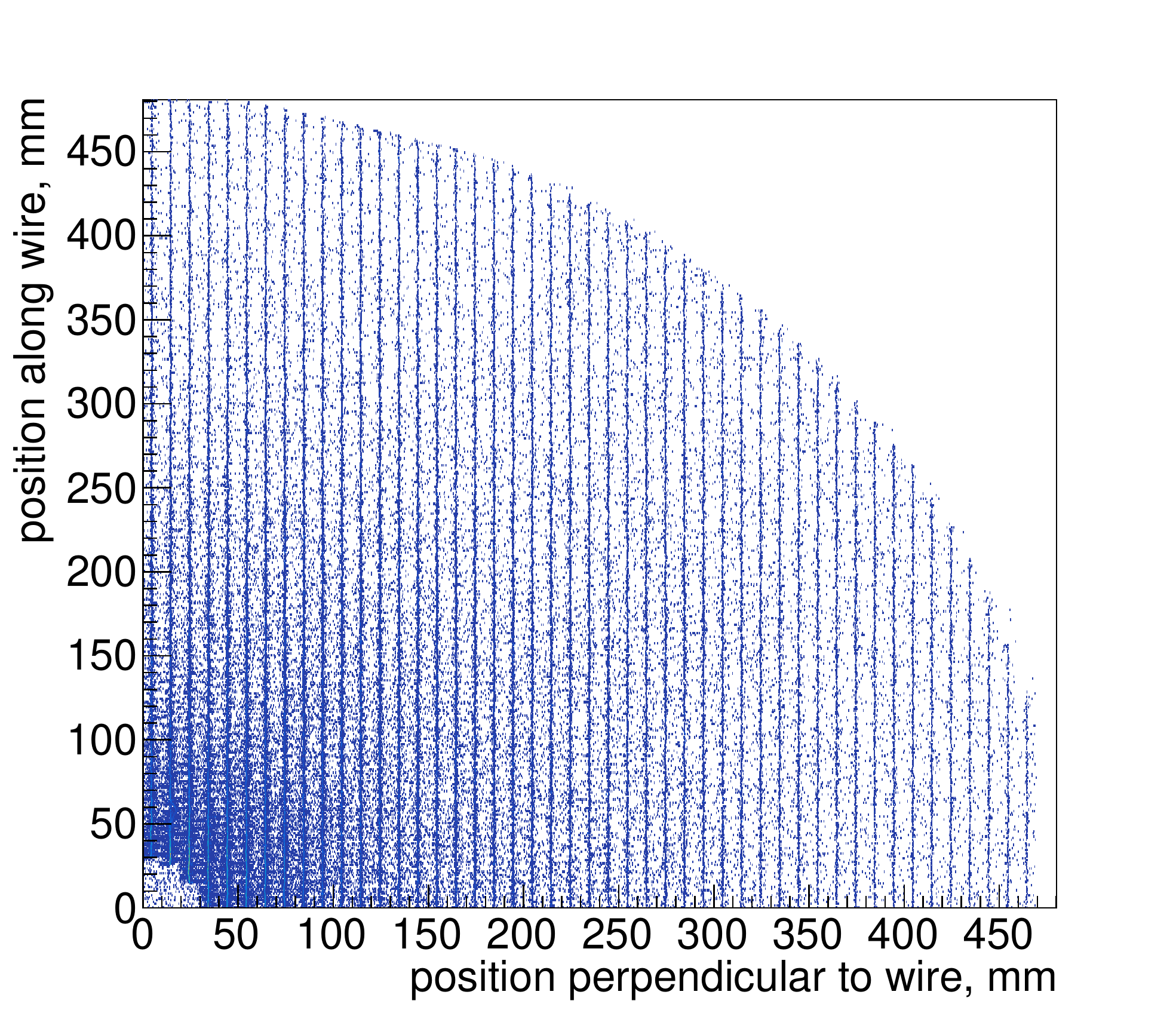}  
\caption{\label{FDC_wires_from_strips} Wire (avalanche) positions reconstructed from the strip information on the two cathodes in one FDC chamber. Only one quarter of the chamber is shown in this figure.
}   
\end{center}  
\end{figure}

For the FDC system, an internal per-chamber calibration process is first performed to optimize the track position accuracy.  
In the FDC the avalanche created around the wire is seen in three projections: on the two cathodes and on the wires.
The drift time information from the wires is used to reconstruct the hit position perpendicular to the wire.
The strip charges from the two cathodes are used to reconstruct the avalanche position along the wire. 
The same strip information can be used to reconstruct the avalanche position perpendicular to the wire,  which, due to the proximity of the avalanche to the wire, is practically the wire position, as illustrated in Fig~\ref{FDC_wires_from_strips}.
This strip information is used to align the strips on the two cathodes with respect to the wires. 
At the same time, the residuals of the reconstructed wire positions are an estimate of the strip resolution.
The resolutions of the detector were reported earlier \cite{FDC_NIM}. 
The strip resolution along the wires, estimated from the wire position reconstruction, varies between $180$ and $80$~$\mu$m, depending on the total charge induced on the strips. The drift distance is reconstructed from the drift time with a resolution between $240$ and $140$~$\mu$m
depending on the distance of the hit to the wire in the $0.5-4.5$~mm range.  

Position offsets and package rotations were determined for both drift chamber systems, first independently, and then together, using the alignment software MILLEPEDE\cite{millepede} in a process described in \cite{GlueXCDCNIM} and in \cite{MikeStaib_thesis}.

Online monitoring software enables shift-takers to check that the number of channels recording data, the distribution of signal arrival times, and the  $dE/dx$ distribution are as expected.


\section[Performance of the charged-particle-tracking system]{Performance of the charged-particle-tracking system \label{sec:trackingperformance}}
\subsection{Track reconstruction}

The first stage in track reconstruction is pattern recognition.  Hits in adjacent
 layers in the FDC in each package are formed into track segments that are 
linked together with other segments in other packages to form FDC track 
candidates using a helical model for the track parameters.
Hits in adjacent rings in the axial layers of the CDC are also associated into 
segments that are linked together with other segments in other axial layers
and fitted with circles in the projection perpendicular to the beam line. Intersections between these circles and the stereo wires are found and a linear fit is performed to find a $z-$position near the beamline and the tangent to the dip
 angle $\lambda=\pi/2-\theta$.  These parameters, in addition to the circle fit 
parameters, form a CDC track candidate for each set of linked axial and stereo 
layers. Candidates that emerge from the target, and pass through both FDC and CDC in the  $5^\circ-20^\circ$ range, are linked together.

The second stage uses a Kalman filter \cite{KalmanFilter, KalmanFilter2} to find the fitted track parameters
\{z,D,$\phi$,$\tan\lambda$,$q/p_T$\}
at the position of closest approach of the track to the beam line. The track candidate parameters are used as an initial guess, where D is the signed distance of closest approach to the beam line.  The Kalman filter proceeds in steps from the hits farthest from the beam line toward the beam line. Energy loss and multiple scattering are taken into account at each step along the way, according to a map of the magnetic field within the bore of the solenoid magnet.
For the initial pass of the filter, the drift time information from the 
wires is not used.  Each particle is assumed to be a pion, except for low momentum track 
candidates ($p<0.8$~GeV/$c$), for which the fits are performed with a proton hypothesis.

The third stage matches each fitted track from the second stage to either
the Start Counter, the Time-of-Flight scintillators, the Barrel Calorimeter, or
the Forward Calorimeter to determine a start time t0 so that the drift time to
each wire associated with the track could be used in the fit. Each track is refitted with
the drift information, separately for each value of mass for particles in the set \{$e^\pm,\pi^\pm,K^\pm,p^\pm$\}.

\subsection{Momentum and vertex resolution}

\begin{figure}[tbp]
\begin{center}
\includegraphics[width=0.45\textwidth]{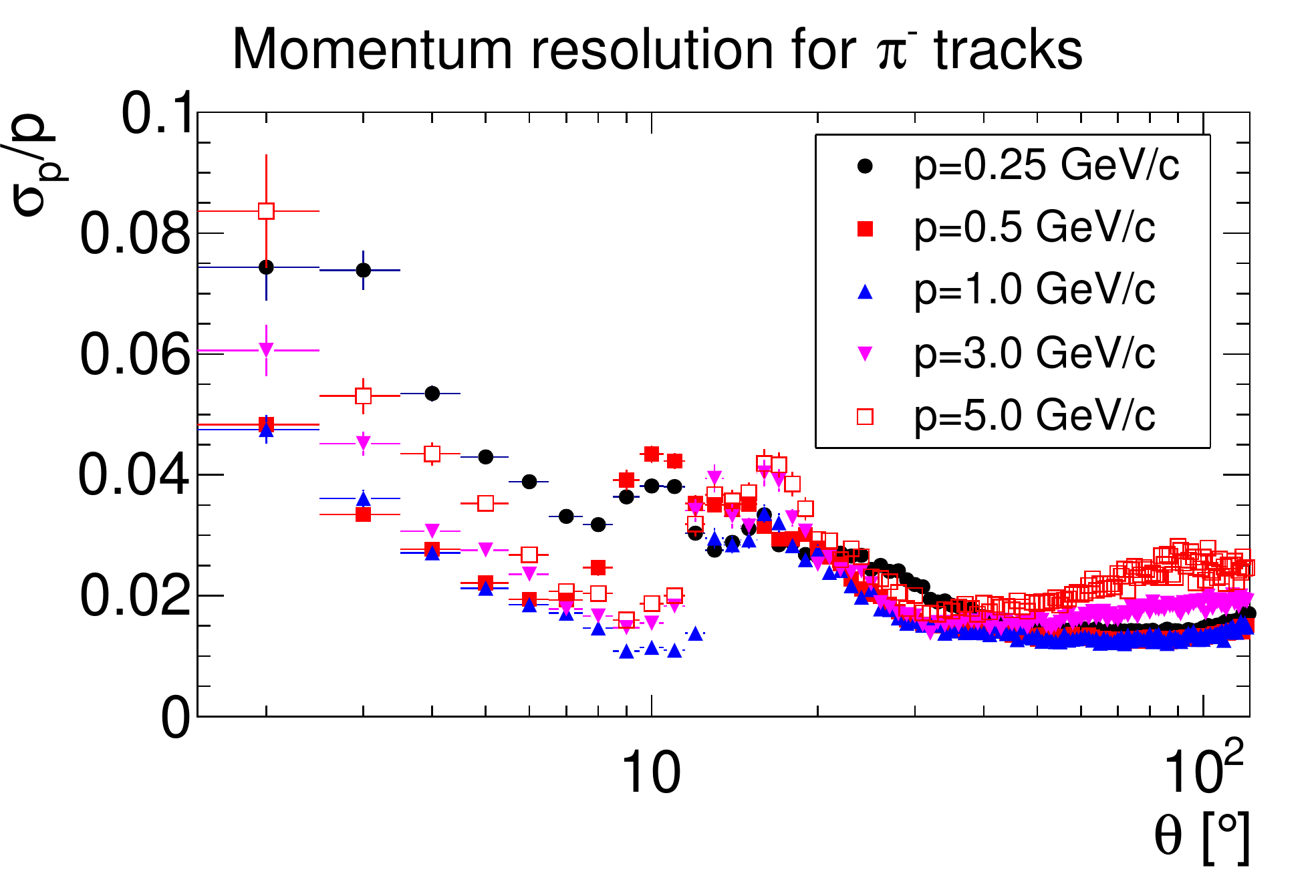}
\includegraphics[width=0.45\textwidth]{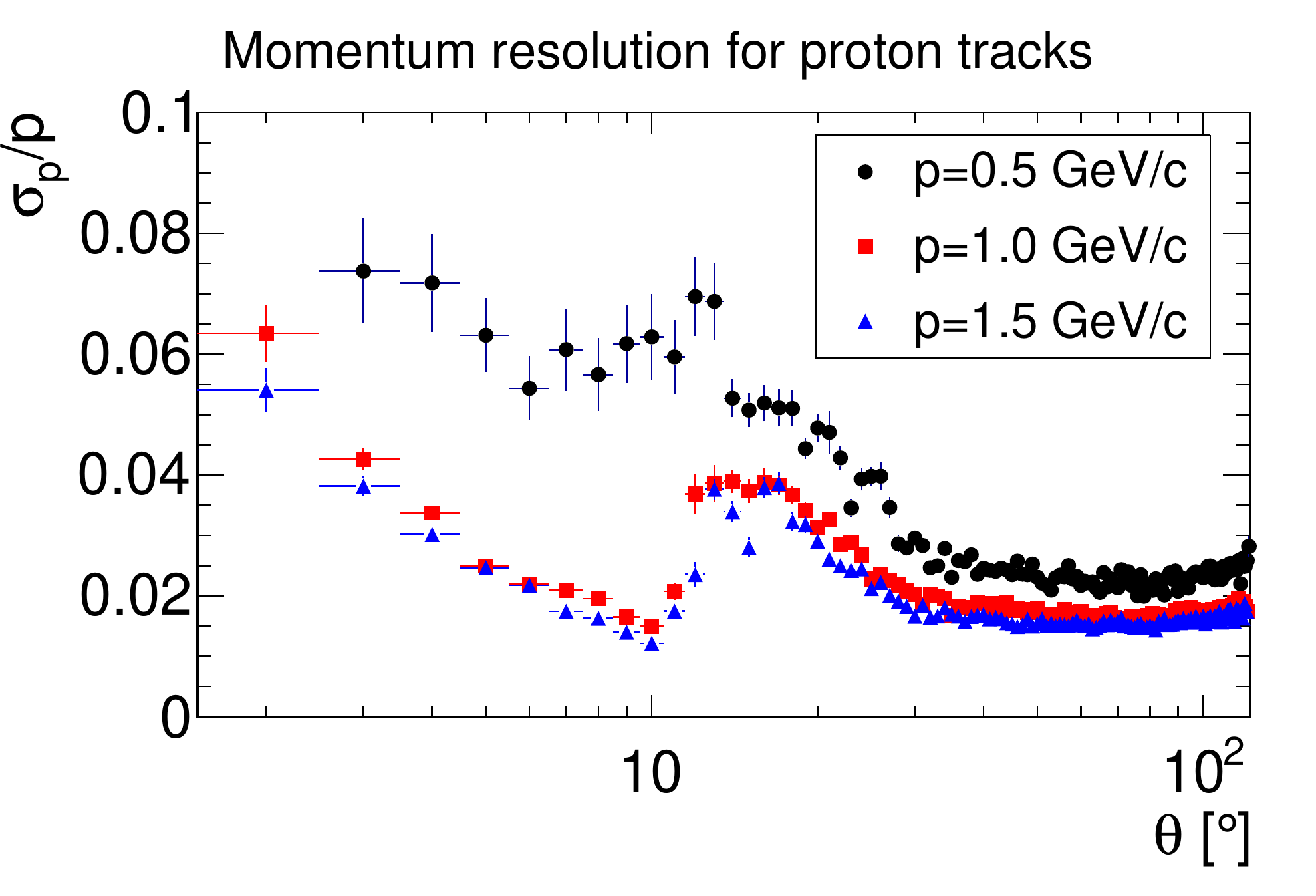}
\caption{\label{fig:dp_p} (Left) Momentum resolution for $\pi^-$ tracks.
(Right) Momentum resolution for proton tracks.}
\end{center}
\end{figure}

\begin{figure}[tbp]
\begin{center}
\includegraphics[width=0.45\textwidth]{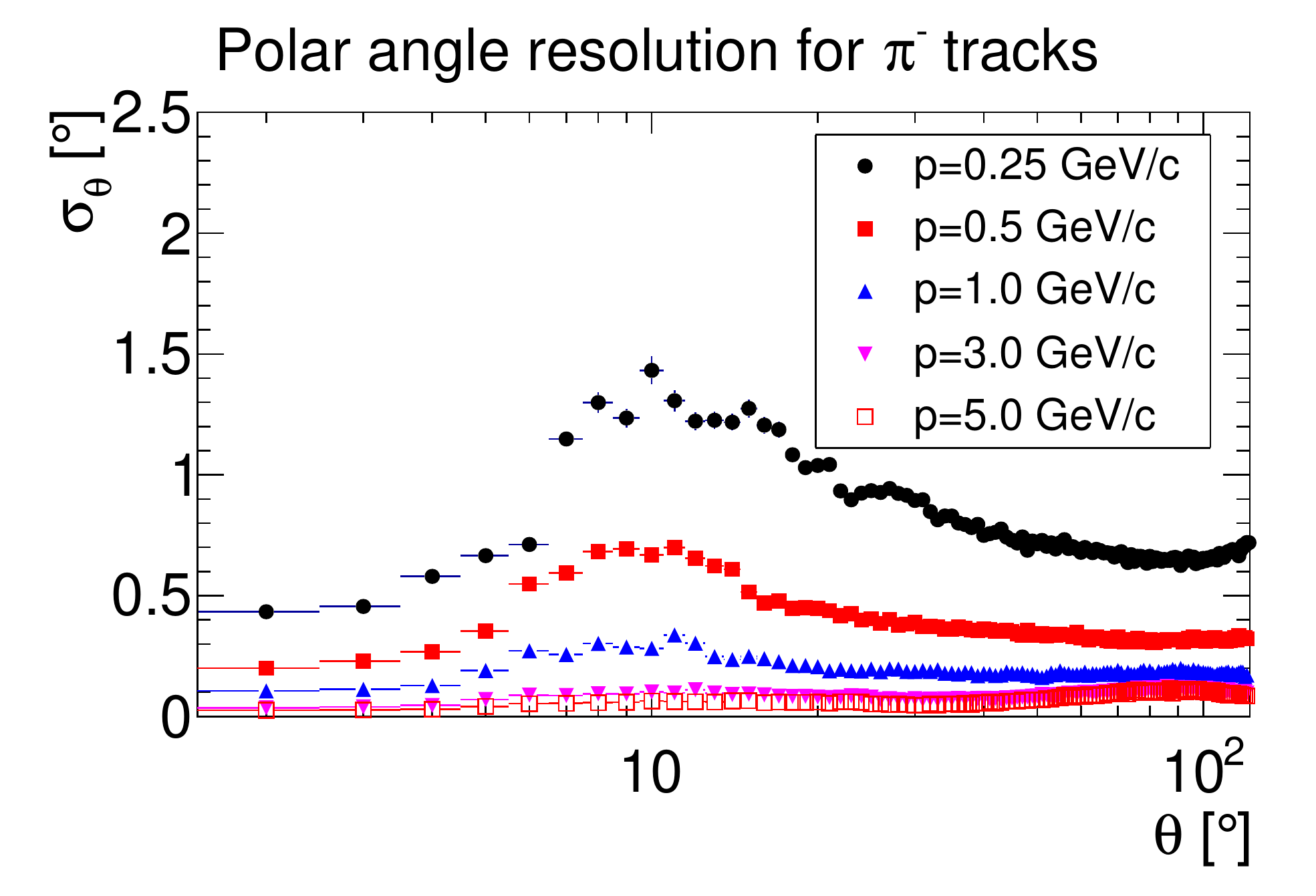}
\includegraphics[width=0.45\textwidth]{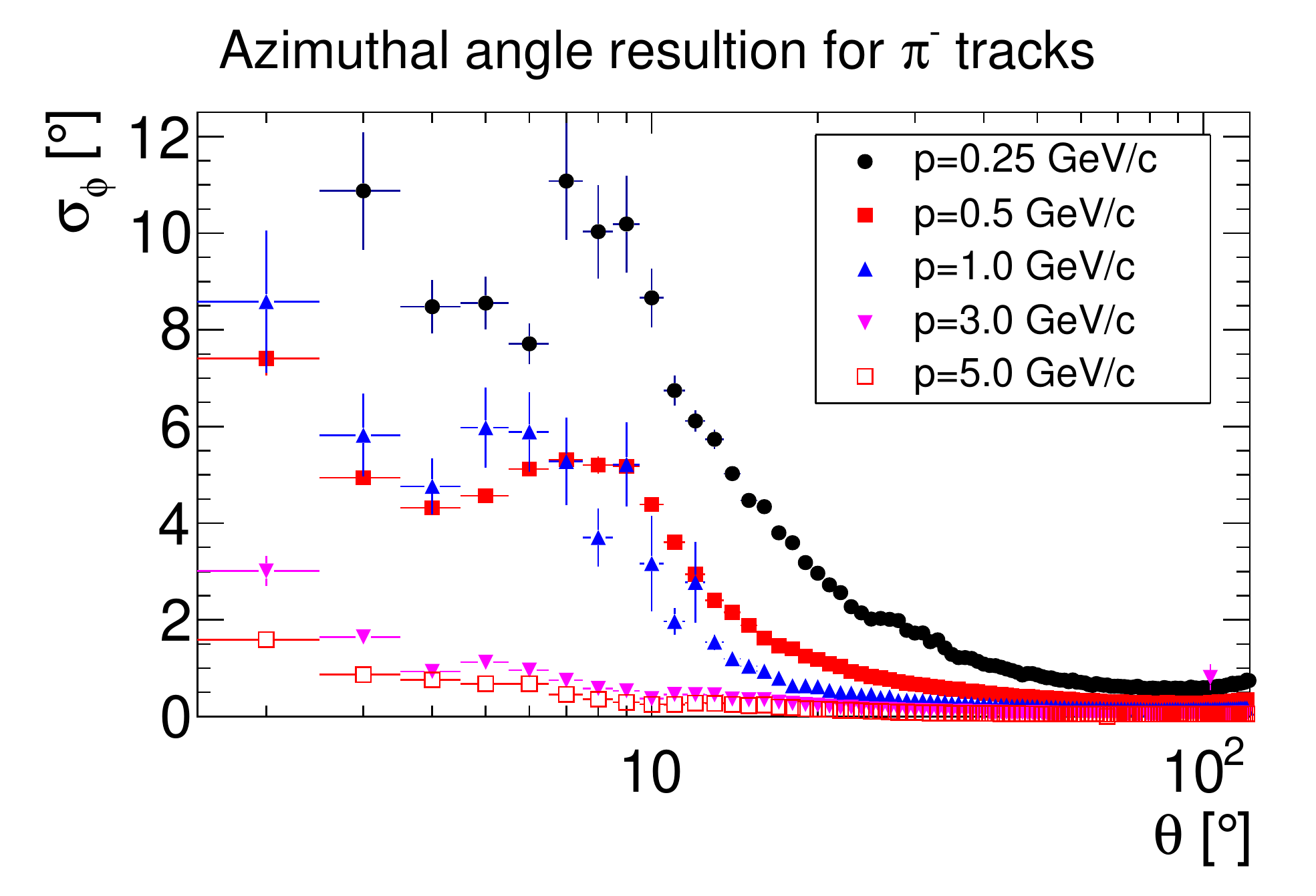}
\caption{\label{fig:angle res} (Left) Polar angle resolution for $\pi^-$ tracks.
(Right) Azimuthal angle resolution for $\pi^-$ tracks.
The resolutions are plotted as a function of the polar angle, $\theta$.}
\end{center}
\end{figure}

The momentum resolution as a function of angle and magnitude for pions and 
protons is shown in Fig.~\ref{fig:dp_p}.  The angular resolution is shown in 
Fig.~\ref{fig:angle res}.

\begin{figure}[tbp]
\begin{center}
\includegraphics[width=0.7\textwidth]{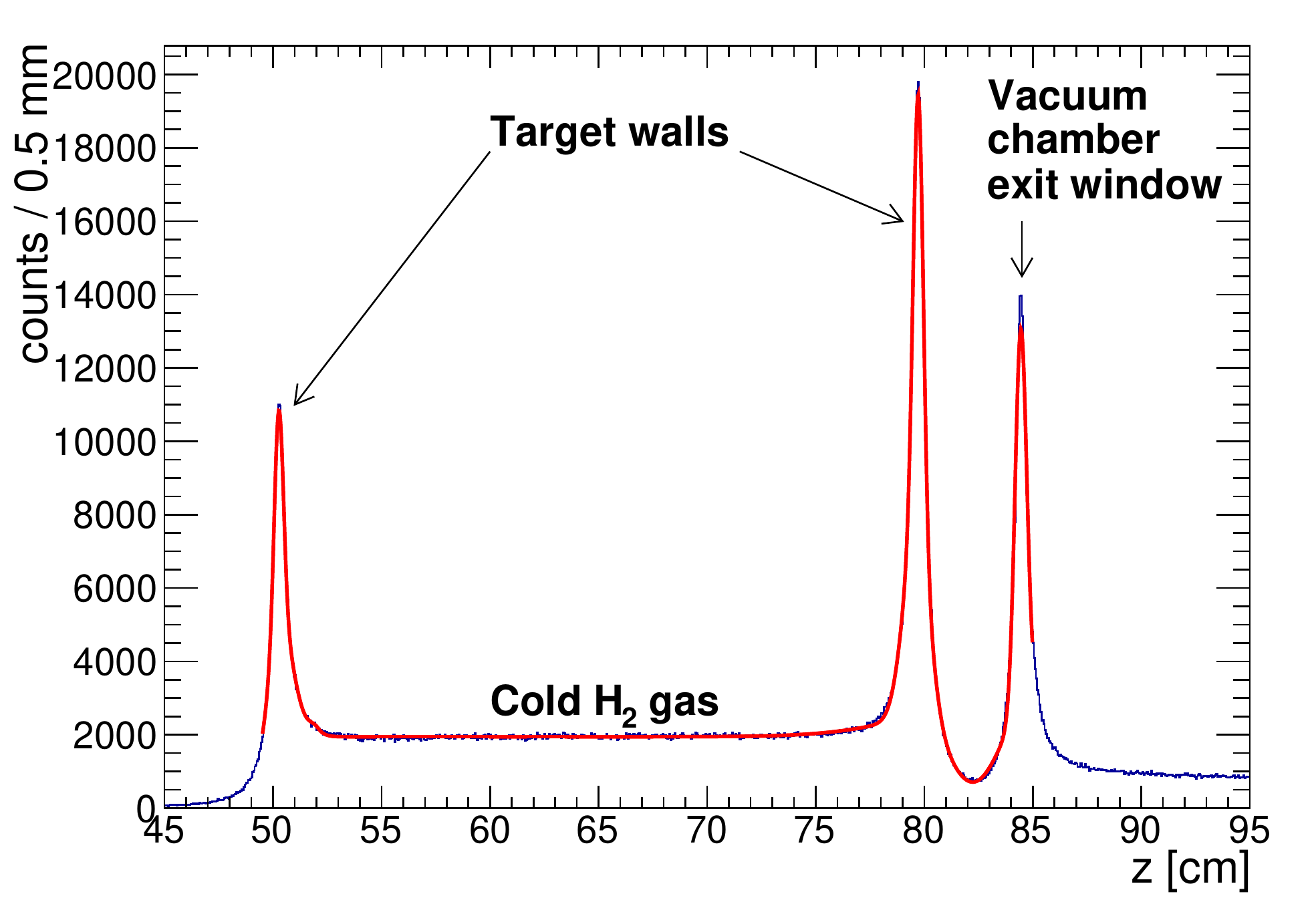}  
\caption{\label{fig:z-vertex} Reconstructed vertex positions within 1 cm radial
 distance with respect to the beam line for an empty target measurement.  The curve shows the result of a fit to the vertex distribution used to determine the vertex
resolution. 
}   
\end{center}  
\end{figure}

The thin windows of the cryogenic target and the exit window of the target
vacuum chamber provide a means to estimate the 
vertex resolution of the tracking system.  Pairs of tracks from empty target measurements are used to reconstruct these windows as illustrated in 
Fig.~\ref{fig:z-vertex}. The distance of closest approach between two tracks, $d$, was required to be less than 1~cm. The vertex position 
is at the mid-point of the line segment (of length $d$) defined by the points of closest approach for each track.
The estimated $z$-position resolution is 3~mm.

\section[Electromagnetic calorimeters] {Electromagnetic calorimeters \label{sec:calorimeters}}

\subsection[Barrel Calorimeter ]{Barrel Calorimeter \label{sec:bcal}}
The Barrel Calorimeter (BCAL) is an electromagnetic sampling calorimeter in the shape of an open cylinder. Photon showers with energies between 0.05~GeV and several GeV, $11^{\circ}$--$126^{\circ}$ in polar angle, and $0^{\circ}$--$360^{\circ}$ in azimuthal angle are detected. The geometry is fairly unique with the production target located in the backward part of the cylinder, as shown in Fig.\,\ref{fig:gluex_cut-away}. The containment of showers depends on the angle of photon incidence, with a thickness of $15.3$ radiation lengths for particles entering normal to the calorimeter face and reaching up to 67 radiation lengths at $14^{\circ}$. Details of the design, construction and performance of the BCAL can be found in Ref.\cite{BEATTIE201824}.

\begin{figure}[tbp]\centering
\includegraphics[scale=0.4]{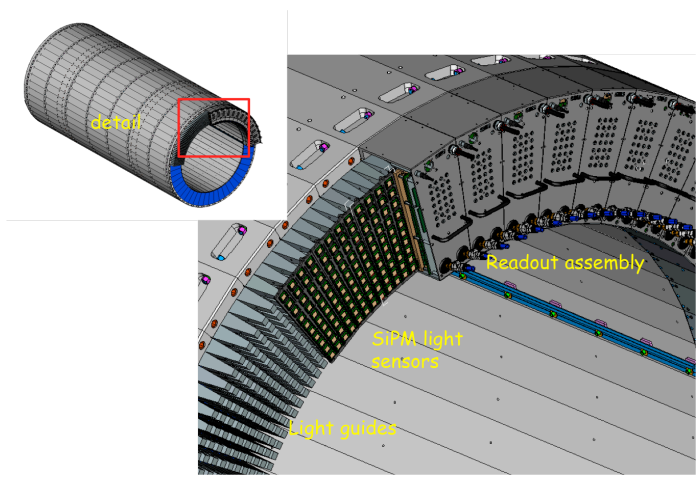}
\caption{\label{fig:bcal:bcal_assemblies}
   Three-dimensional rendition of the light guides mounted at the end of the 
   BCAL, as well as the readout assemblies mounted over them. The 
   readout assemblies contain the 
   SiPMs and their electronics.  (Color online)
  }
\end{figure}

The BCAL is constructed as a lead and  scintillating-fiber matrix, consisting of 0.5~mm-thick corrugated lead sheets and 1.0~mm-diameter Kuraray SCSF-78MJ multi-clad scintillating fibers. The fibers run parallel to the cylindrical axis of the detector. Each module has approximately 185 layers and 15,000 fibers. The BCAL consists of 48 optically isolated modules, each with a trapezoidal cross section, forming a  3.9-m-long cylindrical shell having inner and outer radii of 65~cm and 90~cm, respectively. The light generated in the fibers is collected via small light guides at each end of the module, which transport the light to silicon photomultipliers (SiPMs), which were chosen due to their insensitivity to magnetic fields. The end of the calorimeter with light guides, light sensors and electronics is shown in  Fig.\,\ref{fig:bcal:bcal_assemblies}.

The SiPM light sensors are Hamamatsu S12045(X) Multi-Pixel-Photon Counter (MPPC) arrays \footnote{Hamamatsu Corporation, Bridgewater, NJ 08807, USA \\ (\url{http://sales.hamamatsu.com/en/home.php)}.}, 
which are $4\times4$ arrays of $3\times3$ mm$^2$ tiles \cite{hdnote2913}. The SiPMs were accepted following extensive testing \cite{Barbosa2012100,Qiang2013234,soto,Soto201489,BeattieIEEE,doi:10.1063/1.4955340}. Four thousand units were purchased and 3840 are installed in the detector. The gain of the SiPM depends on the voltage above the breakdown voltage, about 70~V. These are operated at 1.4~V over the breakdown voltage, selected to reduce the effect of readout thresholds. Even at this relatively high overbias, the noise level is dominated by fluctuations in the electronics baseline and not by single-pixel noise. In order to keep a constant gain, the temperature is maintained within practical limits ($\pm$ 2$^\circ$C) using a chilled-water system. The gain is stabilized using a custom circuit that adjusts the bias voltage based on the measured temperature. Two stages of preamplifiers and summing electronics are attached to the sensors. In order to reduce the number of signals that are digitized, circuits sum the outputs of the preamplifiers in groups of radial columns, with coarser granularity away from the target. The layer closest to the target employs a single SiPM, and the next three layers have two, three, and four SiPMs, respectively. On the end of each module, forty SiPMs generate sixteen signals that are delivered to FADCs and twelve signals that are discriminated and then recorded with pipeline TDCs. The FADCs and TDCs are housed in VXS crates located on the floor close to the detector (see Section\,\ref{sec:trig}).

\begin{figure}[tbp]\centering
\includegraphics[height=5cm]{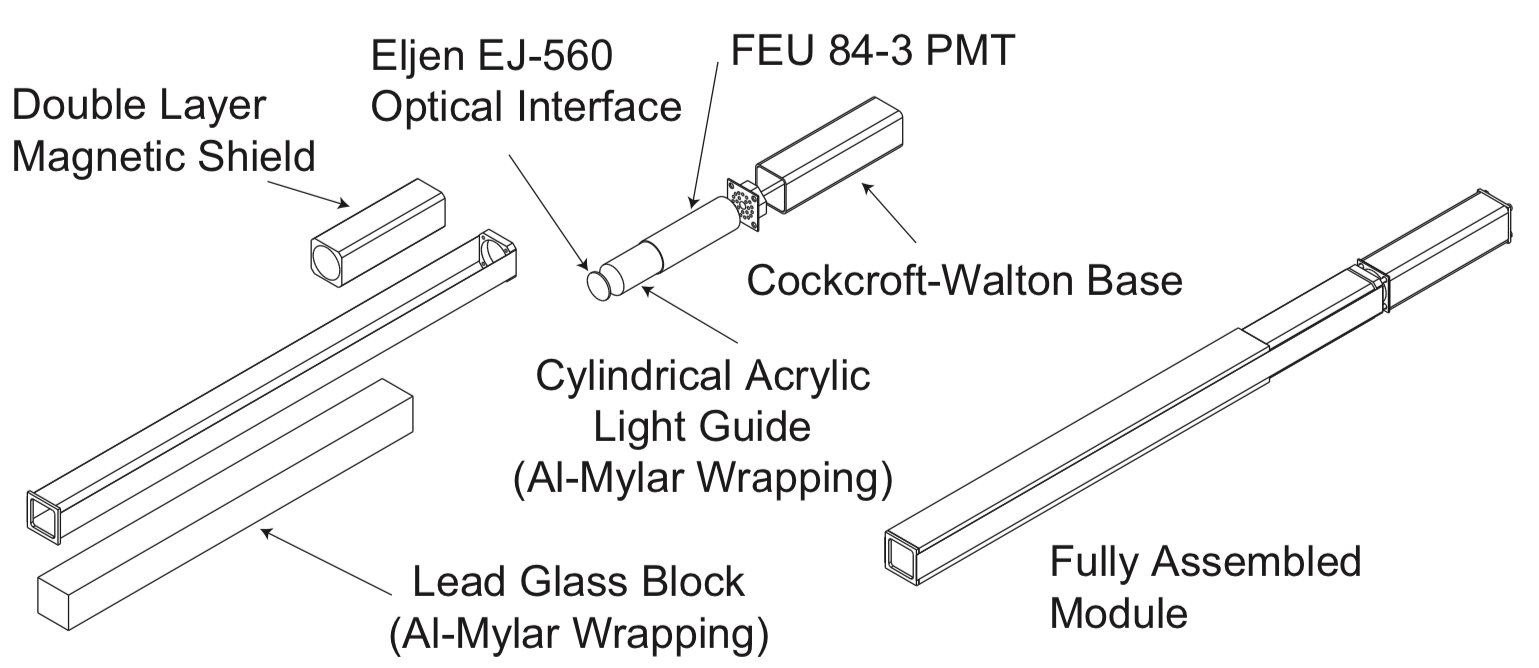}
\caption{\label{fig:fcal:FCAL_single_module}
    Expanded view of a single FCAL module.
  }
\end{figure} 
\subsection{Forward Calorimeter \label{sec:fcal}}
The Forward Calorimeter (FCAL) detects photon showers with energies ranging from 0.1 GeV to several GeV, and  between $1^{\circ}$--$11^{\circ}$ in polar angle. 
The front face of the FCAL is located 5.6~m downstream from the center of the GlueX target and consists of 2800 lead glass blocks stacked in a circular array that has a diameter of 2.4~m. Each lead glass block has transverse dimensions of $4\times4$ cm$^2$ and length of 45 cm.
The material of the lead-glass blocks is equivalent to type F8 manufactured by the Lytkarino Optical Glass Factory.\footnote{http://lzos.ru .} The blocks and most of the PMTs were taken from the decommissioned experiments E852 at Brookhaven National Laboratory \cite{CRITTENDEN1997377} and the RadPhi Experiment at JLab \cite{JONES2007384}. To remove accumulated radiation damage, the glass was annealed by heat treatment prior to installation in \gx{}. The detector is enclosed in a dark room.

The light collection is accomplished via an Eljen EJ-560 optical interface ``cookie'' and a UVT acrylic cylindrical light guide glued to the PMT. The light guide recesses the magnetically sensitive photocathode of the PMT inside a dual layer of soft iron and mu-metal that attenuates the stray field  of the \gx{} solenoid ($\lesssim$200~G). The sensors are FEU 84-3 PMTs with Cockcroft-Walton bases, each consuming 0.2~W.  The design of the PMT base is similar to that noted in Ref.~\cite{Brunner:1998fh}, and eliminates the need for a 2800-channel high-voltage power system. The bases communicate with a controller using the CAN protocol \cite{wiki:CANBus}, with 100 bases on each of 28 CAN buses.  The communication allows continuous monitoring of the PMT voltages, temperatures, and current draw.
A schematic of a single FCAL module is shown in 
Fig.\,\ref{fig:fcal:FCAL_single_module} and more details may be found in Ref.\,\cite{MORIYA201360}. FCAL signals are routed to FADC electronics, situated on a platform, directly behind the FCAL dark room.

\subsection{Electronics \label{sec:calelectronics}}
Custom readout electronics for the two calorimeters are mounted in standard VXS crates and include 
JLab 12-bit 250~MHz FADCs \cite{hdnote1022}, discriminators \cite{hdnote2511} and F1 TDCs \cite{hdnote1021}. The maximum input scale of the FADCs (4095 counts) is set to 2~V.
The FADCs sample each calorimeter channel every 4~ns and generate raw waveforms consisting of 100 samples 
 (400~ns). The samples are available for further processing by the firmware upon a trigger signal, if the waveform exceeds a threshold voltage. The firmware computes several derived quantities of the pulse: pedestal, peak value, integral over a selected window, and time of the halfway point on the leading edge. At most one pulse is extracted from each readout window. These pulse features constitute the raw data that is nominally read out from the FADC.  Optionally, the full waveforms can be read out for diagnostic purposes and to check the firmware output against the offline emulation of the parameter extraction; this is done for less than about 1\% of the production runs.
 
Pulses are identified by the first sample that exceeds a threshold, currently set to 5 (8) counts above the average pedestal for the BCAL (FCAL). These thresholds correspond to approximately 2.5 (12) MeV. The integral is determined using a fixed number of samples relative to the threshold crossing, which was determined by maximizing the ratio of signal to pedestal noise.  The integration window begins one sample before the threshold time and extends to 26 (15) samples after the threshold time for the BCAL (FCAL).  Typical pedestal widths are $\sigma\sim$1.2-1.3 (0.8) counts. For the BCAL, the pedestals are determined for each channel event-by-event, appropriately scaled, and then subtracted from the peak and integral to obtain signals proportional to the energy deposited in the calorimeter. For the FCAL, the average pedestal over a run period is determined offline for each channel and the pedestal contribution to the pulse integral is subtracted when the data are reconstructed.  
 The algorithm that determines the time of the pulse is pulse-height independent and, therefore, time-walk correction is not required for the FADC times~\cite{Bennett:2010nf}.

The outputs of the three inner layers of the BCAL are also fanned out to leading-edge discriminators, which feed the JLab F1 TDCs. The discriminator thresholds are initially set to 35~mV and then adjusted
 channel by channel.  The pulse times are recorded relative to the trigger in a 12-bit word. Multiple hits may be recorded per channel per event (up to eight), but are culled at a later time by comparison to FADC times. The nominal least count is configured to be 58~ps.

\subsection[Calibration and monitoring]{Calibration and monitoring \label{sec:calcalib}}
The relative gains of the calorimeters are monitored using a modular LED-driver system \cite{Anassontzis201441}. The control system is the same for both calorimeters, but the arrangement of LEDs is tailored to the respective detector geometries. In the BCAL, one LED is inserted into each light guide to monitor each individual SiPM and its partner at the far end of the module.
Due to geometry, the illumination varies considerably from channel to channel. 
The average gain stability of the detector over a period of ten days is better than 1\% and the fractional root-mean-square deviation of the mean for each SiPM during a single day from the average over the run period is typically less than 2\%.

For the FCAL, four acrylic panes were installed, each covering the upstream end of one quadrant of the FCAL. Each pane is illuminated by forty LEDs, ten violet, ten blue, and twenty green. In addition to monitoring the stability of the readout, the different colors are used to study the wavelength dependence of the transmission of light though the lead glass blocks.  In particular, radiation damage to lead glass inhibits transmission at the blue end of the spectrum and tends to turn glass a brownish color~\cite{Schaefer:2011gw}. Throughout a several-month experiment, the response to the green LEDs was unchanged. However, the PMT response to violet LEDs degraded by about 10\% in the blocks closest to the beam line, characteristic of radiation damage.  Such damage is only evident in the first two layers of blocks surrounding the 12~cm$\times$12~cm beam hole. This damage is likely confined to the upstream end of the block and does not significantly affect the response to particle showers in the body of the glass.

The energy of a photon or lepton is obtained from the reconstructed electromagnetic shower. Here, a shower is reconstructed using an algorithm that finds a cluster by grouping signals close in time and space, called hits, that have been registered by individual detector elements.
Details of the algorithms to obtain shower energies in the BCAL can be found in Ref.\,\cite{BEATTIE201824}  and in Ref.\,\cite{Jones:2006ru} for the FCAL. The clustering in the FCAL requires that hits register within 15~ns of the primary hit, where the seed threshold is taken to be 35 MeV. Clusters with a single hit are discarded. In the event of overlapping showers, the hit energies are divided among the clusters in proportion to the partition predicted by a typical shower profile. Both detectors have sources of energy-dependent nonlinearities and empirical corrections are developed and applied to minimize the measured energy dependence of the measured $\pi^0$ mass.

\subsection{Performance \label{sec:calperformance}}
The performance of the calorimeter is summarized by its ability to measure the energy, position and timing of electromagnetic showers.

\begin{figure}[tbh]\centering
\includegraphics[width=0.48\textwidth]{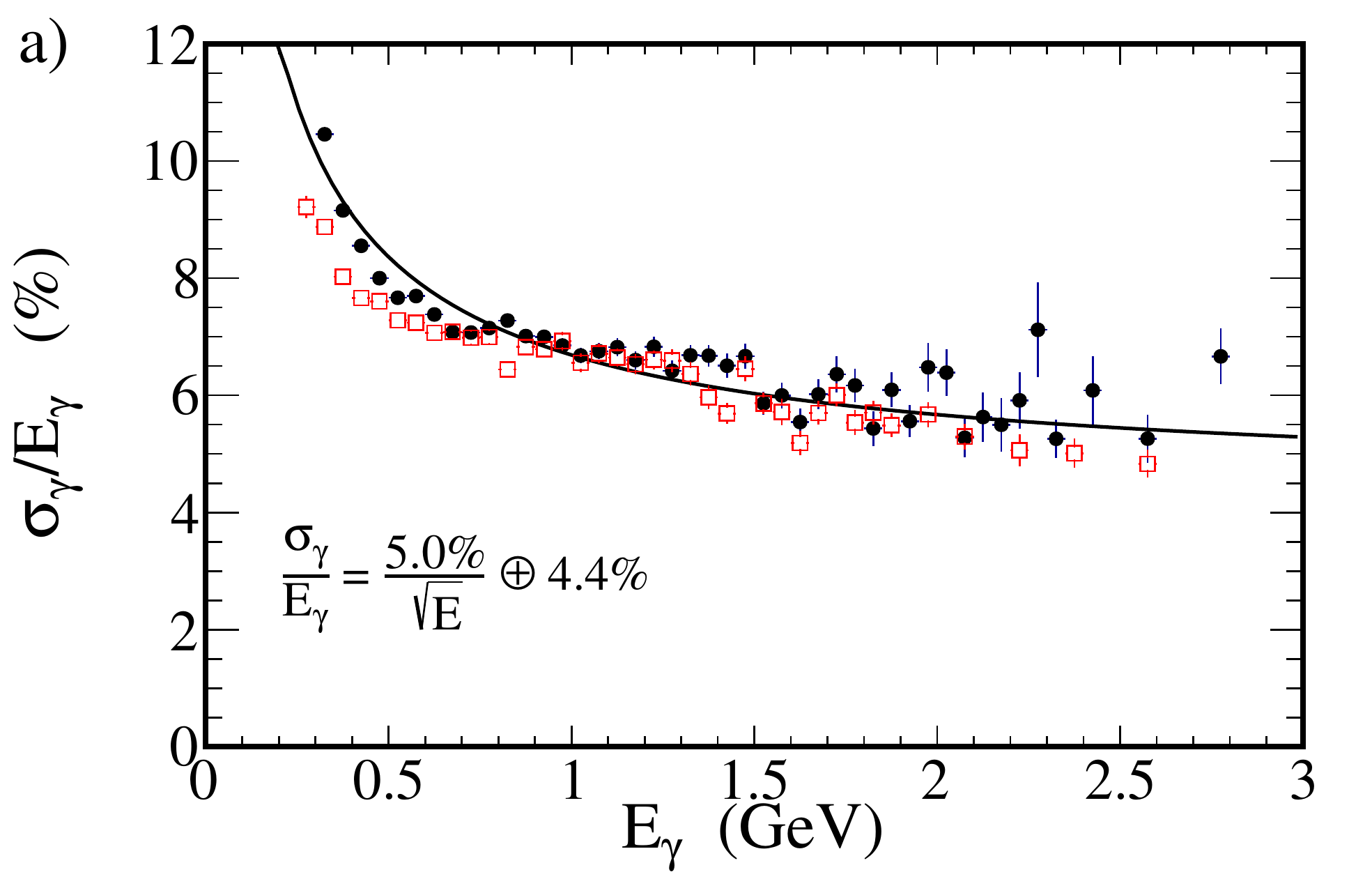} \includegraphics[width=0.48\textwidth]{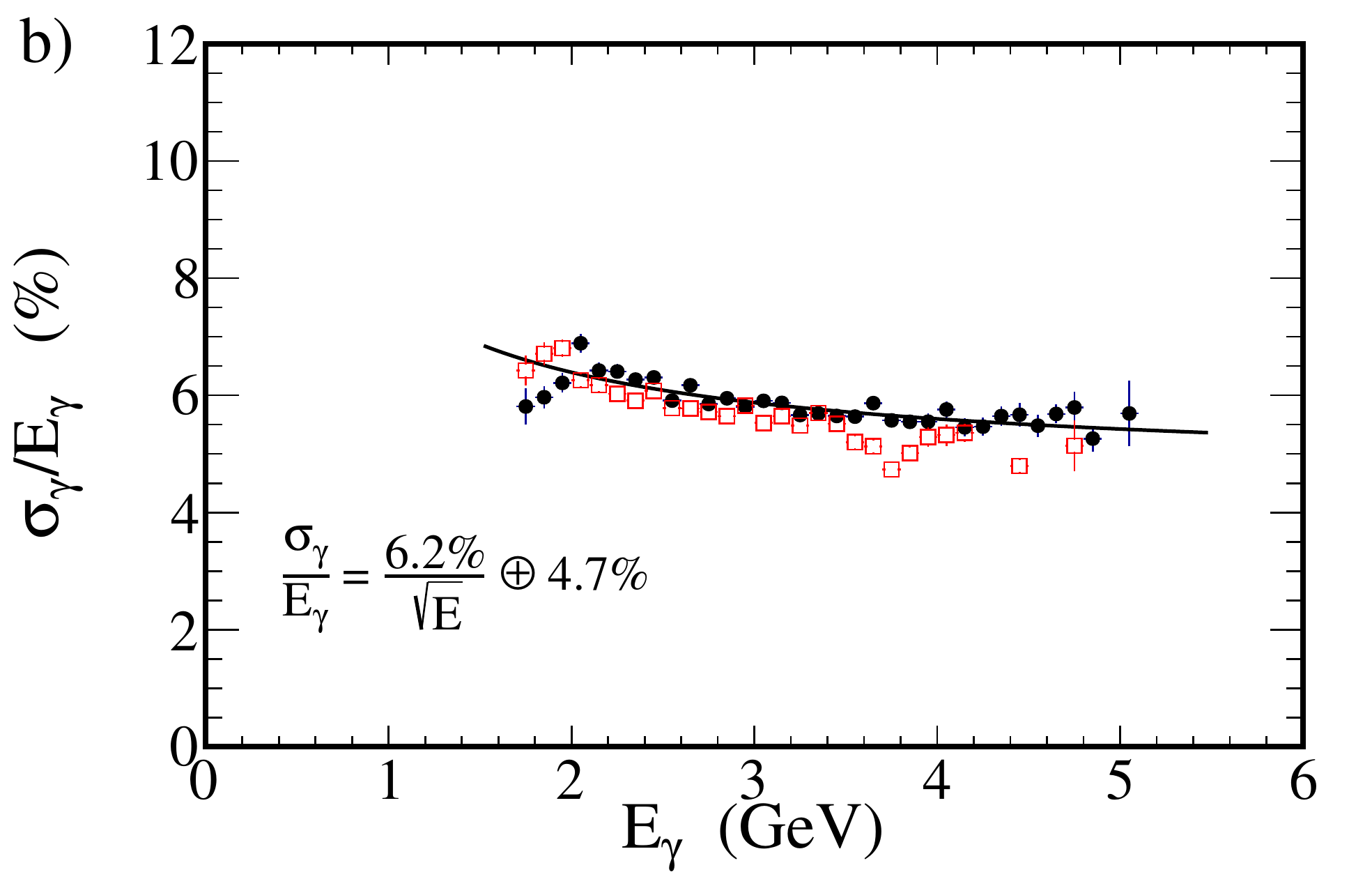}
\caption{\label{fig:bcal:eta_resolution} 
The energy resolution, $\sigma_\gamma/E_\gamma$, for single photons in the a) BCAL and b) FCAL calculated from the $\eta$ mass distribution under the assumption that only the energy resolution contributes to its width.  Solid black circles are data and open red squares are simulation. Fitted curves including the stochastic and constant terms are indicated.
(Color online)
 }
\end{figure}

The energy resolution of each calorimeter was extracted from the measured $\pi^0$ and $\eta$ mass distributions, yielding consistent results. 
To study the $\eta$ mass resolution, events were selected using kinematic fits to $\gamma p \rightarrow p \pi^+ \pi^- \gamma \gamma$, with $\eta\rightarrow \gamma\gamma$ and the photons having the same energies within 10\%.
The proton and pion tracks were used to determine the event vertex, needed to accurately reconstruct the two-photon invariant mass.
This reaction provides a fairly clean sample of $\eta$'s with energy-symmetric photons recorded either both in the BCAL or both in the FCAL. 
The single-photon energy resolution was determined from Gaussian fits to the $\eta$ invariant mass width, neglecting contributions from uncertainty in the opening angle.
Monte Carlo simulation of $\gamma p \rightarrow p \pi^+ \pi^- \eta$ events, with kinematics chosen to approximate the experimental distributions, were used to tune the MC resolution to match the data. 
The single-photon resolutions are shown in Fig.\,\ref{fig:bcal:eta_resolution}(a) for the BCAL and Fig.\,\ref{fig:bcal:eta_resolution}(b) for the FCAL as a function of the mean photon energy, both for data and simulation.
A fit has been performed to the data for each calorimeter to estimate contributions to the width from stochastic and constant processes.  The parameters in the fit are strongly correlated due to the limited range of energy available.\footnote{For the BCAL these data constitute an average over many angles, resulting in a relatively large effective constant term that cannot be extrapolated to higher energy. 
See Ref.\,\cite{BEATTIE201824} Section 11 for details.}

The resolution of the position (Z) along the length of the BCAL ($\sim$\,2.5 cm) is computed from the timing resolution of the system, which was measured to be $\sigma=150$\,ps at 1\,GeV. The transverse position resolution ($\sigma$) obtained from simulation for 1~GeV showers in the FCAL is less than 1.1~cm.

The performance of the calorimeters has been demonstrated in the reconstruction of neutral states including $\pi^0$, $\eta$ and $\eta'$ mesons for the first \gx{} physics publications \cite{AlGhoul:2017nbp,Adhikari:2019gfa}. In addition, although the response of the calorimeters at high energy is still under evaluation, it has provided important electron-pion separation to identify the decays of $J/\psi\rightarrow e^+e^-$ \cite{Ali:2019lzf} where electrons were recorded up to 8~GeV. 

\section[Scintillation detectors]{Scintillation detectors \label{sec:scintillators}}
There are two scintillator-based detectors deployed in the \gx{} spectrometer: a small barrel-shaped detector surrounding the
target, referred to as the Start Counter (ST), and a two-plane hodoscope detector system in the forward direction, referred to as the
Time-of-Flight (TOF) detector. Both detectors provide timing information. Charged-particle identification is derived from
energy loss ($dE/dx$) in the ST and flight time from the TOF.

\subsection{Start Counter \label{sec:st}}

\begin{figure}[!htb]
\centering
\includegraphics[width=1.0\columnwidth]{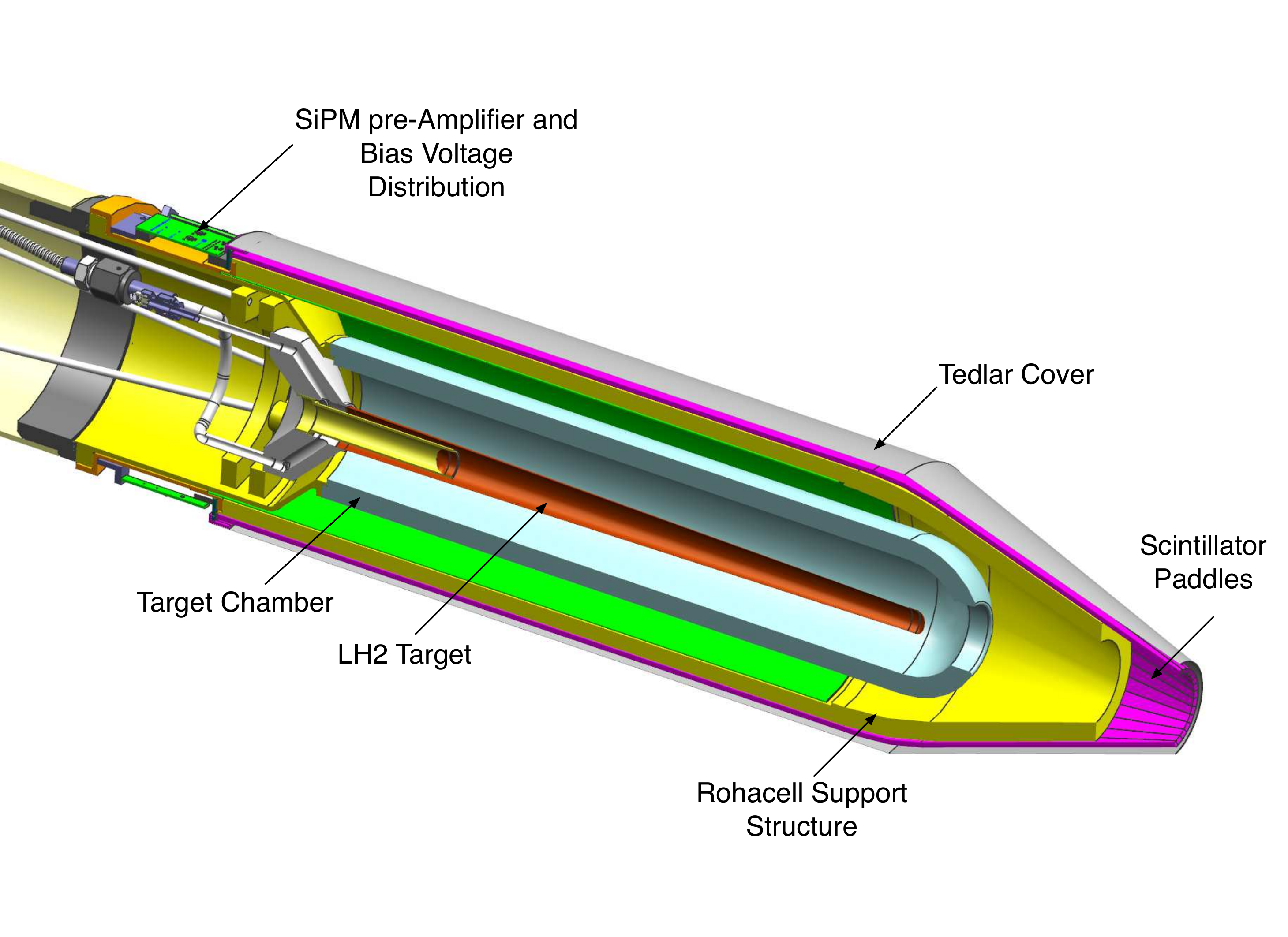}
\caption{The \gx{} Start Counter surrounding the liquid-hydrogen
  target assembly.  The incident beam travels from left to right down the central
  axis.\label{fig:st-overview-drawing}}
\end{figure}

The ST, shown in Fig.~\ref{fig:st-overview-drawing},
surrounds the target
region and covers about 90\% of the solid angle for particles
originating from the center of the target. The ST is designed to operate
at tagged photon beam intensities of up to $10^8$ photons per second
in the coherent peak, and has a high degree of segmentation to limit
the per-paddle rates. The time resolution must be sufficient to resolve the RF beam structure and identify the electron beam bunch from which the event originated
(see Section\,\ref{sec:ebeam}). The ST provides a timing signal that is relatively independent of particle type and trajectory (because of its proximity to the target) and can be used in the Level 1 trigger if necessary. The specific energy deposits $dE/dx$ in ST are used for charged-particle identification in combination with the flight-time from the TOF.
Details of the design, construction and performance of the ST system can be found in 
Ref.\,\cite{Pooser:2019rhu}.

The ST consists of 30 scintillator paddles arranged in a cylinder of radius 78~mm with a ``nose'' section that bends towards the beam
line to a radius of 20~mm at the downstream end. EJ-200 scintillator from Eljen
Technology\footnote{Eljen Technology, https://eljentechnology.com/products/plastic-scintillators.} was
selected for the ST paddles. EJ-200 has a decay time of 2.1~ns with a bulk attenuation length
of 380~cm. 
Each scintillator paddle originated from stock 3~mm thick and 600 mm in
length. The paddles were bent at Eljen to create the nose section, and then machined at McNeal Enterprises Inc.\footnote{McNeal
  Enterprises~Inc., http://www.mcnealplasticmachining.com} to their
final shape, including edges beveled at $6^\circ$ to minimize loss of
acceptance.
The scintillator paddles are supported by a Rohacell closed-cell foam
structure. The Rohacell is 11~mm thick and is rigidly attached to an
aluminum support hub at the upstream end. The downstream support
extends partially into the nose section. The cylindrical length of the Rohacell is further reinforced with three layers of carbon fiber, each layer being 650~$\mu$m thick. The assembly is made light-tight with a Tedlar wrapping, attached to a plastic collar at the upstream end.

Silicon photomultiplier detectors are used as light
sensors, as these are not affected by the magnetic field produced by the solenoid. The SiPMs were placed
at the upstream end of each scintillator element with a 250~$\mu$m air gap.
Each paddle is read out with an array of four 
SiPMs (Hamamatsu S109031-050P multi-pixel photon counters) whose signals are summed. 
The on-board electronics provides two signals per paddle, one delivered to an FADC, and the other to a 5$\times$~amplifier that is sent to a discriminator and then to a TDC. 

\subsection[Time-of-Flight counters]{Time-of-flight counters \label{sec:tof}}
The TOF system delivers fast timing signals from charged particles passing through the detector, thereby providing information for particle identification.
The TOF detector is a wall of scintillators located about 5.5~m downstream from the target, covering 
a polar angular region from 0.6$^{\circ}$ to 13$^{\circ}$. The detector has two planes of
scintillator paddles stacked in the horizontal and vertical direction. Most paddles are 252~cm long and 2.54~cm
thick with a width of 6~cm. 
The scintillator material is EJ-200 from Eljen Technology.
To allow the photon beam to pass through the central region,
an aperture of 12$\times$12\,cm$^2$ is kept
free of any detector material by using four shorter, single-PMT paddle detectors with a length of 120~cm around the beam hole
in each detector plane. These paddles also have a width of 6~cm and a thickness of 2.54~cm. In order to keep the
count rate of the paddles well below 2~MHz the two innermost full-length paddles closest to the beam hole on either side have a reduced width of 3~cm.
Light guides built out of UV transmitting plastic provide the coupling between the scintillator and the PMT and allow the 
magnetic shielding to protect the photocathode by extending about 5~cm past the PMT entrance window. All paddles are wrapped
with a layer of a highly reflective material (DF2000MA from 3M) followed by a layer of strong black Tedlar film for light tightness. 

The scintillator paddles are read out using 
PMTs from Hamamatsu.\footnote{Hamamatsu Photonics, https://www.hamamatsu.com/us/en/index.html.} Full-length paddles
have a PMT at both ends, while the short paddles have a single PMT
at the outer end of the detector. These 2" H10534 tubes have ten stages and are complete assemblies with high voltage base, casing and $\mu$-metal shielding. Additional soft-iron external shielding protects each PMT from significant stray fields from the solenoid magnet.

\subsection{Electronics \label{sec:scelectronics}}
High voltage for the TOF PMTs is provided by CAEN HV modules of type A1535SN, initially controlled by a CAEN SY1527 main frame and
later upgraded to a SY4527.
The PMT outputs are connected to a passive splitter by a 55'-long RG-58 coaxial cables. The signal is split into two equal-amplitude signals. One signal is directly connected to a FADC~\cite{Dong:2007}, while the second signal passes first through a leading-edge discriminator and is then used as an input to a high resolution TDC. The digitizing modules are mounted in VXS crates as described in Section~\ref{sec:trig}.
The threshold of the leading-edge discriminator is controlled separately for each channel and has an intrinsic
deadtime of about 25~ns.

The sparsification threshold for the FADC is set to 120 (160) counts for the ST (TOF), with the nominal pedestal set at 100 counts. The high voltage of each TOF PMT is adjusted to generate a signal amplitude of at least 400 ADC counts above baseline from a minimum-ionizing particle. The data from the FADC are provided by the FPGA algorithm and consist
of two words per channel with information about pedestal, signal amplitude, signal integral, and timing.

The timing signals from the ST system are registered using the JLab F1 TDCs, which have a nominal least count of 58~ps. In order to take advantage of the higher intrinsic resolution of the TOF counters, this system uses the VX1290A TDCs from CAEN\footnote{CAEN, https://www.caen.it/}, which are multi-hit high-resolution TDCs with a buffer of up to 8 words per channel and a nominal least count of 25~ps. Since these TDCs provide the best time measurements in the \gx{} detector, the timing of the accelerator RF signal is also
digitized using these TDCs.

\subsection{Calibration and monitoring \label{sec:sccalib}}
The combined ST and TOF systems are used to determine the flight times of particles, the ST providing a precise start time in combination with the accelerator RF, and the TOF providing the stop time. Both systems may also be used to provide information on particle energy loss. Therefore, the signals in ST and TOF must be 
calibrated to determine corrections for the effects of
time-walk, light propagation time offsets, and light attenuation. The procedures are slightly different for the two detectors because of the different geometries, intrinsic resolutions, and the advantages of the TOF system having two adjacent perpendicular planes. 

For the time-walk correction for each paddle of the ST, the detector signal is sent to both an FADC and a TDC. The time from the FADC, being independent of pulse amplitude, is the reference. The amplitude dependence of the difference between TDC and FDC times is used to measure the time walk; the resulting curve is fit to an empirical function for use in the correction. 
The propagation time is measured as a function of the hit position in a paddle as determined by well-reconstructed charged particle tracks. The propagation velocity is measured in three regions of the counter (``straight,'' ``bend,'' and ``nose'') and is not assumed to be a single value for all hits. The light attenuation is also measured at several positions along the counter using charged particle tracks. The energy-per-unit pathlength in the paddle as a function of distance from the SiPM is fit to a modified exponential, with different parameters allowed for the straight section and the nose section, with continuity enforced at the section boundary.

The calibration procedures for the TOF system take advantage of the two
planes of narrow paddles oriented orthogonal to each other, which permits calibration of the full TOF detector independently
of any other external detector information. The overlap region of two full-length paddles from the two planes defines
a 6$\times$6~cm$^2$ area for most paddles, with a few 3$\times$3~cm$^2$ areas close to the beam hole. The separation between the two detector planes is minimal as they are mounted adjacent to each other, separated only by wrapping
material. While the time-difference (TD) between the two ends of a paddle is related to the hit position along the paddle,
the mean-time (MT) is related to the flight time of a particle from the vertex to the paddle. Therefore, the MT for two overlapping
paddles must be the same when hit by the same particle passing through both paddles, while the hit positions in the horizontal and vertical dimensions are defined by the TD of the two paddles. This relationship results in an internally consistent calibration of all paddles with respect to every other paddle. Prior to finding timing offsets for calibration, all times are corrected for the amplitude-dependent walk. The relation between time at threshold and signal amplitude is parameterized and used to correct for time slewing.

After all full-length paddles have been calibrated, they can be used themselves as references to
calibrate the remaining eight short paddles that only have single-ended readout.  Again we use the fact that any overlap region of two paddles from different
planes has the same particle flight time from the vertex. This coincidence produces peaks in the time difference distributions that can be used to determine the timing offsets of these single-ended readout paddles. 

\begin{figure}[tbp]
\begin{center}
\includegraphics[width=0.6\textwidth]{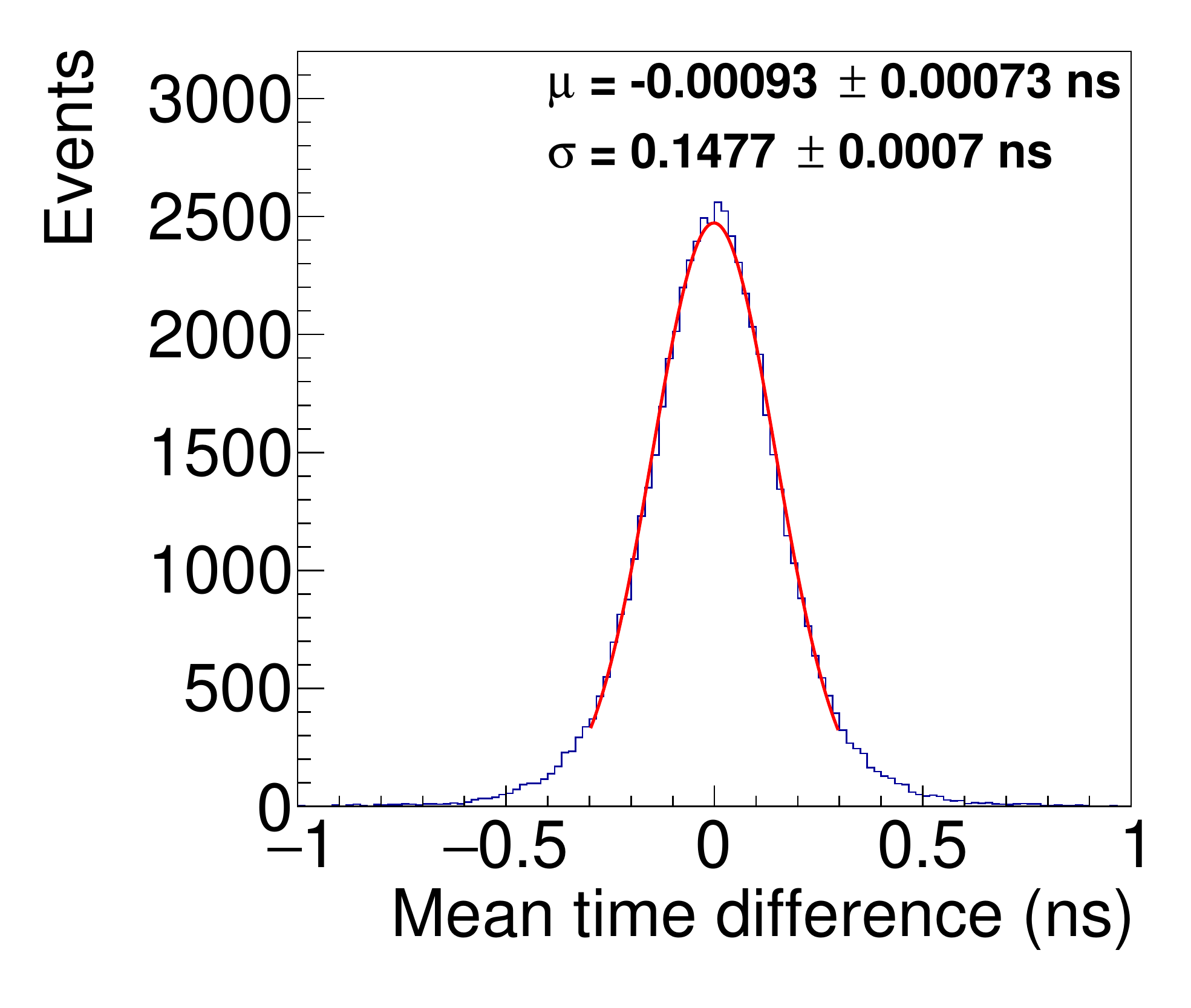}
\caption{\label{fig:mt_diff} Mean time difference between one TOF long paddle of one plane with all other long paddles
of the other plane. (Color online)}
\end{center}
\end{figure}

To test the calibration, we take tracks that are incident on a paddle in one plane and compute the time difference between the MT of that paddle and the MT of every other full-length paddle in the other plane. The resulting distribution of these differences is shown in Fig.~\ref{fig:mt_diff}. Assuming that all paddles have the same timing resolution, we can compute the
average time resolution to be $\sigma$ = 105~ps$=\frac{148}{\sqrt{2}}$~ps, assuming a Gaussian distribution.

\subsection{Performance \label{sec:scperformance}}
The purpose of the ST is to select the electron beam bunch that generated the tagged photon which induced a reaction in the target. The corresponding time derived from a signal from the CEBAF accelerator, which is synchronized with the RF time structure of the machine, is used to determine the event start time. Therefore, the ST resolution does not contribute to the resolution of the flight time as long as the resolution is sufficient to pick out the correct beam bunch with high probability.

\begin{figure}[tbh]
  \centering
  \includegraphics[width=0.6\linewidth]{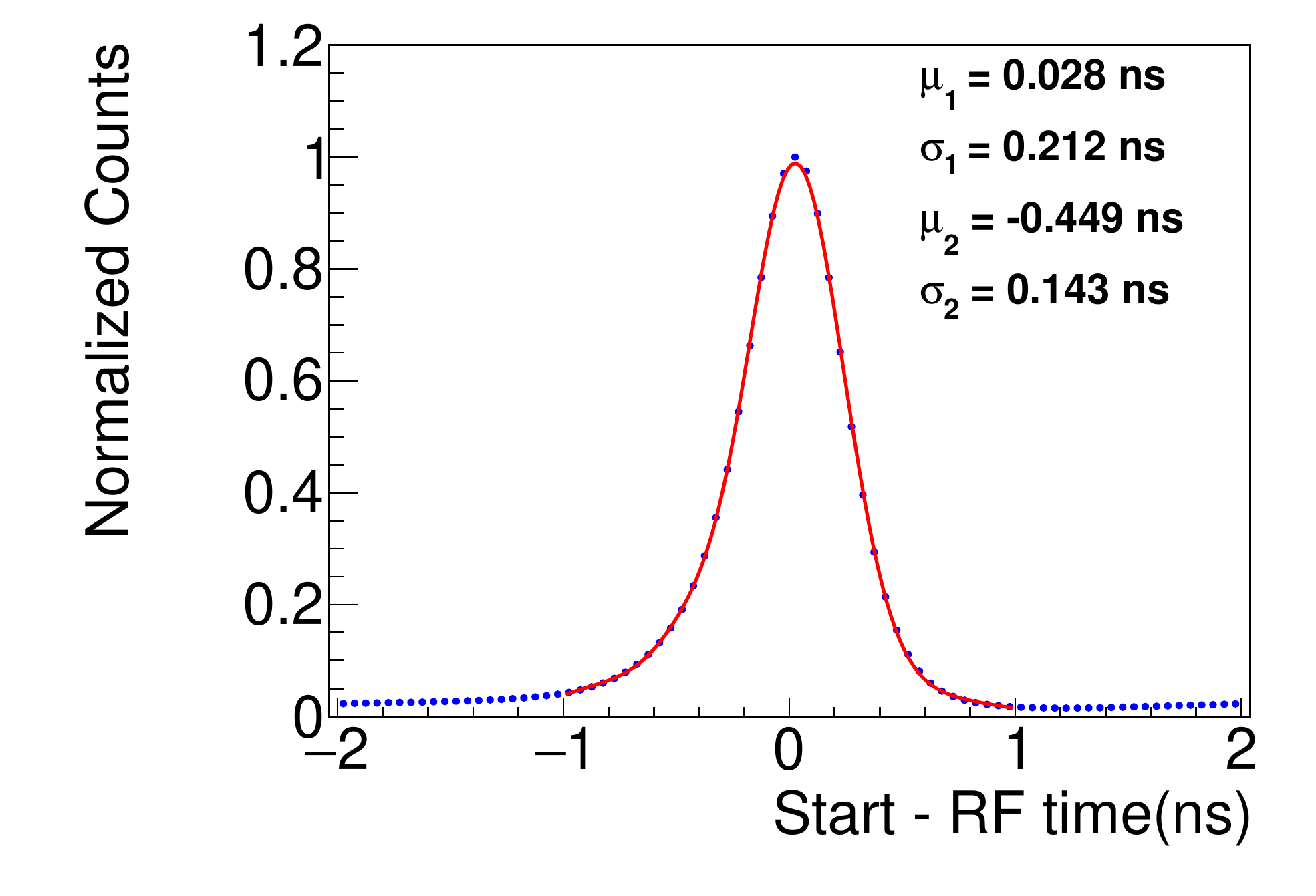}  
  \caption{Time difference distribution between the vertex time computed from the start counter and the accelerator RF. The time from the RF does not contribute significantly to the width of the distribution. The fit function is a double Gaussian plus a third-degree polynomial.
  }
                \label{fig:st-time-resolution}
\end{figure}

The ST timing performance can be determined by comparing the event time at the target measured by the start counter and the accelerator RF time. The start counter time must be corrected for the flight path of the charged particle emerging from the event, and all instrumental corrections mentioned in the previous section must be applied. Fig.~\ref{fig:st-time-resolution} shows the distribution of this time difference. The average time resolution is about $\sigma$=234~ps, where the resolution varies depending on the position of the hit along the counter. 


\begin{figure*}[!htb]
  \centering
  \includegraphics[width=0.6\textwidth]{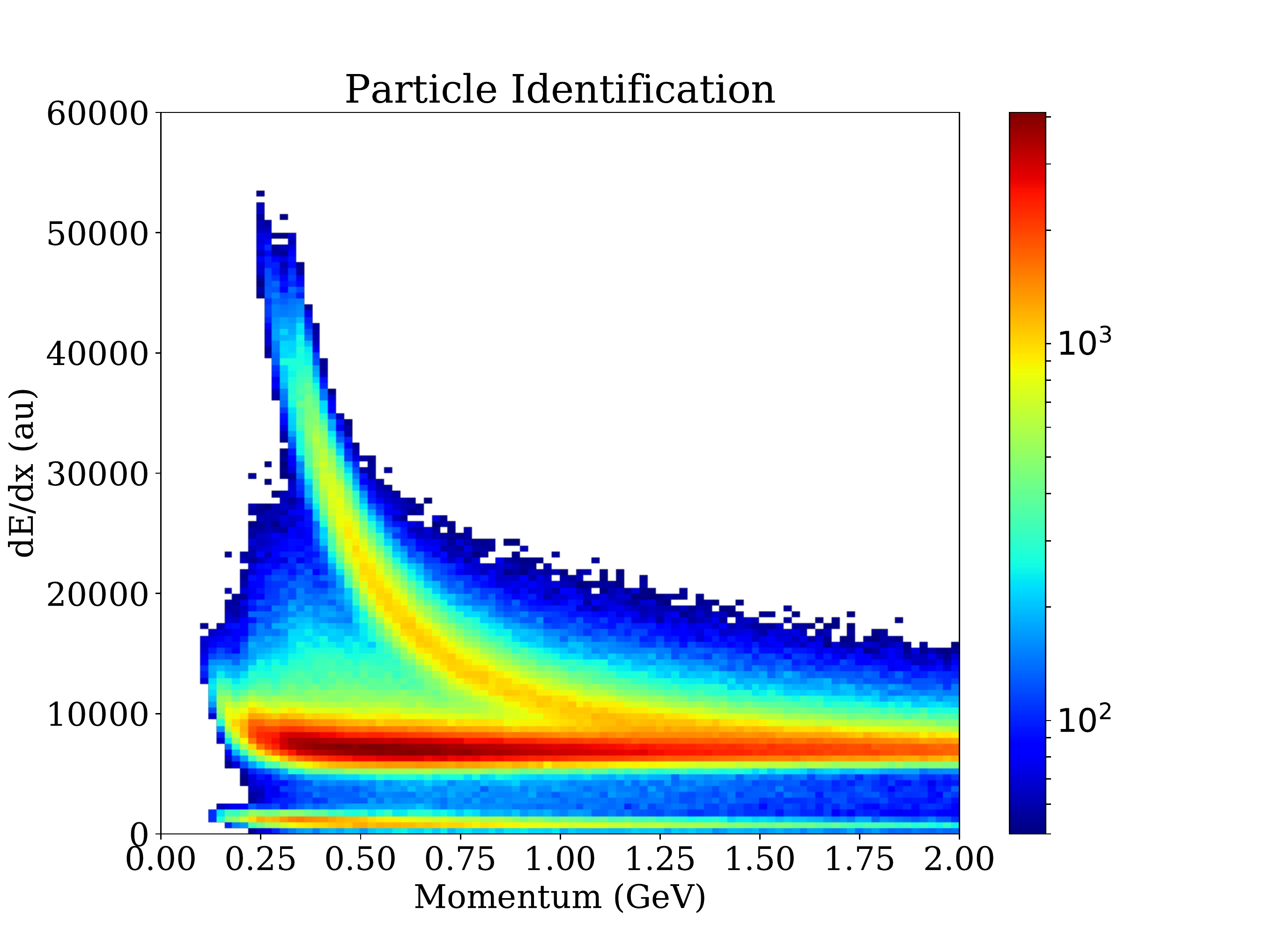}
  \caption{$dE/dx$ vs.\ $p$ for the Start Counter.  The curved band
    corresponds to protons while the horizontal band corresponds to
    electrons, pions, and kaons. Pion/proton separation is achievable
    for tracks with $p < 0.9$~GeV/$c$.}\label{fig:ST_dEdx_vs_p}
\end{figure*}

The ST is also used to identify particles using $dE/dx$. Fig.~\ref{fig:ST_dEdx_vs_p} shows $dE/dx$ versus momentum, $p$,
for charged particles tracked to the Start Counter. Protons can be
separated from pions up to $p=0.9$~GeV/$c$.

\begin{figure}[tbp]
\begin{center}    
\includegraphics[width=0.6\textwidth]{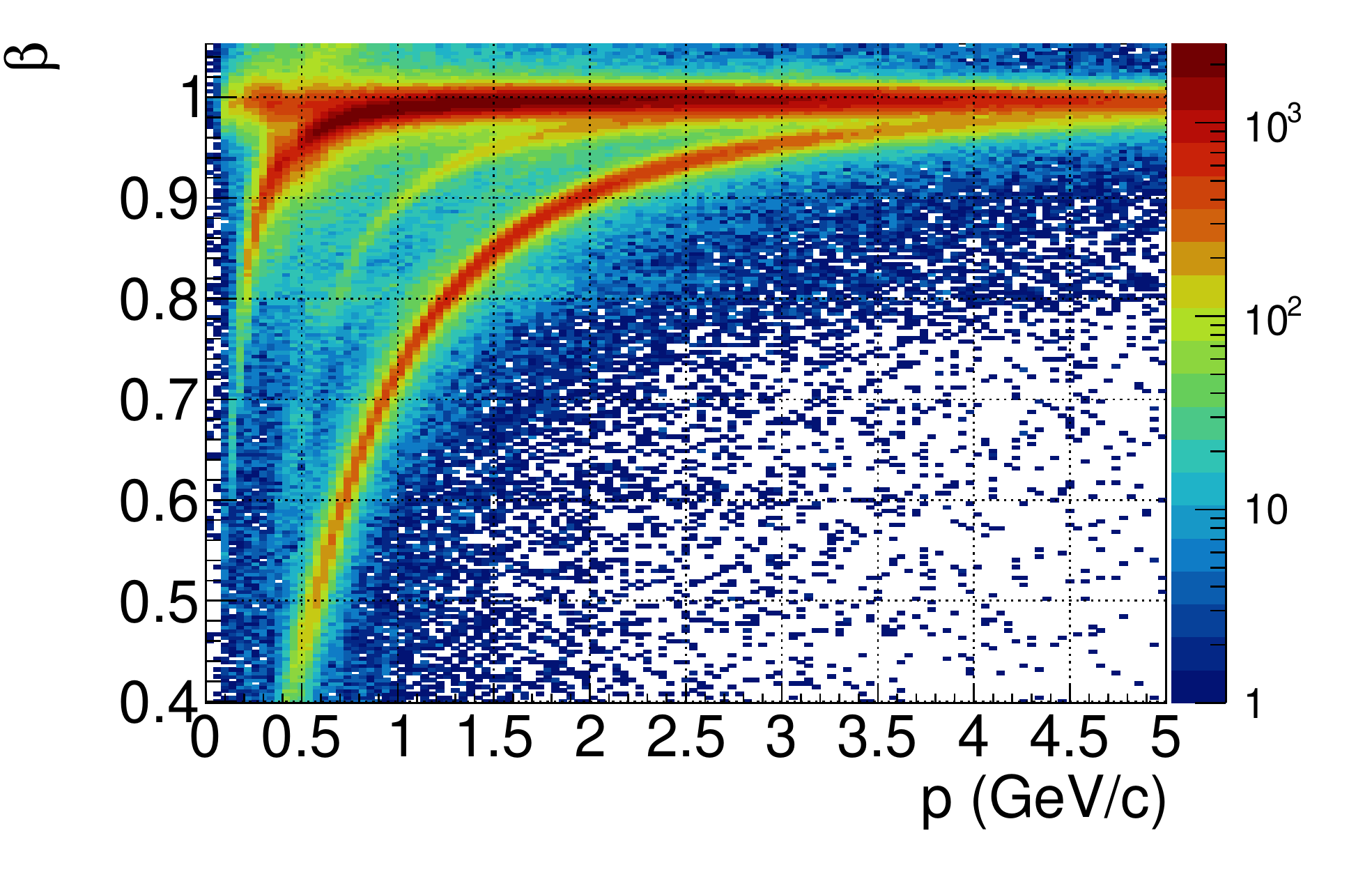}
\caption{\label{fig:betavsp}$\beta$ of positive tracks versus track momentum, showing bands for $e^+$, $\pi^+$, $K^+$ and $p$ for the TOF detector. The color coding of the third dimension
is in logarithmic scale.(Color online)}
\end{center}
\end{figure}

The performance of the TOF detector for particle identification (PID) was investigated by considering the relative number of
particle types within the event sample. Events with at least three fully-reconstructed positively-charged tracks were selected, with at least one of these tracks intersecting the TOF detector. More pions are expected than protons, and more protons than kaons. Looking at the distribution of velocity, $\beta$, of these tracks as a function of momentum, the bands from protons, kaons and pions are identified (see Fig.~\ref{fig:betavsp}). 

\begin{figure}[tbp]
\begin{center}
\includegraphics[width=0.45\textwidth]{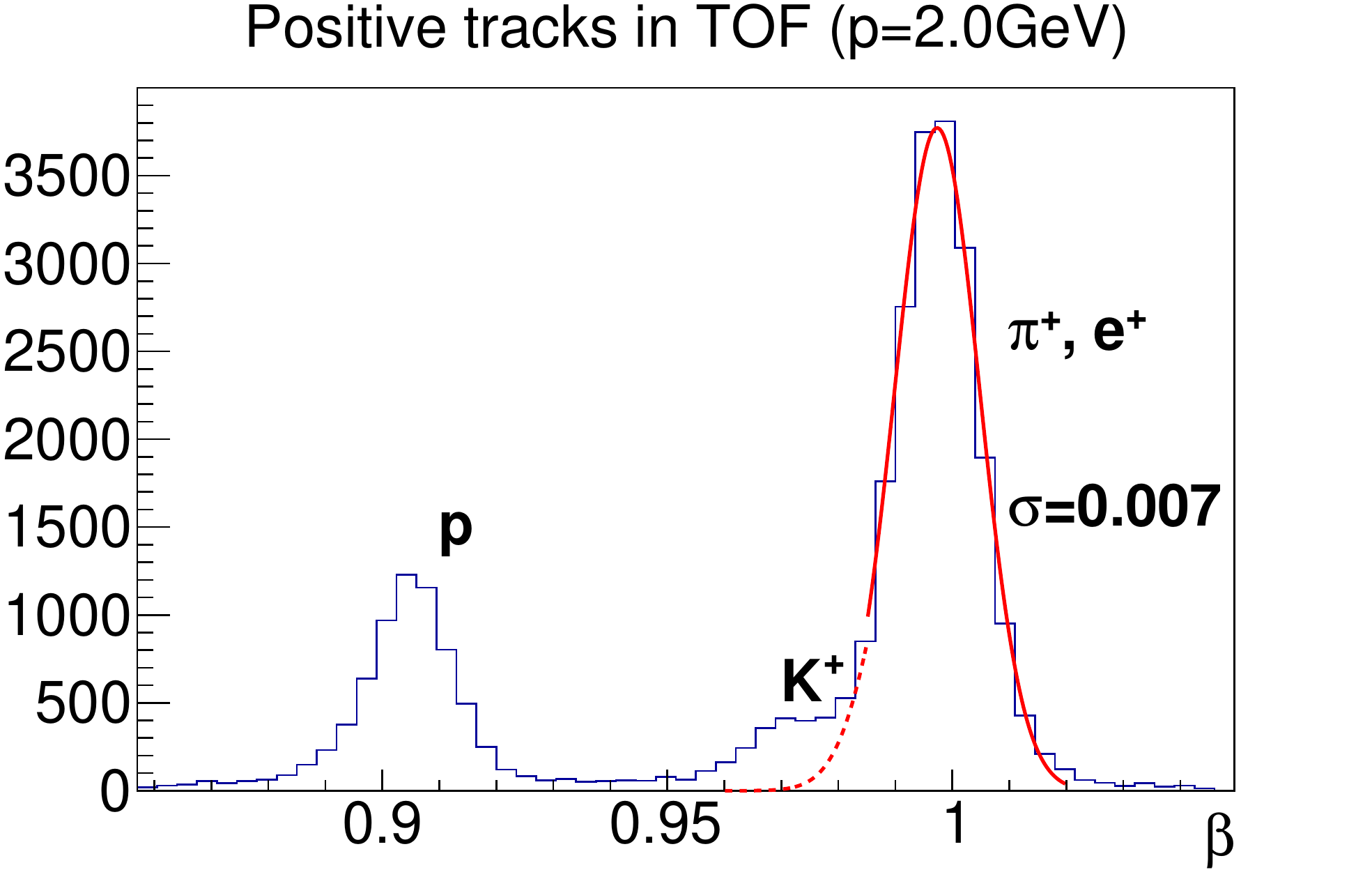}
\includegraphics[width=0.45\textwidth]{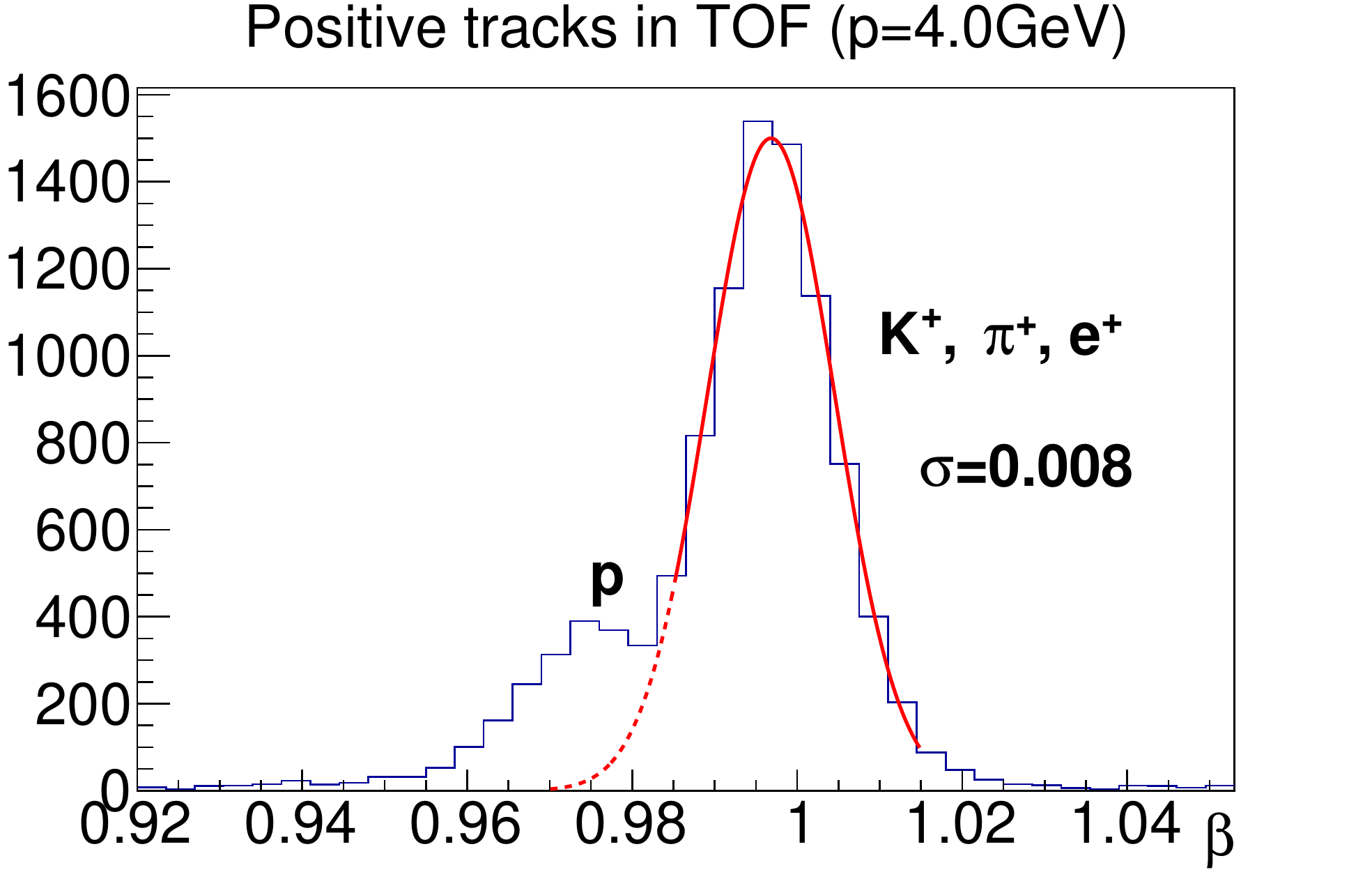}
\caption{\label{fig:betaproj}$\beta$ of positive tracks with 2~GeV/c momentum (left) and with 4~GeV/c (right).}
\end{center}
\end{figure}

The distributions of $\beta$ at two specific track momenta, 2~GeV/c and 4~GeV/c (see Fig.~\ref{fig:betaproj}), are illustrative of the PID capability of the TOF detector. At $p=2$~GeV/c, the TOF detector provides about a 4$\sigma$ separation between
the pion/positron peak and the kaon peak, sufficient to identify tracks as kaons with $\beta=0.97$, or lower, with very
high certainty. However, at  $\beta=0.98$, the probability of the track being a kaon is less than 50\%, due to the abundance of pions that is an order of magnitude larger than kaons. The protons, on the other hand, are very well
separated from the other particle types and can be identified with high confidence over the full range in $\beta$.
At a track momentum of 4~GeV/c, PID becomes much more difficult and represents the limit at which the time-of-flight measurement can identify protons with high confidence. The separation between the large peak containing pions, kaons and positrons from the proton
peak is about 4$\sigma$, while the relative abundance in this case is about a factor of 4. As a consequence, a 4~GeV/c momentum
track with $\beta=0.975$ is most likely a proton, with a small probability of being a pion. At $\beta=0.98$, such
a track has a similar probability for being a proton or a pion.

\section[Trigger]{Trigger \label{sec:trig}}
The goal of the \gx{} trigger is to accept most high-energy hadronic interactions while reducing the background rate induced by electromagnetic and low-energy hadronic interactions to the level acceptable 
by the data acquisition system (DAQ).  The main trigger algorithm is based on measurements of energy depositions in the FCAL and BCAL as described in Ref.~\cite{somov_l1,somov_l11}. Supplementary triggers can also use hits from scintillator detectors, such as the PS, tagging detectors, ST, TOF, and TAC.

\subsection{Architecture \label{sec:trigarchitecture}}
The \gx{} trigger system~\cite{GlueX:2013twa} is implemented on dedicated programmable pipelined electronics modules, designed at JLab using Field-Programmable Gate Arrays (FPGAs).  The \gx{} trigger and readout electronics are hosted in VXS (ANSI/VITA 41.0) crates. VXS is an extension of the VME/VME64x architecture, which uses high-speed backplane lines to transmit trigger information. 

\begin{figure}[tbp]
\begin{center}
\includegraphics[width=0.75\textwidth]{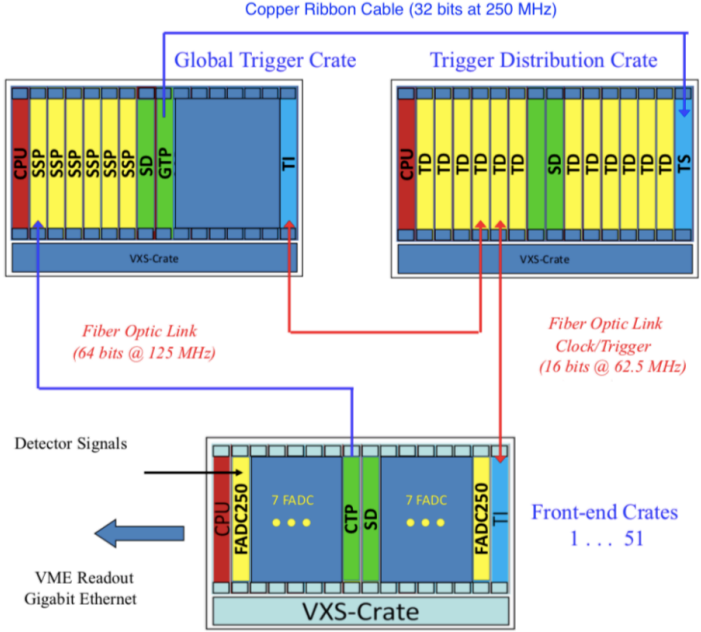}  
\caption{Schematic view of the Level-1 trigger system of the \gx{} experiment. The electronics boards are described in the text.} \label{fig:trig}
\end{center}
\end{figure}

A layout of the trigger system is presented in Fig.~\ref{fig:trig}. Data from the FCAL and BCAL are sent to  FADC modules~\cite{Dong:2007}, situated in 12 and 8 VXS crates, respectively, and are digitized at the sampling rate of 250 MHz. The digitized amplitudes are used for the trigger and are also stored in the FPGA-based pipeline for subsequent readout via VME.
Digitized amplitudes are summed for all 16 FADC250 channels in each 4 ns sampling interval and are transmitted to the crate trigger processor (CTP) module, which sums up amplitudes from all FADC boards in the crate. The sub-system processor (SSP) modules located in the global trigger crate receive amplitudes from all crates and compute the total energy deposited in the FCAL and BCAL. The global trigger processor (GTP) module collects data from the SSPs and makes a trigger decision based on the encoded trigger equations. The core of the trigger system is the trigger supervisor (TS) module, which receives the trigger information from the GTP and distributes triggers to the electronics modules in all readout 
crates in order to initiate the data readout. The \gx{} system has 55 VXS crates in total (26 with FADC250s, 14 with  FADC125s, 14 with F1 TDCs, and 1 CAEN TDC). The TS also provides a synchronization of all crates and provides a 250 MHz clock signal. The triggers and clock are distributed through the trigger distribution (TD) module in the trigger distribution crate. The signals are received by the trigger interface (TI) module and signal distribution (SD) module in each crate. The \gx{} trigger system provides a fixed latency. The longest trigger distribution time of about 3.3 $\mu$s is due to the distance of the tagger hall from Hall D.
The smallest rewritable readout buffer, where hits from the detector are stored, corresponds to about 3.7~$\mu$s for the F1 TDC module. The trigger jitter does not exceed 4 ns.

\subsection{Trigger types \label{sec:triggers}}

The \gx{} experiment uses two main trigger types: the pair spectrometer trigger, and the physics trigger based on energy depositions in the BCAL and FCAL. The 
pair spectrometer trigger is used to measure the flux of beam photons. This trigger requires a time coincidence of hits in the 
two arms of the PS detector, described in Section~\ref{sec:ps}. The physics triggers are generated when the FCAL and BCAL energies  satisfy the following conditions: 
\begin{enumerate}
\item $2\times E_{\rm FCAL} + E_{\rm BCAL}  > 1\;{\rm GeV},  E_{\rm FCAL} > 0\; {\rm GeV}$, \rm{and}  \\
\item $E_{\rm BCAL}  >  1.2\;{\rm GeV}$.
\end{enumerate}
The first condition defines the main trigger that uses the fact that most events produce forward-going energy. The second trigger type is used to accept events with large transverse energy released in the BCAL, such as decays of $J/\psi$ mesons. 

Several other trigger types were implemented for efficiency studies and detector calibration. 
Efficiency of the main production trigger was studied using a trigger based on the coincidence of hits from the ST and TAGH, detectors not used in the main production trigger. A combination of the PS and TAC triggers was used for the acceptance calibration of the PS, described in Section~\ref{sec:ps_flux}. Ancillary minimum-bias random trigger and calorimeter LED triggers were collected concurrently with data taking.

\subsection{Performance \label{sec:trigperformance}}
\begin{figure}[tbp]
\begin{center}
\includegraphics[width=0.5\textwidth]{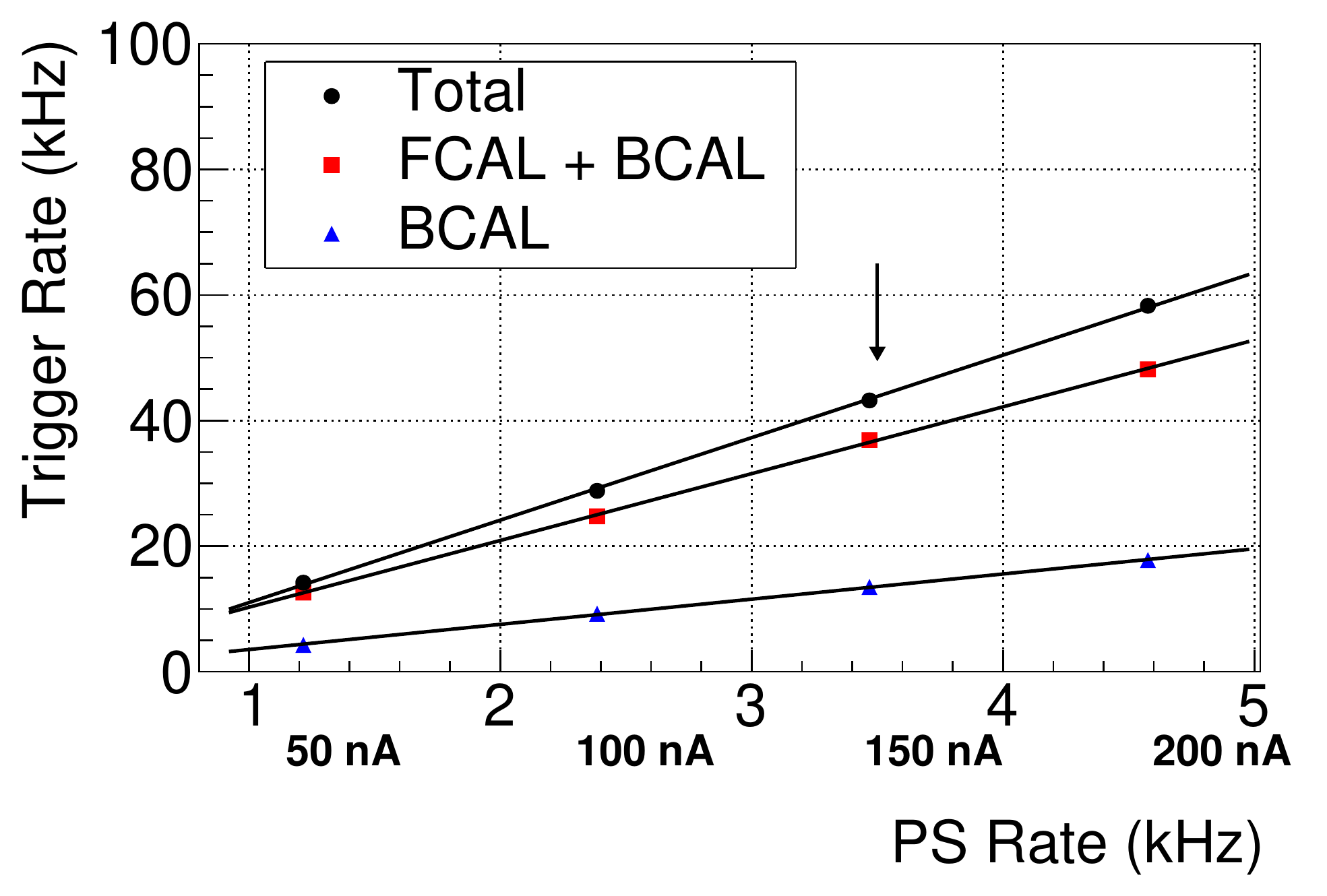}  
\caption{Rates of the main production triggers as a function of the PS rate: FCAL and BCAL trigger (boxes), BCAL trigger (triangles), the total trigger rate (circles). The vertical arrow indicates the run conditions during the spring of 2018 with a diamond
radiator, 5 mm collimator and 75 $\mu$m Be converter.} \label{fig:trig_rate}
\end{center}
\end{figure}

The rate of the main physics triggers as a function of the PS trigger rate is shown in Fig.~\ref{fig:trig_rate}.
The typical rate of the PS trigger in spring 2018 was about 3~kHz, which corresponds to a photon beam flux of $2.5\cdot 10^7\; \gamma/{\rm sec}$ in the coherent peak range. The total trigger rate was about 40 kHz. The rates of the random trigger and each of the LED calorimeter triggers were set to 100 Hz and 10 Hz, respectively. The electronics and DAQ were running with a livetime close to 
$100 \%$, collecting data at a rate of 600 MB per second.
The trigger system can operate at significantly higher rates, considered for the next phase
of the GlueX experiment. The combined dead time of the trigger and DAQ systems at the trigger rate of 80 kHz
was measured to be about $10 \%$. The largest contribution to the dead time comes from the hit processing
time of readout electronics modules. 


\section[Data Acquisition]{Data acquisition \label{sec:daq}}



The \gx{} data acquisition software uses the CEBAF Online Data Acquisition (CODA) framework. CODA is a software toolkit of applications and libraries that allows customized data acquisition systems based on distributed commercial networks. A detailed description of CODA software and hardware can be found in Ref.\,\cite{CLAS12DAQ}. 

\begin{figure}[tbp]
\begin{center}
\includegraphics[width=0.75\textwidth]{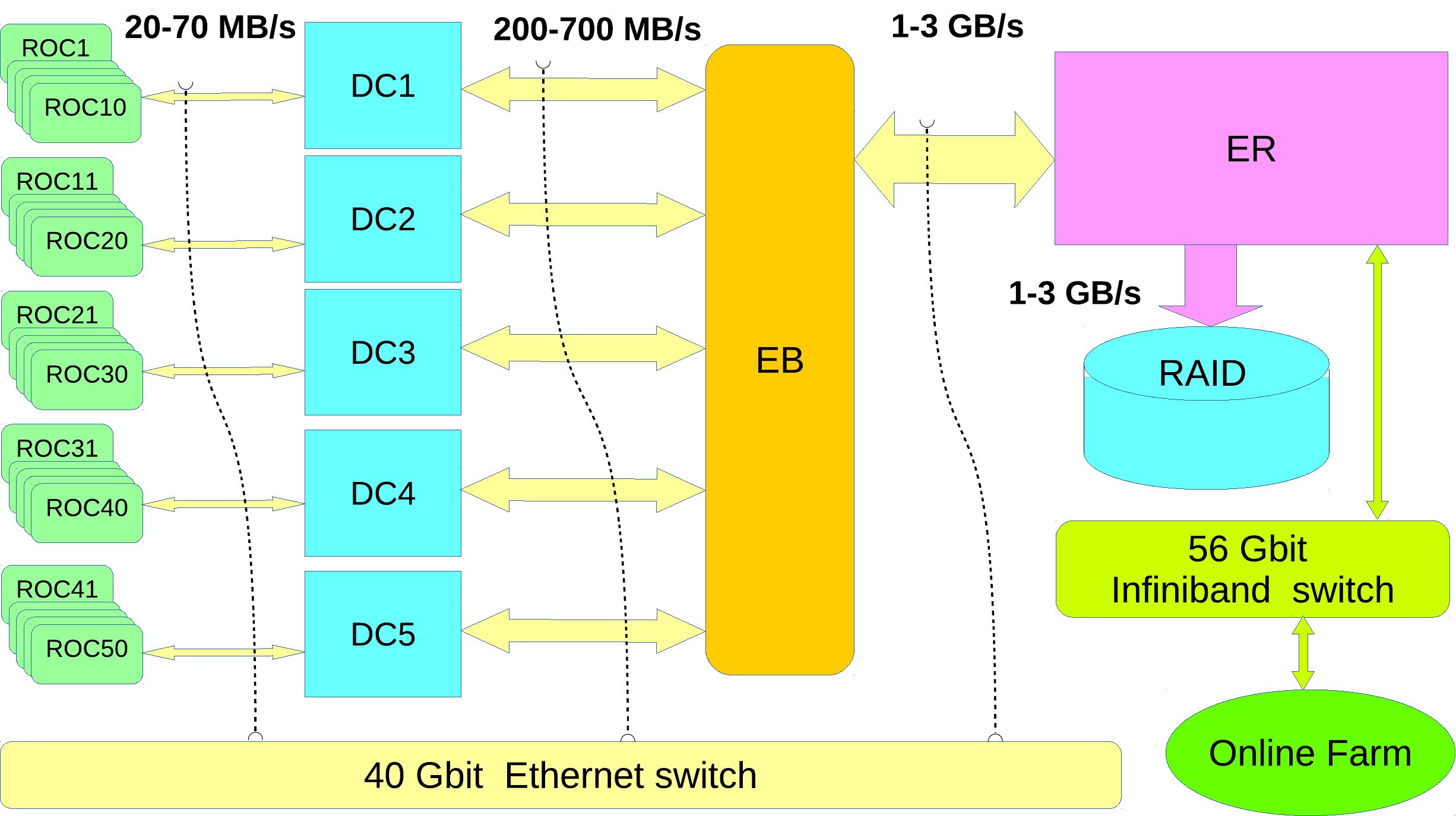}  
\caption{ \label{fig:CODA}
Schematic DAQ configuration for \gx. The high-speed DAQ connections between the ROCs and the ER are contained within an isolated network. The logical data paths are indicated by arrows,
although physically they are routed through the 40 Gbit ethernet switch.  The online monitoring system uses its own separate 56 Infiniband switch.}

\end{center}
\end{figure}

The maximum readout capability of the electronics in the VME/VXS crate is 200 MB/s per crate and the number of crates producing data is about 55.
The data from the electronic modules are read via the VME back-plane (2eSST, parallel bus) by the crate readout controller (ROC), which is a single-board computer running Linux.
The \gx~ network layout and data flow are shown in Fig.~\ref{fig:CODA}.
Typical data rates from a single ROC are in the range of 20--70~MB/s, depending on the detector type and trigger rate.
The ROC transfers data over 1~Gbit Ethernet links to Data Concentrators (DC) using buffers containing event fragments from 40 triggers at a time. Data Concentrators are programs that build partial events received from 10-12 crates and run on a dedicated computer node.
The DC output traffic of 200-600 MB/s is routed to the Event Builder (EB) to build complete events.
The Event Recorder (ER), which is typically running on the same node as an Event Builder, writes data to local data storage.
\gx{} has been collecting data at a rate of 500--900 MB/s, which allows the ER to write out to a single output stream. The system is expandable to handle higher luminosity where rates rise to 1.5--2.5~GB/s. In this case, the ER must write multi-stream data to several files in parallel.
All DAQ computer nodes are connected to both a 40 Gb Ethernet switch and a 56 Gb Infiniband switch.
The Ethernet network is used exclusively for DAQ purposes: receiving data from detectors, building events, and writing data to disk, 
while the Infiniband network is used to transfer events for online data quality monitoring. 
This allows decoupling DAQ and monitoring network traffic.
The livetime of the DAQ is in the range of 92--100\%. The deadtime arises from readout electronics and depends on the trigger rate.  
The DAQ software does not cause dead time during an experimental run, but software-related dead time appears while stopping and starting the run, which takes between 2-8 minutes. 

\begin{landscape}
\begin{figure}[tbp]
\begin{center}

\includegraphics[height=10cm,clip=true]{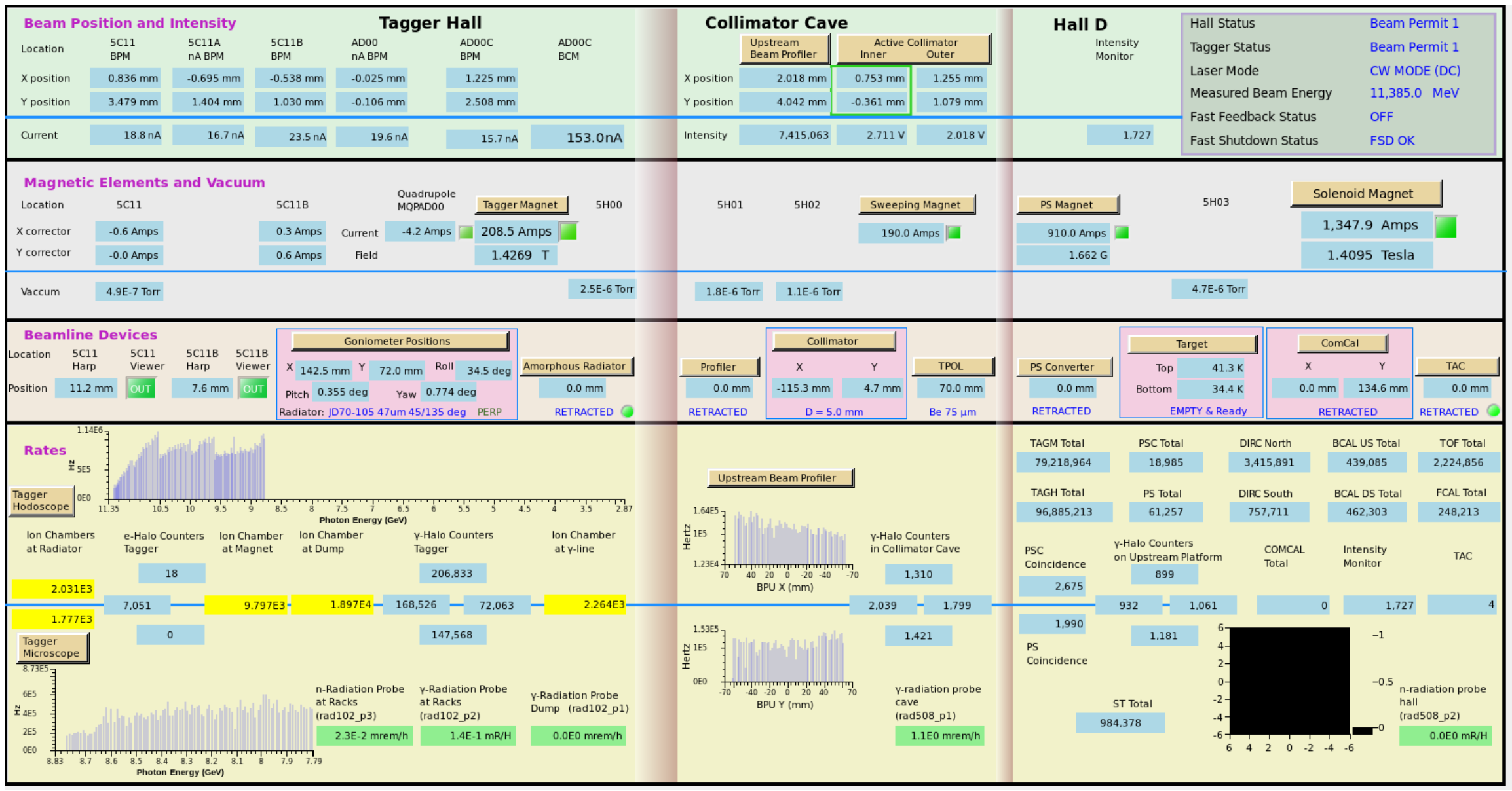}
\caption{Top-level graphical interface for the beamline. This screen provides information on beam currents and rates, radiators, magnet status, target condition, background levels, etc.
\label{fig:GlueX_CSS_overview}
}
\end{center}
\end{figure}
\end{landscape}

\section[Slow controls]{Slow controls \label{sec:controls}}
\GX{} must monitor 
and control tens of thousands of different variables that define the state of the experimental hardware. The values need to be acquired, displayed, archived, and used as inputs to control loops continually with a high degree of reliability. For \gx, approximately 90,000 variables are archived, and many more are monitored.

\subsection{Architecture \label{sec:controlsarchitechture}}
The \gx{} slow control system consists of three layers. The first layer consists of the remote units such as high voltage or low voltage power chassis, magnet power supplies, temperature controller, LabView applications, and PLC-based applications, which directly interact with the hardware and contain almost the all the control loops. The second layer is the Supervisory Control and Data Acquisition (SCADA) layer, which is implemented via approximately 140 EPICS Input/Output Controllers (IOCs). This layer provides the interface between low level applications and higher level applications via the EPICS ChannelAccess protocol. The highest level, referred as the Experiment Control System (ECS), contains applications such as Human-Machine Interfaces, the alarm system, and data archiving system. This structure allows for relatively simple and seamless addition and integration of new components into the overall controls system.    

\subsection{Remote Units \label{sec:controlsinterface}}
\gx{} uses a variety of commercial units to provide control over the hardware used in the experiment. For instance, most detector high voltages are provided by the CAEN SYx527 voltage mainframe,\footnote{https://www.caen.it/subfamilies/mainframes/} while the low and bias voltages are provided by boards residing in a Wiener MPOD chassis\footnote{http://www.wiener-d.com/sc/power-supplies/mpod--lvhv/mpod-crate.html}. These two power supply types provide most voltages for detector elements with the exception of the Tagger Microscope and the Forward Calorimeter. Here custom systems were developed that provide voltage regulation and interact with the EPICS-based layer through higher level interfaces using custom protocols. See Sections~\ref{sec:TAGM} and \ref{sec:fcal} for more details.  

Various beam line devices need to be moved during beam operations. Stepper motors are used to move motorized stages via Newport XPS universal multi-axis motion controllers\footnote{https://www.newport.com/c/xps-universal-multi-axis-motion-controller.} that allow for execution of complex trajectories involving multiple axes. All stage referencing, motion profile computations, and encoder-based closed-loop control occurs within the controller chassis after the basic parameters, such as positions and velocities, are provided by the user via a TCP/IP-based interface to EPICS.   

Custom controls were often developed for each complex installation, such as a superconducting magnet that requires large numbers of input and output channels and sophisticated logic.
For these cases, we used Allen-Bradley CompactLogix and ControlLogix PLC systems\footnote{https://ab.rockwellautomation.com.}. These systems are designed for industrial operations, allow modular design, provide high reliability, and require minimal maintenance. All controls loops are programmed within the PLC application, and are interfaced with EPICS through a TCP/IP-EtherNet/IP-proprietary protocol to allow access by higher level applications to process variables delivered by the PLCs.  

The cryogenic target and the superconducting solenoid employ National Instruments LabView applications. The target controls use both custom-made and vendor-supplied hardware that include built-in remotely-accessible control systems and an NI CompactRIO\footnote{https://www.ni.com/en-us/shop/compactrio.html} chassis. This chassis communicates with the hardware and serves variables using an internal ChannelAccess server and an EPICS IOC running on the CompactRIO controller, as described in Sec.~\ref{sec:target}. A National Instruments PXI high-performance system\footnote{https://www.ni.com/en-us/shop/pxi.html} is used to collect data from different sensors of the solenoid as described in Sec.~\ref{sec:solenoid}. 

\subsection{Supervisory Control and Data Acquisition layer \label{sec:archiver}}
The SCADA layer is the middle layer that distributes the process variables allowing the higher level --and sometimes lower level-- applications to use various process variables of the Hall-D control system. This layer is based on EPICS and uses the ChannelAccess protocol to publish the values of the variables over Ethernet. Efficient exchange of the information between the experiment and accelerator operations is achieved because the accelerator controls also use EPICS.
Several dozen software IOC processes, running on host computers of the experiment control process, collect data from different components of the lowest layer. Each IOC is configured to communicate using the protocol appropriate for the remote units with which data exchange is needed. For instance, the IOC controlling the voltage for the FDC detector needs to be able to communicate with the Wiener MPOD and CAEN SYx527 voltage chassis. The middle layer is primarily used to distribute data between different applications. This layer also contains some EPICS-based applications running on IOCs that provide different control loops and software interlocks.  For instance, the low-voltage power supplies for the FDC detector (see Sec. \ref{sec:fdc}) are shut off if the temperature or the flow of the coolant in the chiller falls outside of required limits. 
\subsection{Experiment Control System \label{sec:alarms}}
The highest level of controls contains applications that archive data, display data in interactive GUIs and as stripcharts, alarm and notify shift personnel and experts when problems occur, and interface with the CODA-based data acquisition system (Sec.~\ref{sec:daq}).
An example of such a GUI is the beamline overview screen, shown in Fig.\,\ref{fig:GlueX_CSS_overview}. Many of the buttons of the GUI are active and allow access to other GUIs.
Display management and the alarm system for \gx{} controls are based on Controls System Studio (CSS),\footnote{http://controlsystemstudio.org/}  which is an Eclipse-based toolkit for operating large systems. CSS is well suited for systems that use EPICS as an integral component. Although CSS provides an archiving engine and stripcharting tools, the MYA archiver,\cite{Slominski:2009icaleps} provided by the JLab accelerator software group, was employed with its tools for displaying the archived data as a time-series. Display management for \gx{} controls is within the CSS BOY~\cite{Chen:2011icaleps} environment, which allows system experts to build sophisticated control screens using standard widgets. The alarm system is based on the CSS BEAST\cite{Kasemir:2009icaleps} alarm handler software, which alerts shift personnel of problems with the detector, and notifies a system expert if the problems are not resolved by shift personnel.   

\section[Online computing system]{Online computing system \label{sec:online}}

This section describes the \GX ~software and computing systems  used for data monitoring and for transport to the tape system for permanent storage.

\subsection{Monitoring \label{sec:onlinemonitoring}}

\begin{figure}[tbp]
\begin{center}
\includegraphics[width=0.99\textwidth, clip,trim=1.5cm 0.9cm 1.7cm 0.8cm]{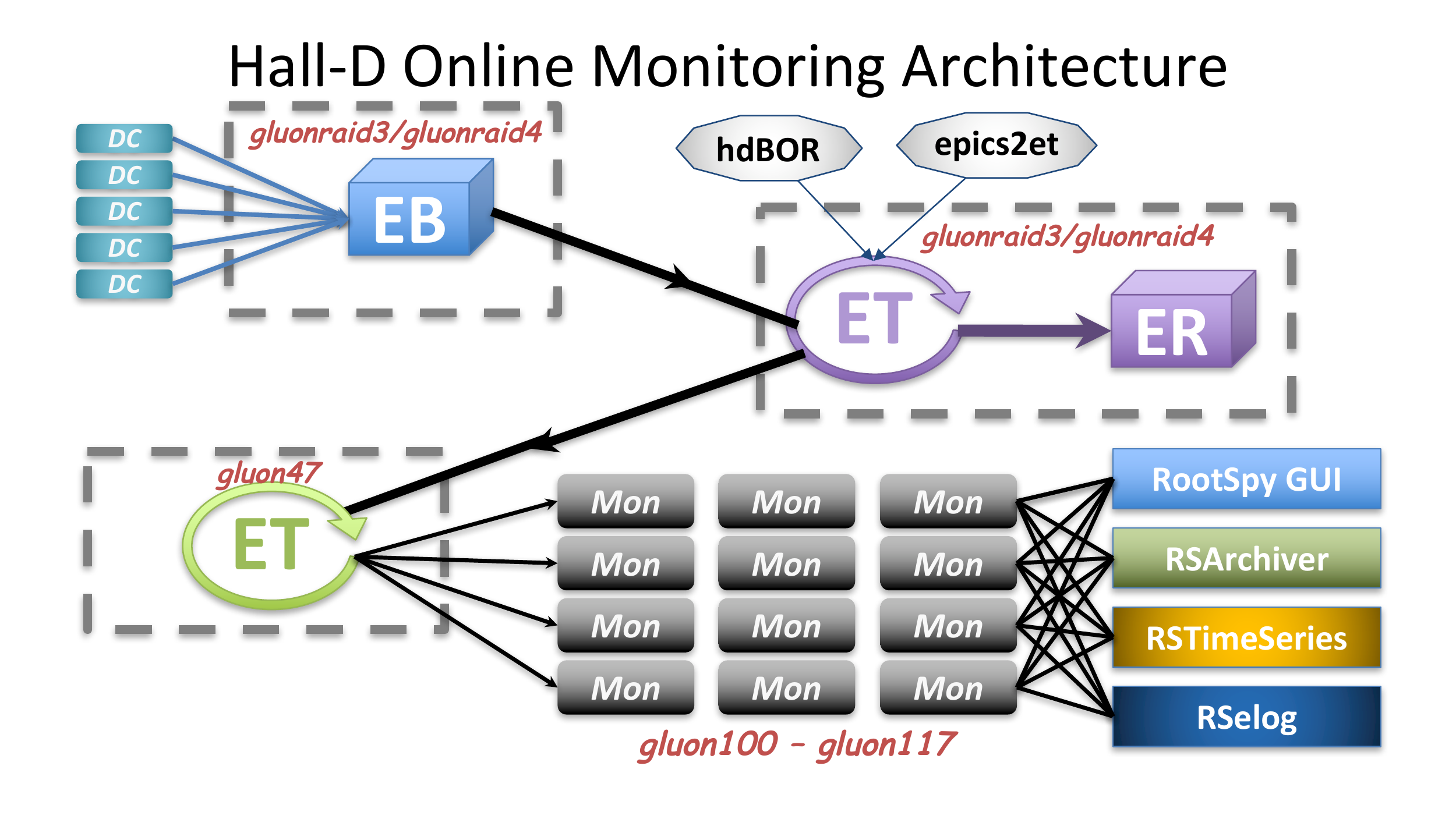}
\caption{\label{fig:online_monitoring_processes}Processes distributed across several computers in the online monitoring system. DC, EB, and ER are the Data Concentrator, Event Builder, and Event Recorder processes, respectively, in the CODA DAQ system.}   
\end{center}  
\end{figure}

The Online Monitoring system consists of multiple stages that provide immediate monitoring of the data, as well as near-term monitoring (a few hours after acquisition). Immediate monitoring is based on the \textit{RootSpy} system\cite{rootspy} written for use in \GX, though its design is not experiment specific. Figure \ref{fig:online_monitoring_processes} shows a diagram of the processes involved in the RootSpy system and how those processes are coupled to the DAQ system. The Event Transfer (ET) process is part of the CODA DAQ system \cite{coda} and is used to extract a copy of a portion of the datastream without interfering with data acquisition. The monitoring system uses a secondary ET to minimize connections to the RAID server running the Event Recorder process.

\begin{figure}[tbp]
\begin{center}
\includegraphics[width=0.99\textwidth]{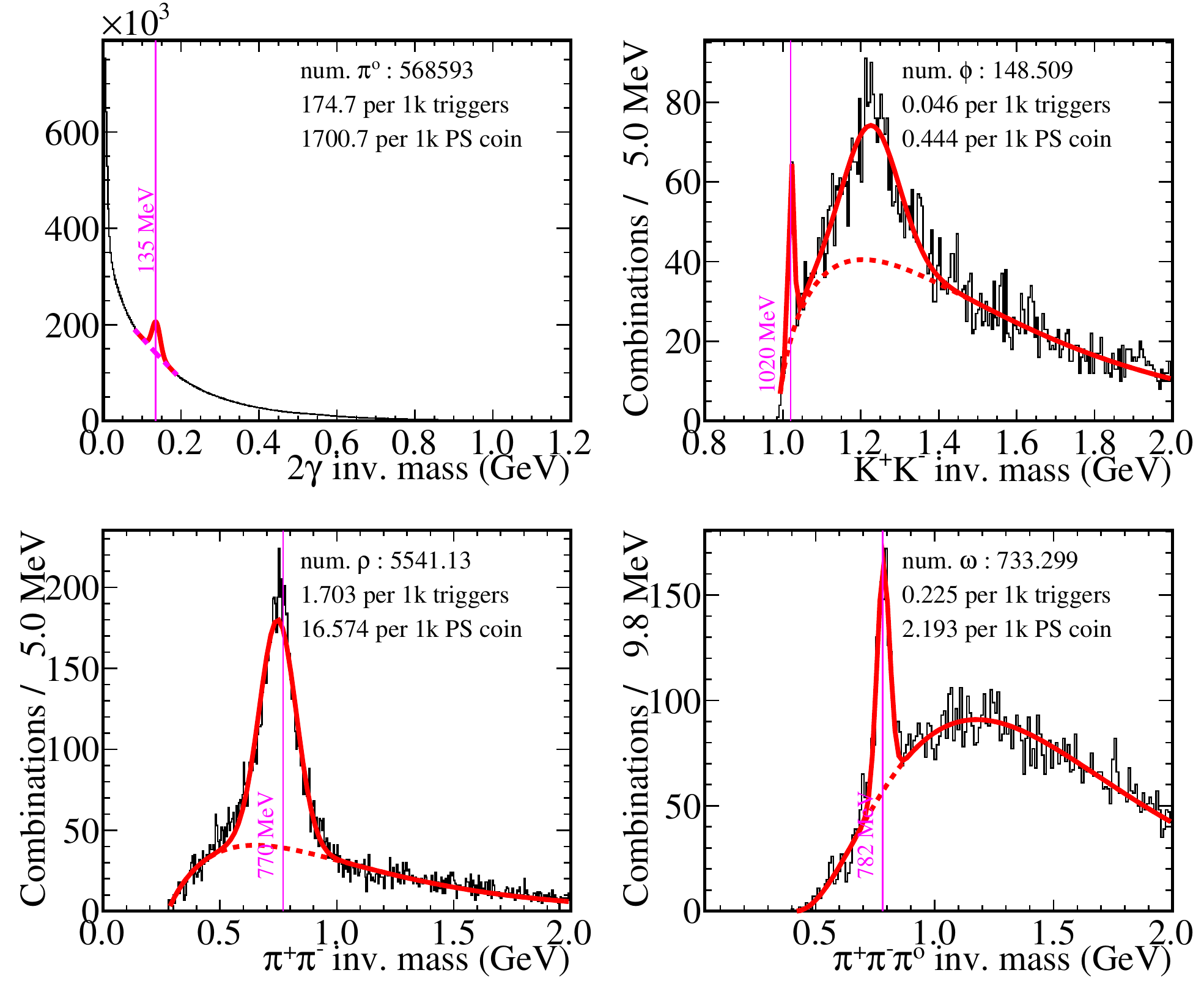}
\caption{\label{fig:online_monitoring_PID}Invariant mass distributions showing $\pi^\circ$, $\omega$, $\rho$, and $\phi$ particles. These plots were generated online in about 1hr 40min by looking at roughly 2\% of the data stream.}   
\end{center}  
\end{figure}

The monitoring system is run on a small computer farm\footnote{The online monitoring farm consists of eight 2012 era Intel x86\_64 computers with 16 cores+16 hyper-threads (ht) plus six 2016 era Intel x86\_64 computers with 36 cores + 36ht. The monitoring farm uses 40 Gbps (QDR) and 56 Gbps(FDR) IB for the primary interconnect. Note that the DAQ system uses a separate 40 Gbps ethernet network that is independent of the farm.}, with each computer processing a small part of the data stream. In total, about 10\% of the data is processed for the low level occupancy plots while roughly 2\% is fully reconstructed for higher level analysis. The CODA ET software system is used to distribute the data among the farm computers. Each farm node generates histograms, which \textit{RootSpy} gathers and combines before display to shift workers in a GUI.
Plots are displayed via a set of ROOT \cite{Brun:1997pa} macros, each responsible for drawing a single page. Most macros divide the page into multiple sections so that multiple plots can be displayed on a single page. Figure \ref{fig:online_monitoring_PID} shows an example of a high-level monitoring plot, where four invariant-mass distributions are shown with fits. Values extracted from the fits are printed on the plots for easy quantitative comparison to a reference plot. 



There are several client programs that summarize the information available in the histograms produced by \textit{RootSpy} and generate output that make it easy to assess the uniformity and quality of the data. One of these is the \textit{RSTimeSeries} program, which periodically inserts data into an InfluxDB time series database. The database provides a web-accessible strip chart of detector hit rates and reconstructed quantities (e.g. number of $\rho$'s per 1k triggers). Another is the \textit{RSArchiver} program that gathers summed histograms to be displayed in the Plot Browser\footnote{https://halldweb.jlab.org/data\_monitoring/Plot\_Browser.html.} website. Plot Browser provides easy comparison of plots between different runs and between different analysis passes. Jobs are automatically submitted to the JLab farm for full reconstruction of the first five files (100GB) of each run. The results are displayed in Plot Browser and may be compared directly with the online analysis of the same run.

\subsection{Data transport and storage \label{sec:onlineprocessing}}

\GX ~Phase I generated production data at rates up to 650MB/s. The data were temporarily stored on large RAID-6 disk arrays, and then copied to an LT0 tape system in the JLab Computer Center for long term storage. Two RAID servers, each with four partitions, were used for staging the data. The partition being written was rotated between runs  to minimize head thrashing on disks by only reading partitions not currently being written. Partitions were kept at approximately 80\% capacity and older files were deleted to maintain this level,  allowing the monitoring farm easy access to files when the beam was down. A copy of the first three files ($\sim1.5\%$) of each run was also kept on the online computers for direct access to samples from each run.      

The data volumes stored to tape are shown in Table \ref{tab:online_data_volumes} in units of petabytes (PB). Entries marked ``actual'' are values taken from the tape storage system. The line marked ``model'' comes from the \GX ~computing model\cite{gx3821}.

\begin{table}[tb]
    \centering
    \begin{tabular}{|l|c|c|c|c|c|}
    \hline
                           & \textbf{2016}  & \textbf{2017}  & \textbf{2018} \\
    \hline
    actual (raw data only) & 0.624 & 0.914 & 3.107 \\
    \hline
     model (raw data only) &       & 0.863 & 3.172 \\
    \hline
    \hline
    actual (production data)    & 0.55  & 1.256 & 1.206 \\
    \hline
    \end{tabular}
    \caption{\GX{} data volumes by year. All values are in petabytes (PB). Most years include two run periods. The line marked ``model'' gives calculated rates from the \GX ~Computing Model\cite{gx3821} based on the detector luminosity. ``Raw data only'' represents data generated by the DAQ system (not including the backup copy). ``Production'' represents all derived data including reconstructed values and ROOT trees. }
    \label{tab:online_data_volumes}
\end{table}


\section[Event reconstruction]{Event reconstruction \label{sec:reconstruction}} 


\GX~uses the computer center batch farm at JLab to perform data monitoring, event reconstruction, and physics analyses.  For data monitoring, detector hit occupancies, calibration and reconstruction quality, and experimental yields and resolutions, are analyzed for several physics channels.  A subset of the data is monitored automatically as it is saved to tape.  Every few weeks, monitoring processes are launched on a subset of the data to study improvements from ongoing calibrations and reconstruction software improvements.  The histograms produced by these monitoring jobs are displayed on a website and ROOT files are available for download, enabling the collaborators to easily study the quality of the data. 

Every few months, a major reconstruction launch over all of the data is performed, linking hits in the various detector systems to reconstruct particles in physics events.  Monitoring plots from these launches are also published to the web. Finally, regular analysis launches over the reconstructed data are performed for the reactions requested by users on a web form. The results of these launches are saved in reaction-specific ROOT TTrees for further analysis.

\begin{figure}[h!]\centering
\includegraphics[width=1.0\textwidth]{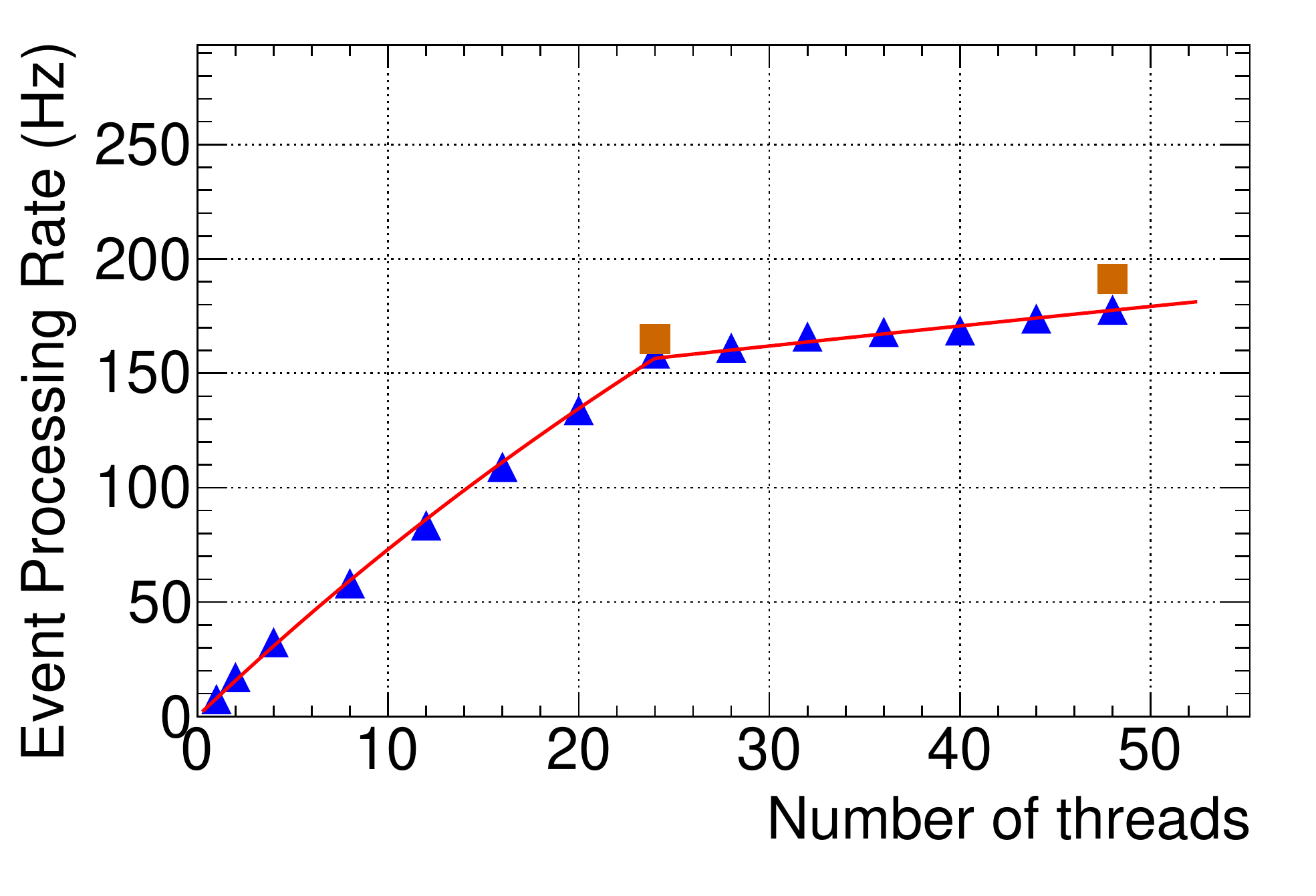}
\caption[]{\label{fig:offline_monitorA}The scaling of program performance as a function of the number of processing threads. The computer used for this test consisted of 24 full cores (Intel x86\_64) plus 24 hyperthreads. The orange squares are from running multiple processes, each with 12 threads.} 
\end{figure}

For all launches, the reconstruction is run in a multi-threaded mode to make efficient use of the available computing resources. Fig.\,\ref{fig:offline_monitorA} shows the multithreaded scaling from our monitoring launches. The program performs near the theoretical limit for jobs that use a number of threads that is less than or equal to the number of physical cores on the processor. By using hyperthreads, a smaller but still significant gain is achieved.
All file outputs are written to a write-through cache system, which is ultimately backed up to tape.

\begin{figure}[hbt]\centering
\includegraphics[width=0.9\textwidth]{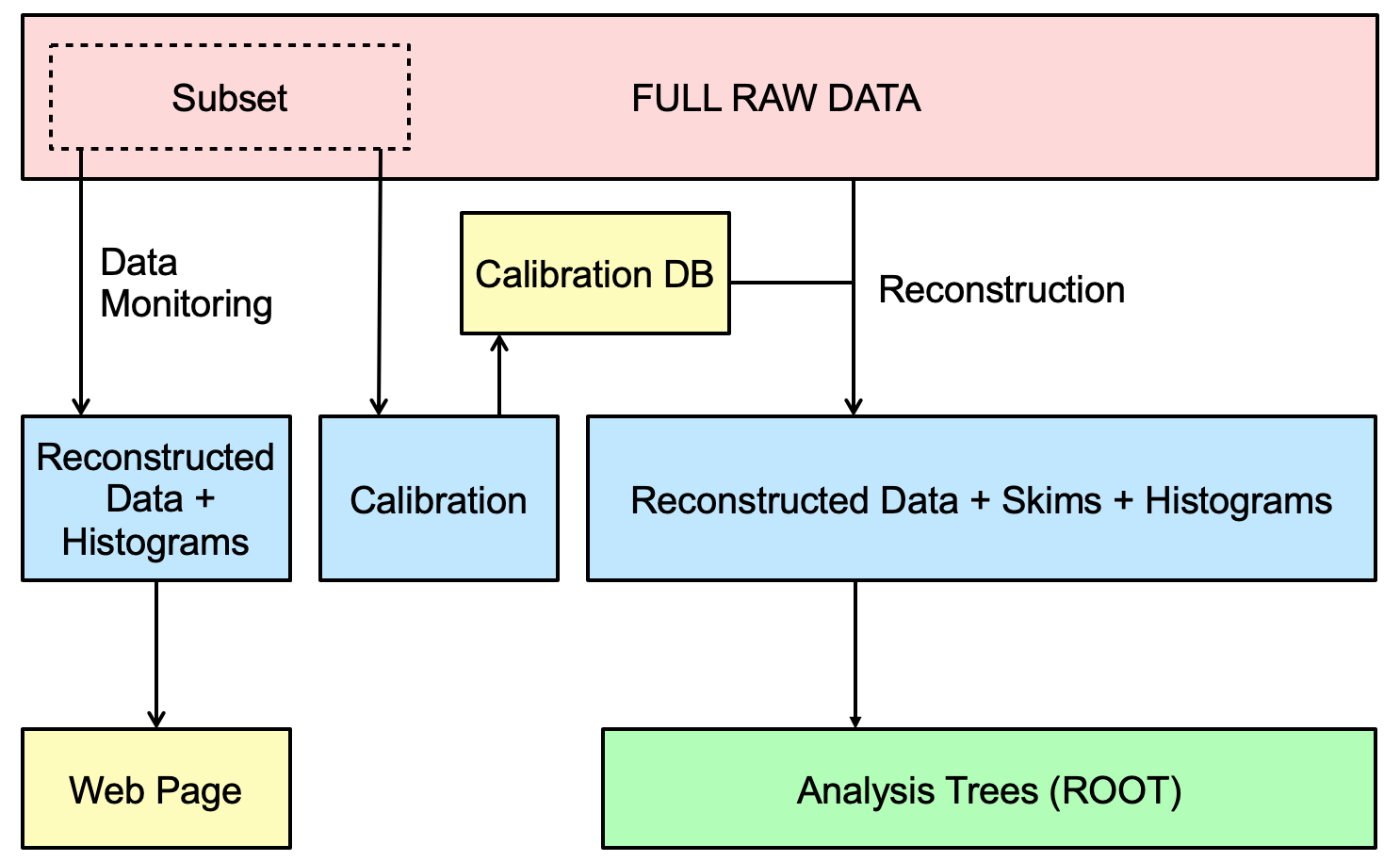}
\caption[]{\label{fig:production_overview}Production flowchart for \GX~data, illustrating analysis steps.} 
\end{figure}

\GX~ Phase I has recorded 1400 separate physics-quality runs, with a total data footprint of about 3 petabytes. Data were saved in 19-GB files, with all runs consisting of multiple files (typically $100$ or more per run). Fig.\,\ref{fig:production_overview} shows an overview of the different production steps for \GX~data, which are described in more detail in the following subsections.

\subsection{Calibration \label{sec:reccalibration}}
During the acquisition of data, a unique run number is assigned to a period of data corresponding to less than about 2 hours of clock time, which may result in writing a couple hundred files. It is assumed that the detector changes very little during this period and therefore there will be no changes in the calibration constants.
Two types of calibration procedures are used, depending on the complexity of the calibration procedures.  Simple, well-understood calibrations such as timing alignment between individual channels and subdetectors or drift chamber gain and time-to-distance calibrations, can be performed with one file of data per run.  These procedures are executed either in the online environment or on the batch farm, and can be repeated as needed following any improvements in reconstruction algorithms or other calibrations.

More complicated calibration procedures, such as calorimeter gain calibration, require more data and are often iterative procedures, requiring several passes through the data.  The raw data are processed upon arrival on the batch farm, resulting in histograms or in selected event data files in EVIO \cite{EVIO} or ROOT-tree format.
Many of these outputs require that charged particle tracks are reconstructed. However, the computationally intensive nature of track reconstruction makes it a challenge to fully reconstruct all raw data immediately. Therefore, the full suite of calibration procedures is only applied to 10 - 20\% of the data.
Processing of the remaining data is mostly focused on separating out, or ``skimming,'' events collected by calibration triggers.

\subsection{Monitoring \label{sec:recmonitoring}}

In Fig.~\ref{fig:production_overview} the ``FULL  RAW DATA'' box represents experimental data that have been backed up to tape. The box labeled ``subset" represents the first five files of each run, which are run through offline monitoring processes. These monitoring jobs are first processed during the run to check the quality of the data, but are also processed after major changes to calibrations or software to validate those changes.
The resulting Reconstructed Events Storage (REST) files and ROOT histogram files are used for checking the detector and reconstruction performance.

\subsection{Reconstruction \label{sec:recreconstruction}}

When the data have been sufficiently well calibrated, a full (production) pass of the reconstructed software on the physics quality data is performed. In the current total \GX~data set, about 1400 runs were deemed ``physics quality." The remaining runs were short runs related to engineering and commissioning tests of the experiment. The 1400 physics quality runs include the majority of the data recorded during the running period, representing about 3 petabytes. All these files were reconstructed using computing resources at several sites, equivalent to more than 20 million core-hours combined. This produced more than $500$ terabytes of REST data files. The large reduction in size from collected event data to physics data files (about a factor of six) permits faster and more efficient physics analyses of the data.

During the REST production, a series of detector studies were performed that required access to raw data and that would not be possible on the reconstructed data alone. Many improvements to software and detector calibration resulted from these studies. Similar studies can be made with simulated data to match and assess the detector acceptance.

\subsection{Offsite reconstruction}
\label{sec:recoffsite}

\begin{figure}[htb]\centering
\includegraphics[width=0.49\textwidth]{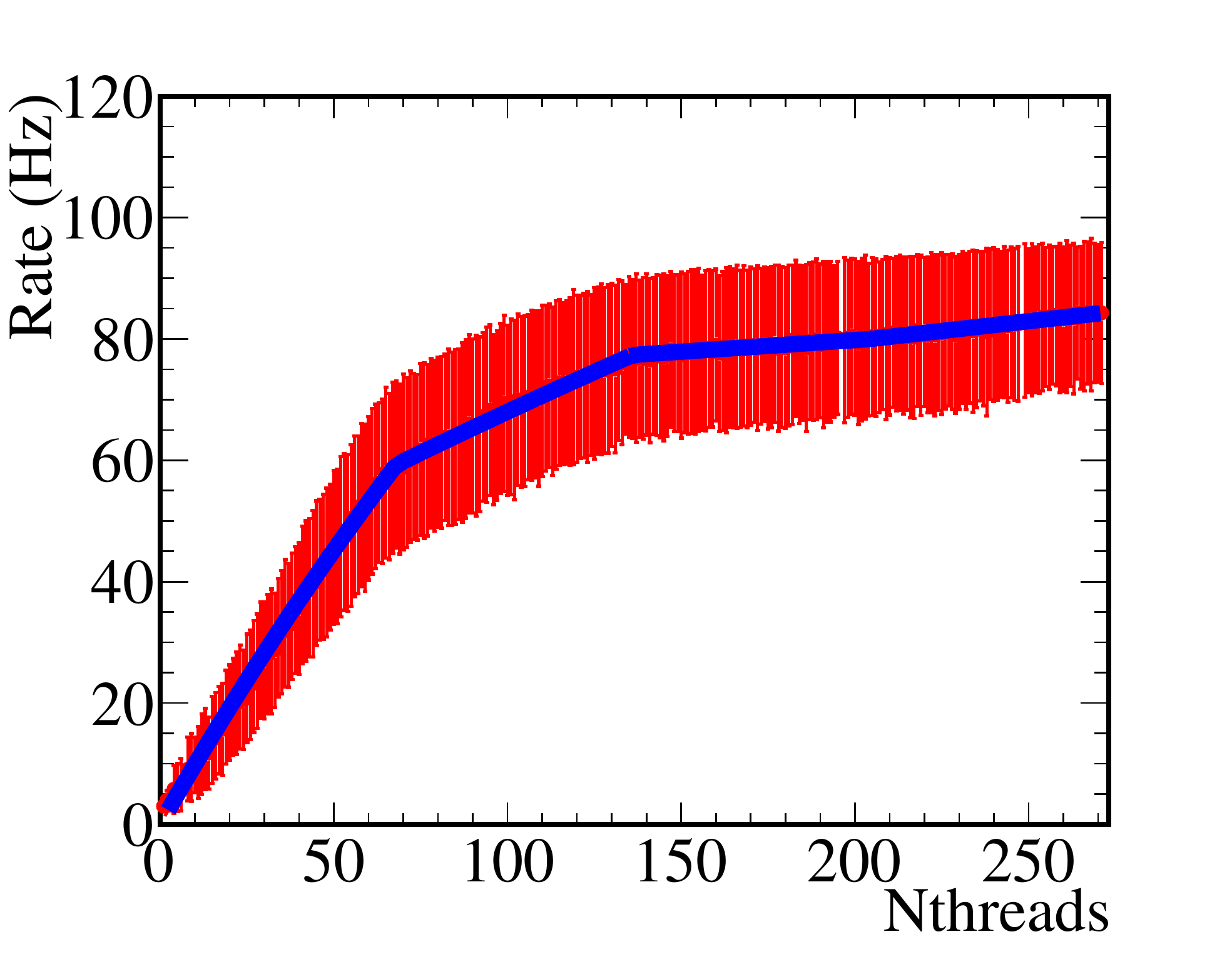}
\includegraphics[width=0.49\textwidth]{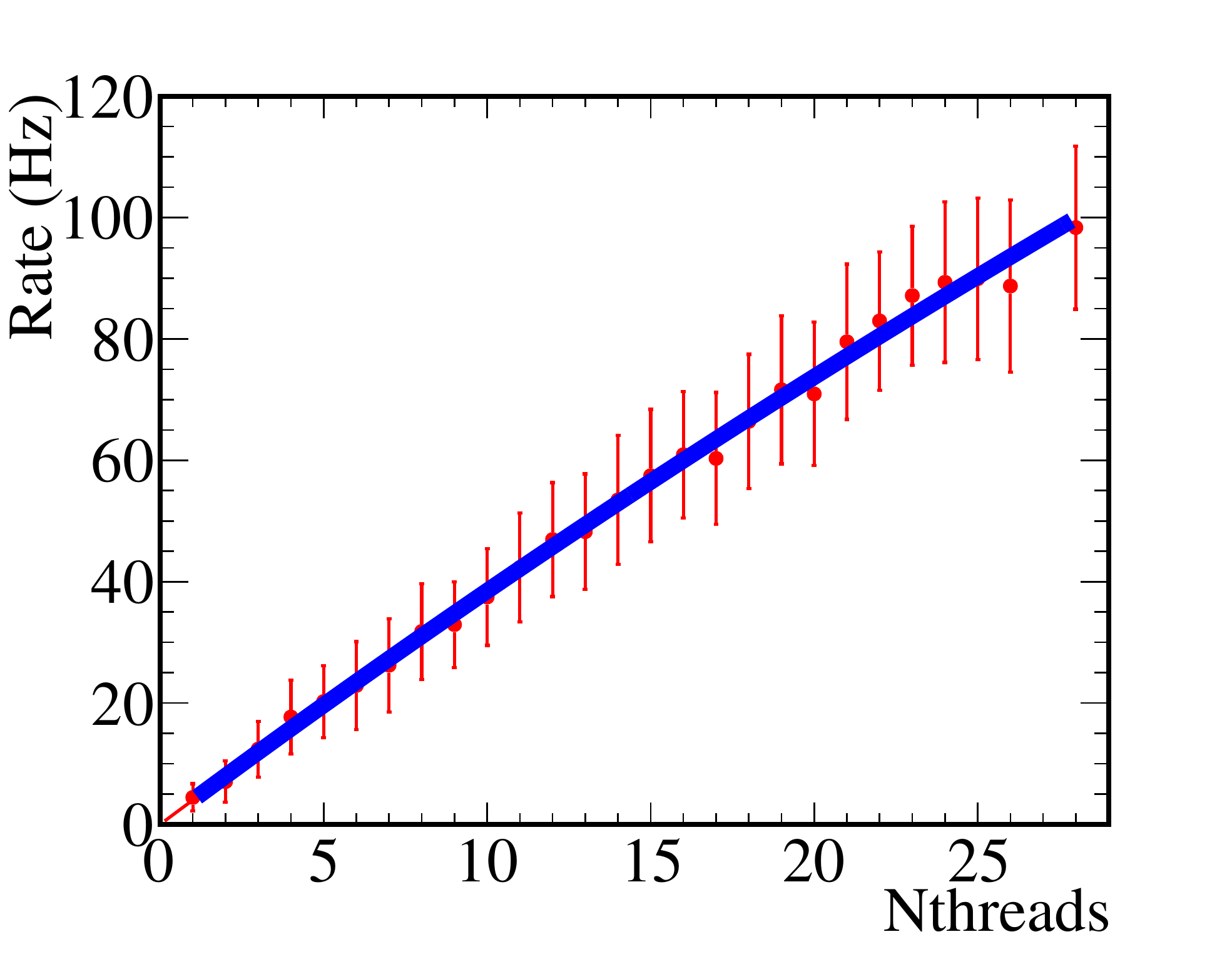}
\caption[]{\label{fig:production_offsite_rate_vs_nthreads_NERSC}Event processing rate versus number of threads for reconstruction jobs on NERSC Cori II (left) and PSC Bridges (right). The slope changes in the NERSC plot are due to the KNL architecture, which had four hardware threads per core. For PSC Bridges, hyper-threading is disabled and the plot shows a single slope.} 
\end{figure}

Production processing of \GX~data uses offsite high-performance computing resources in addition to the onsite computing farm at JLab, specifically, the National Energy Research Supercomputing Center (NERSC) and the Pittsburgh Supercomputing Center (PSC). For NERSC, the total allocation used for the academic year 2018-2019 was 53M NERSC units, which was used to process 70.5k jobs. This is equivalent to approximately 9M core-hours on a Intel x86\_64 processor. The jobs were run on NERSC's Cori II system, which is comprised of KNL (Knight's Landing) processors. The PSC allocation was awarded through the XSEDE\footnote{https://www.xsede.org.} allocation system in the last quarter of calendar year 2019 for 5.9 MSU. Only 0.85M SU were used in 2019 to run 7k jobs on the PSC Bridges system or about 10\% of the number processed at NERSC. Figure~\ref{fig:production_offsite_rate_vs_nthreads_NERSC} shows how the event processing rates scaled with the number of processing threads for both NERSC and PSC. Jobs run at both of those sites were assigned entire nodes so the number of processing threads used was equal to the total number of hardware threads.

Container and distributed file system technologies were used for offsite processing. The software binaries as well as calibration constants, field maps, etc. were distributed using the CERN-VM-file system (CVMFS). 
The binaries were all built at JLab using a CentOS7 system. A very lightweight Docker container was made based on CentOS7 that had only a minimal number of system RPMs\footnote{RedHat Package Management, https://access.redhat.com/documentation/en-us/red\_hat\_enterprise\_linux/5/html/deployment\_guide/ch-rpm} installed. All other software, including third-party packages such as ROOT, were distributed via CVMFS. This meant changes to the container itself were very rare (about once per year). The Docker container was pulled into NERSC's Shifter system without modification. The same container was used to create a Singularity container used at both PSC and on the Open Science Grid (OSG) for simulation jobs.

Raw data ware transferred from JLab to the remote sites using Globus\footnote{https://opensciencegrid.org/technology/policy/globus-toolkit.},  which uses GridFTP. The Globus tasks were submitted and managed by the SWIF2 workflow tool written by the JLab Scientific Computing group. SWIF2 was needed to manage the data retrieval from tape, for transfer to the remote site, for submission of remote jobs, and for transfer of processed data back to JLab. Disk space limitations at both JLab and the remote sites meant only a portion of the data set could be on disk at any one time. Thus, SWIF2 had to manage the jobs through all stages of data transfer and job submission.

\subsection{Analysis \label{sec:recanalysis}}

The full set of reconstructed (REST) data is too large to be easily handled by individual analyzers. For that reason, a system was developed to analyze data at JLab and extract reaction-specific ROOT trees. This step is represented by the right-hand green box at the bottom of Fig.\,\ref{fig:production_overview}.

Users can specify individual reactions via a web interface.
Periodically, the submitted reactions are downloaded into a configuration file, which steers the analysis launch. For each reaction, the \GX~analysis library inside the JANA framework creates possible particle combinations from the reconstructed particle tracks and showers saved in the REST format. Common selection criteria are applied for exclusivity and particle identification before performing a kinematic fit, using vertex and four-momentum constraints. Displaced vertices and inclusive reactions are also supported. Objects representing successful particle combinations (e.g. $\pi^0 \rightarrow \gamma\gamma$) and other objects are managed in memory pools, and can be reused by different channels to reduce the overall memory footprint of the process. With this scheme, up to one hundred different reactions can be combined into one analysis launch processing the reconstructed data.

If the kinematic fit converged for one combination of tracks and showers, the event is stored into a reaction-specific but generic ROOT tree, made accessible to the whole collaboration. The size of the resulting ROOT trees for the full data set strongly depends on the selected reaction, but is usually small enough to be copied to the user's home institution for a more detailed analysis.

\section[Monte Carlo]{Monte Carlo simulation \label{sec:simulation}}
The detailed simulation of events in the Hall-D beamline and \gx{} detector is performed with a GEANT-based software package. The package was originally developed within the GEANT3 framework~\cite{Brun:1987ma} and then migrated to the GEANT4 framework~\cite{Agostinelli:2002hh,Allison:2016lfl}. The simulation framework uses the same geometry definitions and magnetic field maps as used in reconstruction. The geometry includes the full photon beamline, starting at the radiator and ending at the photon beam dump. Both internal and external event generators are supported by the framework.  Internal sources include the coherent bremsstrahlung source and the single particle gun. Events read from any number of external generators are also supported. These input events specify one or more primary vertices to be simulated, which are randomized within the hydrogen target with timing that matches the RF structure of the beam.

\begin{figure}[t]\centering  
\includegraphics[width=0.95\textwidth]{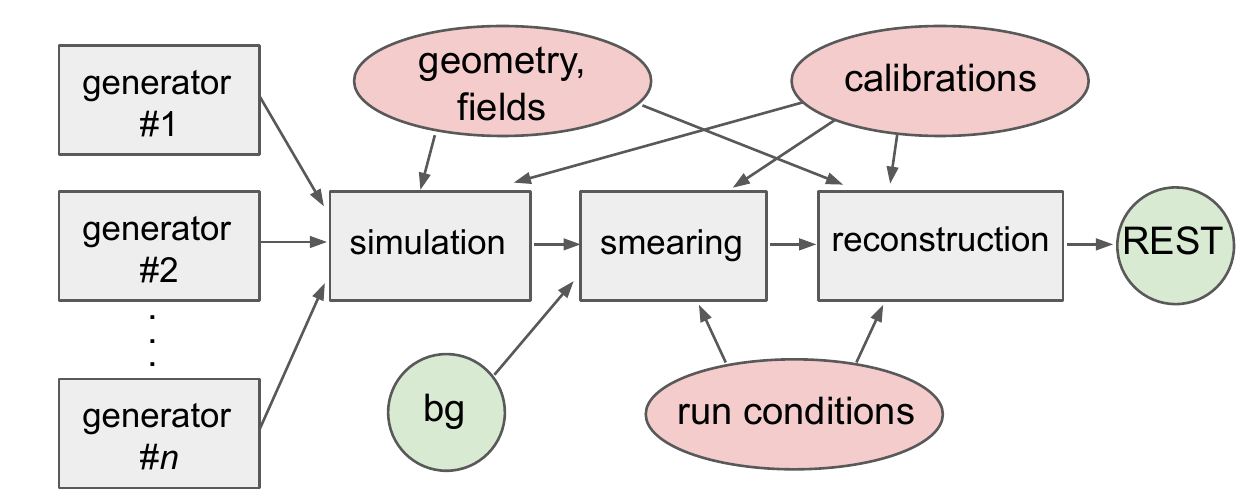}
\caption[]{\label{fig:MC-data-flow}The Monte Carlo data flow from event generators through physics analysis REST files. The ovals represent databases containing tables indexed by run number, providing a common configuration for simulation, smearing, and reconstruction. Background events represented by the circle marked \emph{bg} are real events collected using a random trigger, which are overlaid on the simulated events to account for pile-up in the Monte Carlo.}
\end{figure}

The Monte Carlo data flow is presented in Fig.~\ref{fig:MC-data-flow}. Events of interest are generated using either an internal or user-supplied event generator. The input event specification is fed to the Hall D GEANT simulation code, either {\em hdgeant} or {\em hdgeant4}, which tracks the particles through the experimental setup and records the signals they produce in the active elements of the detector. Behavior of the simulation is conditioned by a run number, which corresponds to a particular set of experimental conditions: beam polarization and intensity, beamline and detector geometry, magnetic field maps, etc. All this information is read by the simulation at run-time from the calibrations database, which functions as the single source for all time-dependent geometry, magnetic field, and calibration data relevant to the simulation.

Events written by the simulation are processed by the detector response package {\em mcsmear}. It applies corrections to the simulated hits to account for detector system inefficiencies and resolution, and overlays additional hits from uncorrelated background events. Loss of hits from detector channels, multi-hit truncation, and electronic deadtime are also applied at this step. Information needed for this processing comes from the databases for calibrations and run-conditions, and from files containing real backgrounds sampled using random triggers. Events emerging from the smearing step are deemed to be faithful representations of what the detector would have produced for the given run in response to the specified input. These Monte Carlo events are then processed with the same reconstruction software as used for the real events, and the output is saved to a REST file. These REST files are then made available for physics analysis.

\subsection[Geometry specification]{\label{sec:materialscan}Geometry specification}
The geometry and material descriptions for the experiment are common across simulation and reconstruction, residing in a family of xml files that follow a common schema called the Hall~D Detector Specification, or \emph{HDDS} \cite{HDDS,gx732}. Run-specific variations of the geometry xml records are maintained in the calibration database. The geometry and magnetic field map are also maintained in the calibration database.

The output events from the simulation are written as a data stream, which may either be piped directly into the next step of the Monte Carlo pipeline or saved to a file. Events are passed between 
all stages of the Monte Carlo processing pipeline, shown in Fig.~\ref{fig:MC-data-flow}, using the common data format of the Hall-D Data Model, HDDM \cite{gx65}. HDDM is used for all intermediate input and output event streams.

\subsection{Event generators \label{sec:generators}}
Simulation starts with the generation of events, which can be specific particles or reactions, or simply unbiased background events. A common toolset has been developed to minimize redundancy. These tools include standard methods to generate the distributions of primary photon beam energies and polarization. An output interface is used to produce files suitable as input to the GEANT simulation.

The photon beam energy distribution can be produced using a coherent bremsstrahlung generator that accounts for the physical properties of the radiator and the photon beamline. This generator allows the user to select the orientation of the diamond radiator, and then calculates the linear polarization for each photon. Photons can also be generated according to the spectrum measured in the pair spectrometer during any actual data run by interfacing to the calibration data base. Here the user inputs the degree of linear polarization and the orientation. Finally, the user can provide a histogram of the photon energy spectrum and a second one of the degree of polarization to be used to generate the photon beam. 

One of the first generators was used to simulate the total photoproduction cross section. It is currently used to study backgrounds to physics reactions as well as develop analysis tools for extracting signals. This event generator, called {\em bggen}, is based on Pythia~\cite{Sjostrand:2006za}, and includes additions that describe the low-energy photoproduction cross sections. Other generators are tied to specific reactions, where the generator needs to describe the underlying physics.

\subsection{HDGEANT \label{sec:hdgeant}}
Both GEANT3 and GEANT4 versions are available for simulation of the experiment. Both versions have been tuned to reproduce the behavior of the experiment, but there are some differences arising from how the two versions decide when to stop tracking particles. In general, the simulation mimics the running conditions found across a range of runs, typically a large part of a single run period. The output from GEANT contains both hit times and energies deposited in detector volumes. 

\subsection[Detector response]{Detector response}
Converting time and energy deposits coming from GEANT into electronic detector responses that match the readout from the experiment is carried out by the detector response package \textit{mcsmear}. The output of this digitization is identical to the real data with the exception that the so-called \emph{truth information} about the data is retained to allow detailed performance studies. In addition to the digitization, at this stage the run-dependent efficiency effects are applied to the data, including both missing electronic channels and reduced efficiency of other channels. Additional smearing of some signals is also applied here to better match the performance of the Monte Carlo to data. 

The \textit{mcsmear} package also folds measured backgrounds into the data stream. During regular data collection, random triggers are collected concurrently with data taking (see Section\,\ref{sec:trig}). These are separated from the actual data and used to provide experimental background signals in the Monte Carlo, with rates based on the actual beam fluxes in the experiment. 

\subsection{Job submission \label{sec:jobsubmission}}
A large number of experimental conditions need to be matched in simulated data. The \emph{MCWrapper} tool was 
developed to streamline the input specifications, implement consistency with corresponding data reconstruction, seamlessly access computer offsite resources, and produce Monte Carlo samples in proportion to the actual data taken. The goal is to model the differences between runs and provide a simulated data set, comparable to the real data. The primary system used for this phase is the Open Science Grid (OSG) in order to leverage resources in addition to the local JLab computing farm. Many automated checks are made to avoid flawed submission, and all aspects of the requests and jobs are monitored during running. Once completed, \emph{MCWrapper} checks for expected output files to be returned as if the jobs were run on the JLab farm. If expected files are not found the system will automatically submit a replacement job. Once the jobs are verified completed and all data from the request have been properly moved, the user receives an automated email alerting them that their request has been fulfilled and providing the location where the user can access the event sample.

Users are able to monitor and control their simulations via an online dashboard. The \emph{MCWrapper} dashboard gives information about active projects and allows users (or administrators) to interact with their requests. Users may cancel, suspend, or declare projects complete. Detailed information is presented about the individual jobs, such as where the jobs are being run, basic usage statistics, and current status.  This information gives individuals a near real-time look into the production of their Monte Carlo samples.

\section[Detector performance]{Detector performance \label{sec:performance}}                                                 

The capability of the \gx{} detector in reconstructing charged and neutral particles and assembling them into fully reconstructed events has been studied in data and simulation using several photoproduction reactions.  The results of these studies are summarized in this section.




\begin{figure}[tbp]
\begin{center}
\includegraphics[width=0.45\textwidth]{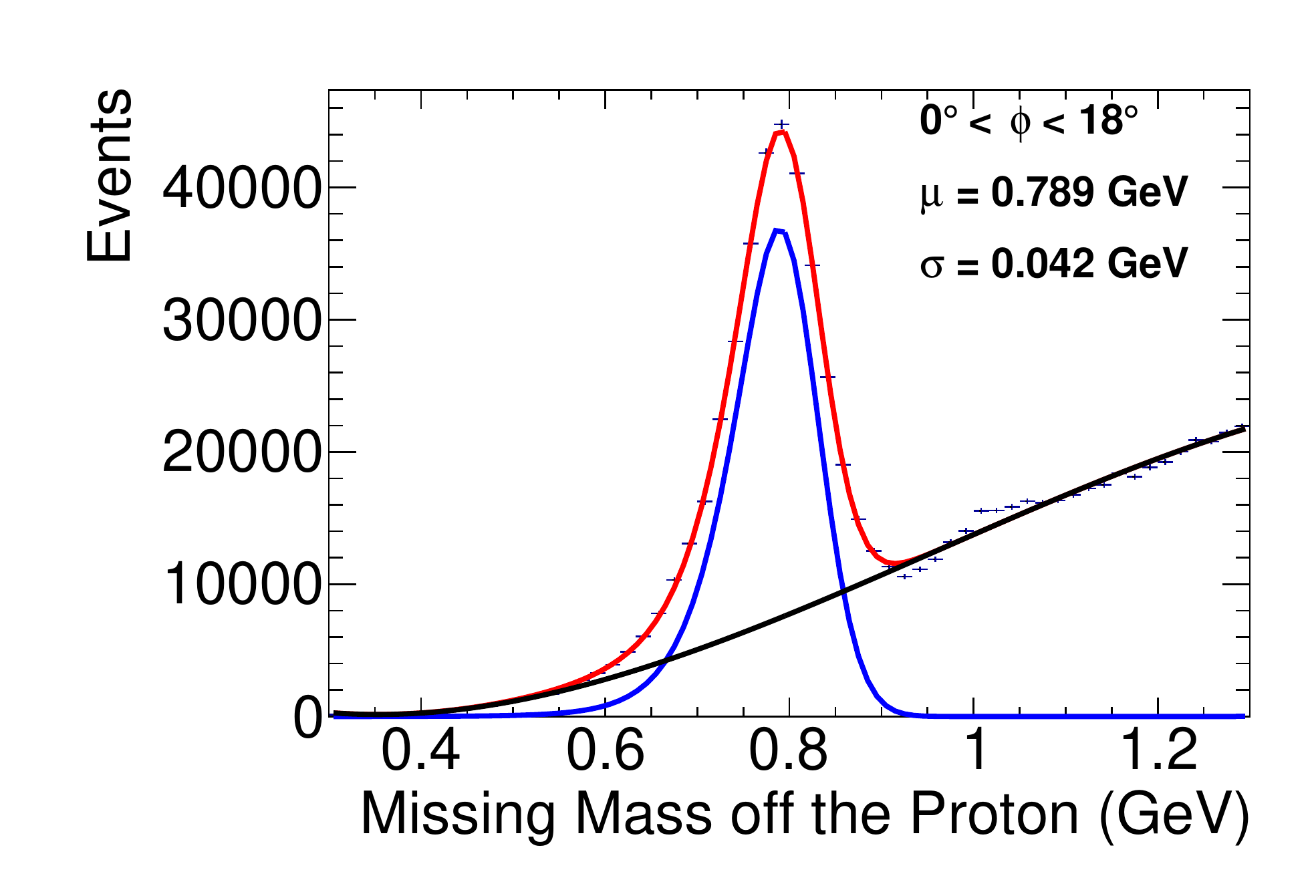}
\includegraphics[width=0.45\textwidth]{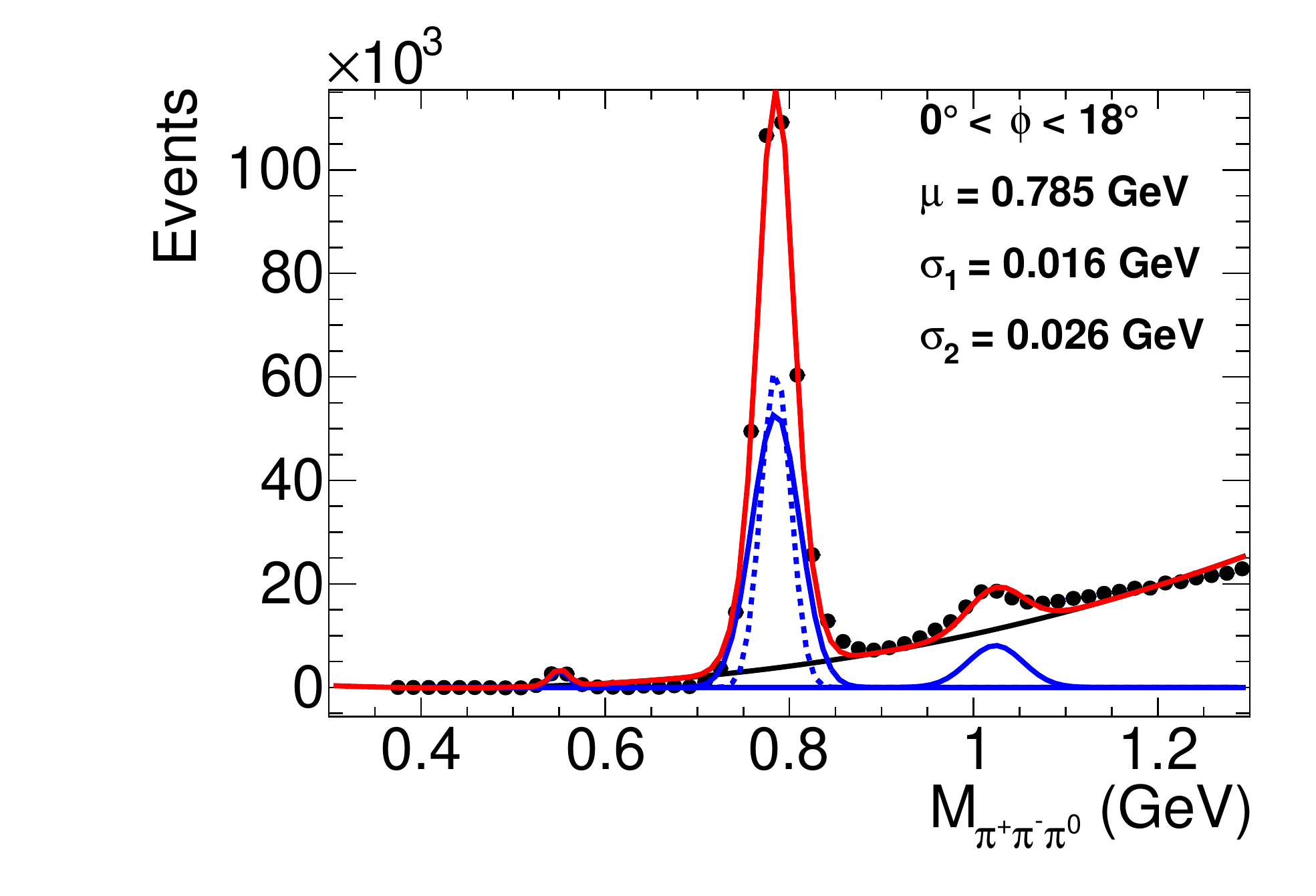}
\caption{\label{fig:omega mass}
Reconstructed mass distributions for the reaction $\gamma p \to p\pi^0\pi^{\pm}(\pi^\mp)$ for a bin in $\phi$.
  (Left) Distribution of the missing mass off the proton.
(Right) Invariant mass distribution for the $\pi^+\pi^-\pi^0$ system.  The blue curves show the resonant contributions, the black
curve show the polynomial backgrounds, and the red curve shows the sum.
 (Color online)}
\end{center}
\end{figure}

\subsection{Charged-particle reconstruction efficiency  \label{sec:trackeff}}  
\begin{figure}[tbp]
\begin{center}
\includegraphics[width=\textwidth]{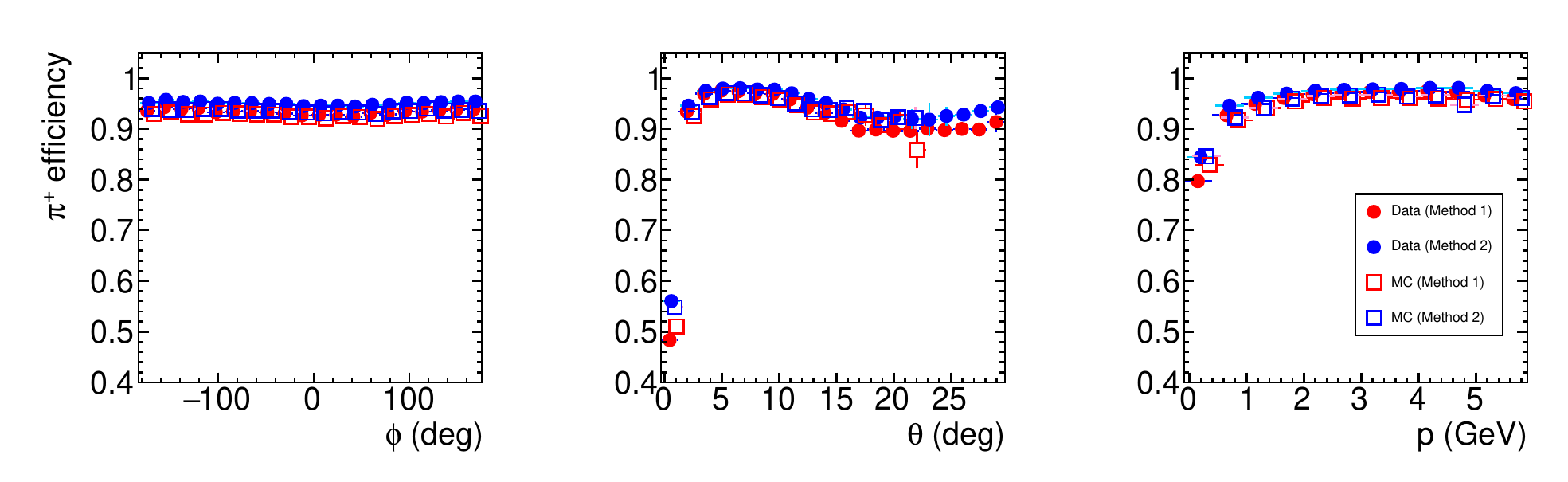} 
\caption{\label{fig:tracking efficiency}
Tracking efficiency for $\pi^+$ tracks, determined by data and simulation using two methods. (Color online)}
\end{center}
\end{figure}

The track reconstruction efficiency was estimated by analyzing $\gamma p \rightarrow p \omega$, $\omega\rightarrow\pi^+\pi^-\pi^0$ events, where the proton, the $\pi^0$, and one of the charged pions were used to predict the three-momentum of the other charged pion. Two methods were used to calculate this efficiency, $\varepsilon=N_{found}/(N_{found}+N_{missing})$.  Events for which no track was reconstructed in the predicted region of 
phase space contributed to $N_{missing}$, while events where the expected track was reconstructed contributed to $N_{found}$.  For the first method, the $\omega$ yields for $N_{found}$ and $N_{missing}$ were estimated from the missing mass off the 
proton; for the second method, the invariant mass of the $\pi^+\pi^-\pi^0$ system was used to find $N_{found}$.  This analysis was performed for individual bins of track momentum, $\theta$, and $\phi$.
Examples of mass histograms for a typical bin in $\phi$ are shown in Fig.~\ref{fig:omega mass}.  The exercise was repeated for a sample of $\omega$ Monte Carlo events.   A comparison of the efficiency for pion reconstruction derived from the 
two methods for both Monte Carlo and experimental data is shown in Fig.~\ref{fig:tracking efficiency}.  The efficiencies for Monte Carlo and experimental data 
agree to within 5\%.

While this reaction only allows the determination of track reconstruction efficiencies for $\theta < 30^\circ$, this covers the majority of charged particles produced in \gx{} due to its fixed-target geometry.  Other reactions are being studied to determine the efficiency at larger angles.

\subsection{Photon efficiency\label{sec:perfneutral}}




Photon-reconstruction efficiency has been studied using different methods for the FCAL and BCAL.  In the FCAL, absolute photon reconstruction efficiencies have been determined using the ``tag-and-probe'' method with a sample of photons from the reaction $\gamma p \to \omega p$, $\omega \to \pi^+\pi^-\pi^0$, $\pi^0 \to \gamma (\gamma)$, where one final photon is allowed but not required to be reconstructed.  The yields with and without the reconstructed photon are determined using two methods.  In the first method, the $\omega$ yield is determined from the missing-mass spectrum, $M_X(\gamma p \rightarrow pX)$, selecting on whether only one or both reconstructed photons are consistent with a final-state $\pi^0$. In the second method,  the count when both photons are found is determined from the $\omega$ yield from the fully reconstructed invariant mass $M(\pi^+\pi^-\gamma\gamma)$. If the photon is not reconstructed, the $\omega$ yield is determined by a fit to the distribution of the missing mass off the proton.  Both methods yield consistent results, with a reconstruction efficiency generally above 90\%, and within 5\% or less agree with the efficiencies determined from simulation.

\begin{figure}[tbp]
\begin{center}
\includegraphics[width=0.45\textwidth]{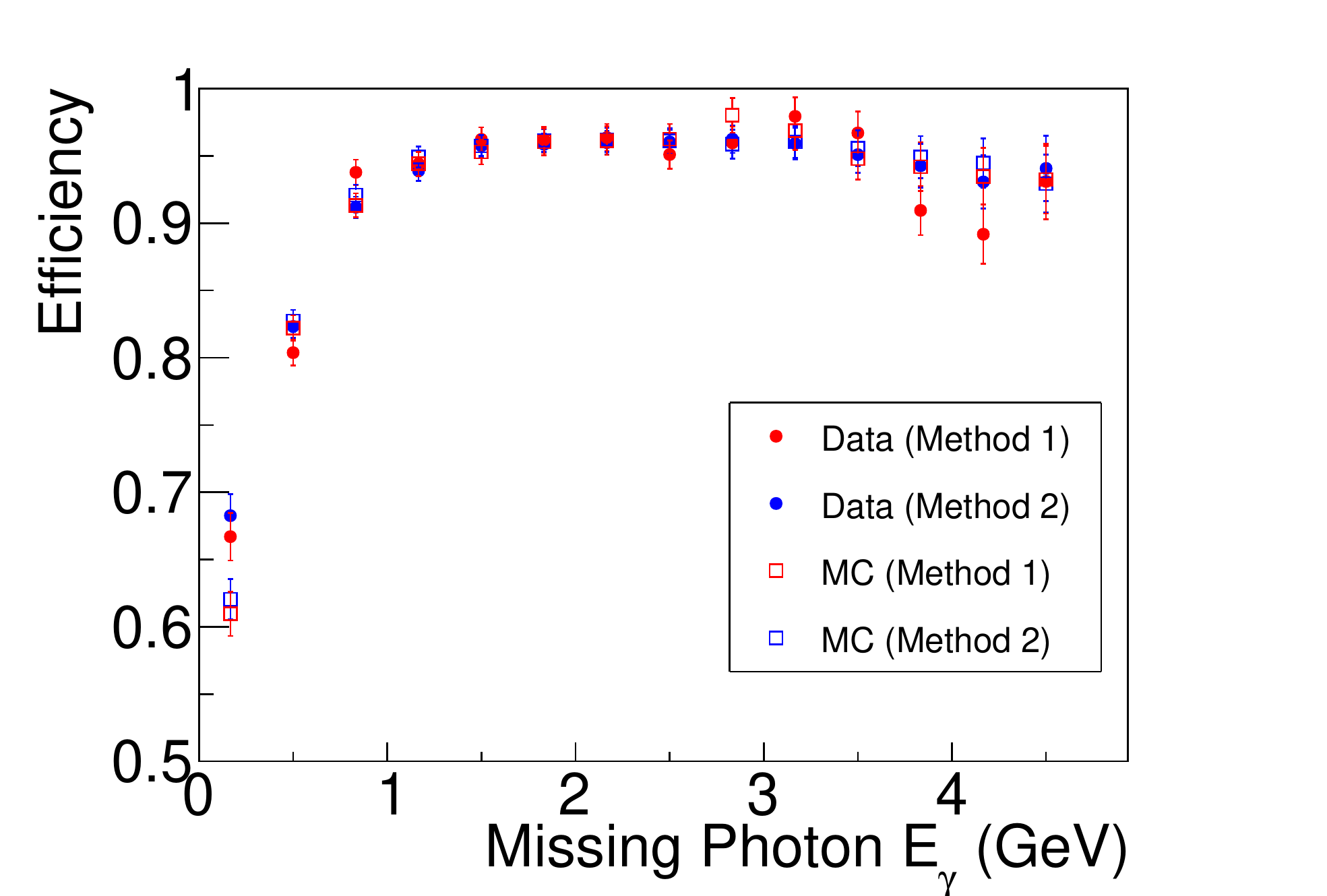}
\includegraphics[width=0.45\textwidth]{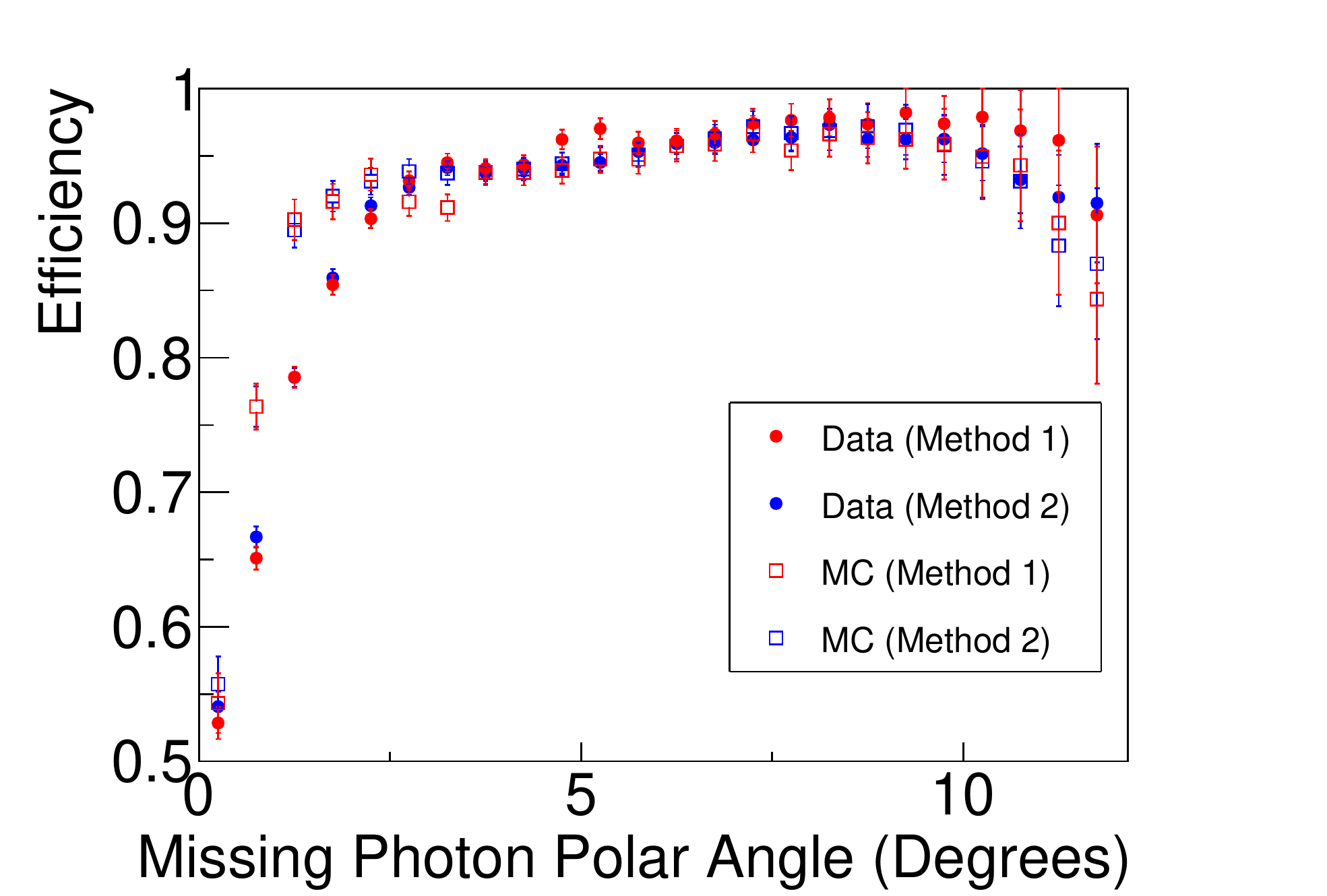}
\caption{\label{fig:fcalphotoneff}
Photon reconstruction efficiency in FCAL determined from $\gamma p \to \omega p$, $\omega \to \pi^+\pi^-\pi^0$, $\pi^0 \to \gamma (\gamma)$ as a function of (left) photon energy and (right) photon polar angle.  Good agreement between data and simulation is observed in the fiducial region $\theta = 2^\circ - 10.6^\circ$. (Color online)
}
\end{center}
\end{figure}

\begin{figure}[tbp]
\begin{center}
\includegraphics[width=\textwidth]{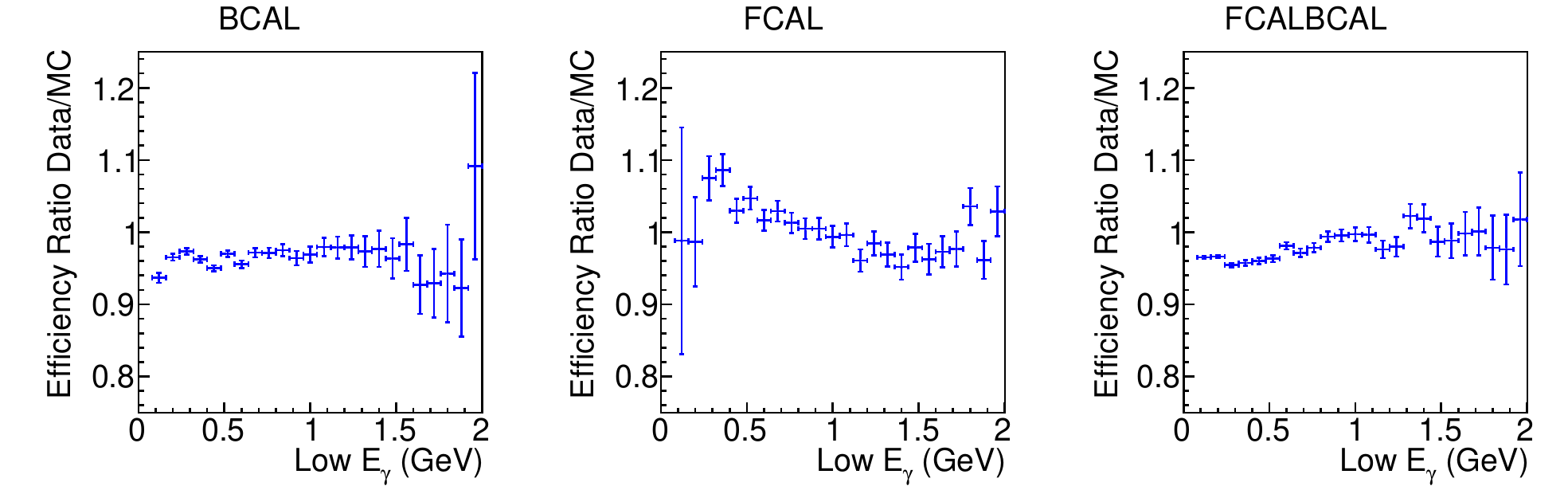}
\caption{\label{fig:bcalpi0photoneff}
Ratios of relative photon reconstruction efficiency between data and simulation determined from $\pi^0\to\gamma \gamma$ decays in $\gamma p \to  \pi^+\pi^-\pi^0 p$ events.  The efficiency ratios are shown for the cases where (left) both photons were measured in the BCAL, (middle) both photons were measured in the FCAL, and (right) one photon was measured in the BCAL and the other in the FCAL.
}
\end{center}
\end{figure}

A relative photon efficiency determination has been performed using $\pi^0\to\gamma\gamma$ decays, which spans the full angular range detected in \gx{}.  A sample of fully reconstructed $\gamma p \to  \pi^+\pi^-\pi^0 p$ events were inspected, taking advantage of the $\pi^0\to\gamma\gamma$ decay isotropy in the center-of-mass frame.  Thus, any anisotropy indicates an inefficiency in the detector. Results from this analysis are illustrated in Fig.~\ref{fig:bcalpi0photoneff}. Generally, this relative efficiency is above 90\%, and agrees within 5\% of that determined from simulation.  

The models for the simulated response of both calorimeters are being updated, and the final agreement between photon efficiency determined in data and simulation is expected to improve.

Detailed studies of detector performance determined the standard fiducial region for most analyses to be $\theta = 2^\circ - 10.6^\circ$ and $\theta > 11.3^\circ$.  These requirements avoid the region dominated by beam-related backgrounds at small $\theta$ and the transition region between the BCAL and FCAL, where shower reconstruction is difficult.

\subsection{Kinematic fitting \label{sec:perffitting}}

Kinematic fitting is a powerful tool to improve the resolution of measured data and to distinguish between different reactions.  In \gx{}, this method takes advantage of the fact that the initial state is very well known, with the target proton at rest, and the incident photon energy measured with very high precision ($<0.1\%$). This knowledge of the initial state gives substantial improvements in the kinematic quantities determined for exclusive reactions.  The most common kinematic fits that are performed are those that impose energy-momentum conservation between the initial and final-state particles.  Additional optional constraints in these fits are for the four-momenta of the daughters of an intermediate particle to add up to a fixed invariant mass, and for all the particles to come from a common vertex (or multiple vertices, in the case of reactions containing long-lived, decaying particles).

\begin{figure}[tbp]
\begin{center}
\includegraphics[width=0.75\textwidth]{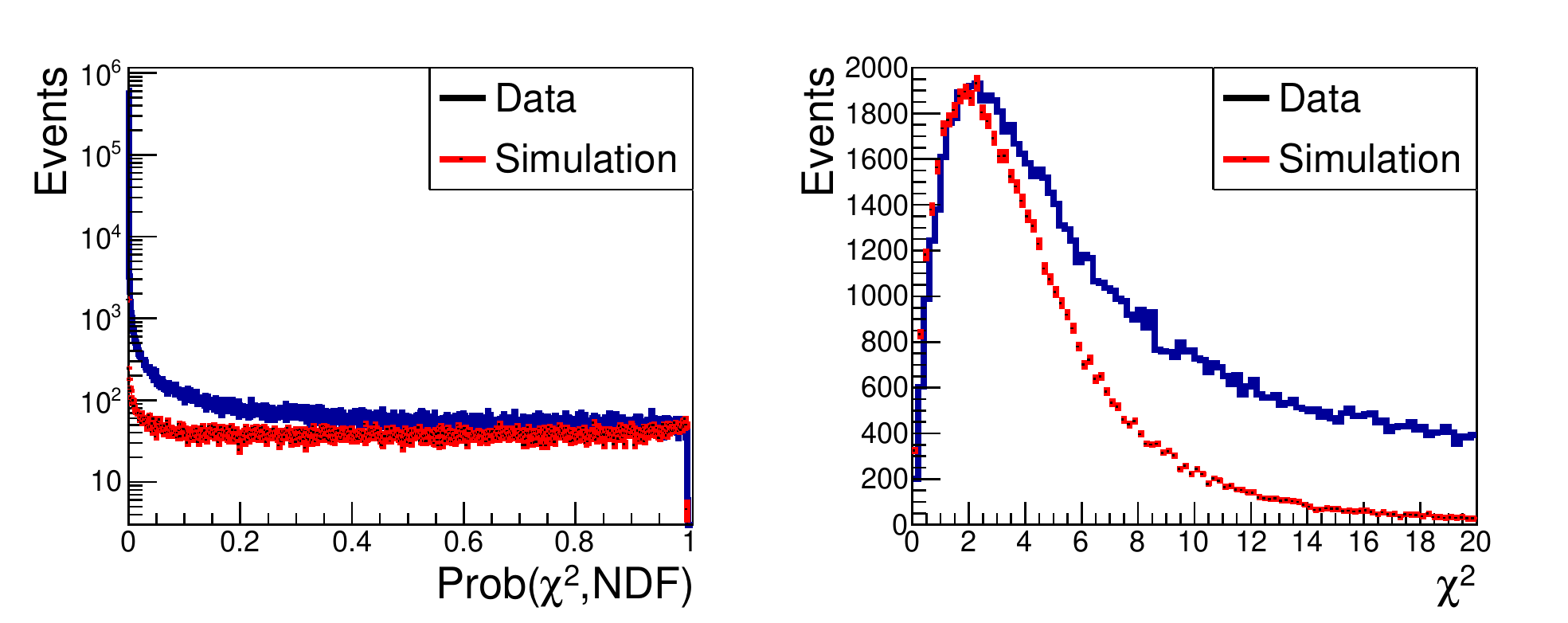}
\caption{\label{fig:kinfitperform}
Distribution of kinematic fit (left) probability and (right) $\chi^2$ for reconstructed $\gamma p \to \eta p$,  $\eta \to \pi^+\pi^-\pi^0$ events in data and simulation.  Both distributions agree reasonably for well-measured events, and diverge due to additional background in data and differences in modeling poorly-measured events.
 (Color online)}
\end{center}
\end{figure}

To illustrate the performance of the kinematic fit, we use a sample of $\gamma p \to \eta p$, $\eta \to \pi^+\pi^-\pi^0$ events selected using a combination of standard particle identification and simple kinematic selections.  
The use of the kinematic fit improves the $\eta$-mass resolution  from  2.6~MeV to 1.7~MeV, which is typical of low-multiplicity meson production reactions.  
The quality of the kinematic fit is determined using either the probability calculated from the $\chi^2$ of the fit and the number of degrees-of-freedom or the $\chi^2$ of the fit itself. 
The distributions of the kinematic fit $\chi^2$ and probability are illustrated in Fig.~\ref{fig:kinfitperform} for both reconstructed and simulated data.  The agreement between the two distributions is good for small $\chi^2$ (large probability), and flat over most of the probability range, indicating good overall performance for most signal events.  The disagreement between the two distributions at larger $\chi^2$ (probability $<0.2$) is due to a combination of background events and deficiencies in the modelling of poorly measured events with large resolution.

The performance of the reconstruction algorithms and kinematic fit can be studied through investigating the ``pull'' distributions, where the pull of a variable $x$ is defined by comparing its measured values and uncertainties and those resulting from the kinematic fit as
\begin{equation}
    \text{pull}_x = \frac{x_\text{fitted} - x_\text{measured}}{\sqrt{\sigma_{x,\text{measured}}^2 - \sigma_{x,\text{fitted}}^2}}.
\end{equation}
If the parameters and covariances of reconstructed particles are Gaussian, are measured accurately, and the fit is performing correctly, then these pull values are expected to have a Gaussian distribution centered at zero with a width $\sigma$ of 1.  If the pull distributions are not centered at zero, this is an indication that there is a bias in the measurements or the fit.  If $\sigma$ varies from unity, this is an indication that the covariance matrix elements are not correctly estimated.  

\begin{figure}[tbp]
\begin{center}          \includegraphics[width=0.29\textwidth]{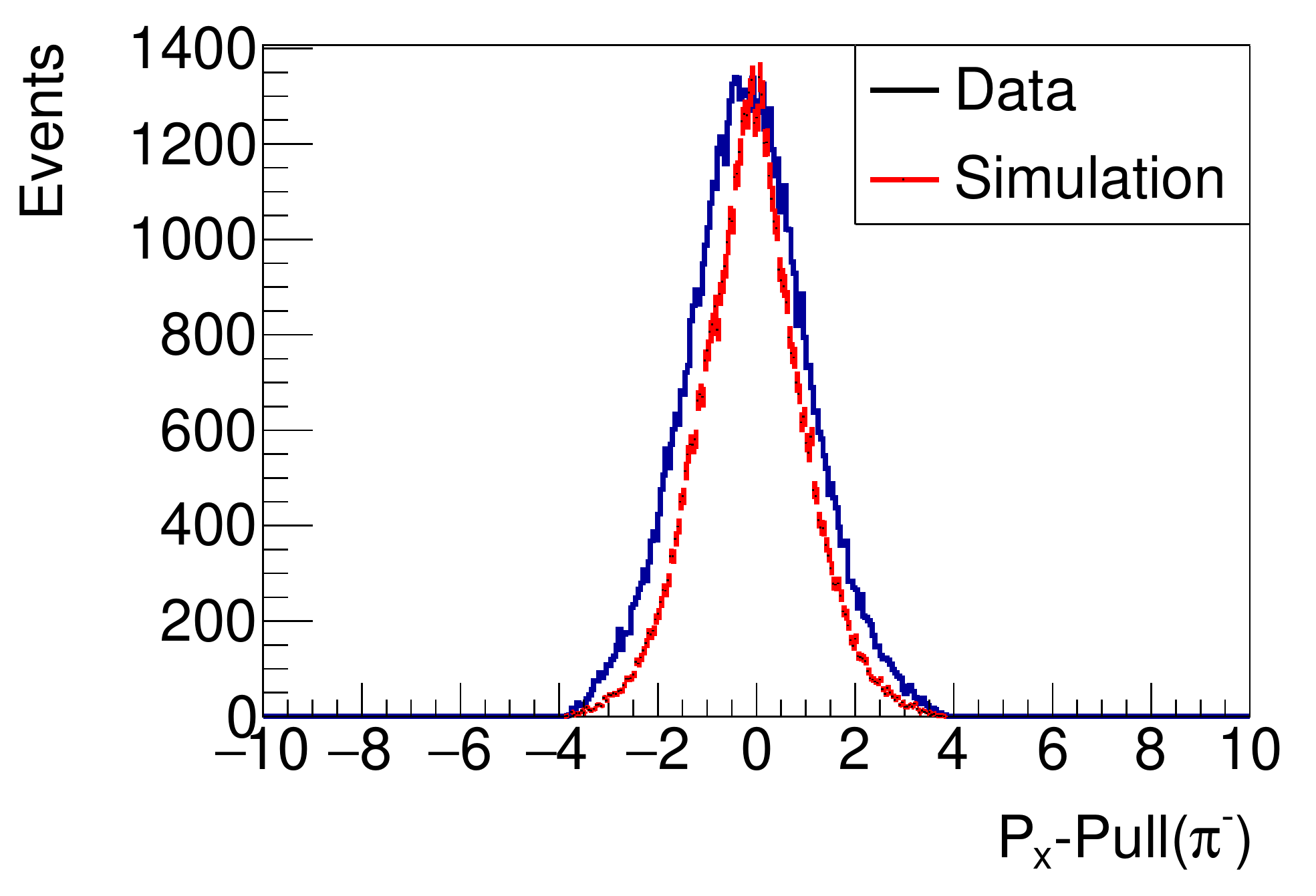}
\includegraphics[width=0.29\textwidth]{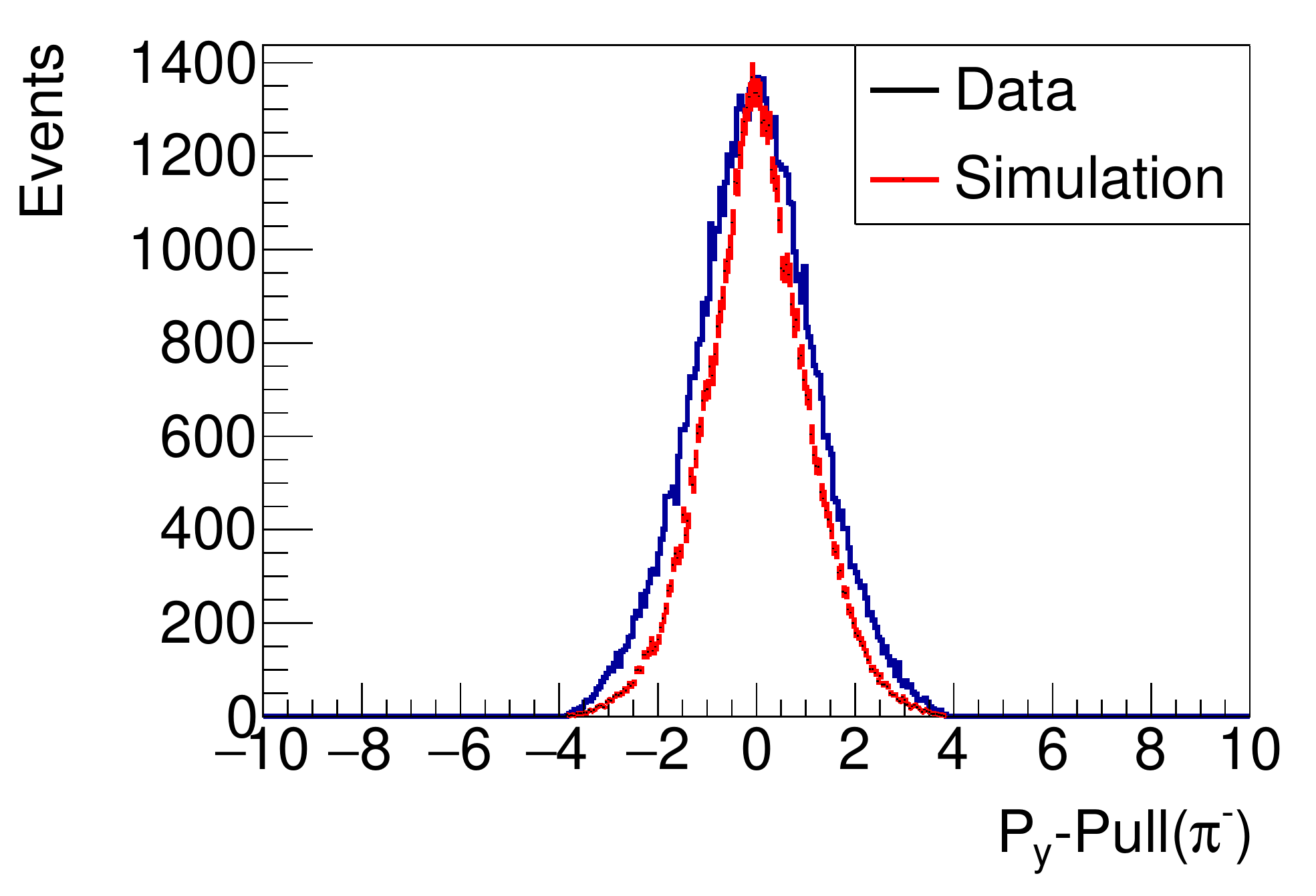}
\includegraphics[width=0.29\textwidth]{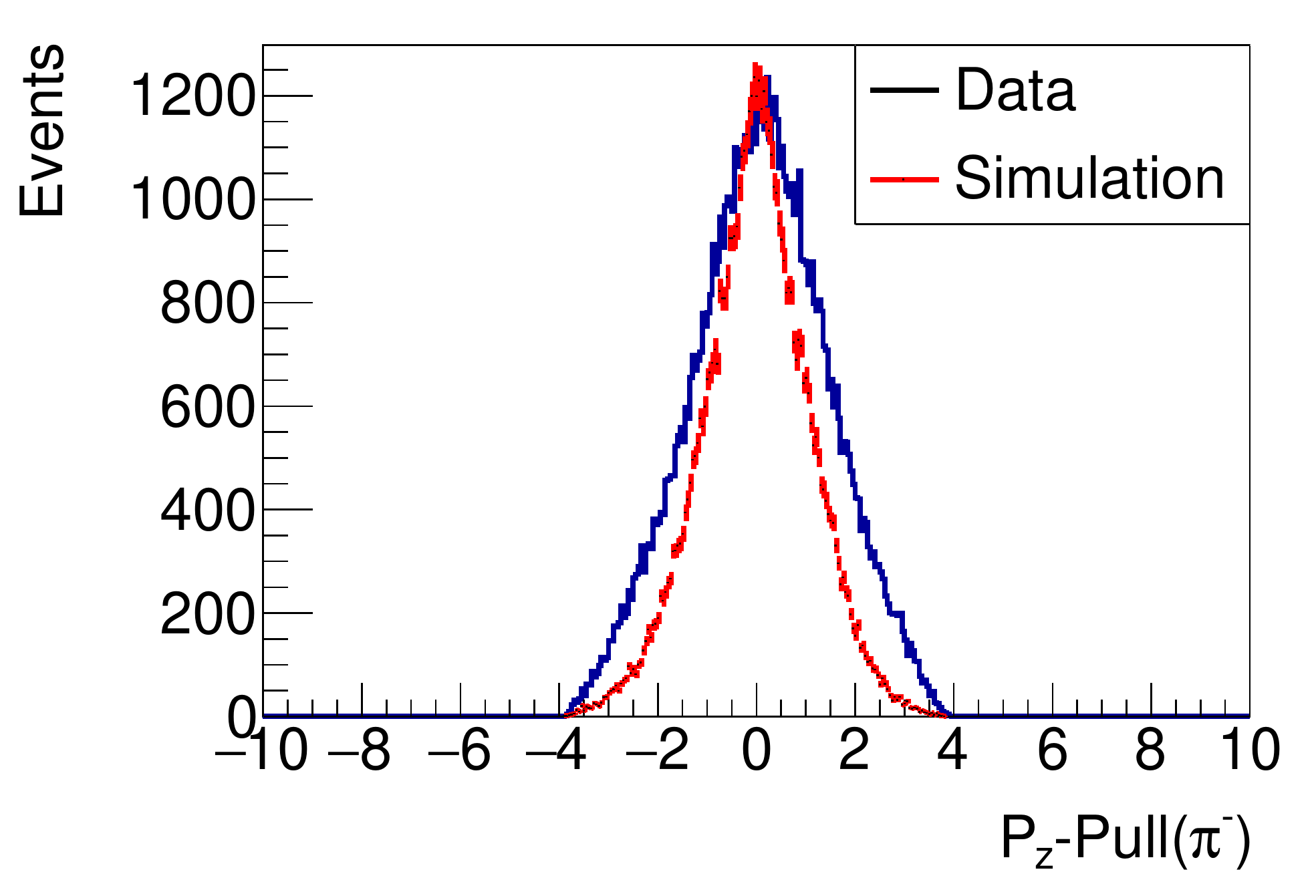}

\caption{\label{fig:kinfitpulls}
Pull distributions for momentum components of the $\pi^-$ from reconstructed $\gamma p \to \eta p$,  $\eta \to \pi^+\pi^-\pi^0$ events in data and simulation for events with fit probability $>0.01$: (left) $p_x$, (center) $p_y$, (right) $p_z$.
 (Color online)}
\end{center}
\end{figure}

\begin{figure}[tbp]
\begin{center}          \includegraphics[width=0.3\textwidth]{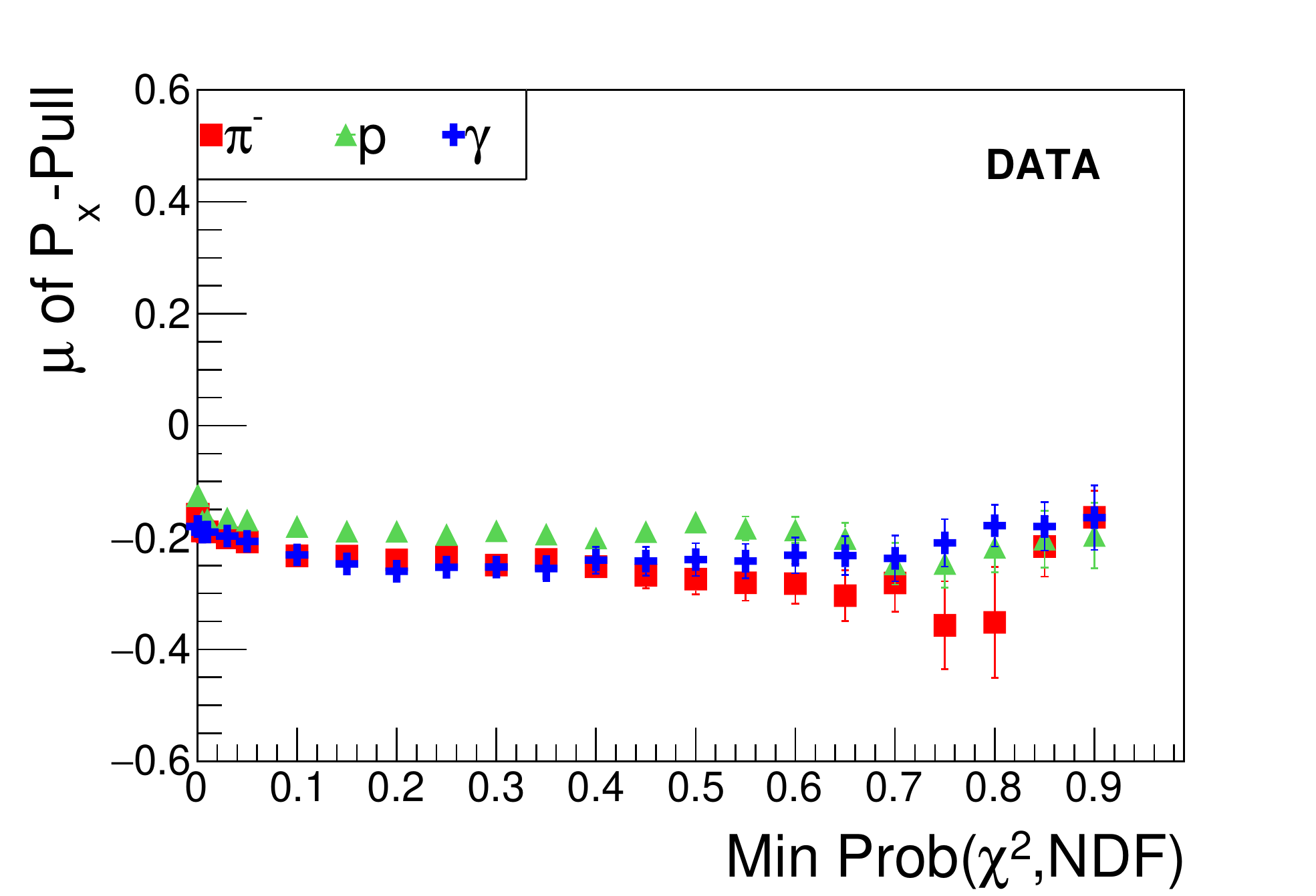}
\includegraphics[width=0.3\textwidth]{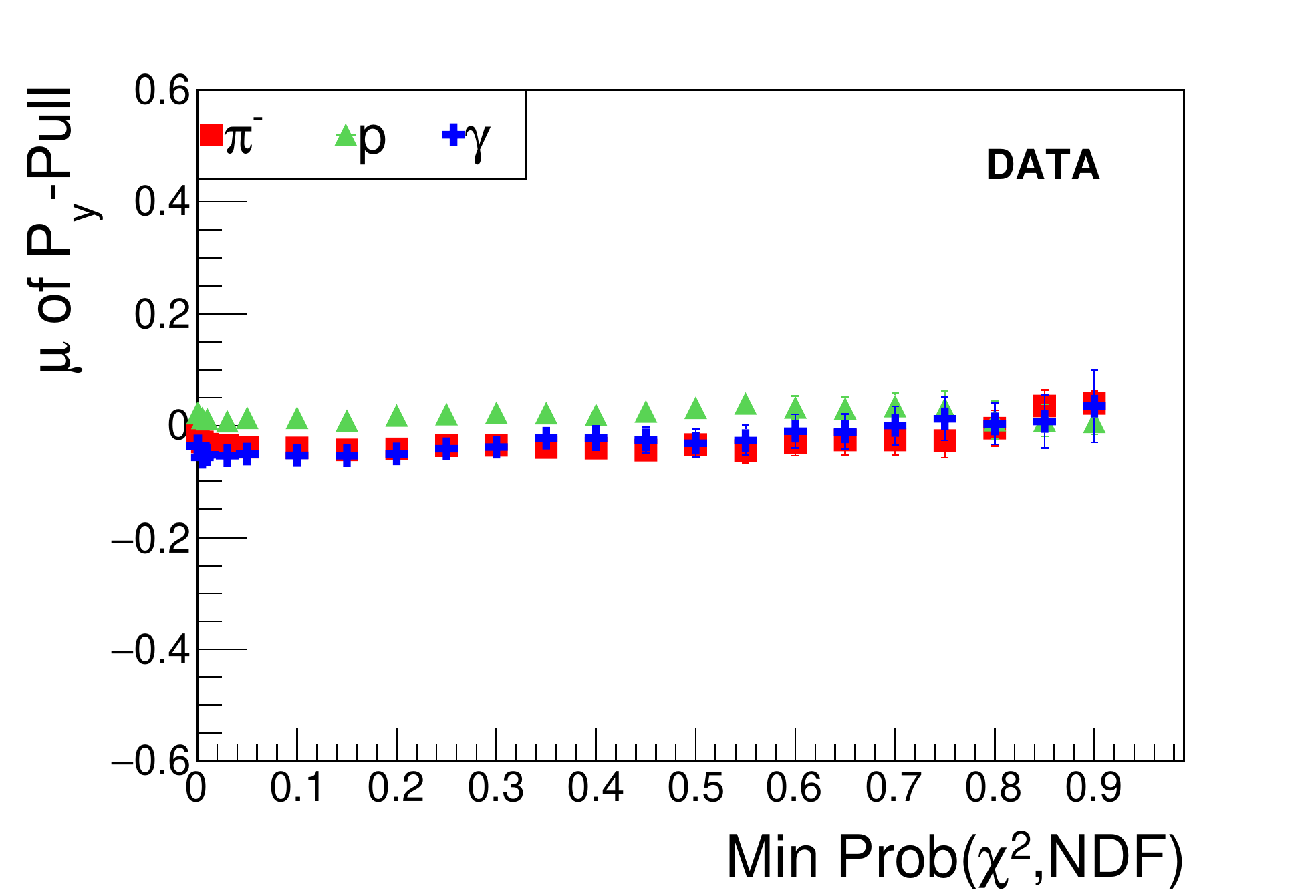}
\includegraphics[width=0.3\textwidth]{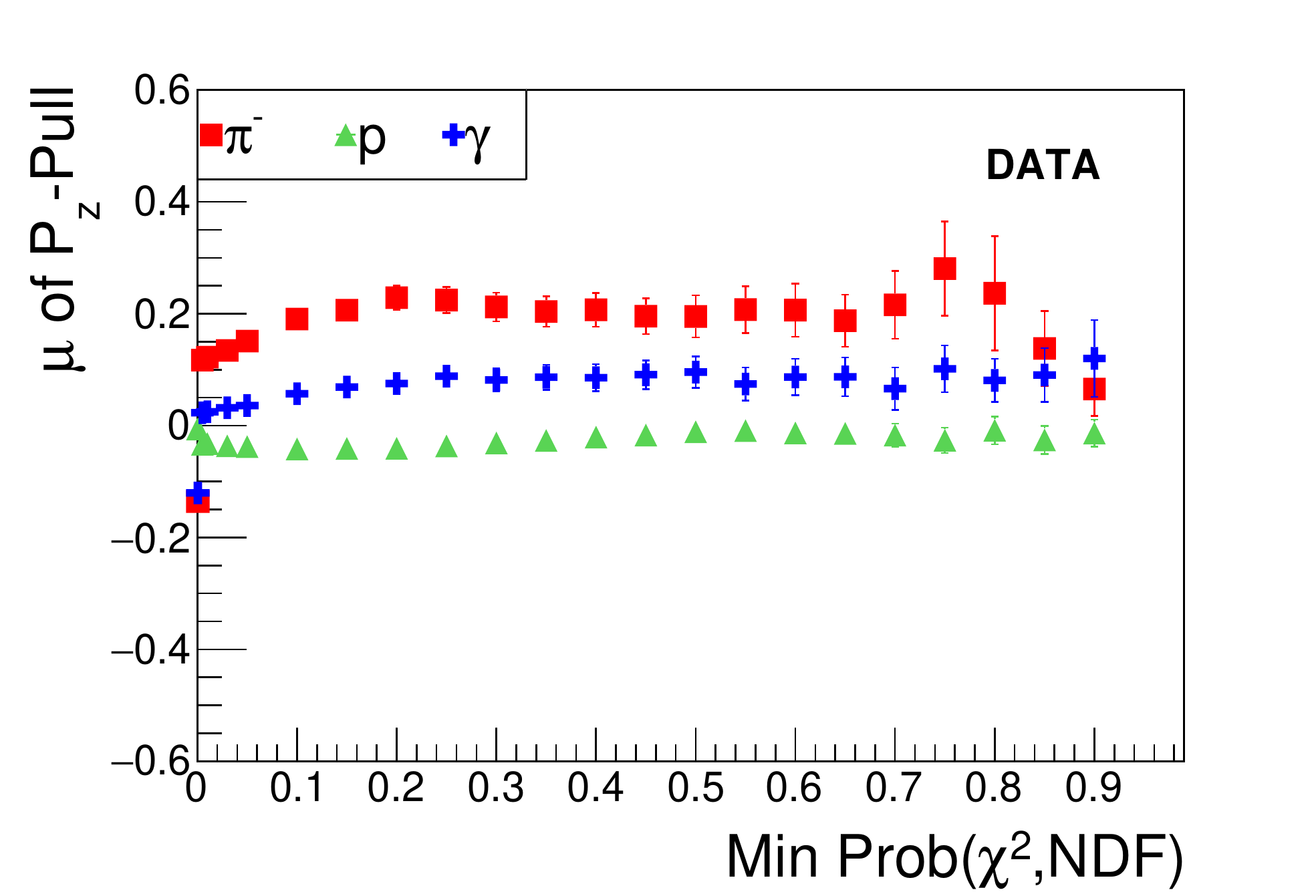}

\includegraphics[width=0.3\textwidth]{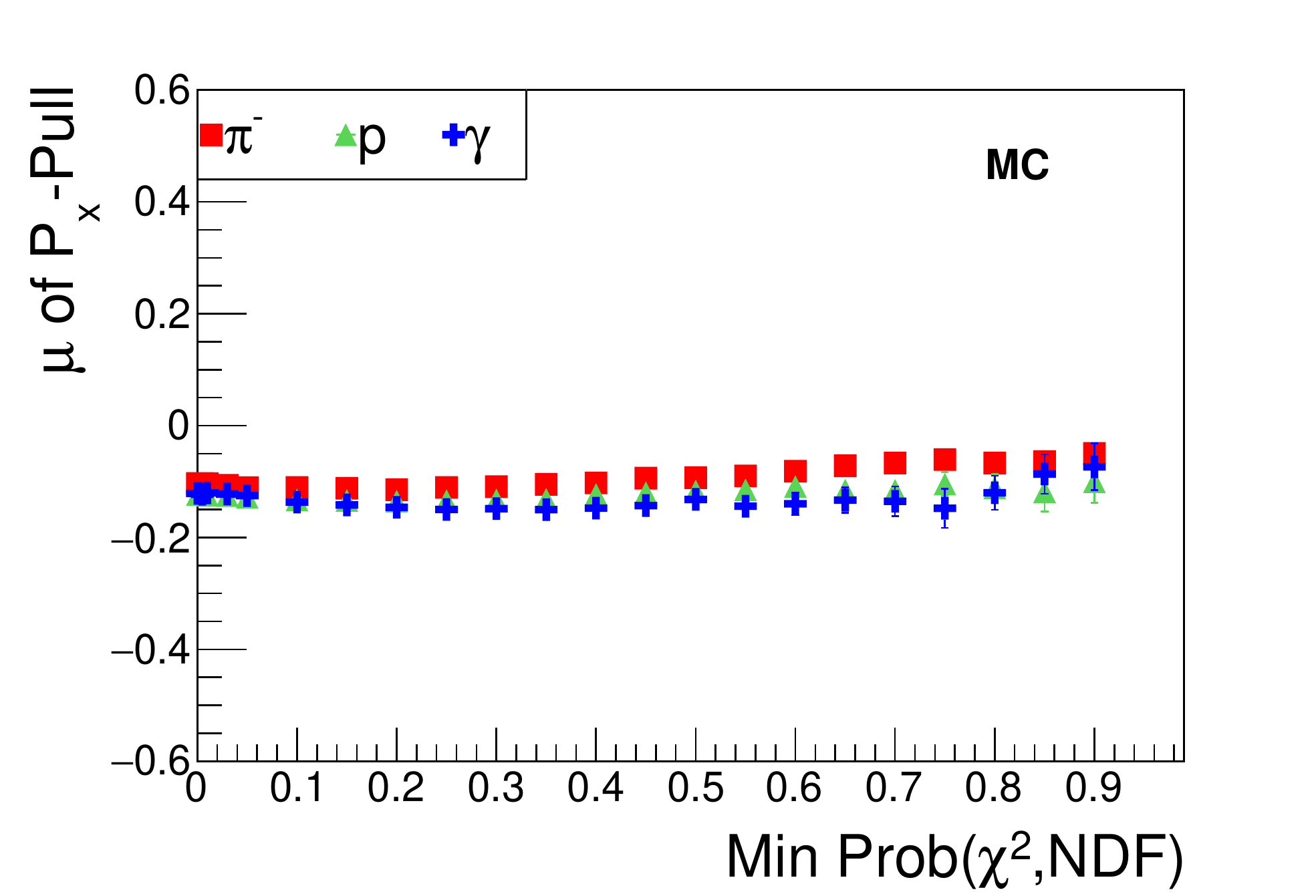}
\includegraphics[width=0.3\textwidth]{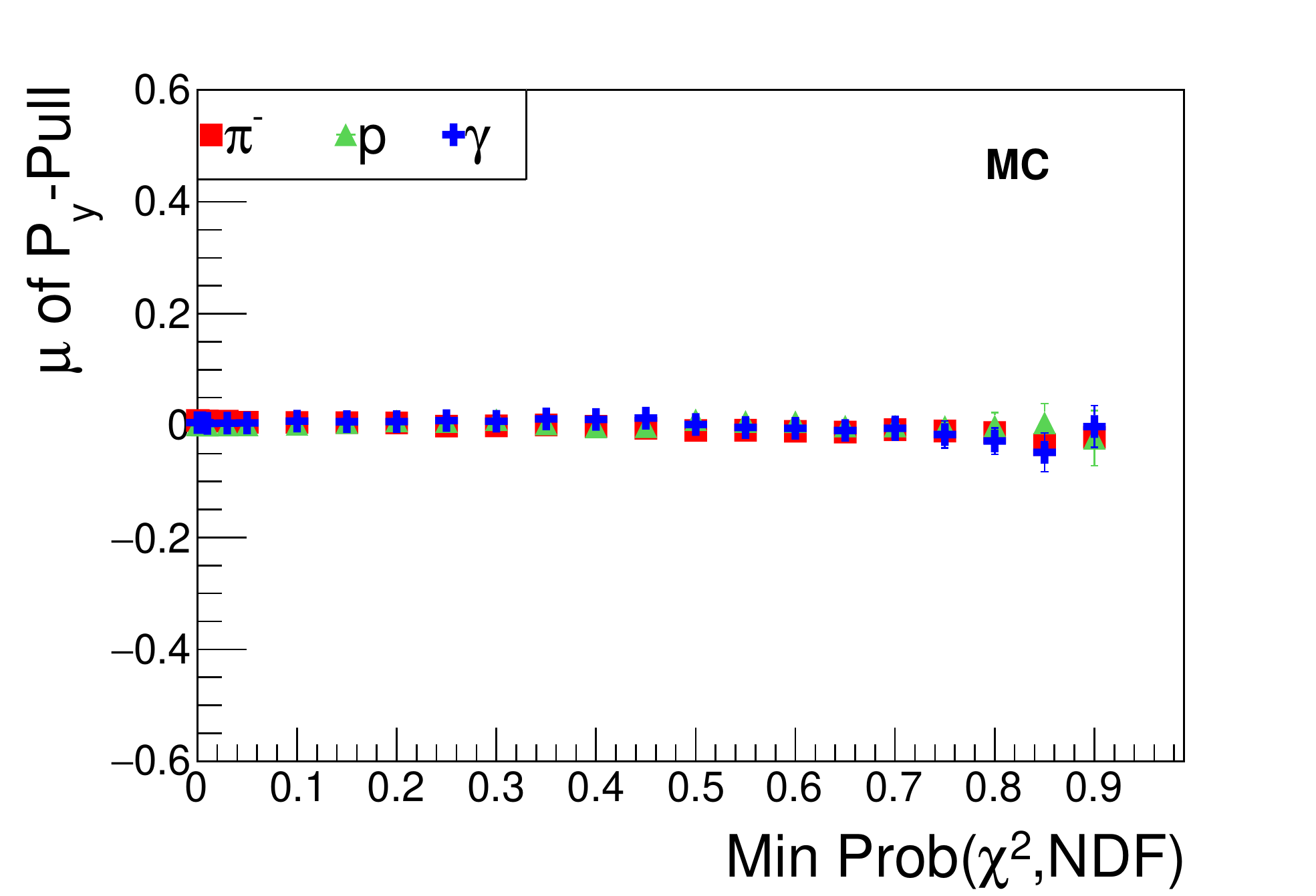}
\includegraphics[width=0.3\textwidth]{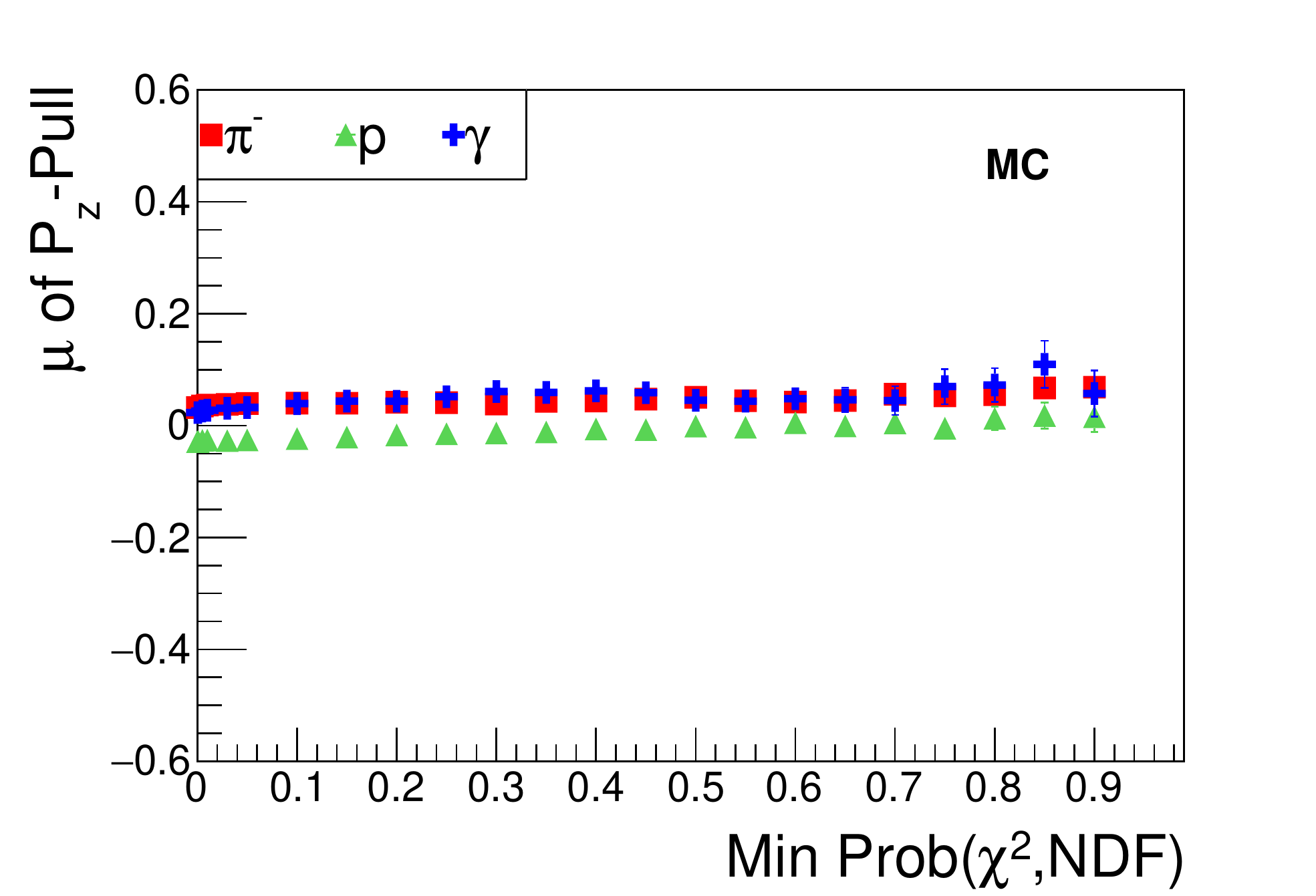}

\includegraphics[width=0.3\textwidth]{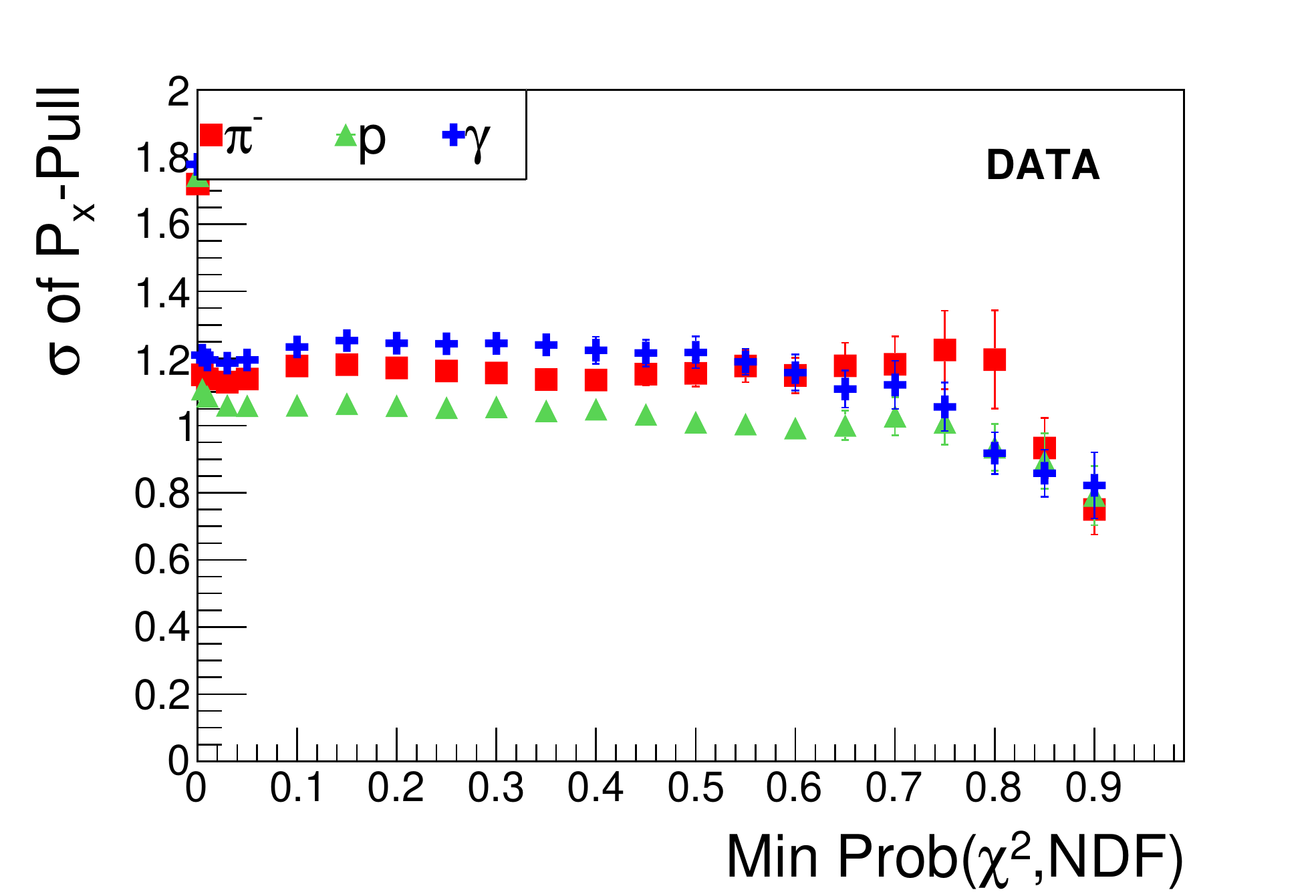}
\includegraphics[width=0.3\textwidth]{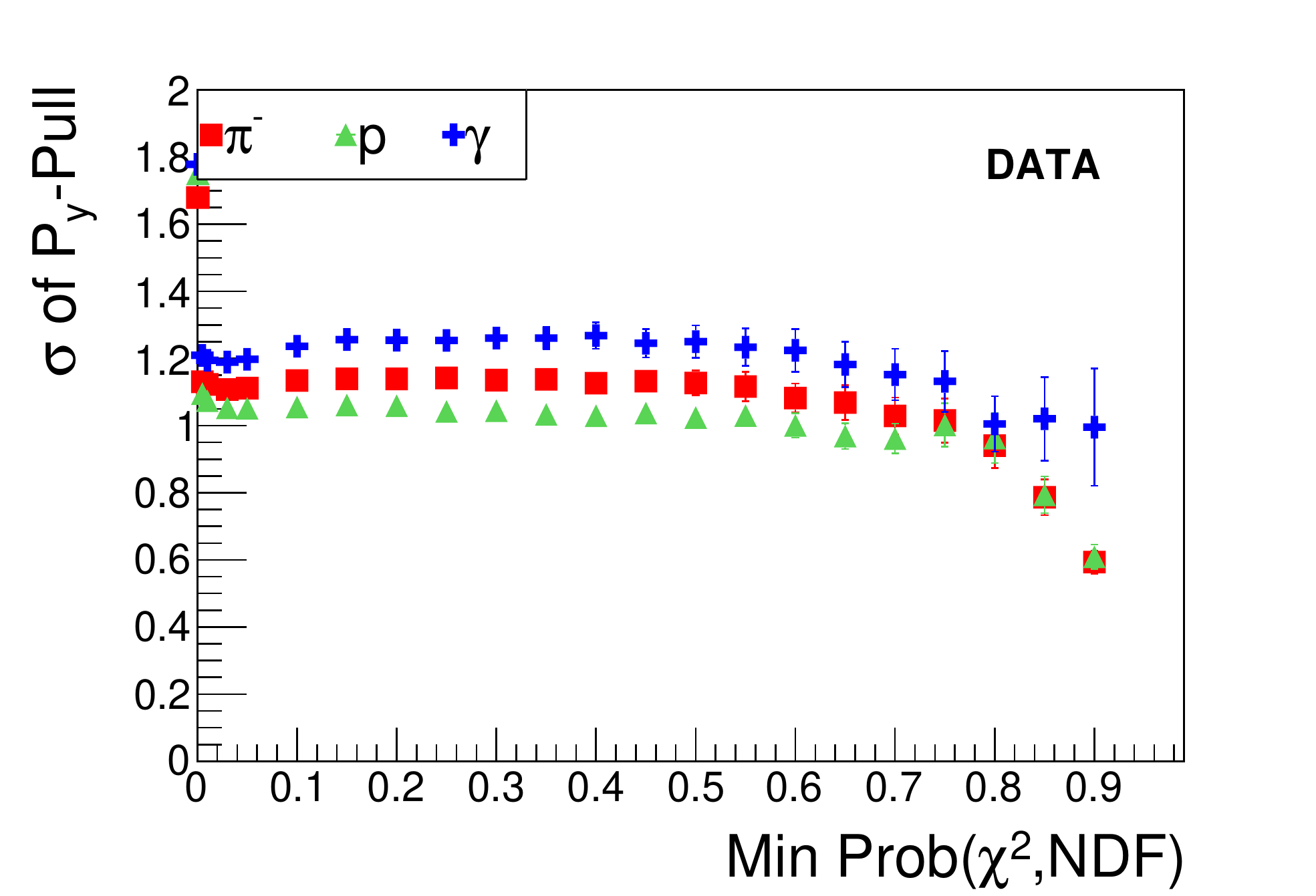}
\includegraphics[width=0.3\textwidth]{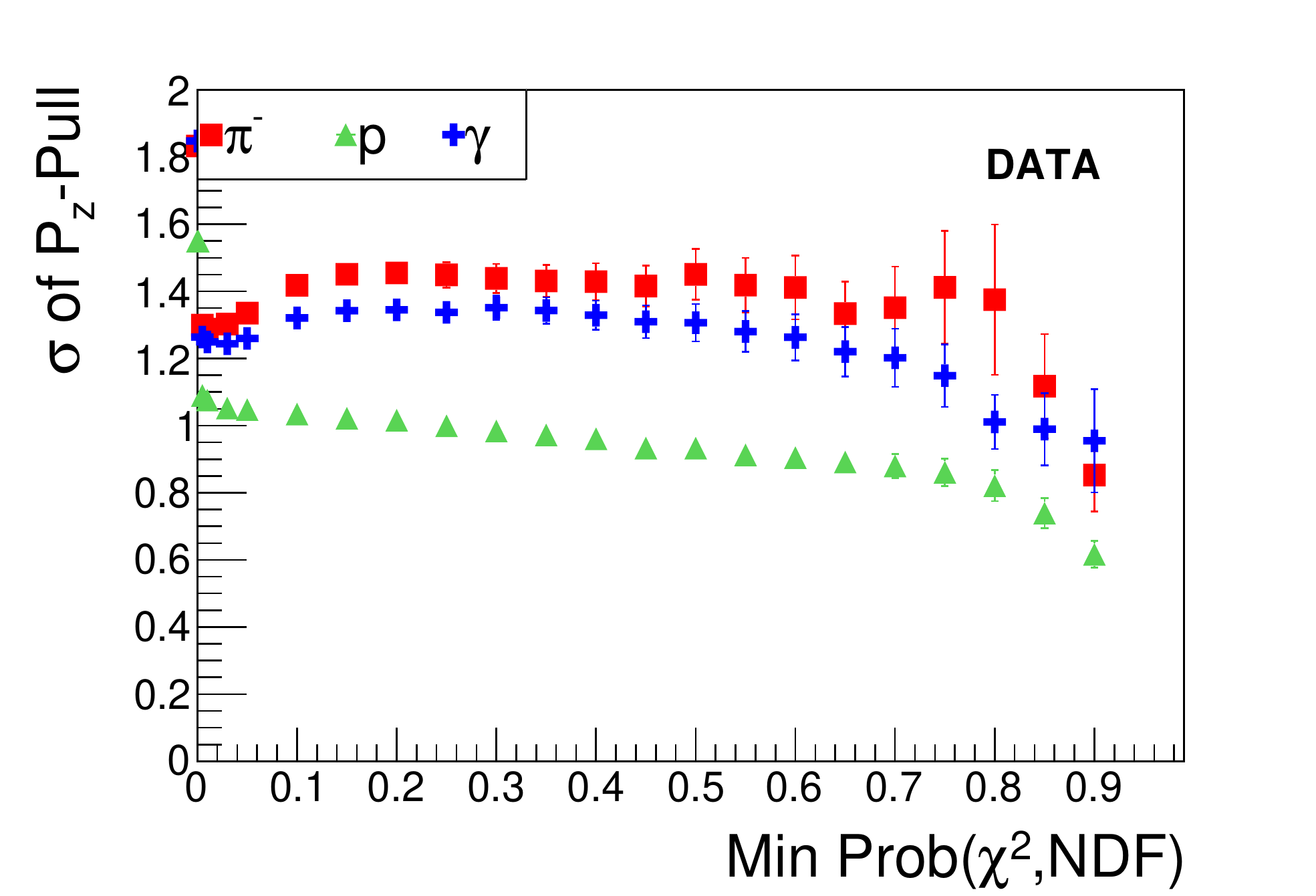}

\includegraphics[width=0.3\textwidth]{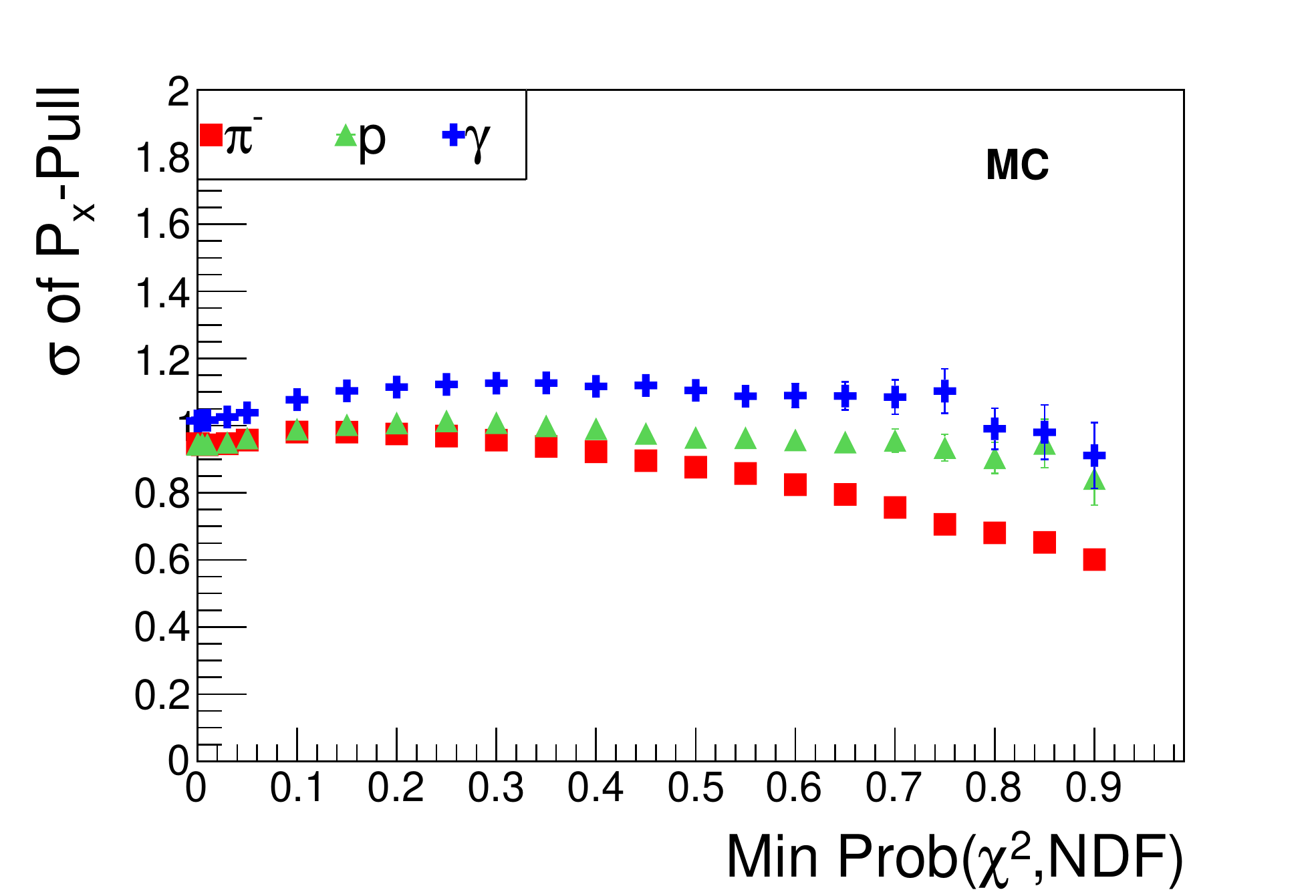}
\includegraphics[width=0.3\textwidth]{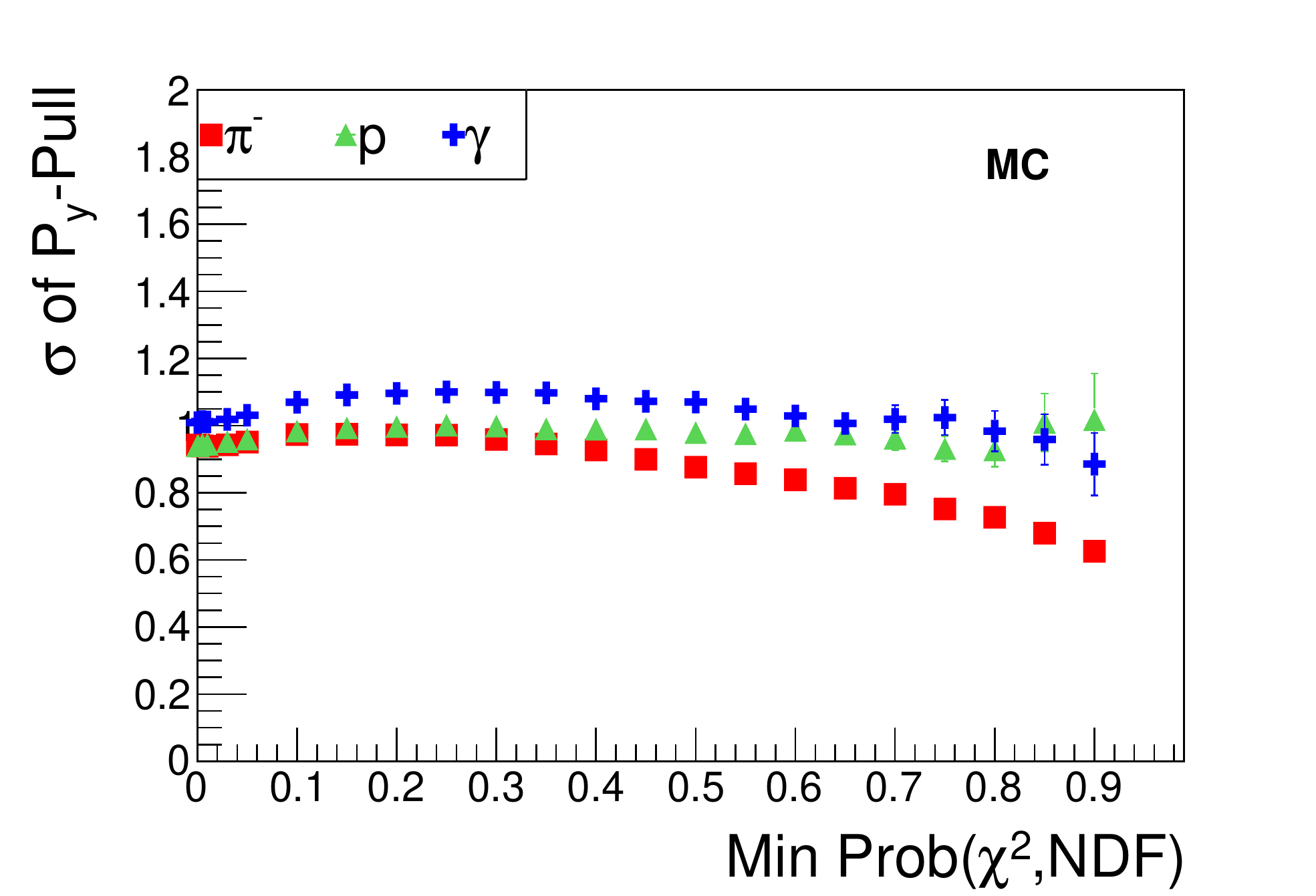}
\includegraphics[width=0.3\textwidth]{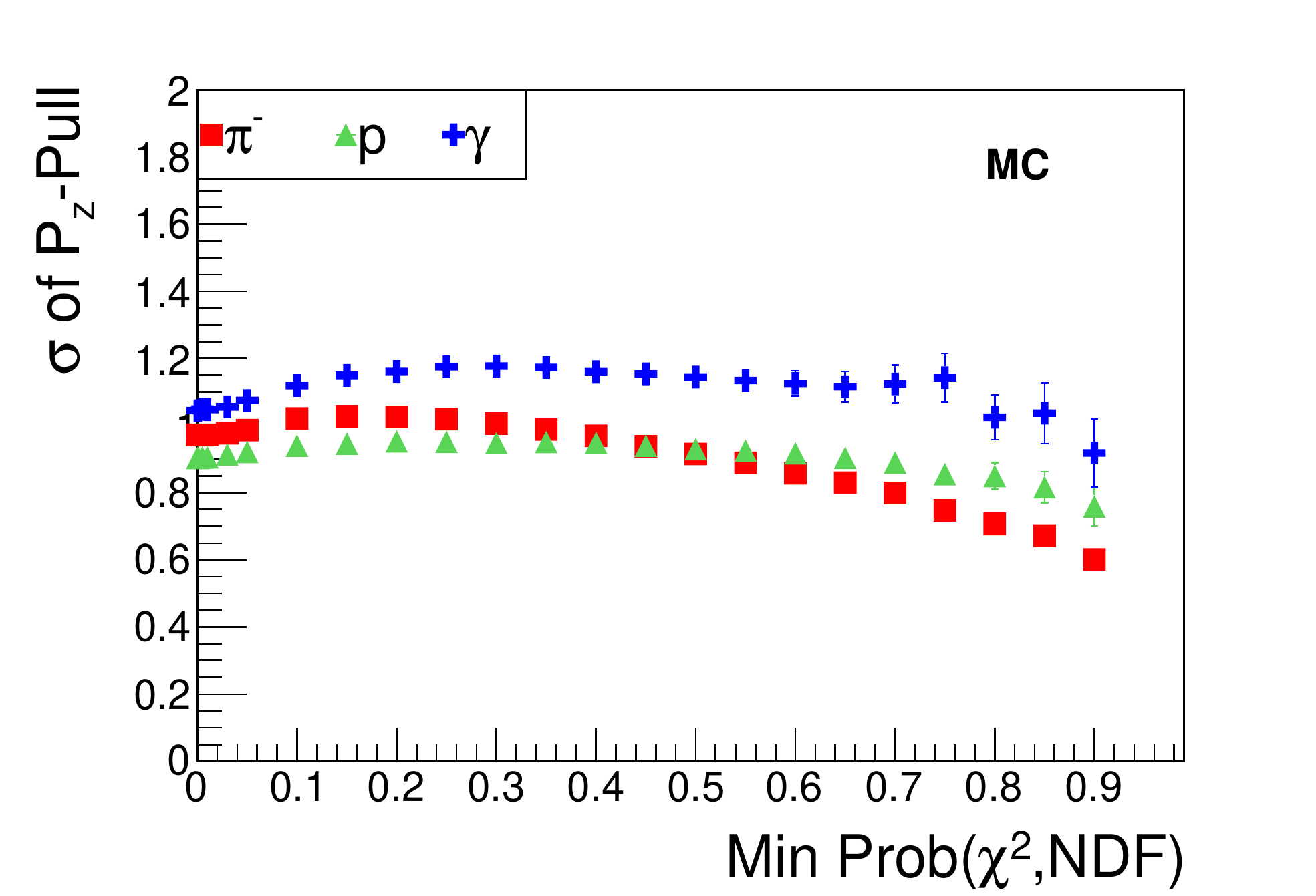}

\caption{\label{fig:kinfitstudy}
Pull means (top) and sigmas (bottom) for the momentum components of each particle as a function of the minimum probability required of the fit from reconstructed $\gamma p \to \eta p$,  $\eta \to \pi^+\pi^-\pi^0$ events.
 (Color online)}
\end{center}  
\end{figure}

As an example, the pull distributions for the momentum components of the $\pi^-$ in reconstructed $\gamma p \to \eta p$, $\eta \to \pi^+\pi^-\pi^0$ events are shown in Fig.~\ref{fig:kinfitpulls}.  Both real and simulated data have roughly Gaussian shapes with similar widths.  More insight into the stability of the results of the kinematic fit can be found by studying the variation of the means and widths of the fit distributions as a function of the fit probability.  The results of such a study are summarized in Fig.~\ref{fig:kinfitstudy}, where broad agreement between the results from real and simulated data is seen.  The means of the pull distributions are generally around zero, except for $p_x$ with a mean of roughly $-0.1$, and the widths within about 20\% of unity.  This level of performance and agreement between data and simulation is acceptable for the initial analysis of data, where very loose cuts on the kinematic fit $\chi^2$ are performed, and steady improvement in the modeling of the covariance matrices of reconstructed particles is expected to continue.

\subsection{Invariant-mass resolution \label{sec:perfchargedresol}}

\begin{figure}[tpb]  
\begin{center}
\includegraphics[width=0.45\textwidth]{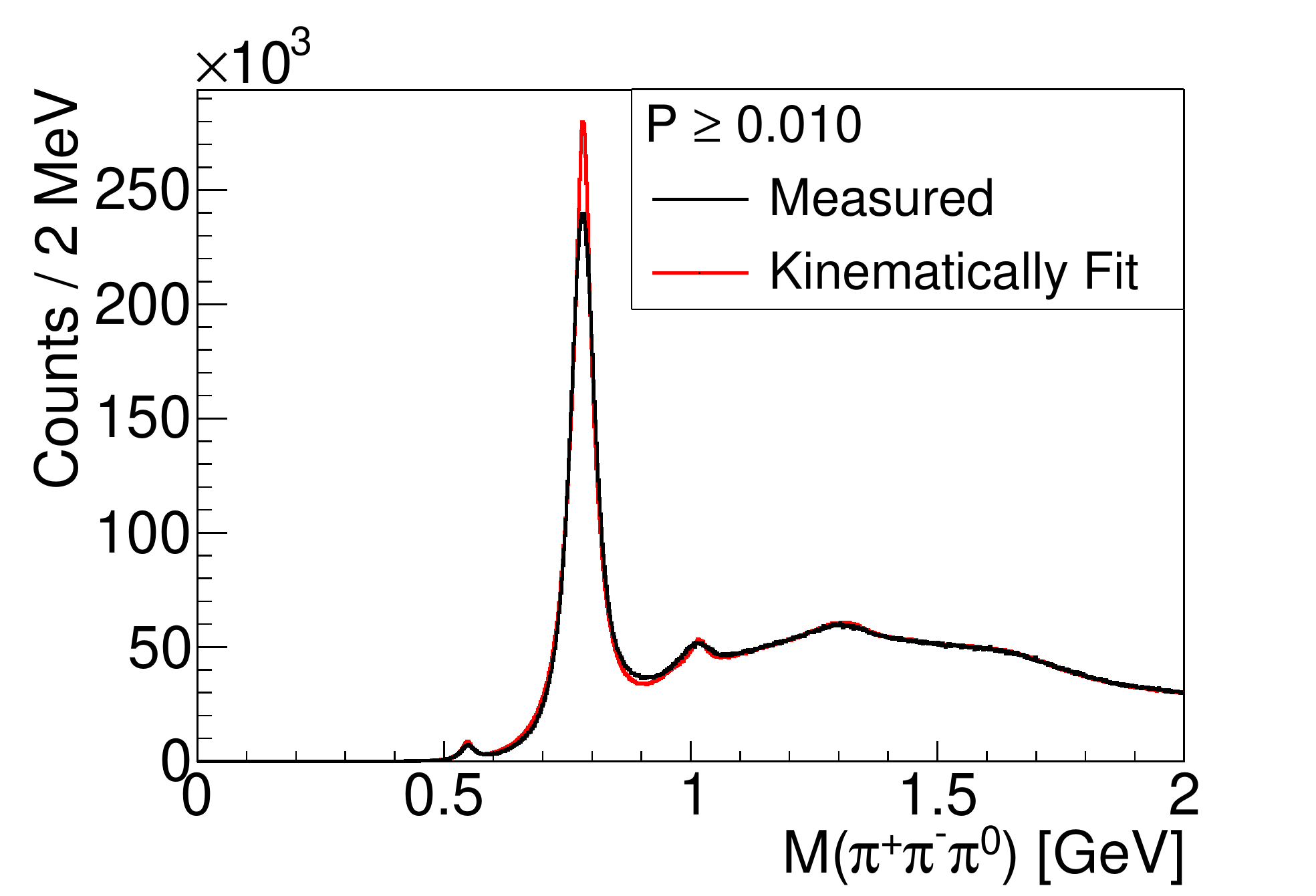}    
\includegraphics[width=0.4\textwidth]{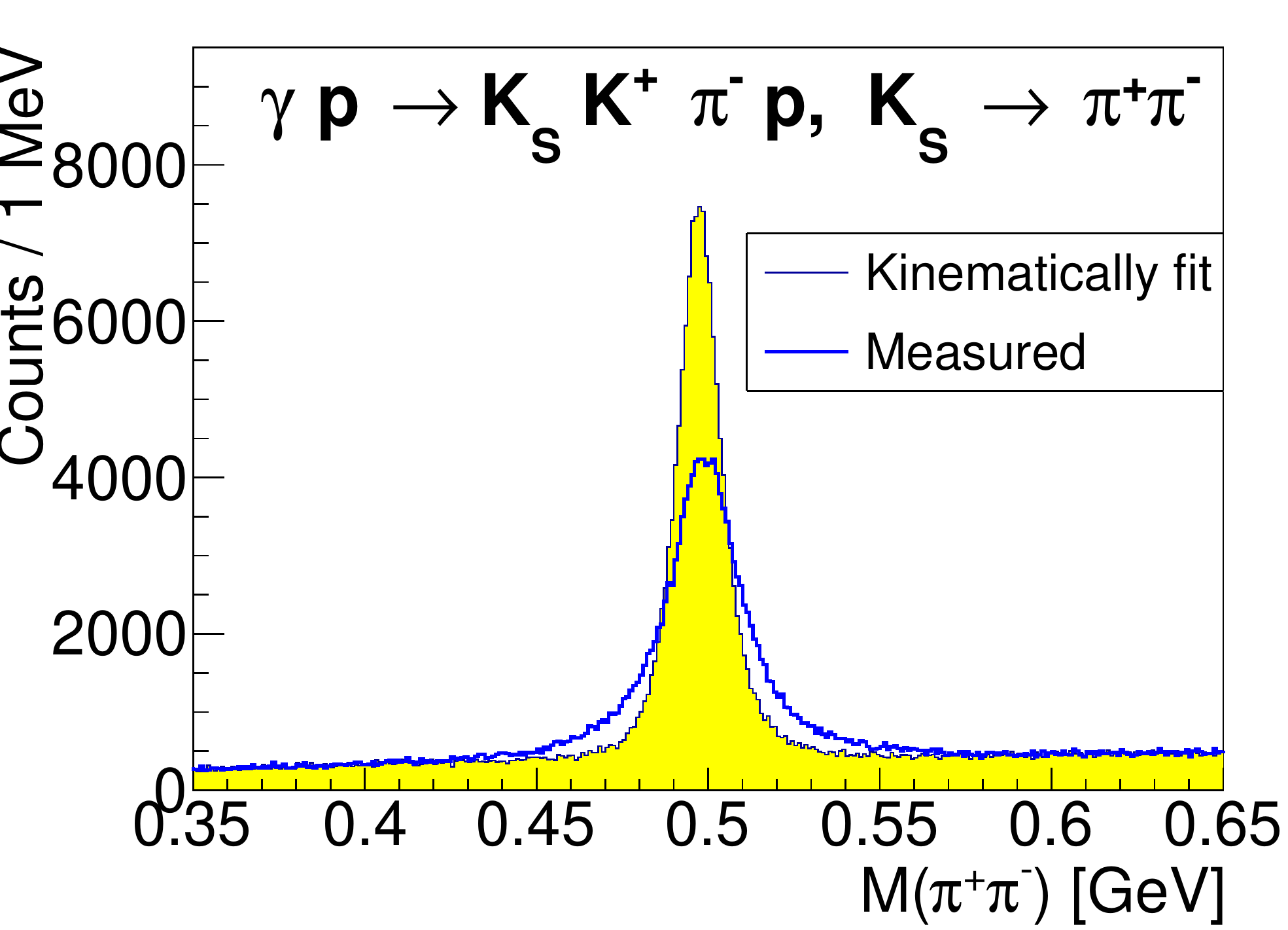}
\caption{\label{fig:invmass1}
(Left) $\pi^+\pi^-\pi^0$ invariant-mass distribution from $\gamma p \to \pi^+\pi^-\pi^0 p$ (Right) $\pi^+\pi^-$ invariant mass distribution from $\gamma p \to K_S K^+ \pi^- p$, $K_S\to\pi^+\pi^-$. (Color online)}
\end{center}
\end{figure}

\begin{figure}[tpb]
\begin{center}\includegraphics[width=0.5\textwidth]{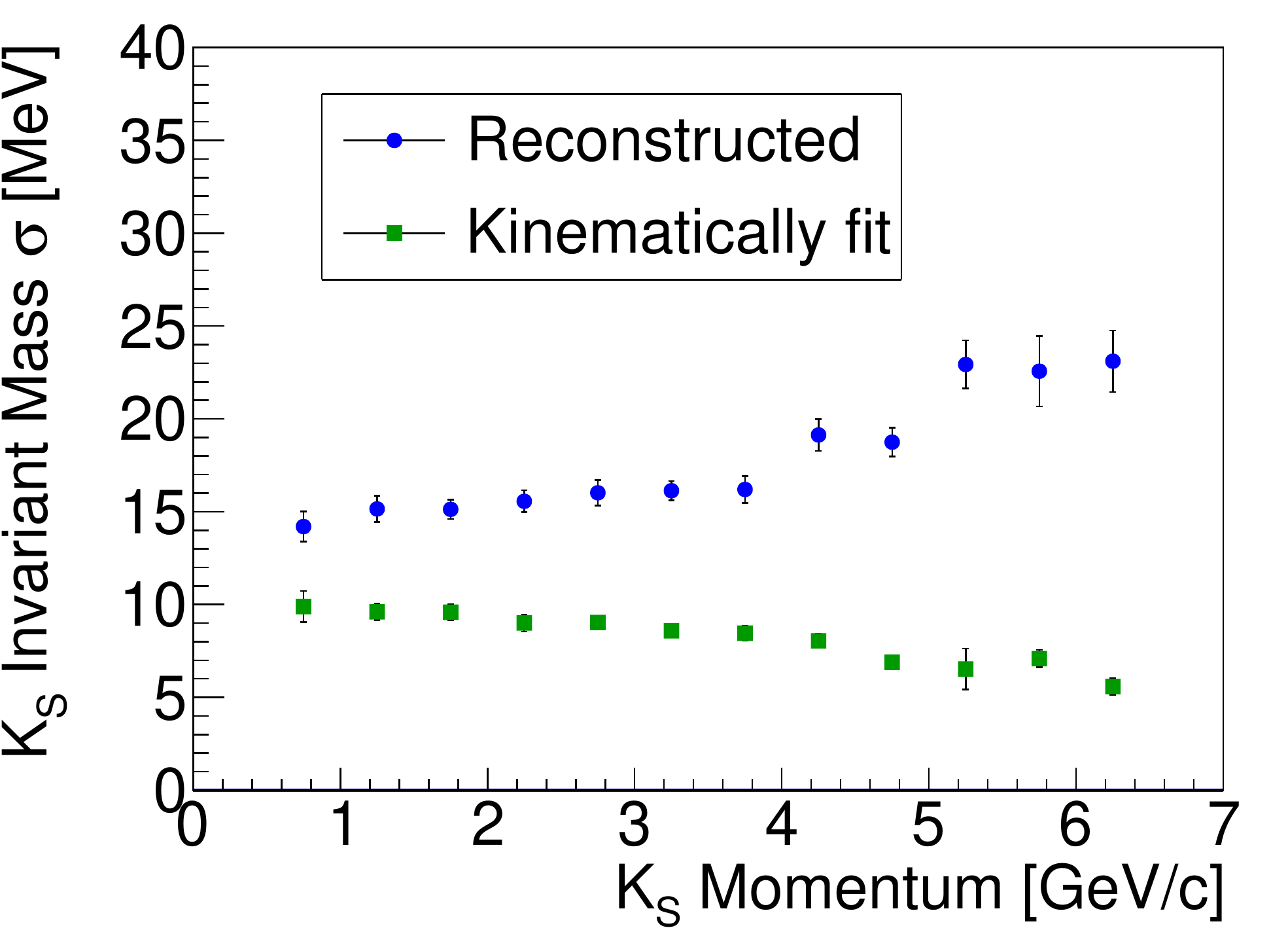}
\caption{\label{fig:invmass1a}
$K_S\to\pi^+\pi^-$ invariant mass resolution for the events shown in Fig.\,\ref{fig:invmass1}, as a function of $K_S$ momentum, both before and after a kinetic fit, which constrains energy and momentum conservation.  
(Color online)}
\end{center}
\end{figure}

\begin{figure}[tpb]
\begin{center}
\includegraphics[width=0.42\textwidth]{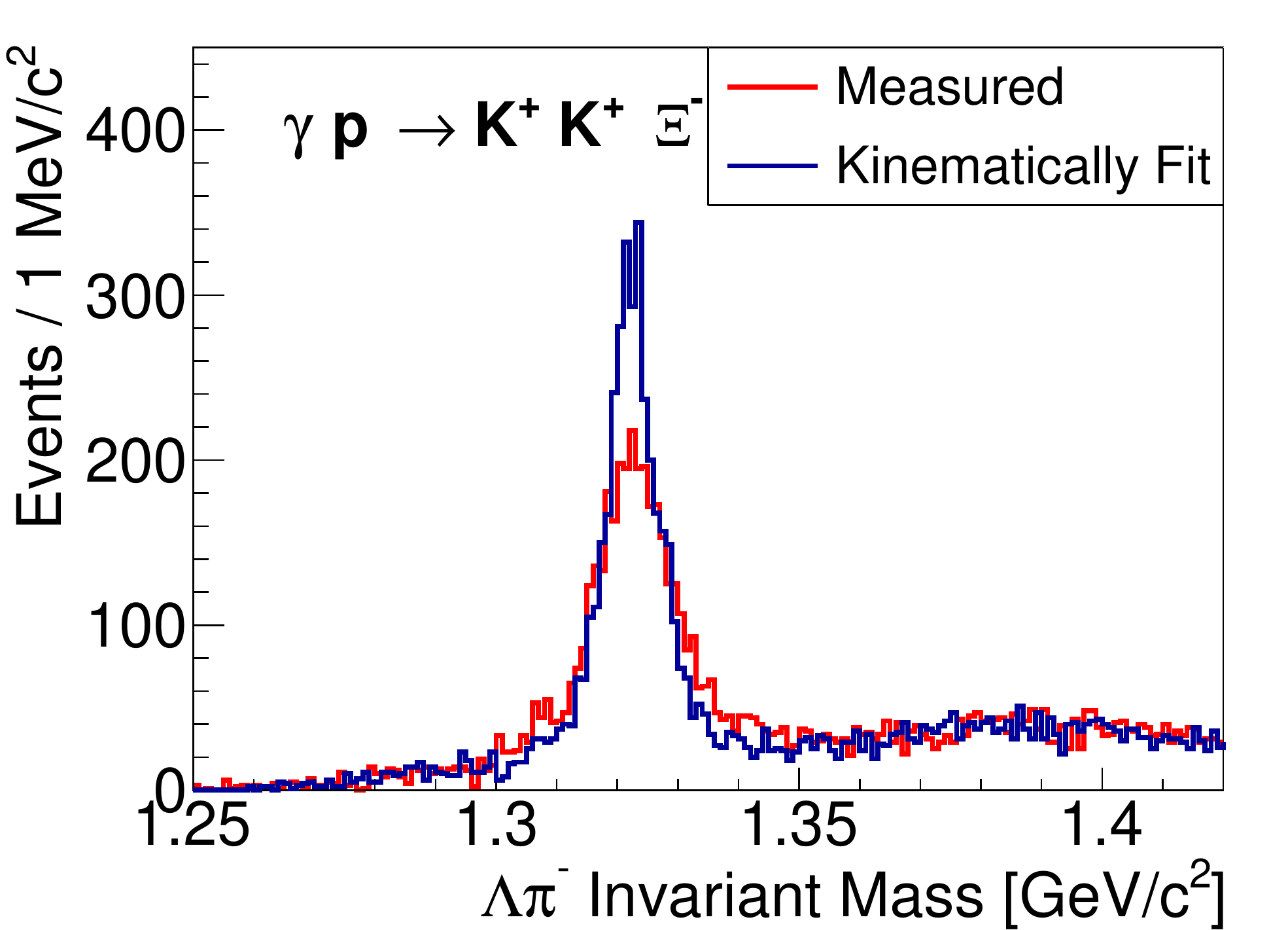}
\includegraphics[width=0.42\textwidth]{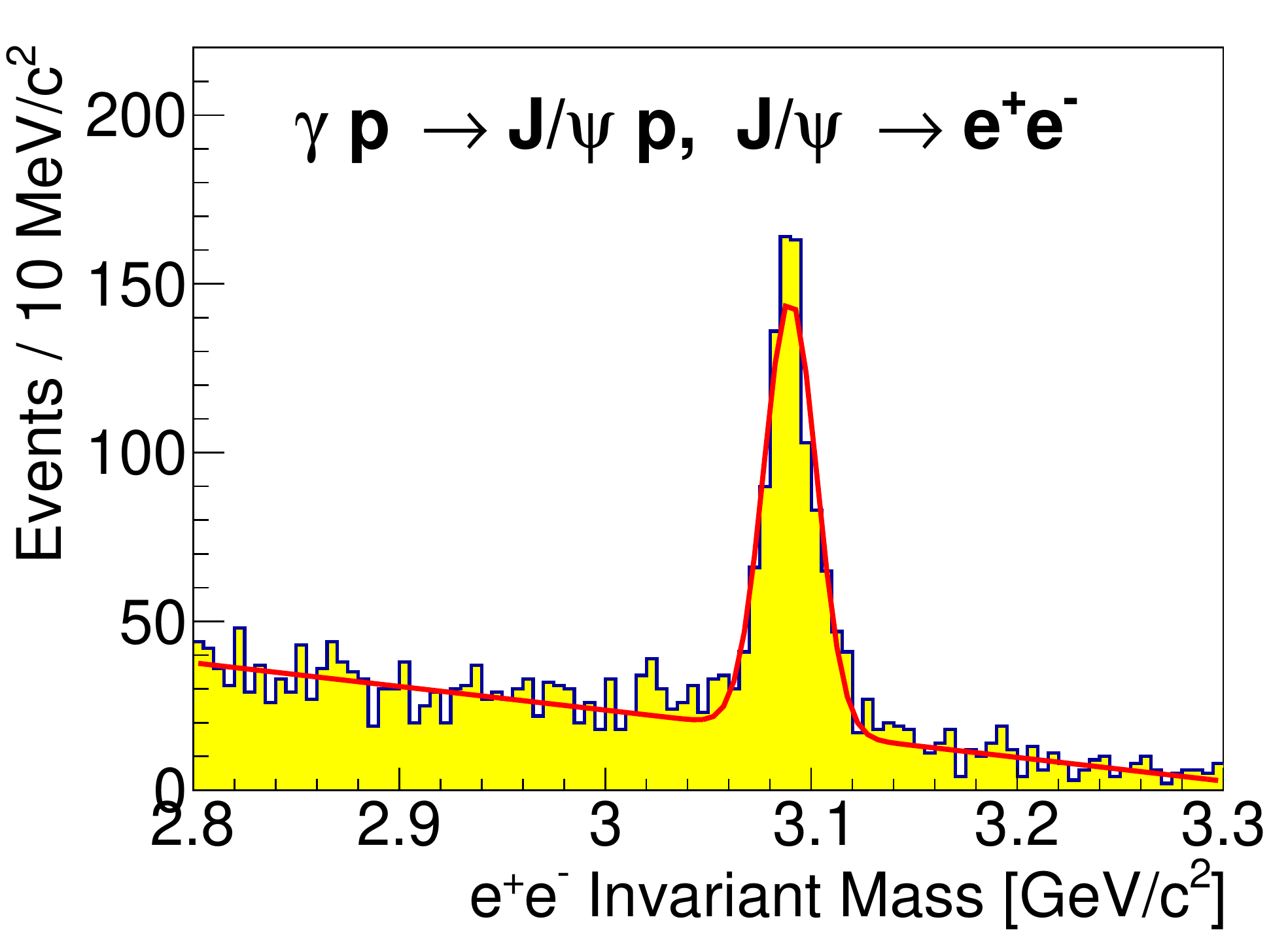}
\caption{\label{fig:invmass2}
(Left) $\Lambda^0\pi^-$ invariant mass distribution from $\gamma p \to K^+ K^+ \pi^- \pi^- p$. (Right) $e^+e^-$ invariant mass distribution from kinematically fit $\gamma p \to e^+e^- p$ events. (Color online)}
\end{center}
\end{figure}

The invariant-mass resolution for resonances depends on the momenta and angles of their decay products.  This resolution has been studied using several different channels, which are illustrated in Figs.~\ref{fig:invmass1} and \ref{fig:invmass2}. A typical meson production channel including both charged particles and photons, $\omega \to \pi^+\pi^-\pi^0$ from $\gamma p \to \omega p$, is shown in the left panel of Fig.~\ref{fig:invmass1}. The distribution shows the strong peak due to $\omega$ meson production.  Other structures are also seen, such as peaks corresponding to the production of $\eta$ and $\phi$ mesons.  The $\omega$ peak resolution obtained is 26.1~MeV when using only the reconstructed  particle 4-vectors, and improves to 16.4~MeV after a kinematic fit. The invariant-mass distribution of $\pi^+\pi^-$ from $\gamma p \to K_S K^+ \pi^- p$, $K_S\to\pi^+\pi^-$ exhibits the peak due to $K_S\to\pi^+\pi^-$ decays (right panel of Fig.\,\ref{fig:invmass1}).  The $K_S$ peak resolution is 17.0~MeV using only the reconstructed charged particle 4-vectors, and improves to  8.6~MeV after a kinematic fit imposing energy and momentum conservation. The dependence of the $K_S\to\pi^+\pi^-$ invariant-mass resolution as a function of $K_S$ momentum is shown in Fig.\,\ref{fig:invmass1a} , both before and after an energy/momentum-constraint kinematic fit.  

The invariant mass of $\Lambda^0\pi^-$ from $\gamma p \to K^+ K^+ \pi^- \pi^- p$ is shown in the left panel of Fig.\,\ref{fig:invmass2},  illustrating the peak due to $\Xi^- \to \pi^- \Lambda^0$, $\Lambda^0 \to p \pi^-$.  The $\Xi^-$ peak resolution obtained is 7.3~MeV when using only the reconstructed charged particle 4-vectors, and improves to 4.6~MeV after a kinematic fit imposing energy and momentum conservation and the additional constraint that the mass of the $p \pi^-$ pairs must be that of the $\Lambda^0$ mass.  The $e^+e^-$ invariant mass distribution from kinematically fit $\gamma p \to e^+e^- p$ events is shown in the right panel of Fig.\,\ref{fig:invmass2}, illustrating the peak due to $J/\psi\to e^+e^-$.  The resolution of the peak is 13.7~MeV.            
 
\subsection{Particle identification \label{sec:perfpid}}

Particle identification in \gx{} uses information from both energy loss in different detector systems and time-of-flight measurements.  This information can be used for identification in several ways.  The simplest method is to apply selections directly on the relevant PID variables.  To include detector resolution information, one can create a $\chi^2$ variable comparing a measured value to the expected value for a particular hypothesis, that is
\begin{equation}
    \chi^2(p) = \left(  \frac{ X(\mathrm{measured}) - X(\mathrm{expected})_p}{\sigma_X} \right)^2
\end{equation}
where $X$ is the given PID variable, $p$ is the particle hypothesis, and $\sigma_X$ is the resolution of this variable.  Multiple PID variables can be combined into one probability, or a figure-of-merit.   Standard, loose selections on time-of-flight and energy loss are sufficient for initial physics analyses, while the performance of more complicated selections is being actively studied.

\begin{figure}[tbp]
\begin{center}
\includegraphics[width=0.6\textwidth]{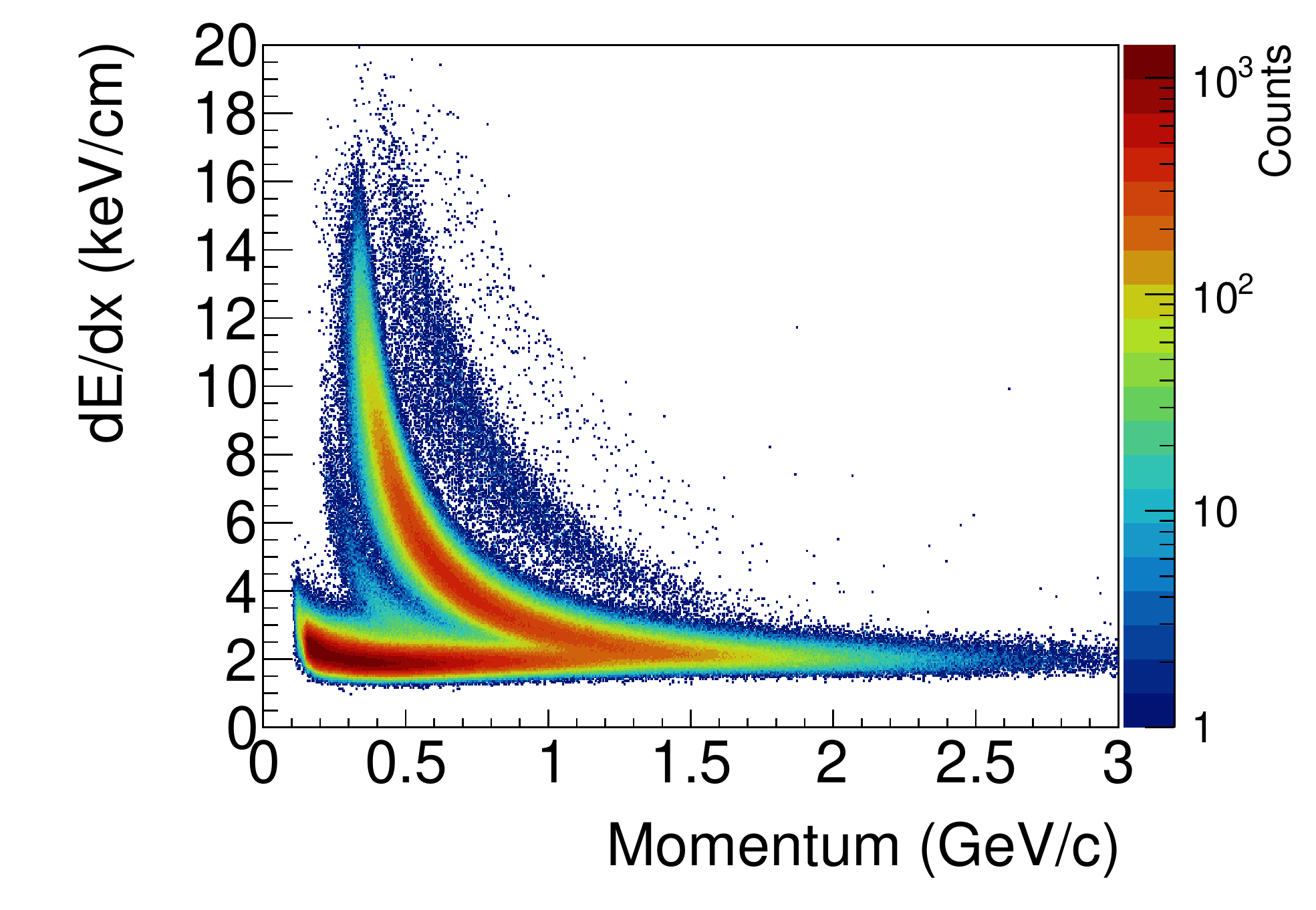}
\caption{\label{fig:performcdcdedx}
CDC energy loss ($dE/dx$) for positively charged particles that have at least 20 hits in the detector, as a function of measured particle momentum.  The band corresponding to protons curves upwards, showing a larger energy loss than pions and other lighter particles at low momentum.  The two bands show a clear separation for momenta  $\lesssim 1$~GeV.  A faint kaon band can be seen between them.
}
\end{center}
\end{figure}

At sufficiently large $\theta$, the energy loss for charged particles in the central drift chamber $dE/dx$ can be used.   Fig.~\ref{fig:performcdcdedx} illustrates these distributions for positively charged particles, showing a clear separation of pions and protons in the momentum range $\lesssim 1$~GeV. 
The $dE/dx$ resolution is approximately 27\%, with the separation between the pion and proton bands dropping from about $8\sigma$ at $p=0.5$~GeV/$c$ to about $2\sigma$ at $p=1.0$~GeV/$c$, with both bands fully merged by $p=1.5$~GeV/$c$.

\begin{figure}[tbp]
\begin{center}          
\includegraphics[width=0.29\textwidth]{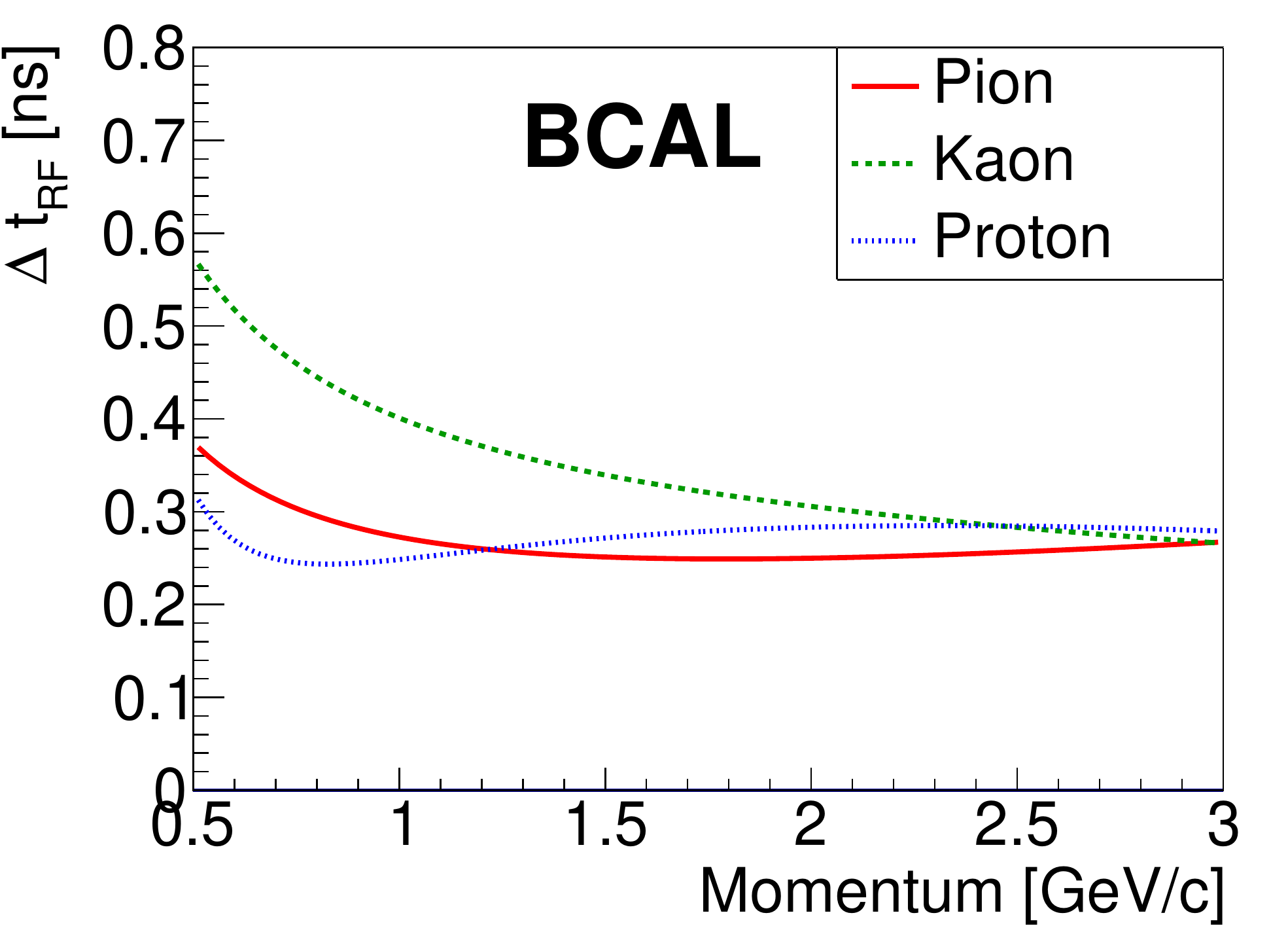}
\includegraphics[width=0.29\textwidth]{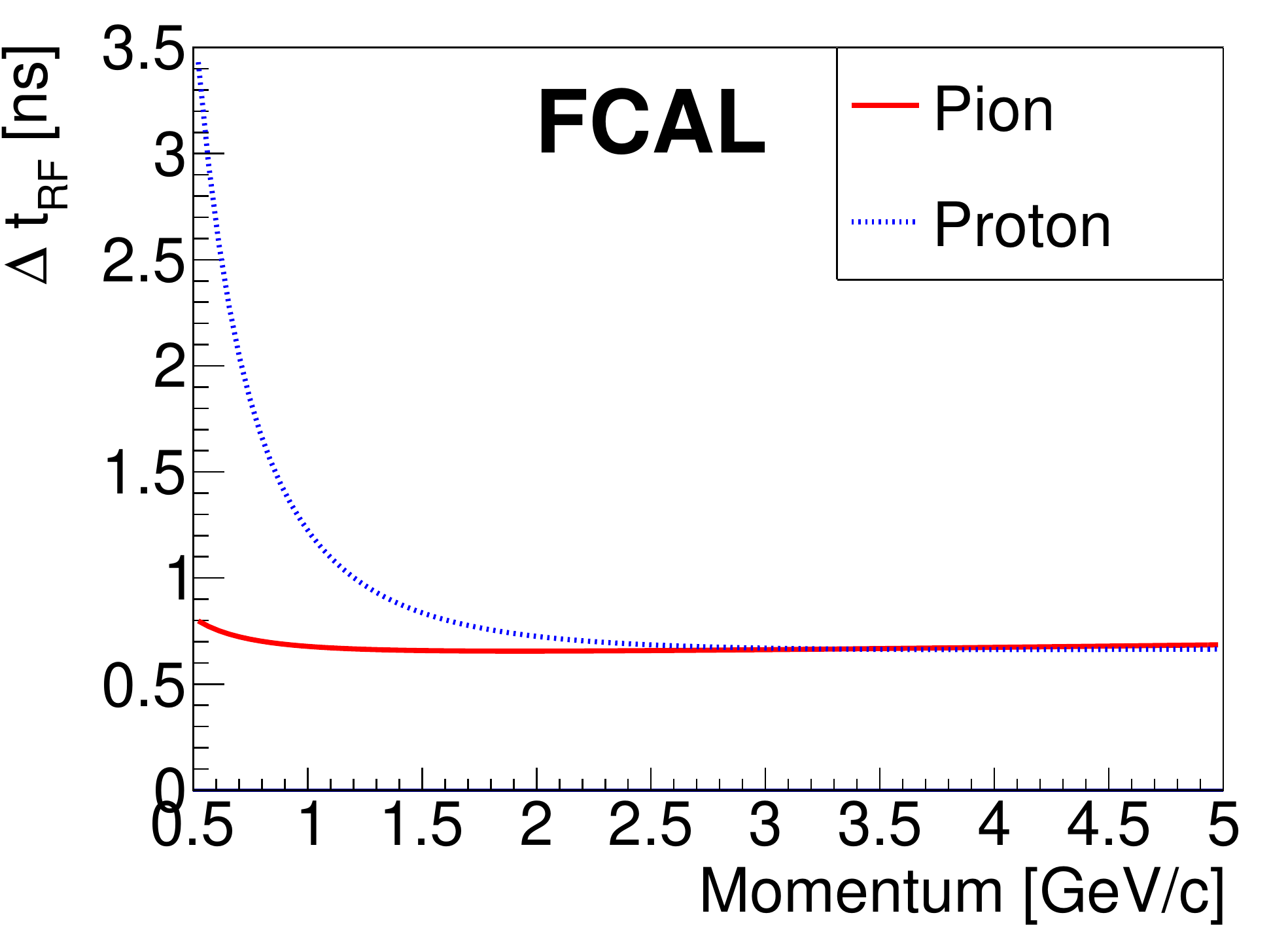}
\includegraphics[width=0.29\textwidth]{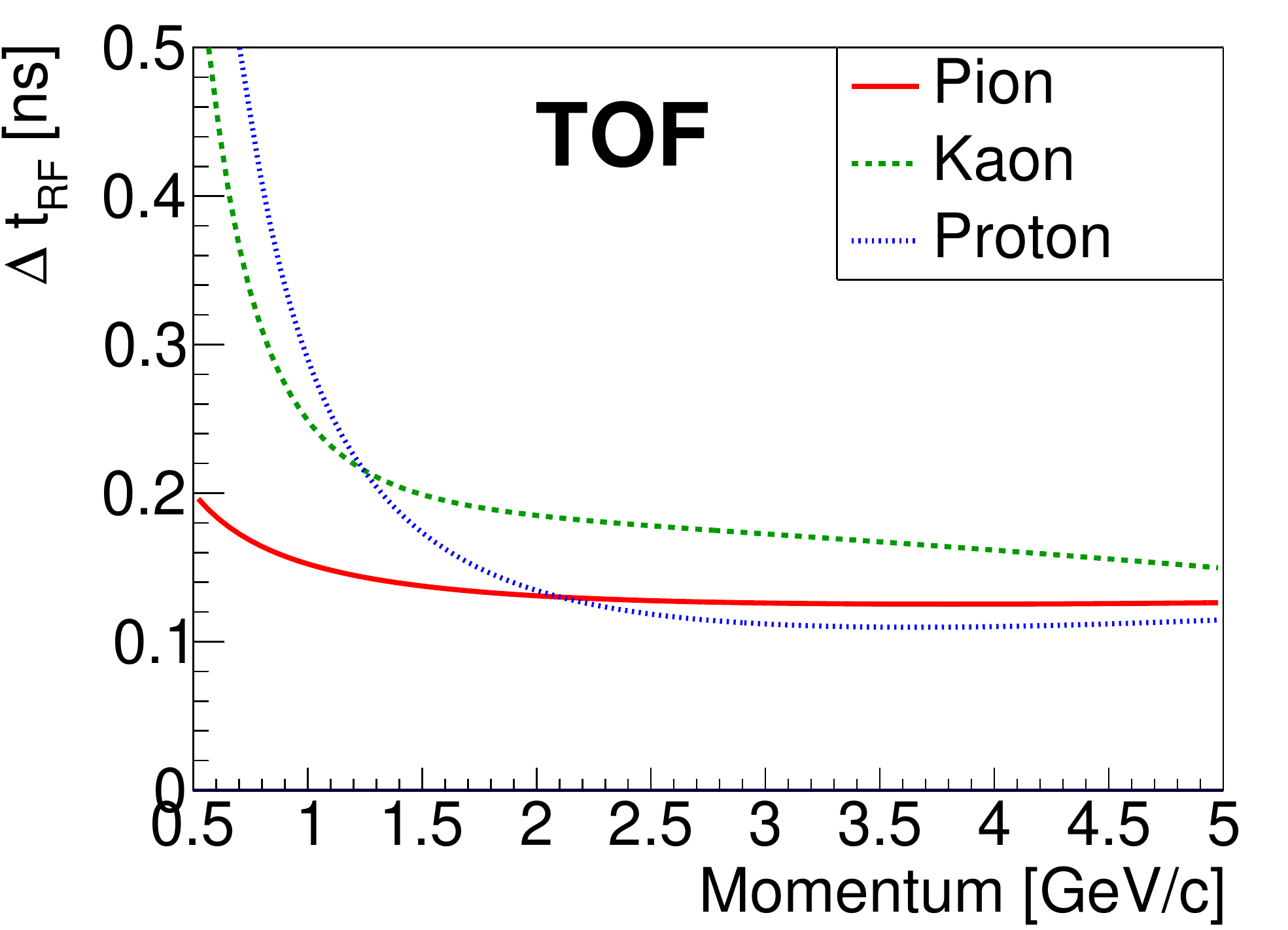}

\caption{\label{fig:timingresol}
Resolution as a function of particle momentum for  $\Delta t_\mathrm{RF}$ in various subdetectors: (left) BCAL, (center) FCAL, (right) TOF
 (Color online)}
\end{center}
\end{figure}

The primary means of particle identification is through time-of-flight measurements, and information from several sources is combined to make the most accurate determination.  The RF reference signal from the accelerator is used to define the time when each photon bunch enters the target.  The reconstructed final-state particles are used to determine which photon bunch most likely generated the detected reaction, with the primary determination coming from the signals from the Start Counter associated with the charged particle tracks.  The photon bunch determination has a resolution of $<10$~ps. Each charged particle is associated with additional timing information based on the hit in the highest resolution detector (for example the BCAL or TOF).  The flight  time to this measured hit $t_\mathrm{meas}$ relative to the time of the photon bunch that generated the event $t_\mathrm{RF}$ can be used to distinguish between particles of different mass.  Two common variables that are used are the velocity ($\beta$) determined using the measured time-of-flight and the momentum of the particle, and $\Delta t_\mathrm{RF}$, the difference between the measured and RF times after they both have been extrapolated back to the center of the target, assuming some particle-mass hypothesis.
An example of the separation between different particle types can be seen in Fig.~\ref{fig:betavsp}.
The loose selections used for initial analyses of this data placed on the $\Delta t_\mathrm{RF}$ distributions and the momentum dependence of the resolution of this variable in different detectors are shown in Fig.~\ref{fig:timingresol}.  
Requiring reconstructed particles to have  $\Delta t_\mathrm{RF} \lesssim 1-2$~ns has been found to be sufficient for analyses of high-yield channels which are the focus of initial analysis.  The study of the selections required for more demanding channels is ongoing.

\begin{figure}[tbp]
\begin{center}
\includegraphics[width=0.4\textwidth]{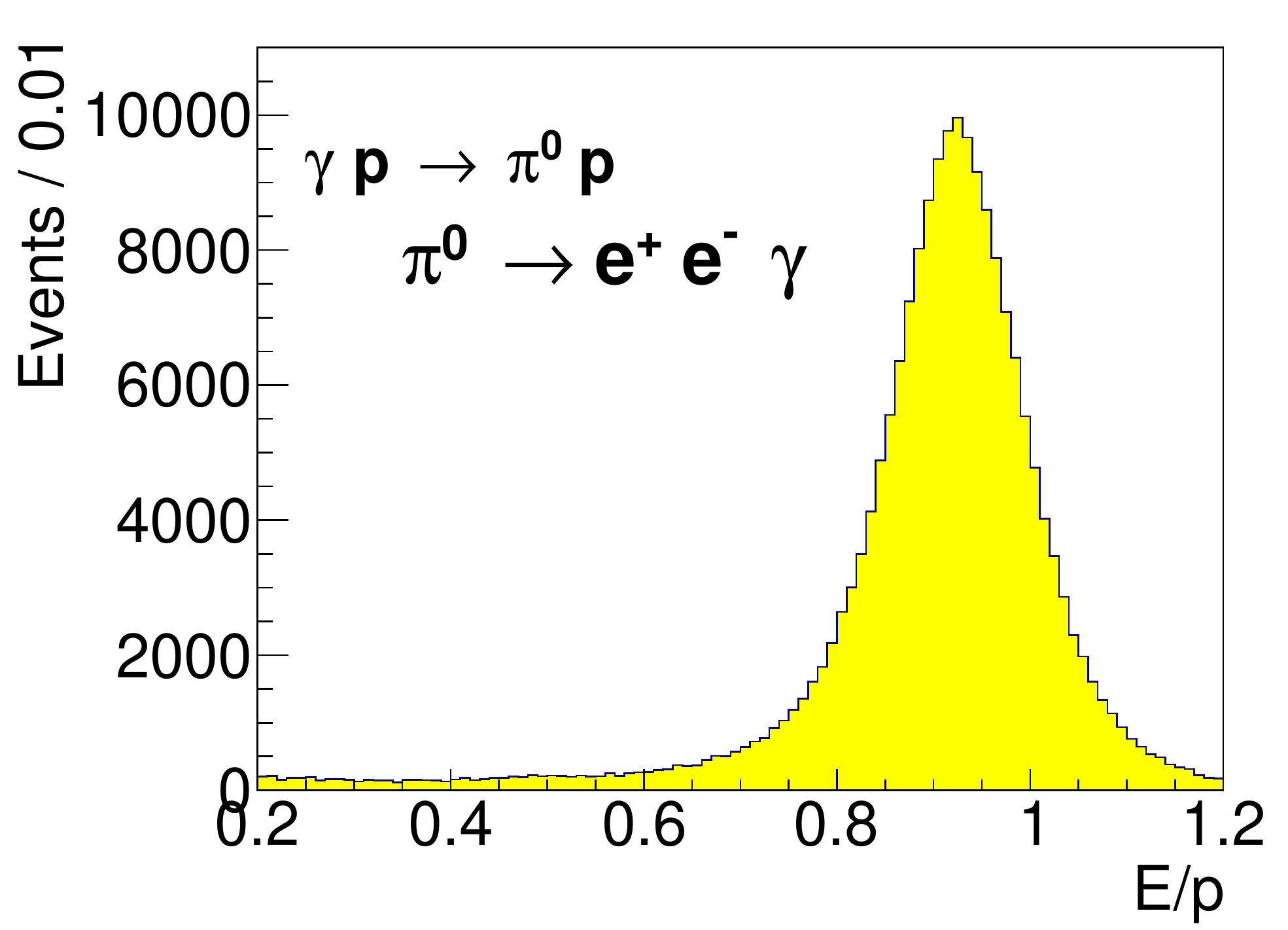}
\includegraphics[width=0.4\textwidth]{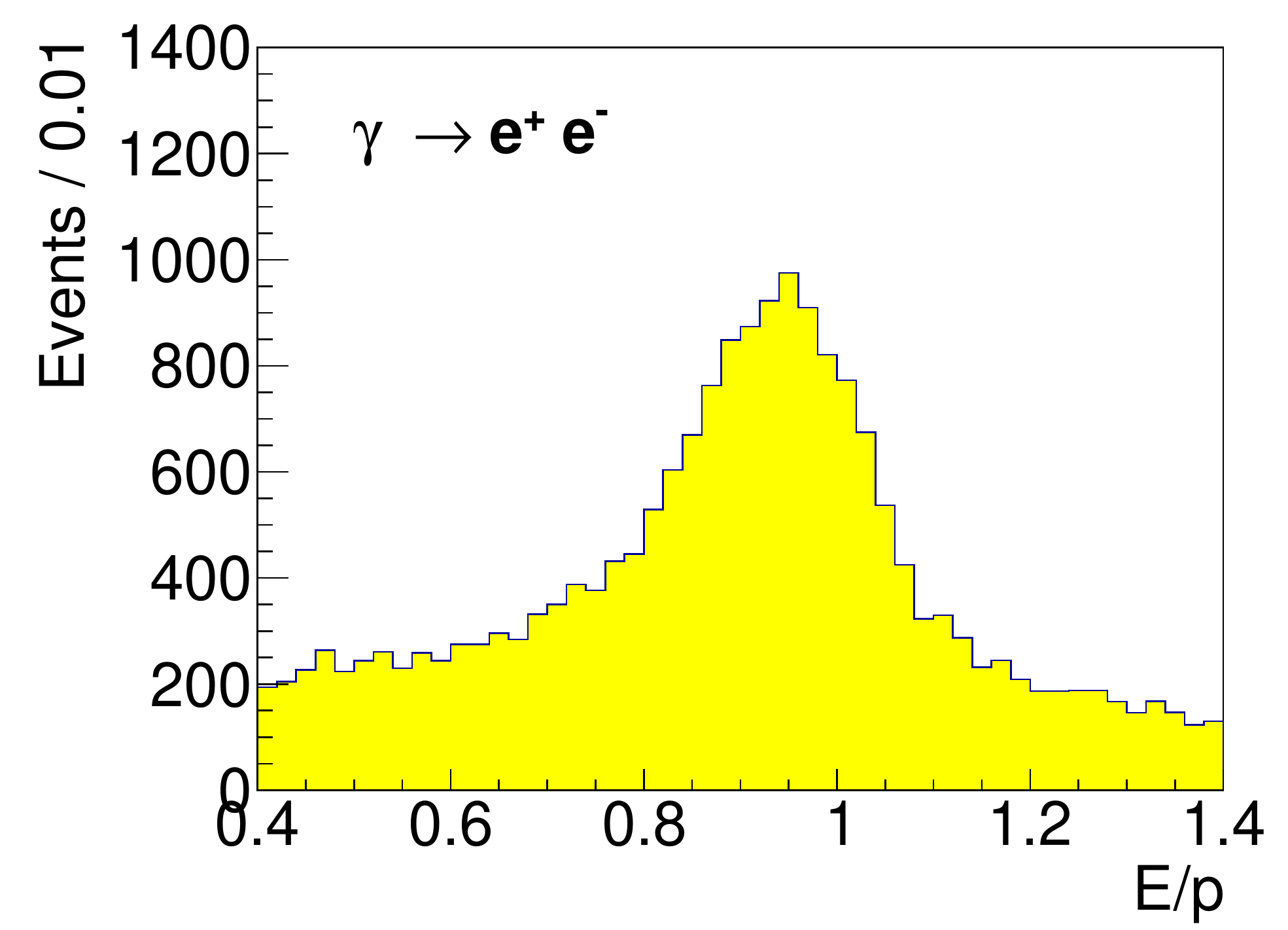}
\caption{\label{fig:performeop}
Electron identification in the calorimeters is performed using the $E/p$ variable, the ratio of the energy loss in the electromagnetic calorimeters ($E$) to the momentum reconstructed in the drift chambers ($p$).  Left) This distribution was obtained using $e^{\pm}$ showers reconstructed in the FCAL from the reaction $\gamma p \to \pi^0 p$, $\pi^0\to e^+e^-\gamma$. Right) $e^{\pm}$ showers reconstructed in the BCAL from photon conversions.
}
\end{center}
\end{figure}

Electrons are identified using the ratio of their energy loss in the electromagnetic calorimeters $E$ to the momentum reconstructed in the drift chambers~$p$.  This $E/p$ ratio should be approximately unity for electrons and less for hadrons.  The overall distributions of this variable are illustrated 
in  Fig.~\ref{fig:performeop}.  Other variables, such as the shape of the showers generated by the charged particles in the calorimeter, promise to provide additional information to separate electron and hadron showers.

\section{Summary and outlook\label{sec:summary} }
\begin{figure}[tbh]\centering
\includegraphics[width=0.48\textwidth]{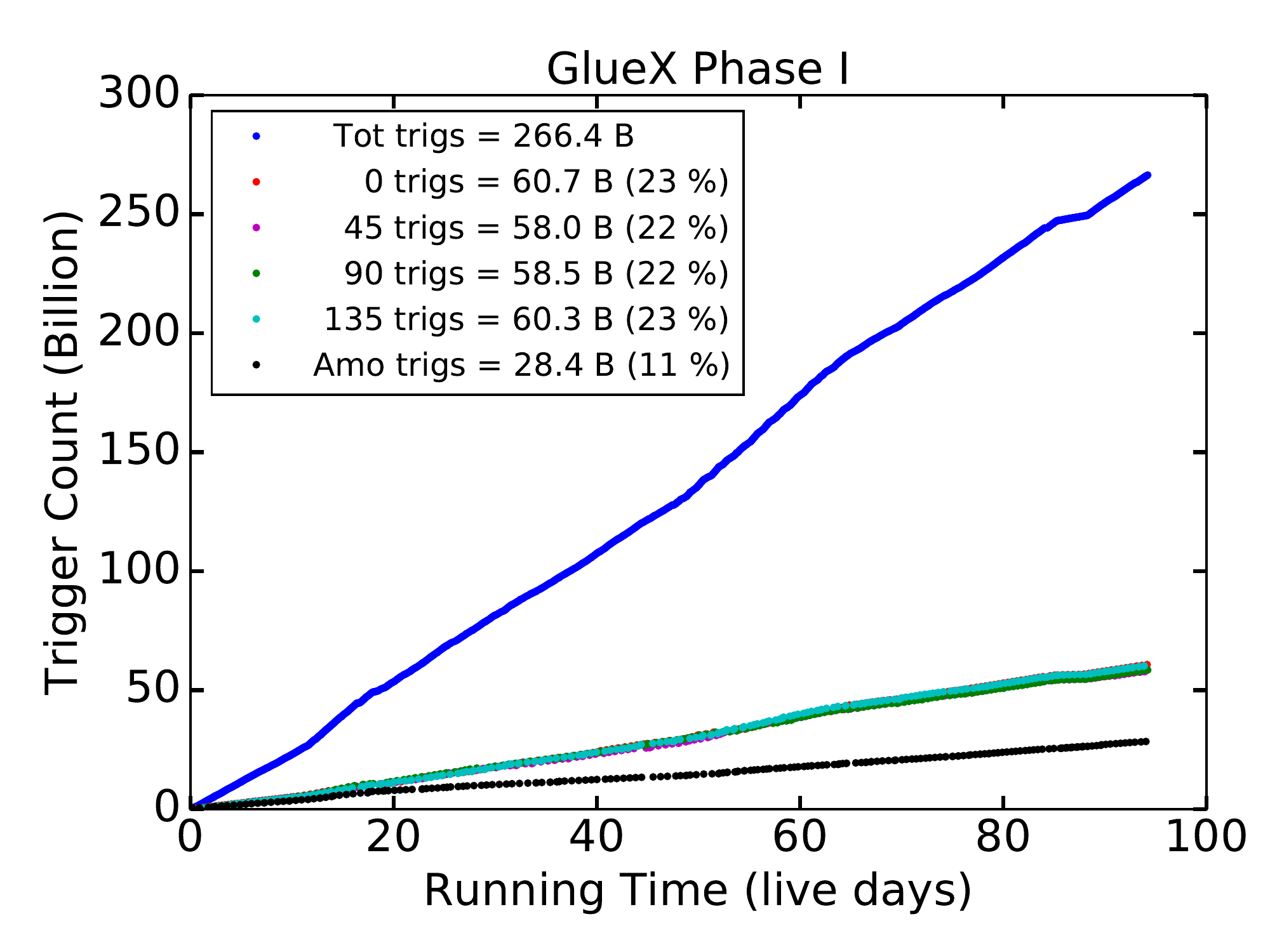}
\caption{\label{fig:plot_rcdb3_phaseI} 
Plot of integrated number of triggers versus the number of live days in 2017 and 2018. The legend provides the number of triggers for the four diamond orientations relative to the horizontal (0, 45, 90, 135$^\circ$) and the amorphous radiator. The trigger curves of the four diamond configurations fall on top of one another, as we attempted to match the amount of data taken for each configuration. 
(Color online)    
 }   
\end{figure}

We have presented the design, construction, and performance, of the beamline and detector of the \gx{} experiment in Hall D at Jefferson Lab during its first phase of operation. The experiment operated routinely at an incident photon flux of $2\times 10^{7}$ photons/s in the coherent peak with an open trigger, taking data at 40 kHz, and recording 600 MB/s to tape with live time $>$95\%. During this period the experiment accumulated  121.4 pb$^{-1}$ in the coherent peak and 319.4 pb$^{-1}$ total for $E_\gamma>$8.1 GeV. Data were collected in two sets of orthogonal linear polarizations of the incident photons, with $\sim$23\% of the data in each of the four orientations. The remaining $\sim$11\% was collected with unpolarized photons. Approximately 270 billion triggers ($\sim$ 3PB) were accumulated during this period, as shown in Fig.\,\ref{fig:plot_rcdb3_phaseI}.  

The operational characteristics of the charged and neutral particle detectors, trigger, DAQ, online and offline systems have been verified, and individual components performed as designed. The detector is able to reconstruct exclusive final states, reconstruction efficiencies have been determined, and Monte Carlo simulations compare well with experimental data. The infrastructure is in place to process our high volume of data both on the JLab computing farm as well on other offsite facilities, providing the ability to process the data in a timely fashion.

Future running will include taking data at higher luminosity  and with improved particle identification capability. The \gx~experiment has already implemented the necessary infrastructure to allow the experiment to operate at a flux of $5\times10^{7}$ photons/s in the coherent peak for the upcoming run periods and has added a new DIRC detector\footnote{Four ``bar boxes" from the BaBar DIRC\cite{Aubert:2001tu} detector have been installed and tested.} to extend particle identification of kaons to higher momenta.


\section{Acknowledgments}  
We gratefully acknowledge the outstanding efforts of technical support at all the collaborating institutions and the support groups at Jefferson Lab that completed the assembly, installation,
and maintenance of the detector. We acknowledge the contributions of D. Bennett, M. Lara, A. Subedi and P. Smith to the construction and commissioning of the Forward Calorimeter. We thank E.C. Aschenauer, G. Young and all members of the JLab 12 GeV Project for guidance and direction during the design and construction phases of the project. This work was supported in part by the U.S. Department of Energy, the U.S. National Science Foundation, the Natural Sciences and Engineering Research Council of Canada (NSERC),the German Research Foundation, Forschungszentrum J\"ulich GmbH, GSI Helmholtzzentrum f\"{u}r Schwerionenforschung GmbH, the Russian Foundation for Basic Research, the UK Science and Technology Facilities Council, the Chilean Comisi\'{o}n Nacional de Investigaci\'{o}n Cient\'{i}fica y Tecnol\'{o}gica, the National Natural Science Foundation of China, and the China Scholarship Council. This material is based upon work supported by the U.S. Department of Energy, Office of Science, Office of Nuclear Physics under contract DE-AC05-06OR23177. 

\newpage

\section*{References}
   
\bibliography{GlueX_nim}

\begin{thebibliography}{10}
\expandafter\ifx\csname url\endcsname\relax
  \def\url#1{\texttt{#1}}\fi
\expandafter\ifx\csname urlprefix\endcsname\relax\def\urlprefix{URL }\fi
\expandafter\ifx\csname href\endcsname\relax
  \def\href#1#2{#2} \def\path#1{#1}\fi

\bibitem{Crede:2008vw}
V.~Crede, C.~A. Meyer, {The Experimental Status of Glueballs}, Prog. Part.
  Nucl. Phys. 63 (2009) 74--116.
\newblock \href {http://arxiv.org/abs/0812.0600} {\path{arXiv:0812.0600}},
  \href {https://doi.org/10.1016/j.ppnp.2009.03.001}
  {\path{doi:10.1016/j.ppnp.2009.03.001}}.

\bibitem{Meyer:2010ku}
C.~A. Meyer, Y.~Van~Haarlem, {The Status of Exotic-quantum-number Mesons},
  Phys. Rev. C82 (2010) 025208.
\newblock \href {http://arxiv.org/abs/1004.5516} {\path{arXiv:1004.5516}},
  \href {https://doi.org/10.1103/PhysRevC.82.025208}
  {\path{doi:10.1103/PhysRevC.82.025208}}.

\bibitem{Meyer:2015eta}
C.~A. Meyer, E.~S. Swanson, {Hybrid Mesons}, Prog. Part. Nucl. Phys. 82 (2015)
  21--58.
\newblock \href {https://doi.org/10.1016/j.ppnp.2015.03.001}
  {\path{doi:10.1016/j.ppnp.2015.03.001}}.

\bibitem{gluex-ref}
{The GlueX Collaboration}, {The GlueX Experiment in Hall~D},
  \href{https://halldweb1.jlab.org/wiki/index.php/GlueX_Project_Overviews}{GlueX
  Project Overviews} (hyperlink) (2010).

\bibitem{Leemann:2001dg}
C.~W. Leemann, D.~R. Douglas, G.~A. Krafft, {The Continuous Electron Beam
  Accelerator Facility: CEBAF at the Jefferson Laboratory}, Ann. Rev. Nucl.
  Part. Sci. 51 (2001) 413--450.
\newblock \href {https://doi.org/10.1146/annurev.nucl.51.101701.132327}
  {\path{doi:10.1146/annurev.nucl.51.101701.132327}}.

\bibitem{CEBAF12GeV}
H.~Areti, et~al., {CEBAF at 12 GeV}, in preparation for submission to Phys.
  Rev. AB (2020).

\bibitem{timm1969}
U.~Timm, {Coherent Bremsstrahlung of Electrons in Crystals}, Fortschritt der
  Physik 17 (1969) 765--808.
\newblock \href {https://doi.org/10.1002/prop.19690171202}
  {\path{doi:10.1002/prop.19690171202}}.

\bibitem{LIVINGSTON2009205}
K.~Livingston, {The Stonehenge technique. A method for aligning coherent
  bremsstrahlung radiators}, Nucl. Instrum. Meth. A 603~(3) (2009) 205 -- 213.
\newblock Available from:
  \url{http://www.sciencedirect.com/science/article/pii/S0168900209003477},
  \href {https://doi.org/https://doi.org/10.1016/j.nima.2009.02.010}
  {\path{doi:https://doi.org/10.1016/j.nima.2009.02.010}}.

\bibitem{gx3076}
C.~Meyer, A review of asymmetry measurements in vector meson photoproduction
  experiments, Tech. Rep. 3076, Carnegie Mellon University,
  https://halldweb.jlab.org/doc-private/DocDB/ShowDocument?docid=3076 (August
  2016).

\bibitem{Bilokon:1983}
H.~Bilokon, et~al., {Coherent bremsstrahlung in crystals as a tool for
  producing high energy photon beams to be used in photoproduction experiments
  at CERN SPS}, Nucl. Instrum. Meth. 204 (1983) 299--310.
\newblock Available from:
  \url{https://www.sciencedirect.com/science/article/pii/0167508783900613},
  \href {https://doi.org/10.1016/0167-5087(83)90061-3}
  {\path{doi:10.1016/0167-5087(83)90061-3}}.

\bibitem{YANG2010719}
G.~Yang, et~al., Rocking curve imaging for diamond radiator crystal selection,
  Diamond and Related Materials 19~(7) (2010) 719 -- 722, proceedings of
  Diamond 2009, The 20th European Conference on Diamond, Diamond-Like
  Materials, Carbon Nanotubes and Nitrides, Part 2.
\newblock Available from:
  \url{http://www.sciencedirect.com/science/article/pii/S0925963510000063},
  \href {https://doi.org/https://doi.org/10.1016/j.diamond.2009.12.017}
  {\path{doi:https://doi.org/10.1016/j.diamond.2009.12.017}}.

\bibitem{YANG2012}
G.~Yang, et~al., {High resolution X-ray diffraction study of single crystal
  diamond radiators}, physica status solidi (a) 209~(9) (2012) 1786--1791.
\newblock Available from:
  \url{https://onlinelibrary.wiley.com/doi/abs/10.1002/pssa.201200017}, \href
  {https://doi.org/10.1002/pssa.201200017} {\path{doi:10.1002/pssa.201200017}}.

\bibitem{BORGGREEN19631}
J.~Borggreen, B.~Elbek, L.~P. Nielsen, {A proposed spectrograph for heavy
  particles}, Nuclear Instruments and Methods 24 (1963) 1 -- 12.
\newblock Available from:
  \url{http://www.sciencedirect.com/science/article/pii/0029554X63902763},
  \href {https://doi.org/https://doi.org/10.1016/0029-554X(63)90276-3}
  {\path{doi:https://doi.org/10.1016/0029-554X(63)90276-3}}.

\bibitem{Sober2000263}
D.~Sober, et~al., {The bremsstrahlung tagged photon beam in Hall B at JLab},
  Nucl. Instrum. and Meth. A 440~(2) (2000) 263 -- 284.
\newblock Available from:
  \url{http://www.sciencedirect.com/science/article/pii/S0168900299007846},
  \href {https://doi.org/http://dx.doi.org/10.1016/S0168-9002(99)00784-6}
  {\path{doi:http://dx.doi.org/10.1016/S0168-9002(99)00784-6}}.

\bibitem{DIPOLE_YANG}
G.~L. Yang, {A summary of the optics design for the \gx{} single dipole tagger
  spectrometer}, Tech. Rep. GlueX-doc-1186, Glasgow University,
  https://halldweb.jlab.org/doc-public/DocDB/ShowDocument?docid=1186 (January
  2009).

\bibitem{DIPOLE_SOMOV}
A.~Somov, Resolution studies of a dipole tagger magnet: response to the magnet
  review referees, Tech. Rep. GlueX-doc-1368, Jefferson Lab,
  https://halldweb.jlab.org/doc-public/DocDB/ShowDocument?docid=1368 (January
  2010).

\bibitem{gx4271}
D.~I. Sober, {Analysis of the Hall D Tagger Dipole Magnet Field Maps}, Tech.
  Rep. GlueX-doc-4271, The Catholic University of America,
  https://halldweb.jlab.org/doc-private/DocDB/ShowDocument?docid=4271 (July
  2015).

\bibitem{Fischer:2000zu}
H.~Fischer, et~al., {Implementation of the dead time free F1 TDC in the COMPASS
  detector readout}, Nucl. Instrum. Meth. A461 (2001) 507--510.
\newblock \href {http://arxiv.org/abs/hep-ex/0010065}
  {\path{arXiv:hep-ex/0010065}}, \href
  {https://doi.org/10.1016/S0168-9002(00)01285-7}
  {\path{doi:10.1016/S0168-9002(00)01285-7}}.

\bibitem{tagh:base}
V.~{Popov}, et~al., {Performance studies of Hamamatsu R9800 photomultiplier
  tube with a new active base designed for use in the Hall D Broadband tagger
  Hodoscope}, in: 2014 IEEE Nuclear Science Symposium and Medical Imaging
  Conference (NSS/MIC), Seattle, WA, 2014, pp. 1--4.
\newblock \href {https://doi.org/10.1109/NSSMIC.2014.7431075}
  {\path{doi:10.1109/NSSMIC.2014.7431075}}.

\bibitem{nist_xrays}
J.~Hubbell, S.~Seltzer, {Tables of X-Ray Mass Attenuation Coefficients and Mass
  Energy-Absorption Coefficients from 1 keV to 20 MeV for Elements Z=1 to 92
  and 48 Additional Substances of Dosimetric Interest}, Tech. rep., Radiation
  Physics Division, Physical Measurement Laboratory, National Institute of
  Standards and Technology,
  \href{https://www.nist.gov/pml/x-ray-mass-attenuation-coefficients}{NIST
  Standard Reference Database 126} (hyperlink) (2009).

\bibitem{Miller:1973yi}
G.~Miller, D.~R. Walz, {A Tungsten Pin Cushion Photon Beam Monitor}, Nucl.
  Instrum. Meth. 117 (1974) 33.
\newblock \href {https://doi.org/10.1016/0029-554X(74)90380-2}
  {\path{doi:10.1016/0029-554X(74)90380-2}}.

\bibitem{DUGGER2017115}
M.~Dugger, et~al., Design and construction of a high-energy photon polarimeter,
  Nucl. Instrum. Meth. A867 (2017) 115 -- 127.
\newblock Available from:
  \url{http://www.sciencedirect.com/science/article/pii/S0168900217305715},
  \href {https://doi.org/10.1016/j.nima.2017.05.026}
  {\path{doi:10.1016/j.nima.2017.05.026}}.

\bibitem{BARBOSA2015376}
F.~Barbosa, et~al., {Pair spectrometer hodoscope for Hall D at Jefferson Lab},
  Nucl. Instrum. Meth. A 795 (2015) 376 -- 380.
\newblock Available from:
  \url{http://www.sciencedirect.com/science/article/pii/S0168900215007573},
  \href {https://doi.org/https://doi.org/10.1016/j.nima.2015.06.012}
  {\path{doi:https://doi.org/10.1016/j.nima.2015.06.012}}.

\bibitem{Barbosa:2017zzw}
F.~Barbosa, et~al., {{Time characteristics of detectors based on silicon
  photomultipliers for the GlueX experiment}}, Instrum. Exp. Tech. 60 (2017)
  322--329.
\newblock \href {https://doi.org/10.1134/S0020441217030022}
  {\path{doi:10.1134/S0020441217030022}}.

\bibitem{Somov:2017kif}
A.~Somov, others., {{The silicon photomultipliers in the detector subsystems of
  the GlueX experiment}}, J. Phys. Conf. Ser. 798 (2017) 012223.
\newblock \href {https://doi.org/10.1088/1742-6596/798/1/012223}
  {\path{doi:10.1088/1742-6596/798/1/012223}}.

\bibitem{Tolstukhin:2014zsa}
I.~A. Tolstukhin, et~al., {Recording of relativistic particles in thin
  scintillators}, Instrum. Exp. Tech. 57~(6) (2014) 658--661.
\newblock \href {https://doi.org/10.1134/S0020441214060153}
  {\path{doi:10.1134/S0020441214060153}}.

\bibitem{Somov:2017vhp}
A.~Somov, et~al., {{Commissioning of the Pair Spectrometer of the GlueX
  experiment}}, J. Phys. Conf. Ser. 798 (2017).
\newblock \href {https://doi.org/10.1088/1742-6596/798/1/012175}
  {\path{doi:10.1088/1742-6596/798/1/012175}}.

\bibitem{Somov:2016bgb}
A.~Somov, et~al., {{Performance of the pair spectrometer of the GlueX
  experiment}}, J. Phys. Conf. Ser. 675~(4) (2016) 042022.
\newblock \href {https://doi.org/10.1088/1742-6596/675/4/042022}
  {\path{doi:10.1088/1742-6596/675/4/042022}}.

\bibitem{somov_flux}
A.~Somov, {Pair Spectrometer acceptance determination (Spring 2019)}, Tech.
  rep., Jefferson Lab,
  \href{https://halldweb.jlab.org/doc-public/DocDB/ShowDocument?docid=3924}{Technical
  Report GlueX-doc-3924} (hyperlink) (Feb. 2019).

\bibitem{clasnote1992014}
D.~Sober, {Calibration of the Tagged Photon Beam: Normalization Methods, Shower
  Counter and Pair Spectrometer}, Tech. rep., Catholic University of America,
  \href{https://www.jlab.org/Hall-B/notes/clas_notes92/note92-014.pdf}{Technical
  Report CLAS-NOTE-92-014} (hyperlink) (1992).

\bibitem{clasnote1993011}
A.~Eppich, R.~Sealock, {Studies of a Lead Glass Total Absorption Counter},
  Tech. rep., Jefferson Lab,
  \href{https://www.jlab.org/Hall-B/notes/clas_notes93/note93-011.pdf}{Technical
  Report CLAS-NOTE-93-011} (hyperlink) (1993).

\bibitem{clasnote1999002}
E.~Anciant, et~al., {Photon Flux Normalization for CLAS}, Tech. rep.,
  CEA-Saclay,
  \href{http://www.jlab.org/Hall-B/notes/clas_notes99/norma.ps}{Technical
  Report CLAS-NOTE-1999-002} (hyperlink) (1992).

\bibitem{Alcorn-confer-1972}
J.~S. Alcorn, H.~Peterson, S.~S. Lorant, {SLAC two-meter diameter,
  25-kilogauss, superconducting solenoid, {UAMH BINN}}, in: Applied
  Superconductivity Conference, Inst. of Electrical and Electronics Engineers,
  Inc., New York; Stanford Univ., CA, 1972, p. 273.

\bibitem{Aston:1987uc}
D.~Aston, et~al., {The LASS spectrometer}, Tech. Rep. {SLAC-298}, SLAC,
  Stanford, CA, {Technical report SLAC-R-298} (1987).
\newblock Available from:
  \url{https://www-public.slac.stanford.edu/scidoc/docMeta.aspx?slacPubNumber=slac-R-298}.

\bibitem{Ballard:2011tm}
J.~Ballard, et~al., {{Refurbishment and testing of the 1970's era LASS solenoid
  coils for JLab's Hall D}}, AIP Conf. Proc. 1434 (2012) 861--868.
\newblock \href {https://doi.org/10.1063/1.4707001}
  {\path{doi:10.1063/1.4707001}}.

\bibitem{Ballard:2015wma}
J.~Ballard, et~al., {Commissioning and Testing the 1970's Era LASS Solenoid
  Magnet in JLab's Hall D}, IEEE Trans. Appl. Supercond. 25~(3) (2015) 4500805.
\newblock \href {https://doi.org/10.1109/TASC.2014.2385152}
  {\path{doi:10.1109/TASC.2014.2385152}}.

\bibitem{Lavendure:2014:refrig}
N.~Laverdure, et~al., {The Hall D solenoid helium refrigeration system at
  JLab}, AIP Conf. Proc. 1573~(1) (2014) 329--336.
\newblock \href {https://doi.org/10.1063/1.4860719}
  {\path{doi:10.1063/1.4860719}}.

\bibitem{HAKOBYAN2008218}
H.~Hakobyan, et~al., {A double-target system for precision measurements of
  nuclear medium effects}, Nucl. Instrum. Meth. A: 592~(3) (2008) 218 -- 223.
\newblock \href {https://doi.org/https://doi.org/10.1016/j.nima.2008.04.055}
  {\path{doi:https://doi.org/10.1016/j.nima.2008.04.055}}.

\bibitem{VanHaarlem:2010yq}
Y.~V. Haarlem, et~al., {The GlueX Central Drift Chamber: Design and
  Performance}, Nucl. Instrum. Meth. A622 (2010) 142--156.
\newblock \href {https://doi.org/10.1016/j.nima.2010.06.272}
  {\path{doi:10.1016/j.nima.2010.06.272}}.

\bibitem{GlueXCDCNIM}
N.~S. Jarvis, et~al., {The Central Drift Chamber for \gx{}}, Nucl. Instrum.
  Meth. A962 (2020) 163727.
\newblock \href {https://doi.org/https://doi.org/10.1016/j.nima.2020.163727}
  {\path{doi:https://doi.org/10.1016/j.nima.2020.163727}}.

\bibitem{KADYK1991436}
J.~A. Kadyk, Wire chamber aging, Nucl. Instrum. Meth. 300~(3) (1991) 436 --
  479.
\newblock \href {https://doi.org/10.1016/0168-9002(91)90381-Y}
  {\path{doi:10.1016/0168-9002(91)90381-Y}}.

\bibitem{VAVRA20031}
J.~Va'vra, Physics and chemistry of aging - early developments, Nucl. Instrum.
  Meth. A515 (2003) 1 -- 14, proceedings of the International Workshop on Aging
  Phenomena in Gaseous Detectors.
\newblock \href {https://doi.org/https://doi.org/10.1016/j.nima.2003.08.124}
  {\path{doi:https://doi.org/10.1016/j.nima.2003.08.124}}.

\bibitem{hdnote2515}
F.~Barbosa, {Electronics overview}, Tech. rep., Jefferson Lab,
  \href{https://halldweb.jlab.org/doc-public/DocDB/ShowDocument?docid=2515}{Technical
  Report GlueX-doc-2515} (hyperlink) (Jun. 2014).

\bibitem{FDC_NIM}
L.~Pentchev, et~al., {Studies with cathode drift chambers for the GlueX
  experiment at Jefferson Lab}, Nucl. Instrum. Meth. A845 (2017) 281--284.
\newblock \href {https://doi.org/10.1016/j.nima.2016.04.076}
  {\path{doi:10.1016/j.nima.2016.04.076}}.

\bibitem{Visser2008}
G.~Visser, {High Density 125 MSPS Differential Input ADC Module Specifications
  – for GlueX Drift Chamber Application } (2008).
\newblock Available from:
  \url{https://halldweb.jlab.org/DocDB/0008/000855/002/Drifts_ADC_Specification_Document.pdf}.

\bibitem{5873864}
G.~{Visser}, et~al., {A 72 channel 125 MSPS analog-to-digital converter module
  for drift chamber readout for the GlueX detector}, in: IEEE Nuclear Science
  Symposuim Medical Imaging Conference, 2010, pp. 777--781.
\newblock \href {https://doi.org/10.1109/NSSMIC.2010.5873864}
  {\path{doi:10.1109/NSSMIC.2010.5873864}}.

\bibitem{hdnote1021}
F.~Barbosa, et~al., {The Jefferson Lab High Resolution Time-to-Digital
  Converter (TDC)}, Tech. rep., Jefferson Lab,
  \href{https://halldweb.jlab.org/doc-public/DocDB/ShowDocument?docid=1021}{Technical
  Report GlueX-doc-1021} (hyperlink) (Apr. 2008).

\bibitem{millepede}
V.~Blobel, {Millipede II} (2007).
\newblock Available from:
  \url{https://www.desy.de/\~kleinwrt/MP2/doc/html/index.html}.

\bibitem{MikeStaib_thesis}
M.~Staib, {Calibrations for charged particle tracking and the measurements of
  $\omega$ photoproduction with the GlueX Detector}, Ph.D. thesis, Carnegie
  Mellon University, Department of Physics,
  \href{https://halldweb.jlab.org/doc-public/DocDB/ShowDocument?docid=3393}{Technical
  Report GlueX-doc-3393} (hyperlink) (September 2017).

\bibitem{KalmanFilter}
R.~E. Kalman, {A New Approach to Linear Filtering and Prediction Problems},
  ASME Journal of Basic Engineering 82~(1) (1960) 35--45.
\newblock \href {https://doi.org/10.1115/1.3662552}
  {\path{doi:10.1115/1.3662552}}.

\bibitem{KalmanFilter2}
R.~E. Kalman, R.~S. Bucy, {New Results in Linear Filtering and Prediction
  Theory}, ASME Journal of Basic Engineering 83~(1) (1961) 95--108.
\newblock \href {https://doi.org/10.1115/1.3658902}
  {\path{doi:10.1115/1.3658902}}.

\bibitem{BEATTIE201824}
T.~Beattie, et~al., {Construction and performance of the barrel electromagnetic
  calorimeter for the \gx{} experiment}, Nucl. Instrum. Meth. A896 (2018) 24 --
  42.
\newblock \href {https://doi.org/10.1016/j.nima.2018.04.006}
  {\path{doi:10.1016/j.nima.2018.04.006}}.

\bibitem{hdnote2913}
E.~Smith, {Development of Silicon Photomultipliers and their Applications to
  GlueX}, Tech. rep., Jefferson Lab, {AIP Proceedings 1753 -- XI Latin American
  Symposium on Nuclear Physics and Applications, Medell\'in, Colombia}.
  \href{https://halldweb.jlab.org/doc-public/DocDB/ShowDocument?docid=2913}{Technical
  Report GlueX-doc-2913} (hyperlink) (Dec. 2015).

\bibitem{Barbosa2012100}
F.~Barbosa, et~al., {Silicon photomultiplier characterization for the GlueX
  barrel calorimeter}, Nucl. Instrum. Meth. A695 (2012) 100 -- 104.
\newblock \href {https://doi.org/10.1016/j.nima.2011.11.059}
  {\path{doi:10.1016/j.nima.2011.11.059}}.

\bibitem{Qiang2013234}
Y.~Qiang, et~al., {Radiation hardness tests of SiPMs for the JLab Hall D Barrel
  calorimeter}, Nucl. Instrum. Meth. A698 (2013) 234 -- 241.
\newblock \href {https://doi.org/10.1016/j.nima.2012.10.015}
  {\path{doi:10.1016/j.nima.2012.10.015}}.

\bibitem{soto}
O.~Soto, et~al., {Characterization of novel Hamamatsu Multi Pixel Photon
  Counter (MPPC) arrays for the GlueX experiment}, Nucl. Instrum. Meth. A732
  (2013) 431--436.
\newblock \href {https://doi.org/10.1016/j.nima.2013.06.071}
  {\path{doi:10.1016/j.nima.2013.06.071}}.

\bibitem{Soto201489}
O.~Soto, et~al., {Novel Hamamatsu Multi-Pixel Photon Counter (MPPC) array
  studies for the GlueX experiment: New results}, Nucl. Instrum. Methods A 739
  (2014) 89--97.
\newblock \href {https://doi.org/10.1016/j.nima.2013.12.032}
  {\path{doi:10.1016/j.nima.2013.12.032}}.

\bibitem{BeattieIEEE}
T.~Beattie, et~al., {Methodology for the Determination of the Photon Detection
  Efficiency of Large-Area Multi-Pixel Photon Counters}, IEEE Transactions on
  Nuclear Science 62 (2015) 1865--1872.
\newblock \href {https://doi.org/10.1109/TNS.2015.2442262}
  {\path{doi:10.1109/TNS.2015.2442262}}.

\bibitem{doi:10.1063/1.4955340}
E.~Smith, {Development of Silicon Photomultipliers and their Applications to
  GlueX Calorimetry}, AIP Conference Proceedings 1753~(1) (2016) 010001.
\newblock \href {https://doi.org/10.1063/1.4955340}
  {\path{doi:10.1063/1.4955340}}.

\bibitem{CRITTENDEN1997377}
R.~Crittenden, et~al., A 3000 element lead-glass electromagnetic calorimeter,
  Nucl. Instrum. Meth. A387~(3) (1997) 377 -- 394.
\newblock \href {https://doi.org/10.1016/S0168-9002(97)00101-0}
  {\path{doi:10.1016/S0168-9002(97)00101-0}}.

\bibitem{JONES2007384}
R.~Jones, et~al., Performance of the {RadPhi} detector and trigger in a high
  rate tagged photon beam, Nucl. Instrum. Meth. A570~(3) (2007) 384 -- 398.
\newblock \href {https://doi.org/10.1016/j.nima.2006.09.039}
  {\path{doi:10.1016/j.nima.2006.09.039}}.

\bibitem{Brunner:1998fh}
A.~Brunner, et~al., {A Cockcroft-Walton base for the FEU84-3 photomultiplier
  tube}, Nucl. Instrum. Meth. A414 (1998) 466--476.
\newblock \href {https://doi.org/10.1016/S0168-9002(98)00651-2}
  {\path{doi:10.1016/S0168-9002(98)00651-2}}.

\bibitem{wiki:CANBus}
{Wikipedia contributors}, {CAN} bus --- {Wikipedia}{,} the free encyclopedia,
  [Online; accessed 28-October-2019] (2019).
\newblock Available from:
  \url{https://en.wikipedia.org/w/index.php?title=CAN_bus&oldid=922757529}.

\bibitem{MORIYA201360}
K.~Moriya, et~al., {A measurement of the energy and timing resolution of the
  GlueX Forward Calorimeter using an electron beam}, Nucl. Instrum. Meth. 726
  (2013) 60 -- 66.
\newblock \href {https://doi.org/10.1016/j.nima.2013.05.109}
  {\path{doi:10.1016/j.nima.2013.05.109}}.

\bibitem{hdnote1022}
F.~Barbosa, et~al., {A VME64x, 16-Channel, Pipelined 250 MSPS Flash ADC With
  Switched Serial (VXS) Extension}, Tech. rep., Jefferson Lab,
  \href{https://halldweb.jlab.org/doc-public/DocDB/ShowDocument?docid=1022}{Technical
  Report GlueX-doc-1022} (hyperlink) (Apr. 2007).

\bibitem{hdnote2511}
M.~Dugger, et~al., {Hall D / GlueX Technical Construction Report, Chapter
  3.10}, Tech. rep., Jefferson Lab,
  \href{https://halldweb.jlab.org/doc-public/DocDB/ShowDocument?docid=2511}{Technical
  Report GlueX-doc-2511} (hyperlink) (Jul. 2017).

\bibitem{Bennett:2010nf}
J.~V. Bennett, et~al., {Precision timing measurement of phototube pulses using
  a flash analog-to-digital converter}, Nucl. Instrum. Meth. A622 (2010)
  225--230.
\newblock \href {https://doi.org/10.1016/j.nima.2010.06.216}
  {\path{doi:10.1016/j.nima.2010.06.216}}.

\bibitem{Anassontzis201441}
E.~Anassontzis, et~al., {Relative gain monitoring of the \gx{} calorimeters},
  Nucl. Instrum. Meth. A738 (2014) 41 -- 49.
\newblock \href {https://doi.org/10.1016/j.nima.2013.11.054}
  {\path{doi:10.1016/j.nima.2013.11.054}}.

\bibitem{Schaefer:2011gw}
B.~D. Schaefer, et~al., {Radiation Damage of F8 Lead Glass with 20 MeV
  Electrons}, Nucl. Instrum. Meth. B274 (2012) 111--114.
\newblock \href {https://doi.org/10.1016/j.nimb.2011.12.005}
  {\path{doi:10.1016/j.nimb.2011.12.005}}.

\bibitem{Jones:2006ru}
R.~T. Jones, et~al., {A bootstrap method for gain calibration and resolution
  determination of a lead-glass calorimeter}, Nucl. Instrum. Meth. A566 (2006)
  366--374.
\newblock \href {https://doi.org/10.1016/j.nima.2006.07.061}
  {\path{doi:10.1016/j.nima.2006.07.061}}.

\bibitem{AlGhoul:2017nbp}
H.~A. Ghoul, et~al., {Measurement of the beam asymmetry $\Sigma$ for $\pi^0$
  and $\eta$ photoproduction on the proton at $E_\gamma = 9$ GeV}, Phys. Rev.
  C95~(4) (2017) 042201(R).
\newblock \href {https://doi.org/10.1103/PhysRevC.95.042201}
  {\path{doi:10.1103/PhysRevC.95.042201}}.

\bibitem{Adhikari:2019gfa}
S.~Adhikari, et~al., {Beam Asymmetry $\mathbf{\Sigma}$ for the Photoproduction
  of $\mathbf{\eta}$ and $\mathbf{\eta^{\prime}}$ Mesons at
  $\mathbf{E_{\gamma}=8.8}$ GeV}, Phys. Rev. C100~(5) (2019) 052201(R).
\newblock \href {https://doi.org/10.1103/PhysRevC.100.052201}
  {\path{doi:10.1103/PhysRevC.100.052201}}.

\bibitem{Ali:2019lzf}
A.~Ali, et~al., {First Measurement of Near-Threshold J/$\psi$ Exclusive
  Photoproduction off the Proton}, Phys. Rev. Lett. 123~(7) (2019) 072001.
\newblock \href {https://doi.org/10.1103/PhysRevLett.123.072001}
  {\path{doi:10.1103/PhysRevLett.123.072001}}.

\bibitem{Pooser:2019rhu}
E.~Pooser, et~al., {The GlueX Start Counter Detector}, Nucl. Instrum. Meth.
  A927 (2019) 330--342.
\newblock \href {https://doi.org/10.1016/j.nima.2019.02.029}
  {\path{doi:10.1016/j.nima.2019.02.029}}.

\bibitem{Dong:2007}
H.~{Dong}, et~al., {Integrated tests of a high speed VXS switch card and 250
  MSPS flash ADCs}, in: 2007 IEEE Nuclear Science Symposium Conference Record,
  Vol.~1, 2007, pp. 831--833.
\newblock \href {https://doi.org/10.1109/NSSMIC.2007.4436457}
  {\path{doi:10.1109/NSSMIC.2007.4436457}}.

\bibitem{somov_l1}
A.~Somov, {Level-1 Trigger of the GlueX Experiment}, Tech. rep., Jefferson Lab,
  \href{https://halldweb.jlab.org/doc-public/DocDB/ShowDocument?docid=1137}{Technical
  Report GlueX-doc-1137} (hyperlink) (Jul. 2008).

\bibitem{somov_l11}
A.~Somov, {Update on the trigger simulation}, Tech. rep., Jefferson Lab,
  \href{https://halldweb.jlab.org/doc-public/DocDB/ShowDocument?docid=1272}{Technical
  Report GlueX-doc-1272} (hyperlink) (Jul. 2009).

\bibitem{GlueX:2013twa}
A.~Somov, {Development of level-1 triggers for experiments at Jefferson Lab},
  AIP Conf. Proc. 1560~(1) (2013) 700--702.
\newblock \href {https://doi.org/10.1063/1.4826876}
  {\path{doi:10.1063/1.4826876}}.

\bibitem{CLAS12DAQ}
S.~Boyarinov, et~al., {The CLAS12 Data Acquisition System}, Nucl. Instrum.
  Meth.In press (2020).
\newblock \href {https://doi.org/10.1016/j.nima.2020.163698}
  {\path{doi:10.1016/j.nima.2020.163698}}.

\bibitem{Slominski:2009icaleps}
C.~Slominski, et~al., {A MySQL based EPICS archiver}, {Proceedings,
  ICALEPCS2009} (2010) 447--449Available from:
  \url{http://accelconf.web.cern.ch/AccelConf/icalepcs2009/papers/wep021.pdf}.

\bibitem{Chen:2011icaleps}
X.~Chen, K.~Kasemir, {BOY, a modern graphical operator interface editor and
  runtime}, {Proceedings, ICALEPCS2011} (2011) 1404--1406Available from:
  \url{http://accelconf.web.cern.ch/AccelConf/PAC2011/papers/weobn3.pdf}.

\bibitem{Kasemir:2009icaleps}
K.~Kasemir, et~al., {The Best Ever Alarm System Toolkit”}, {Proceedings,
  ICALEPCS2009} (2010) 1062--1065Available from:
  \url{http://accelconf.web.cern.ch/AccelConf/icalepcs2009/papers/tua001.pdf}.

\bibitem{rootspy}
{D. Lawrence and others}, {RootSpy Data Quality Monitoring System}.
\newblock Available from: \url{https://www.jlab.org/RootSpy/}.

\bibitem{coda}
{JLab Data Acquisition Group}, {CODA},
  \href{https://coda.jlab.org}{coda.jlab.org}.
\newblock Available from: \url{https://coda.jlab.org}.

\bibitem{Brun:1997pa}
R.~Brun, F.~Rademakers, {ROOT: An object oriented data analysis framework},
  Nucl. Instrum. Meth. A389 (1997) 81--86, {See also http://root.cern.ch/}.
\newblock \href {https://doi.org/10.1016/S0168-9002(97)00048-X}
  {\path{doi:10.1016/S0168-9002(97)00048-X}}.

\bibitem{gx3821}
M.~Ito, D.~Lawrence, {GlueX Computing Model for RunPeriod-2017-01}, Tech. rep.,
  Jefferson Lab,
  \href{https://halldweb.jlab.org/doc-public/DocDB/ShowDocument?docid=3821}{Technical
  Report GlueX-doc-3821} (hyperlink) (Jun. 2018).

\bibitem{EVIO}
{JLab Data Acquisition Group}, {CODA Online Data Formats},
  \href{https://coda.jlab.org/drupal/system/files/eventbuilding.pdf}{https://coda.jlab.org/drupal/system/files/eventbuilding.pdf}.

\bibitem{Brun:1987ma}
R.~Brun, F.~Bruyant, M.~Maire, A.~C. McPherson, P.~Zanarini, {GEANT3} (1987).

\bibitem{Agostinelli:2002hh}
S.~Agostinelli, et~al., {GEANT4: A Simulation toolkit}, Nucl. Instrum. Meth.
  A506 (2003) 250--303.
\newblock \href {https://doi.org/10.1016/S0168-9002(03)01368-8}
  {\path{doi:10.1016/S0168-9002(03)01368-8}}.

\bibitem{Allison:2016lfl}
J.~Allison, et~al., {Recent developments in Geant4}, Nucl. Instrum. Meth. A835
  (2016) 186--225.
\newblock \href {https://doi.org/10.1016/j.nima.2016.06.125}
  {\path{doi:10.1016/j.nima.2016.06.125}}.

\bibitem{HDDS}
R.~Jones, {HDDS Schema}.
\newblock Available from:
  \url{https://halldsvn.jlab.org/repos/trunk/hdds/HDDS-1\_1.xsd}.

\bibitem{gx732}
R.~Jones, {Detector Models for GlueX Monte Carlo Simulation: the CD2 Baseline},
  Tech. rep., University of Connecticut,
  \href{https://halldweb.jlab.org/doc-public/DocDB/ShowDocument?docid=732}{Technical
  Report GlueX-doc-732} (hyperlink) (Jan. 2007).

\bibitem{gx65}
R.~Jones, {HDDM -- Hall D Data Model}, Tech. rep., University of Connecticut,
  \href{https://halldweb.jlab.org/doc-public/DocDB/ShowDocument?docid=65}{Technical
  Report GlueX-doc-65} (hyperlink) (Sep. 2003).

\bibitem{Sjostrand:2006za}
T.~Sjostrand, S.~Mrenna, P.~Z. Skands, {PYTHIA 6.4 Physics and Manual}, JHEP 05
  (2006) 026.
\newblock \href {https://doi.org/10.1088/1126-6708/2006/05/026}
  {\path{doi:10.1088/1126-6708/2006/05/026}}.

\bibitem{Aubert:2001tu}
B.~Aubert, et~al., {The BaBar detector}, Nucl. Instrum. Meth. A479 (2002)
  1--116.
\newblock \href {https://doi.org/10.1016/S0168-9002(01)02012-5}
  {\path{doi:10.1016/S0168-9002(01)02012-5}}.

\end{thebibliography}
\bibliographystyle{elsarticle-num-modified}

\end{document}